%% file: main.tex
\RequirePackage[l2tabu, orthodox]{nag} 

\documentclass[twoside]{um}  

\title{Variational Quantum \\ Algorithms for \\ Many-Body Systems}  
\tagline{A Prelude to Fault-Tolerant Quantum Computation}      
\author{Mirko Consiglio}            
\authorID{194898M}                   
\supervisor{Prof. Tony J.\ G.\ Apollaro}              
\department{Department of Physics}   
\faculty{Faculty of Science}       
\degreeName{Ph.D. (Melit.) in Physics}       
\doctype{thesis}               
\degreeDate{\monthyeardate\today}    

\newcommand{\DHS}[2]{D_\text{HS}\left(#1, #2\right)}
\newcommand{\EHS}[1]{E_\text{HS}\left(#1\right)}

\newcommand{\PTr}[2]{\text{Tr}_{#1}\left\{#2\right\}}

\newcommand{\cF}{\mathcal{F}}
\newcommand{\cZ}{\mathcal{Z}}
\newcommand{\cS}{\mathcal{S}}
\newcommand{\cW}{\mathcal{W}}
\newcommand{\cC}{\mathcal{C}}
\newcommand{\cO}{\mathcal{O}}
\newcommand{\cH}{\mathcal{H}}
\newcommand{\cU}{\mathcal{U}}
\newcommand{\cP}{\mathcal{P}}
\newcommand{\cA}{\mathcal{A}}
\newcommand{\cG}{\mathcal{G}}
\newcommand{\cK}{\mathcal{K}}

\newcommand{\bE}{\mathbb{E}}
\newcommand{\bV}{\mathbb{V}}

\newcommand{\dI}{\mathds{1}}
\newcommand{\dU}{\mathds{U}}

\newcommand{\GHZ}{\ket{\text{\acs{GHZ}}}}

\DeclarePairedDelimiter{\ceil}{\lceil}{\rceil}
\DeclarePairedDelimiter{\floor}{\lfloor}{\rfloor}

\renewcommand{\eqref}[1]{\hypersetup{linkcolor=DissertationColor}\textup{(\ref{#1})}\hypersetup{linkcolor=black}}

\newcommand{\nref}[1]{\hypersetup{linkcolor=DissertationColor}\textup{\nameref{#1}}\hypersetup{linkcolor=black}}

\newcommand{\figref}[1]{\hypersetup{linkcolor=DissertationColor}\textup{\ref{#1}}\hypersetup{linkcolor=black}}

\newcommand{\secref}[1]{\hypersetup{linkcolor=DissertationColor}\textup{\ref{#1}}\hypersetup{linkcolor=black}}

\newcommand{\tabref}[1]{\hypersetup{linkcolor=DissertationColor}\textup{\ref{#1}}\hypersetup{linkcolor=black}}

\hypersetup{
    colorlinks=true, 
    linkcolor=black,
    citecolor=DissertationColor,
    urlcolor=DissertationColor
}

\newcommand*{\doi}[1]{DOI \texttt{\href{http://dx.doi.org/#1}{#1}}}

\begin{document}

\frontmatter
\maketitle
\input{frontmatter/copyright}\cleardoublepage
\input{frontmatter/dedication}\cleardoublepage
\input{frontmatter/acknowledgements}\cleardoublepage
\input{frontmatter/abstract}\cleardoublepage
\tableofcontents*\cleardoublepage
\input{frontmatter/publications}\cleardoublepage
\listoffigures\cleardoublepage
\listoftables\cleardoublepage
\input{frontmatter/abbreviations}\cleardoublepage

\pagestyle{umpage}
\floatpagestyle{umpage}
\acresetall
\mainmatter
\input{chapter1/introduction}
\input{chapter2/lit_review}
\input{chapter3/vqe}
\input{chapter4/vsv}
\input{chapter5/gibbs}
\input{chapter6/conclusion}
\appendix
\input{appendixA/vqe_appendix} 
\input{appendixB/swap_test_appendix}
\input{appendixC/css_appendix}
\input{appendixD/gilbert_appendix}
\input{appendixE/entangling_gates}
\input{appendixF/barren_plateau}
\input{appendixG/xy}
\input{addendum/addendum}

\pagestyle{umpageback}
\let\thebibliography\stdthebibliography
{\backmatter
    \renewcommand*\bibname{References}
    \if@openright\cleardoublepage\else\clearpage\fi
    \bibliographystyle{unsrtnat}
    \bibliography{ref}
    \printindex
}

\end{document}

%% file: frontmatter/copyright.tex
\begin{copyrightenv}
    
\end{copyrightenv}

%% file: frontmatter/dedication.tex
\begin{dedication}
    {\Large{G$\hbar$al Nicole, g$\hbar$all-familja, u g$\hbar$al log$\hbar$ba 40k}}\\[3mm]
    Spe\.cjalment g$\hbar$al ommi, talli dejjem kont hemm g$\hbar$alija
\end{dedication}

%% file: frontmatter/acknowledgements.tex
\begin{acknowledgements}
    The research work disclosed in this thesis is funded by the Tertiary Education Scholarships Scheme. The author acknowledges the use of IBM Quantum services for this work; the views expressed are those of the author, and do not reflect the official policy or position of IBM or the IBM Quantum team.
    
    First and foremost, I wish to express my sincere gratitude to Prof. Tony John George Apollaro, my supervisor and friend, for his consistent support and the wealth of knowledge he has imparted to me throughout this journey. His expertise and guidance, both on an academic and personal level, have greatly influenced this thesis and contributed to my growth as a researcher.

    I am also deeply thankful to the Department of Physics at the University of Malta for their support in addressing any queries I had. Special thanks go to Prof. Charles V. Sammut, whose inspiration and guidance led me to pursue a path in academia, as well as to the members of the Quantumalta group (past and present)\footnote{Particularly Andr\'e, Jake, Karl, and Claudio} for their support and motivation.

    I would like to acknowledge the hospitality of the Quantum Group at Trinity College Dublin where I spent three months\footnote{A heartfelt thank you goes to Steve and Fiona for welcoming me into their lovely home during this time.} during the third year of my Ph.D. I thoroughly enjoyed collaborating with Prof. John Goold on preparing Gibbs states on a quantum computer, Sl\'ainte!
    
    I extend my deepest gratitude to my parents, Mark and Sonia, for their constant support, love, and countless sacrifices that have enabled me to pursue my dreams. Their unwavering faith in me has been a constant source of inspiration and strength.

    I am grateful to my friends\footnote{Andreas, Alex, Jake, Jeremy, Kyle, Emma, Claudia, Tania, Carola, Gabriel, Adrian, Thomas, Daniel, and Gabb, who have all been a part of my journey in some way.} and family, who have patiently listened to my profuse ramblings (and occasional endless grumbles) about the intricacies and depth of quantum mechanics. A special shout-out goes to the people at the Realm of Pure Fantasy\footnote{In particular, Brian and Daniele, as well as the Team Malta members, Gabriel, Finian, Ryan, Luke, Adam, James, Marshal, William and Enrico, it has been both a privilege and an honour to serve as your captain.} for providing a safe haven and a second home, where I could spend countless hours playing, talking, and laughing away my frustrations.
    
    Finally, I want to express my deepest appreciation to Nicole, my girlfriend, for her continual understanding, support, and patience during this challenging journey. I am incredibly fortunate to have her by my side, her love and support have been my beacon during the most trying times.
\end{acknowledgements}

%% file: frontmatter/abstract.tex

\begin{abstract}
    \Acp{VQA} incorporate hybrid quantum--classical computation aimed at harnessing the power of \ac{NISQ} computers to solve challenging computational problems. In this thesis, three main \acp{VQA} are presented, each tackling a different facet of many-body physics.
    
    The first is the \ac{VQE}, which is designed to determine the ground-state of the extended Fermi-Hubbard model. The \ac{VQE} was applied to study the ground-state properties of $N$-component interacting fermions. To this end, an SU($N$) fermion-to-qubit encoding was devised, based on an extension of the \acl{JW} mapping. The ground-state of the Hubbard model, with different dynamical parameters, was specifically obtained by using a number-conserving \ac{PQC}. The persistent current, having applications in the emergent field of atomtronics, was then investigated and numerically obtained by varying the magnetic flux and adiabatically assisting the \ac{VQE}. This approach lays out the basis for a current-based quantum simulator of many-body systems that can be implemented on \ac{NISQ} computers. 

    The second \ac{VQA} is the \ac{VSV}, which is a novel approach to determining the \ac{CSS} of an arbitrary quantum state, with respect to the \ac{HSD}. The  performance of the \ac{VSV} is first assessed by investigating the convergence of the optimisation procedure for \acs{GHZ} states of up to seven qubits, using both statevector and shot-based simulations. The results indicate that current \ac{NISQ} devices may be useful in addressing the \acs{NP}-hard full separability problem using the \ac{VSV}, due to the shallow quantum circuit imposed by employing the destructive SWAP test to evaluate the \ac{HSD}.
    
    The final \ac{VQA} was designed for the preparation of thermal states. The preparation of an equilibrium thermal state of a quantum many-body system on \ac{NISQ} devices is an important task in order to extend the range of applications of quantum computation. Faithful Gibbs state preparation would pave the way to investigate protocols such as thermalisation and out-of-equilibrium thermodynamics, as well as providing useful resources for quantum algorithms, where sampling from Gibbs states constitutes a key subroutine. The novelty of the \ac{VQA} consists in implementing a \ac{PQC} acting on two distinct, yet connected, quantum registers. The \ac{VQA} evaluates the Helmholtz free energy, where the von Neumann entropy is obtained via post-processing of computational basis measurements on one register, while the Gibbs state is prepared on the other register, via a unitary rotation in the energy basis. Finally, the \ac{VQA} is benchmarked by preparing Gibbs states of several spin-$1/2$ models and achieving remarkably high fidelities across a broad range of temperatures in statevector simulations. The performance of the \ac{VQA} was assessed on IBM quantum computers, showcasing its feasibility on current \ac{NISQ} devices.
\end{abstract}

%% file: frontmatter/publications.tex
\chapter{List of Publications}

\let\oldaddcontentsline\addcontentsline
\renewcommand{\addcontentsline}[3]{}  

\let\stdthebibliography\thebibliography
\def\thebibliography{%
  \let\chapter\subsection%
  \let\section\subsection%
  \stdthebibliography%
}

\makeatletter
\let\original@biblabel\@biblabel  
\makeatother

\makeatletter
\renewcommand\@biblabel[1]{}  
\makeatother

\renewcommand*\bibname{Publications Supporting the Thesis}

\renewcommand*\bibname{Additional Research Contributions}

\renewcommand*\bibname{Newspaper Articles}

\makeatletter
\let\@biblabel\original@biblabel  
\makeatother

\let\addcontentsline\oldaddcontentsline

%% file: frontmatter/abbreviations.tex

\chapter{List of Abbreviations}

\begin{acronym}[X-MEMS]
\acro{BFGS}{Broyden--Fletcher--Goldfarb--Shanno\acroextra{ (optimiser)}}
\acro{BP}{barren plateau}
\acro{BQP}{bounded-error quantum polynomial time\acroextra{ (complexity)}}
\acro{CPTP}{completely-positive trace-preserving}
\acro{CPU}{classical processing unit}
\acro{CSS}{closest separable state}
\acro{CVaR}{conditional value-at-risk}
\acro{DMRG}{density matrix renormalisation group}
\acro{GHZ}{Greenberger--Horne--Zeilinger\acroextra{ (state)}}
\acro{GME}{genuine multipartite entanglement}
\acro{GR}{Grover--Rudolph\acroextra{ (ansatz)}}
\acro{GSA}{generalised simulated annealing}
\acro{hc}[h.c.]{Hermitian conjugate}
\acro{HSD}{Hilbert--Schmidt distance}
\acro{HSE}{Hilbert--Schmidt entanglement}
\acro{JW}{Jordan--Wigner\acroextra{ (mapping)}}
\acro{k-CSS}[$k$-CSS]{closest $k$-separable state}
\acro{KKT}{Karush--Kuhn--Tucker\acroextra{ (conditions)}}
\acro{LOCC}{local operations and classical communication}
\acro{ML}{maximum likelihood}
\acro{NFT}{Nakanishi--Fujii--Todo\acroextra{ (optimiser)}}
\acro{NISQ}{noisy intermediate-scale quantum}
\acro{NP}{non-deterministic polynomial time\acroextra{ (complexity)}}
\acro{NSB}{Nemenman--Shafee--Bialek\acroextra{ (estimator)}}
\acro{PQC}{parametrised quantum circuit}
\acro{QAOA}{quantum approximate optimisation algorithm\acroextra{ (or quantum alternating operator ansatz)}}
\acro{QEC}{quantum error correction}
\acro{QGA}{quantum Gilbert algorithm}
\acro{QIP}{quantum information processing}
\acro{QMA}{quantum Merlin--Arthur time\acroextra{ (complexity)}}
\acro{QPU}{quantum processing unit}
\acro{QST}{quantum state transfer}
\acro{SLOCC}{stochastic local operations and classical communication}
\acro{SLSQP}{sequential least squares programming\acroextra{ (optimiser)}}
\acro{SPAM}{State preparation and measurement\acroextra{ (errors)}}
\acro{SPSA}{simultaneous perturbation stochastic approximation\acroextra{ (optimiser)}}
\acro{SSH}{Su--Schrieffer--Heeger\acroextra{ (model)}}
\acro{TFD}{thermofield double}
\acro{UCC}{unitary coupled cluster\acroextra{ ansatz}}
\acro{HVA}{Hamiltonian variational ansatz}
\acro{VQA}{variational quantum algorithm}
\acro{VQE}{variational quantum eigensolver}
\acro{VSV}{variational separability verifier}
\acro{X-MEMS}[$X$-MEMS]{maximally-entangled mixed $X$-states}
\end{acronym}

%% file: chapter1/introduction.tex
\chapter{Introduction}

\epigraph{\textit{Nature isn’t classical, dammit, and if you want to make a simulation of nature, you’d better make it quantum mechanical, and by golly it's a wonderful problem, because it doesn't look so easy.}}{--- Feynman, when referring to simulating physics with\\ quantum computers.}

Quantum computing emerged as a nascent multidisciplinary field inspired by Benioff's conceptualisation of a quantum-mechanical counterpart to Turing machines~\cite{Benioff1980}. Subsequently, Deutsch~\cite{Deutsch1985}, and Feynman~\cite{Feynman1982, Feynman1986}, along with numerous others, further propelled its development. Research in quantum computing flourished when Shor devised a quantum algorithm that efficiently determines prime factors of composite integers, which could result in the collapse of many modern cryptographic protocols~\cite{Shor1994}. Since then, many quantum algorithms have been designed with varying applications, ranging from optimisation, machine learning, quantum many-body physics and chemistry, cryptography, communication and pharmaceuticals~\cite{Montanaro2016,Bharti2022}. 

\section{(Quantum) Computational Complexity Theory}

Computational complexity theory is a means of assigning computational problems to different complexity classes, each with its own set of characteristics and limitations in terms of some resource, most commonly time and memory. For the sake of conciseness, the asymptotic notation of function complexity will be briefly described, and only the relevant complexity classes will be mentioned in this thesis.\footnote{Interested readers can refer to \citet{Arora2009} for a more comprehensive review.} The most important three definitions for asymptotic complexity (as a function the size of the system $n$) are:
\begin{itemize}
    \item $f(n) = \cO(T(n))$, where $f$ is said to be asymptotically bounded above by $T$ (up to a multiplicative constant).
    \item $f(n) = \Omega(T(n))$, where $f$ is said to be asymptotically bounded below by $T$ (up to a multiplicative constant).
    \item $f(n) = \Theta(T(n))$, where $f$ is said to be asymptotically bounded above and below by $T$ (up to two multiplicative constants).
\end{itemize}

\begin{figure}[t]
    \centering
    \includegraphics[width=0.6\textwidth]{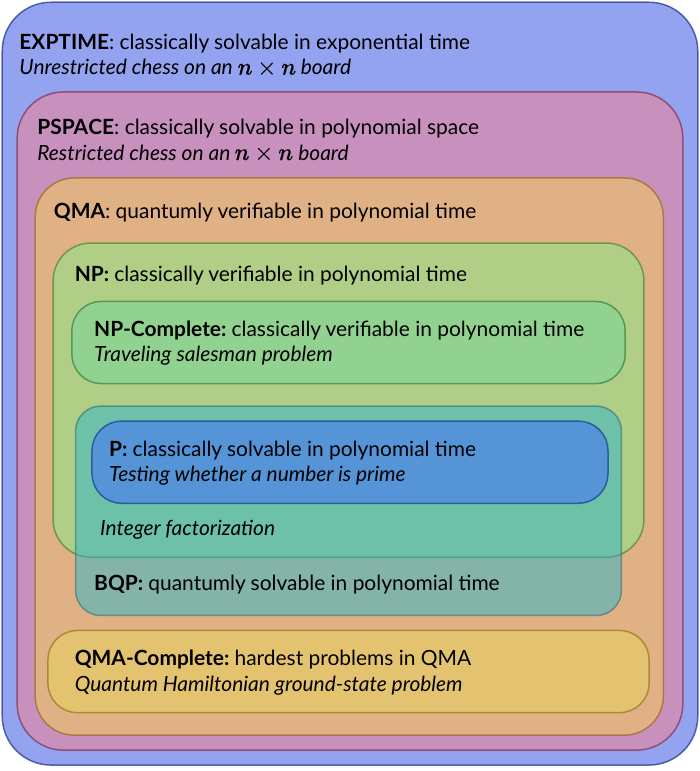}
    \caption[Illustration of computational complexity classes.]{Illustration of computational complexity classes along with a few examples of problems contained within some of the classes. Note that the containers are suggestive, and have not been mathematically proven for many of the classes in the figure. One important conjecture is in fact whether P $=$ \ac{NP}. Figure adapted from \citet{Bharti2022}.}
    \label{fig:complexity_classes}
\end{figure}

Now let us turn our attention to complexity classes which are relevant to the algorithms discussed and mentioned in this thesis:
\begin{itemize}
    \item P: problems that can be solved in a polynomial time by a deterministic classical computer. This means that the problem can be solved in $\cO\left(n^k\right)~\exists~k > 0$. Note that $k = 1$ denotes linear time whilst $0 < k < 1$ denotes fractional time.
    \item \acs{NP}: stands for \acl{NP}, that is problems that can be verified in a polynomial time by a deterministic classical computer, but not necessarily solved in such a time. This implies that the problem can be verified in $\cO\left(n^k\right)~\exists~k > 1$, however the solution cannot be determined faster than $\Omega\left(n^k\right)~\forall~k > 1$.
    \item \acs{BQP}: stands for \acl{BQP}. Essentially equivalent to P but for a quantum computer. 
    \item PSPACE: stands for Polynomial Space. Problems that can be solved in a polynomial space by a deterministic classical computer.
    \item EXPTIME: stands for Exponential Time. Problems that can be solved in an exponential time by a deterministic classical computer. This means that the problem can be solved in $\cO\left(k^n\right)~\exists~k > 1$.
    \item \acs{QMA}: stands for \acl{QMA} and is the quantum analogy for \ac{NP}. A problem is said to be in \ac{QMA} if the solution can be verified in a polynomial time by a quantum computer, if it is given ``yes'' as an answer.
\end{itemize}
The classes mentioned above can be neatly placed in containers as shown in Fig.~\figref{fig:complexity_classes}. Furthermore, the last few important definitions are the concepts of reducibility, hardness and completeness~\cite{Bharti2022}. A problem $A$ is said to be reducible to a problem $B$, if a method for solving $B$ can be used to construct a method for solving $A$ with a comparable amount of computational resources, denoted as $A \leq B$. This can also be interpreted to mean that solving $B$ is at least as hard as solving $A$. Given a complexity class $C$, a problem $X$ is said to be $C$-hard if every problem in $C$ reduces to $X$, and finally, a problem $X$ is said to be $C$-complete if $X$ is $C$-hard and also $X \in C$.

A typical example of a \ac{BQP} problem is integer factorisation, which is polynomially solvable by a quantum computer using Shor's algorithm~\cite{Shor1994}. On the other hand, there is no known classical algorithm that can factor integers in polynomial time. However, it is crucial to carry out tests for quantum algorithmic complexity, as it is obscure whether quantum computers are able to solve \ac{NP}-Complete problems in polynomial time. In fact, most of the algorithms which will be discussed in this work are related to the preparation of a desired initial state. It is established that preparing --- and also \textit{finding} specific states, such as ground-states of Hamiltonians --- is \ac{QMA}-complete~\cite{Kempe2005, Watrous2008, Bookatz2013}.

\section{The Era of Noise. Is there a Solution?}

While the fault-tolerant quantum computer era might herald in a new age of unparalleled efficiency through quantum algorithms, we are still a far cry away from achieving \textit{true} quantum advantage~\cite{Preskill2018}. In essence, quantum advantage\footnote{Formally, it consists of two fundamental tasks: the engineering feat of constructing a robust quantum computer, and the computational complexity challenge of identifying a problem that can be efficiently solved by such a quantum computer, showcasing a superpolynomial speedup compared to the most efficient classical algorithms available for that specific task~\cite{Harrow2017}.} refers to the ability of quantum computers to solve certain problems significantly faster than classical computers.

Initially, most of the proposed quantum algorithms required millions of physical qubits to operate successfully and efficiently, along with the implementation of \ac{QEC} protocols, to reduce the inherent decoherence of qubits, improper gate control and measurement errors --- collectively known as noise. Currently, we are in the era of \ac{NISQ} devices, where we only have access to around the order of 100 noisy qubits at most. As a result, most of the current quantum algorithms have to be devised with the current severe limitations in mind. In fact, the main goal in the \ac{NISQ} era is to develop techniques and algorithms which are able to extract the maximum available quantum computational power, while also being suitable to scale for the future long-term goal of fault-tolerant quantum computation~\cite{Terhal2015}.

Nevertheless, attempts at attaining quantum advantage, or quantum supremacy, have been made in the \ac{NISQ} era. In particular, the main proposals for demonstrating quantum advantage typically fall under one of the following algorithms: factoring integers, notably using Shor's algorithm~\cite{Shor1994}; boson sampling, a computing paradigm centred on the transmission of identical photons through a linear-optical network can effectively tackle specific sampling and search problems~\cite{Aaronson2011}; and cross-entropy benchmarking~\cite{Boixo2018}, which consists of sampling the output distribution of random quantum circuits.

For instance, Google claimed to have achieved quantum supremacy in 2019 with its 53-qubit quantum computer, Sycamore, demonstrating that it could solve a specific problem faster than the most powerful classical supercomputers at the time~\cite{Arute2019}. In response, IBM indicated that certain claims were overstated and proposed a potential time frame of 2.5 days rather than 10,000 years for solving the same problem, detailing various techniques that a classical supercomputer could employ to enhance computational efficiency. IBM's input holds significance, especially considering that the most potent supercomputer at that time, Summit, was developed by IBM~\cite{NPR2019}. Subsequent to this, researchers have refined algorithms for the sampling problem pivotal in asserting quantum supremacy, leading to significant reductions in the disparity between Google's Sycamore processor and classical supercomputers~\cite{Liu2021, Bulmer2022}, even surpassing it in certain instances~\cite{Pan2022}.

In December 2020, researchers from the University of Science and Technology of China, led by Jian-Wei Pan, similarly, claimed to have achieved quantum supremacy by employing Gaussian boson sampling on 76 photons with their photonic quantum computer named Jiuzhang~\cite{Ball2020}. According to the paper, the quantum computer generated a number of samples in 200 seconds that would necessitate a classical supercomputer 2.5 billion years to compute. In October 2021, researchers from the University of Science and Technology of China once more demonstrated quantum advantage by constructing two supercomputers named Jiuzhang 2.0 and Zuchongzhi. The Jiuzhang 2.0, which operates on light-based principles, employed Gaussian boson sampling to detect 113 photons from a 144-mode optical interferometer, achieving a sampling rate acceleration of 1024 --- representing an advancement of 37 photons and 10 orders of magnitude compared to the previous Jiuzhang system~\cite{Zhong2021}. Similarly, Zuchongzhi, a superconducting quantum computer with 66 qubits in a tunable coupling architecture, performed random quantum circuit sampling for benchmarking. It was estimated that Zuchongzhi's task, completed in 1.2 hours, would take the most powerful supercomputer at least 8 years to achieve~\cite{Wu2021}.

The current state-of-the-art in experimentation and the growing need for \ac{QEC} have spurred the development of inventive algorithms aimed at achieving the long-anticipated quantum advantage. The term ``near-term quantum computation'' has been introduced to encompass these quantum algorithms specifically tailored for quantum computing hardware expected to emerge in the coming years. Conversely, \ac{NISQ} devices can only execute quantum circuits structured according to a specified graph topology, with each node representing qubits and gates typically operating on one or two qubits. Due to the inherent noise in gate operations, \ac{NISQ} algorithms are inherently limited to shallow depths~\cite{Barak2022}. It is important to clarify that \ac{NISQ} is a hardware-centric concept, and does not inherently imply a temporal aspect. Nevertheless, it is important to identify that the term ``near-term'' is subjective and varies among researchers, as predictions regarding experimental progress are prone to human biases~\cite{Bharti2022}. Algorithms tailored for ``near-term'' hardware may become infeasible if hardware advancements fail to align with the experimental requirements of the algorithm.\footnote{Readers interested in understanding better the current quantum technologies and their implementations can refer to \citet{Ezratty2023}.}

So, what are the means of harnessing the power given by \ac{NISQ} computers? Since the capacity of a modern quantum computer is limited, many algorithms are devised in a hybrid quantum--classical manner. Such algorithms assign a classically demanding component to a quantum computer, while the tractable part is still carried out on a (sufficiently powerful) classical computer. The concept of these algorithms typically involves some variational updates to parameters in a \ac{PQC}, subsequently termed \acp{VQA} (which is a subset of hybrid quantum--classical algorithms)~\cite{Cao2019, Cerezo2021b, Endo2021}. Originally, the first two designs for \acp{VQA} were: the \ac{VQE}~\cite{Peruzzo2014, Wecker2015, McClean2016}, initially proposed to solve quantum chemistry problems; and the \ac{QAOA}~\cite{Farhi2014}, intended to solve combinatorial optimisation problems. Both of these algorithms can be considered as the parents of the entire \ac{VQA} family~\cite{Bharti2022}.

\begin{figure}[t]
    \centering
    \includegraphics[width=\textwidth]{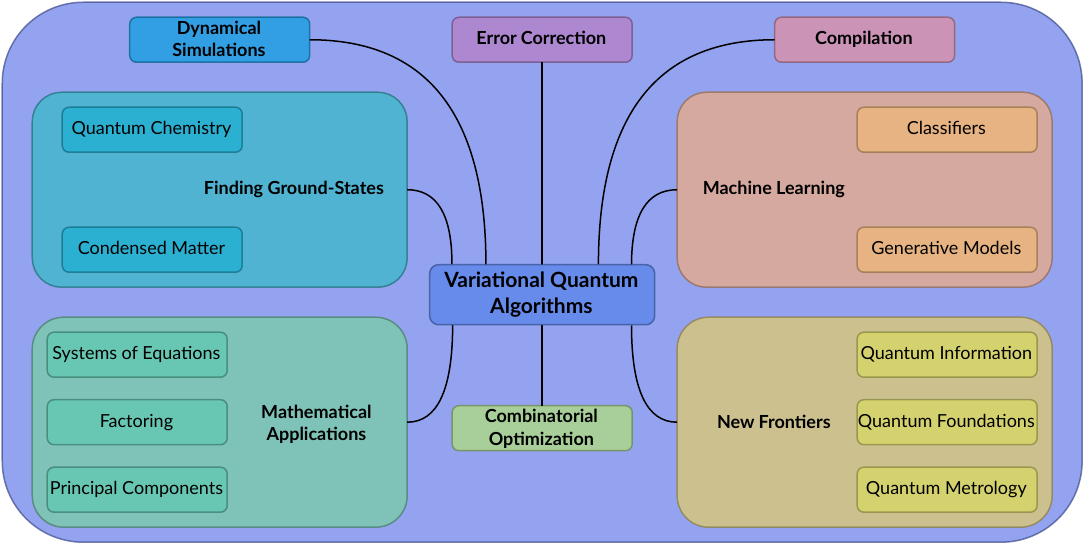}
    \caption[Applications of \acsp{VQA}.]{Applications of \aclp{VQA}. Figure adapted from \citet{Cerezo2021b}.}
    \label{fig:applications}
\end{figure}

\acp{VQA} attempt to bridge the gap between quantum and classical, by leveraging the peak capabilities of current classical computers, and the potential of \ac{NISQ} and near-term devices to obtain a significant computational advantage. Accounting for all the limitations imposed by \ac{NISQ} computers necessitates employing either an optimisation-based or learning-based strategy, a task effectively tackled by \acp{VQA}. They serve as the quantum counterpart to highly successful machine learning techniques, such as neural networks. Furthermore, \acp{VQA} utilise classical optimisation techniques by employing \acp{PQC}, which are then optimised by classical optimisers. This method offers the benefit of maintaining a shallow quantum circuit depth, thereby reducing noise, unlike quantum algorithms designed for fault-tolerant systems~\cite{Cerezo2021a}. \acp{VQA} have been extensively explored across a wide array of applications, as illustrated in Fig.~\figref{fig:applications}, virtually encompassing most of the envisioned uses for quantum computing. While they hold promise as a pathway to achieving quantum advantage in the near term, \acp{VQA} encounter significant challenges related to their trainability, accuracy, and efficiency.

\section{Overview of the Thesis}

This thesis consists of a compendium of both existing and original \acp{VQA} which have been investigated while undertaking this study, including the exploration of novel applications for such algorithms. The \acp{VQA} investigated here include the prototypical \ac{VQE}, despite this being one of the earliest and most investigated algorithms in the \ac{NISQ} era. The next \ac{VQA} that was developed was the \ac{VSV}, which is a completely novel hybrid quantum--classical algorithm that utilises quantum computers to investigate the defining feature of quantum mechanics itself: quantum entanglement. The intended use of this algorithm is to advance the theory describing the geometry of quantum states, which has been a crucial problem of quantum mechanics for decades. Last, but not least, a new \ac{VQA} was designed to prepare Gibbs states on a quantum computer, whereby it alleviates one crucial issue in state-of-the-art Gibbs state preparation algorithms: the efficient computation of the von Neumann entropy. A proof-of-principle implementation of the algorithm on real quantum hardware demonstrates considerable promise for near-term applications in quantum thermodynamics.

Chapter~\secref{chap:2} will first delve into the modular components that make up a \ac{VQA}, and how one can devise and construct a hybrid quantum--classical algorithm. Each of the different modules including the objective function, \ac{PQC}, measurement technique and choice of classical optimiser, will be described and discussed. Although the level of detail in each of the components is restricted to the use cases presented in this work, the references mentioned in each of the relevant sections provide a means of pointing interested readers to more specific components and aspects of \acp{VQA}.

The first \ac{VQA} is then properly introduced in Chapter~\secref{chap:3}, where the \ac{VQE} is used to solve instances of the challenging SU($N$) Hubbard model, using a number-preserving and hardware-efficient \acl{HVA}. The chapter presents the relevant information in a concise way so that the reader can easily interpret the results of the \ac{VQE} simulations. However, the reader is invited to peruse Appendix~\secref{app:vqe}, where exhaustive detail is provided on deriving the qubit Hamiltonian via the $N$-component \ac{JW} mapping, and extending the model to cover more complex scenarios, while the implementation of the \ac{VQE} itself is comprehensively explained.

Chapter~\secref{chap:4} will further introduce deeper quantum-mechanical concepts such as quantum entanglement. This is followed by delving into the design of the \ac{VQA} used to characterise the level of entanglement, while possibly constructing new witnesses of entanglement. The cost function employed in the \ac{VSV} is the \ac{HSD}, which requires evaluating overlaps of quantum states on a quantum device. This is carried out by using the so-called \textsc{SWAP} test (or its destructive version), which is highlighted in Appendix~\secref{app:swap}. Moreover, Appendix~\secref{app:css} contains the derivation of the analytical form of the \ac{CSS} for $n$-qubit \ac{X-MEMS}, which are explained in more detail in the chapter. Lastly, in Appendix~\secref{app:gilbert}, the \ac{VSV} is compared and contrasted with the \ac{QGA}~\cite{Gilbert1966}, which is designed to be run on a classical computer. The \ac{QGA} is shown to be extremely inefficient when implemented on a quantum computer, which further solidifies the \ac{VSV} as a beneficial \ac{NISQ} algorithm.

Chapter~\secref{chap:5} presents a novel \ac{VQA} for both determining and preparing Gibbs states on a quantum computer, along with presenting the results of applying the algorithm to the Ising, XY and Heisenberg XXZ models. In this \ac{VQA}, the generalised Helmholtz free energy is used as a cost function, along with utilising a specific \ac{PQC} that enables the efficient computation of the von Neumann entropy. Furthermore, the algorithm is carried out on IBM quantum hardware. Appendices~\secref{app:entangling_gates} and~\secref{app:bp} are related to the details and implementation of the \ac{VQA} for Gibbs state preparation.

The conclusion of this work is presented in Chapter~\secref{chap:6}, where a summary of the thesis encapsulates a brief review of the results. Furthermore, the future prospects of the work carried out pertaining to the thesis shall be discussed. Finally, research that relates to some of topics discussed, but which falls outside the scope of this thesis, is also discussed after the appendices in~\nref{add:related_research}.

%% file: chapter2/lit_review.tex
\chapter{Building a Variational Quantum Algorithm} \label{chap:2}

\epigraph{\textit{I think I can safely say that nobody understands quantum mechanics.}}{--- \citet{Feynman1965}}

\begin{figure}[t]
    \centering
    \includegraphics[width=\textwidth]{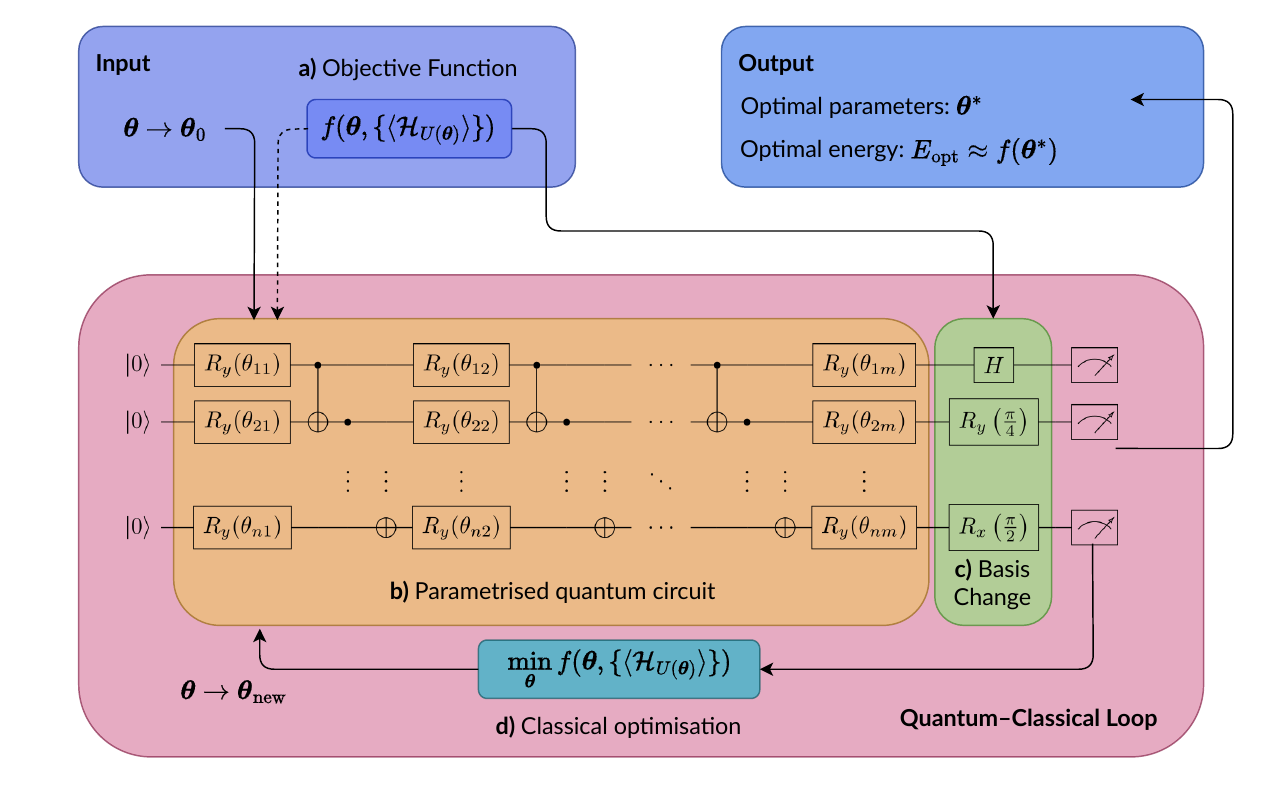}
    \caption[Diagrammatic representation of a \acs{VQA}.]{A diagram showcasing the setup of a \ac{VQA}, with the four main modules: \textbf{a)} The objective function $f$, which encodes the problem to be solved; \textbf{b)} the \acf{PQC}, in which the parameters $\bm{\theta}$ are variationally updated to minimise the objective function; \textbf{c)} the measurement technique, which involves basis changes and measurements needed to compute the objective function; and \textbf{d)} the classical optimiser that minimises the objective function while proposing a new set of variational parameters. Figure adapted from \citet{Bharti2022}.}
    \label{fig:VQA}
\end{figure}

A \ac{VQA} is made up of separate modules that can be easily combined, improved and extended according to the developments in quantum technology. The different modules of a \ac{VQA} are the:
\begin{enumerate}
    \item Objective function
    \item \Acl{PQC}
    \item Measurement technique
    \item Classical optimiser
\end{enumerate}
In this chapter, the above modules are briefly discussed, further detail on constructing \acp{VQA} can be found in \citet{Bharti2022}. A diagram of the setup of a \ac{VQA} can be seen in Fig.~\figref{fig:VQA}.

\section{Objective Function}

When considering a quantum system, the Hamiltonian typically contains all of the relevant information of that system and generally forms its description. In quantum chemistry or many-body physics, the expectation value of the Hamiltonian can lead to the energy of the system, which is often used as the objective function to be optimised for a \ac{VQA}. On the other hand, for combinatorial problems, and others which do not fall in the realm of real physical systems, the respective problem can be encoded into a Hamiltonian, allowing them to be solved on a quantum computer.

The objective function $f$, defined in terms of an expectation value of a Hamiltonian, can be computed as
\begin{equation}
    \label{eq:general_cost_function}
    \min_{\bm{\theta}} f\left(\bm{\theta}, \left\{\langle \cH \rangle_{U(\bm{\theta})}\right\}\right),
\end{equation}
where $\langle \cH \rangle_{U(\bm{\theta})}$ denotes the expectation of the Hamiltonian $\cH$ in the quantum state generated by the unitary $U(\bm{\theta})$, which can be expanded as
\begin{equation}
    \langle \cH \rangle_{U(\bm{\theta})} \equiv \bra{0}^{\otimes n}U^\dagger(\bm{\theta}) \cH U(\bm{\theta})\ket{0}^{\otimes n}.
\end{equation}
The cost function~\eqref{eq:general_cost_function} is ubiquitously associated with the \ac{VQE} algorithm, as witnessed in Chapter~\secref{chap:3}. Once the Hamiltonian of the problem is determined --- along with the associated objective function --- the next step is to decompose it into a set of operators which are measurable on the quantum device. The most common of which is Pauli string decomposition, where such a decomposed Hamiltonian is given as
\begin{equation}
    \cH = \sum\limits_{i=0}^{\cP-1} w_i P_i,
    \label{eq:pauli_decomp}
\end{equation}
where $w_i$ is the weight of the Pauli string $P_i$. A Pauli string can be written as a tensor product of Pauli operators $\sigma^x, \sigma^y, \sigma^z$ --- $P_i = \bigotimes_{j = 1}^{n} \sigma_i^{(j)}$, where $\sigma_i^{(j)} \in \{\dI, \sigma^x, \sigma^y, \sigma^z\}$, with $\dI$ denoting the identity operator. As a result, the expectation value of the Hamiltonian can then be measured by evaluating the expectation value of each of the decomposed Pauli strings:
\begin{equation}
    \langle \cH \rangle_{U(\bm{\theta})} = \sum\limits_{i=0}^{\cP-1} w_i \langle P_i \rangle_{U(\bm{\theta})}.
\end{equation}
This form of decomposed qubit Hamiltonians include molecules, spin chains, or other encoded models~\cite{Bharti2022}.

Pauli string decomposition is both practical and useful because Pauli matrices form a complete basis for the space of Hermitian operators. Additionally, since quantum computers measure in the computational basis (as described in Sec.~\secref{sec:measurement_technique}), Pauli strings can be rotated into this basis using single-qubit gates. This ensures compatibility with hardware constraints, such as native gate sets and qubit connectivity. Moreover, this decomposition often results in sparse Hamiltonian representations, reducing the overhead of experimental implementations of \acp{VQA}.

Other objective functions can be used as well, such as the infidelity and the \ac{CVaR}. In fact, any cost function that can be written in an operational form can be evaluated using a quantum computer, and thus used as an objective function. Objective functions such as the \acl{HSD} and the generalised Helmholtz free energy were indeed used to obtain the results presented in Chapters~\secref{chap:4} and~\secref{chap:5}, respectively.

\section{Parametrised Quantum Circuit}

Once the objective function is determined, the next component of the \ac{VQA} is the quantum circuit that is capable of preparing the quantum state that matches with the minimum of the objective function. The quantum state is prepared via a unitary operation that depends on a set of parameters, called a \acf{PQC}.

The \ac{PQC} is generally applied after preparing an initial state $\ket{\psi_0}$, such that the quantum state is then
\begin{equation}
    \ket{\psi(\bm{\theta})} = U(\bm{\theta})\ket{\psi_0},
\end{equation}
where $\bm{\theta}$ are the variational parameters. Typically, the initial state is simply the all-zero state $\ket{0}^{\otimes n} = \ket{00\cdots0}$, where $n$ is the number of qubits. However, as an example, the \acf{QAOA} uses an initial state starting from a uniform superposition of all qubits in the computational basis~\cite{Farhi2014}, such that 
\begin{equation}
    \ket{D} = H^{\otimes n} \ket{0}^{\otimes n},
    \label{eq:uniform_superposition}
\end{equation}
where $H = (\sigma^x + \sigma^z)/\sqrt{2}$ is the Hadamard gate~\cite{Nielsen2010}. In quantum chemistry models, the initial state is generally chosen to correspond to the Hartree--Fock approximation~\cite{Fischer1987}. The purpose of preparing a good initial state is so that the algorithm is able to optimise the parameters $\bm{\theta}$ close to an optimal point, helping the overall convergence of the algorithm. By starting from a more favourable region, the likelihood of escaping local minima and avoiding \acp{BP} is increased, leading to more efficient optimisation.

Following initial state preparation, the choice of the unitary ansatz $U(\bm{\theta})$ considerably affects the performance of the \ac{VQA}. A necessary condition to find the minimum is that the quantum state that minimises the objective function lies within the subspace of the Hilbert space traversable by the \ac{PQC}. However, more complex unitaries generally lead to deeper and non-local circuits, which are more susceptible to errors. Deeper circuits require a larger number of quantum gates, which increases the overall execution time and the likelihood of errors due to decoherence and gate imperfections. Non-local operations, which involve interactions between qubits that are not physically adjacent, often necessitate the use of SWAP gates to facilitate qubit movement. This further increases circuit depth and introduces additional sources of error. As a result, the accumulation of gate errors, qubit decoherence, and cross-talk between qubits becomes more pronounced, ultimately degrading the fidelity of the computation. Thus, unitary ansatz selection normally falls into two main categories, problem-inspired or hardware-efficient, depending on their structure.

In this chapter, we will mainly discuss fixed-structure ans\"{a}tze,\footnote{For a detailed review on different types of ans\"{a}tze, refer to  \citet{Tilly2022} and \citet{Bharti2022}.} which will be briefly mentioned in Sec.~\secref{sec:other_ansatze}, since these will be employed in the discussed works. However, it is important to note that there are adaptive-structure ans\"{a}tze.

\subsection{Problem-Inspired Ans\"{a}tze}

Given a Hermitian operator $g$, one can construct a unitary operator generated by $g$, describing the time evolution of a system as a function of time $t$:
\begin{equation}
    U(t) = e^{-\imath g t}.
\end{equation}
While $g$ can be any Hermitian operator, such as a Pauli operator, it is often taken to be the Hamiltonian which describes the unitary evolution of the system. Therefore,
\begin{equation}
    U(t) = e^{-\imath \cH t},
    \label{eq:unitary_evolution}
\end{equation}
which can be decomposed into Pauli operators as in Eq.~\eqref{eq:pauli_decomp}. However, it is generally not straightforward to implement the unitary given by Eq.~\eqref{eq:unitary_evolution} on a quantum device, and as a consequence, a Suzuki--Trotter decomposition~\cite{Suzuki1976} is needed to implement an approximation of the unitary operator. If we once again look at Eq.~\eqref{eq:pauli_decomp}, and the sets of operators $\left\{P_i\right\}$ are chosen such that they are commuting, then $e^{-\imath P_i t}$ can be easily implemented. The full evolution over $t$ of such a decomposition can then be implemented as $k$ equal-sized steps
\begin{equation}
    e^{-\imath \cH t} = \lim\limits_{k \rightarrow \infty} \left( \prod\limits_{i=0}^{\cP-1} e^{-\imath \frac{w_i P_i t}{k}} \right)^k.
\end{equation}
An approximation is then carried out by taking $k$ finite steps, such that
\begin{equation}
    e^{-\imath \cH t} \approx \left( \prod\limits_{i=0}^{\cP-1} e^{-\imath \frac{w_i P_i t}{k}} \right)^k.
\end{equation}
This is known as \textit{Trotterisation}, however in practice, this generally results in challenging experimental implementations of such ans\"{a}tze. Knowledge about the physics of the Hamiltonian can substantially reduce the complexity, depth and number of gates of the \ac{PQC} using Trotterised ans\"{a}tze~\cite{Kivlichan2018}. A problem-inspired ansatz --- specifically parity-preserving --- is extensively used in Chapter~\secref{chap:5}.

The \acf{QAOA} is one of the original algorithms designed to perform well in the \ac{NISQ} era to solve combinatorial optimisation problems~\cite{Farhi2014}, that is also inspired by adiabatic approaches. While it can be thought of as a type of \ac{VQA}, it can also be considered as a type of \ac{PQC} --- that is typically referred to as the quantum alternating operator ansatz. Given a cost function $C$ that encodes a combinatorial problem in bit strings forming the computational basis vectors $\ket{i}$, then the problem Hamiltonian $\cH_P$ is defined as
\begin{equation}
    \cH_P \equiv \sum\limits_{i=0}^{n-1} C(i)\ketbra{i},
\end{equation}
and the mixing Hamiltonian $\cH_M$ as
\begin{equation}
    \cH_M \equiv \sum\limits_{i=0}^{n-1} \sigma_i^x,
\end{equation}
with the initial state given as the uniform superposition state $\ket{D}$ from Eq.~\eqref{eq:uniform_superposition}. The final quantum state is given by applying alternating unitaries generated by $\cH_P$ and $\cH_M$ on the initial state $p$-times, such that
\begin{equation}
    \ket{\psi(\bm{\gamma}, \bm{\beta})} \equiv e^{-\imath \beta_p \cH_M}e^{-\imath \gamma_p \cH_P} \cdots e^{-\imath \beta_1 \cH_M}e^{-\imath \gamma_1 \cH_P}\ket{D}.
\end{equation}
The cost function is then given by
\begin{equation}
    C(\bm{\gamma}, \bm{\beta}) \equiv \bra{\psi(\bm{\gamma}, \bm{\beta})} \cH_P \ket{\psi(\bm{\gamma}, \bm{\beta})},
\end{equation}
where $\bm{\gamma}$ and $\bm{\beta}$ are the variational parameters, and $p$ represents the number of layers of the \ac{QAOA}. The minimisation at level $p-1$ is a constrained version of the minimisation at level $p$, meaning that the algorithm improves monotonically with $p$~\cite{Bharti2022}.

Although \ac{QAOA} is not directly explored in these studies, some of them implement an ansatz inspired by it. The \ac{HVA} was introduced to enhance convergence and reduce the number of variational parameters~\cite{Wecker2015, McClean2016}. Given a Hamiltonian that is described in a Pauli decomposition as in Eq.~\eqref{eq:pauli_decomp}, then the unitary can be described by
\begin{equation}
    U(\bm{\theta}) = \prod\limits_i e^{-\imath \theta_i P_i},
\end{equation}
which can be repeated multiple times in the \ac{PQC}, corresponding to multiple \textit{layers} of the unitary operator in the \ac{PQC}.

\subsection{Hardware-Efficient Ans\"{a}tze} \label{sec:hardware_efficient}

The problem-inspired ans\"{a}tze are constructed from the underlying physics of the model to be solved. However, given that we are in the \ac{NISQ} era, we are required to address experimental limitations that consist of qubit connectivity, restricted gate set, decoherence and gate errors, among others. Thus, the concept of hardware-efficient ans\"{a}tze was then introduced~\cite{Kandala2017}, with the idea that they respect limited qubit connectivity, with simple and relatively local gates, which are set up so as to form a sequence of one-qubit gates, followed by an entangling sequence of two-qubit gates. These layers are then repeated enough times to generate a sufficient amount of entanglement and enable the exploration of a large portion of the Hilbert space.

\begin{figure}[t]
    \centering
    \includegraphics[width=\textwidth]{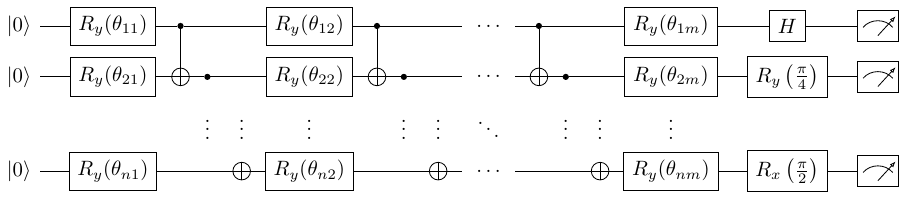}
    \caption[Example of a hardware-efficient \acs{PQC}.]{Example of an $n$-qubit hardware-efficient \ac{PQC} with $m$ layers, where $R_y$ gates are used to hold the parameters with \textsc{CNOT} gates used as the entangling operators.}
    \label{fig:he_ansatz}
\end{figure}

A hardware-efficient unitary operator with $L$ layers can be written as
\begin{equation}
\label{eq:ala}
    U(\bm{\theta}) = \prod\limits_{k=0}^{L-1} U_k(\bm{\theta}_k)W_k,
\end{equation}
where $U_k(\bm{\theta}_k) = \exp(-\imath \bm{\theta}_k V_k)$ is a unitary generated by a non-parametrised Hermitian operator $V_k$, generally being single-qubit rotation gates, and $W_k$ is an entangling gate, which is ordinarily non-parametrised. The entangling gate generally depends on the native gate set of the quantum device, for example, for superconducting architectures the typical native two-qubit entangling gate is the \textsc{CNOT} or \textsc{C}$Z$ gate~\cite{Krantz2019}, or $XX$ gates for trapped ions~\cite{Wright2019}. Different types of hardware-efficient ans\"{a}tze are used in Chapter~\secref{chap:5}. An example of a hardware-efficient ansatz is shown in Fig.~\figref{fig:he_ansatz}.

\subsection{Other Ans\"{a}tze} \label{sec:other_ansatze}

In the context of quantum chemistry, the \ac{UCC} ansatz\footnote{More detail on the \ac{UCC} ansatz can be found in \citet{Taube2006} and \citet{Yung2014}.} is in fact the ansatz of choice in the first proposed \ac{VQE}~\cite{Peruzzo2014}. Coupled cluster theory is a post Hartree--Fock method that aims at recovering a portion of electron correlation energy by evolving an initial wave function. 

Moreover, rather than choosing between problem-inspired and hardware-efficient ans\"{a}tze, one can find a middle ground. In fact, in Chapter~\secref{chap:3}, we will investigate the midway approach of a \ac{HVA} that utilises hardware-efficient gates, which can be implemented natively in transmon devices~\cite{Ganzhorn2019,Sagastizabal2019,Consiglio2022a}. This approach is also a symmetry-preserving method, which has two main benefits: it restricts the size of the Hilbert space, reducing the risk of \acp{BP} (which will be discussed in Section~\secref{sec:barren_plateaus}) and potentially speeding up optimisation; and may avoid degeneracies and localised errors~\cite{Kandala2017, Ryabinkin2019}. The particular symmetry-preservation that will be considered is particle number conservation, with the most commonly employed entangling gate being
\begin{equation}
    A(\theta, \phi) = \left( 
    \begin{array}{cccc}
        1 & 0 & 0 & 0 \\
        0 & \cos(\theta) & -e^{-\imath \phi}\sin(\theta) & 0 \\
        0 & e^{\imath \phi}\sin(\theta) & \cos{\theta} & 0 \\
        0 & 0 & 0 & 1 \\
    \end{array} 
    \right),
\end{equation}
which can be decomposed into three \textsc{CNOT} gates and two rotation gates~\cite{Consiglio2022a}. Note that symmetry constraints can also be achieved by modifying the cost function, typically in the form of including Lagrangian multipliers~\cite{Ryabinkin2019}.

While only fixed-structure ans\"{a}tze have been mentioned, it is important to acknowledge the concept of adaptive-structure ans\"{a}tze.\footnote{For a review on adaptive-structure ans\"{a}tze, see Section 6.3 of \citet{Tilly2022}.} The main idea is to construct a circuit that adapts to the problem at hand. The most common procedure is to iteratively add new operators to the circuit based on their contribution to the overall energy, generally by minimising an alternative cost function after completing each iteration of a regular \ac{VQA}.

\section{Measurement Technique} \label{sec:measurement_technique}

A quantum computer can only supply information via measuring quantum states. Typically, the expectation value of a Hamiltonian is required to minimise an objective function. This requires a unitary transformation on the quantum state to the diagonal basis of the Hamiltonian. However, in principle, this is hard to achieve on a \ac{NISQ} device. The purpose of transforming a Hamiltonian into Pauli strings is because their expectation value can be easily determined on a quantum computer. The computational basis states $\ket{0}$ and $\ket{1}$ can be measured to give the expectation value of the $\sigma^z$ operator,
\begin{equation}
    \langle \sigma^z \rangle = p_0 - p_1 = 2p_0 - 1,
\end{equation}
where $p_0$, $p_1$ are the probabilities to measure the qubit in state $\ket{0}$, $\ket{1}$, respectively. The last equality is due to $p_0 + p_1 = 1$. Subsequently, measurements of $\sigma^x$ and $\sigma^y$ can be carried out by transforming the operators
\begin{equation}
    \sigma^x = H \sigma^z H,
    \label{eq:x_transform}
\end{equation}
\begin{equation}
    \sigma^y = S H \sigma^z H S^\dagger,
    \label{eq:y_transform}
\end{equation}
where $S = \sqrt{\sigma^z}$ and $H$ is the Hadamard gate. The expectation values can be determined by applying the transforming gates on the states before measuring in the computational basis, so that
\begin{equation}
    \langle \sigma^x \rangle = \bra{\psi}\sigma^x\ket{\psi} = \bra{\psi}H \sigma^z H\ket{\psi} = 2p_0 - 1,
\end{equation}
\begin{equation}
    \langle \sigma^y \rangle = \bra{\psi}\sigma^y\ket{\psi} = \bra{\psi}S H \sigma^z H S^\dagger\ket{\psi} = 2p_0 - 1.
\end{equation}
Thus, any Pauli string can be evaluated by, if necessary, transforming each individual qubit $k$ into the $\sigma^z$ basis and then measuring:
\begin{align}
    \langle P \rangle_{U} &= \left\langle \prod\limits_{i=0}^{\cP-1} \sigma^{f(i)}(i) \right\rangle_U = \left\langle \prod\limits_{i=0}^{\cP-1} \sigma^z(i) \right\rangle_{VU} = \prod\limits_{i=0}^{\cP-1} (2p_0(i) - 1),
\end{align}
where $V$ is the product of one-qubit rotations given by Eqs.~\eqref{eq:x_transform} and~\eqref{eq:y_transform}, depending on the Pauli terms $\sigma^{f(i)}$ at qubit $i$. 

In many cases, the number of Pauli strings can be so large that it becomes costly to evaluate each one of them individually. Recently, various approaches for efficiently measuring the expectation value of a Hamiltonian have been proposed~\cite{Bonet_monroig2020}, such as grouping different Pauli strings together to enable simultaneous measurement~\cite{Jena2019, Zhao2020b}. Grouping strategies are discussed in Section~\secref{sec:shots_and_grouping}, and are implicitly used in Chapters~\secref{chap:3},~\secref{chap:4} and~\secref{chap:5}.

Other techniques exist with their respective advantages and disadvantages, such as: measuring the overlap of a quantum state with a unitary operator; using approaches such as the Hadamard test~\cite{Miquel2002}; or direct implementation techniques~\cite{Mitarai2019}. Another method is shadow tomography~\cite{Huang2020}, which involves measuring the classical shadow of a quantum state to extract the expectation value of observables. The destructive \textsc{SWAP}~\cite{Garcia_escartin2013, Cincio2018} test is extensively used in Chapter~\secref{chap:4}.

\section{Classical Optimiser}

So far, only the components of a \ac{VQA} that operate on a quantum device have been discussed. The final piece is the classical optimisation loop, which updates the variational parameters of the \ac{PQC}, tying everything together to create a hybrid quantum--classical loop. The specific choice of an optimiser greatly impacts the performance of the \ac{VQA}. In principle, one can apply any method available in the entire field of multivariate optimisation~\cite{Lavrijsen2020}. However, the \ac{NISQ} era continues to be a limitation due to short coherence times that prevent the efficient implementation of analytical gradient circuits, and result in lengthy computation times for numerous function evaluations needed to calculate the objective function. At the same time, the results from a quantum computer are noisy, both from statistical and quantum device errors, necessitating the use of an optimiser that is noise-resistant. As a result, most of the frequently employed classical optimisers are not adequate for \acp{VQA}, and as a result, new optimisation algorithms and techniques are being developed to optimise \acp{PQC} and their corresponding objective functions.

Classical optimisers can generally be categorised into two classes: gradient-based and gradient free. We shall go through a brief description of the common approaches to both types.\footnote{More details can be found in \citet{Lavrijsen2020} and \citet{Bonet_monroig2023}.}

\subsection{Gradient-Based Approach} \label{sec:gradient_based}

The most common approach to minimise an objective function $f(\bm{\theta})$ is through evaluating its gradient. The gradient of a function indicates the direction in which it exhibits the largest rate of change, also called the direction of steepest \textit{ascent}. Suppose we start with an initial set of $K$ parameters $\bm{\theta}^{(0)}$, and iteratively update $\bm{\theta}^{(n)}$ over $n$ iterations. One such update rule is
\begin{equation}
    \bm{\theta}^{(n+1)} = \bm{\theta}^{(n)} - \eta \nabla f(\bm{\theta}^{(n)}),
\end{equation}
where $\eta$ is a hyperparameter, denoted as the learning rate (which can be a positive series $\eta_n$ converging to zero), and $\nabla = (\partial_1,\dots,\partial_K)$ representing the gradient of the objective function, with $\partial_i \equiv \frac{\partial}{\partial \theta_i}$. This translates to optimising each parameter $\theta_i$ via
\begin{equation}
    \theta_i^{(n+1)} = \theta_i^{(n)} - \eta \partial_i f(\bm{\theta}^{(n)}),
\end{equation}
using Einstein notation, whereby each parameter is nudged in the direction of steepest \textit{descent} due to the negative sign. 

The gradient can be estimated on a quantum computer by using different methods, with finite difference techniques and parameter shift rules being the most common methods. The finite-difference technique can be expressed as
\begin{equation}
\label{eq:finite_difference}
    \partial_i f(\bm{\theta}^{(n)}) \approx \frac{f(\bm{\theta}^{(n)} + \epsilon \bm{e}_i) - f(\bm{\theta}^{(n)} - \epsilon \bm{e}_i)}{2\epsilon},
\end{equation}
where $\epsilon$ is a small number and $\bm{e}_i$ is the unit vector in the $i^\text{th}$ direction. The smaller the $\epsilon$, the closer the approximation is to the true value of the gradient. On the other hand, a smaller $\epsilon$ corresponds to a smaller difference $f(\bm{\theta}^{(n)} + \epsilon \bm{e}_i) - f(\bm{\theta}^{(n)} - \epsilon \bm{e}_i)$, which would in turn entail a larger number of samples from the quantum computer to achieve a good estimation of the gradient~\cite{Bharti2022}. Furthermore, $\epsilon$ can be a function of $n$, i.e. $\epsilon = \epsilon^{(n)}$, that typically decreases with $n$, to obtain a finer resolution. Note that with this method, each iteration of the optimiser requires $2K$ evaluations of the cost function $f$.

The gradients of \acp{PQC} that are constructed via single-parameter unitary gates can be analytically determined using what are known as parameter shift rules~\cite{Romero2018, Mitarai2018, Schuld2019}, which is another technique that is commonly employed. Consider a sequence of $N$ parametrised unitary operators, as derived from Eq.~\eqref{eq:ala},
\begin{equation}
    U(\bm{\theta}) = \prod_{i=0}^{N-1} U_i(\theta_i),
\end{equation}
and
where without loss of generality we assume that the fixed gates are represented by $U_k(\theta_k)$ such that $\theta_k$ is fixed. Since each of these gates are unitary, then they can be represented by a generator $g_i$ such that
\begin{equation}
    U_i(\theta_i) = e^{-\imath \theta_i g_i},
\end{equation}
and
\begin{equation}
    \partial_i U_i(\theta_i) = -\imath g_i e^{-\imath \theta_i g_i} = -\imath e^{-\imath \theta_i g_i} g_i.
\end{equation}
With a slight abuse of notation $\ket{0} \equiv \ket{0}^{\otimes n}$, a common cost function in most \acp{VQA} is
\begin{equation}
    f(\bm{\theta}) = \bra{0} U(\bm{\theta})^\dagger \cH U(\bm{\theta}) \ket{0},
\end{equation}
in which the derivative with respect to $\theta_i$ can be rewritten as
\begin{align}
\label{eq:pre_param_shift}
    \partial_i f(\bm{\theta}) &= \partial_i \bra{0} U(\bm{\theta})^\dagger \cH U(\bm{\theta}) \ket{0} \nonumber \\
    &= \bra{0} \partial_i[U(\bm{\theta})^\dagger \cH U(\bm{\theta})] \ket{0} \nonumber \\
    &= \bra{0} \partial_i[U(\bm{\theta})^\dagger] \cH U(\bm{\theta}) + U(\bm{\theta})^\dagger \cH \partial_i[(]U(\bm{\theta})] \ket{0} \nonumber \\
    &= \bra{\psi_{i-1}} \partial_i[U_i(\theta_i)^\dagger] \cH_{i+1} U_i(\theta_i) + U_i(\theta_i)^\dagger \cH_{i+1} \partial_i[U_i(\theta_i)] \ket{\psi_{i-1}} \nonumber \\
    &= \imath \bra{\psi_{i-1}} U_i(\theta_i)^\dagger g_i^\dagger \cH_{i+1} U_i(\theta_i) - U_i(\theta_i)^\dagger \cH_{i+1}g_i U_i(\theta_i) \ket{\psi_{i-1}} \nonumber \\
    &= \imath \bra{\psi_{i-1}} U_i(\theta_i)^\dagger (g_i\cH_{i+1} - \cH_{i+1} g_i) U_i(\theta_i) \ket{\psi_{i-1}} \nonumber \\
    &= \imath \bra{\psi_{i-1}} U_i(\theta_i)^\dagger [g_i, \cH_{i+1}] U_i(\theta_i) \ket{\psi_{i-1}},
\end{align}
where we have used the fact that $g_i = g_i^\dagger$, the commutator of two operators $A$ and $B$ is defined as $[A, B] = AB - BA$, and define $\ket{\psi_{i-1}} = \left(\prod_{j=0}^{i-1}U_j(\theta_j)\right) \ket{0}$, $\cH_{i+1} = \left(\prod_{j=i+1}^{N-1} U_j\right)^\dagger \cH \left(\prod_{j=i+1}^{N-1} U_j\right)$. \citet{Schuld2019} demonstrated that if the generator $g_i$ has only two distinct eigenvalues $\{\pm r\}$\footnote{This inherently includes all single-qubit gates.} (that can be repeated), then
\begin{equation}
\label{eq:commutator}
    [g_i, \cH] = -\imath r \left[ U_i^\dagger\left(\frac{\pi}{4r}\right) \cH U_i\left(\frac{\pi}{4r}\right) -U_i^\dagger\left(-\frac{\pi}{4r}\right) \cH U_i\left(-\frac{\pi}{4r}\right) \right].
\end{equation}
Putting Eq.~\eqref{eq:commutator} in Eq.~\eqref{eq:pre_param_shift}, we get
\begin{align}
\label{eq:param_shift}
    \partial_i f(\bm{\theta}) &= r\bra{\psi_{i-1}} U_i\left(\theta_i + \frac{\pi}{4r}\right)^\dagger \cH_{i+1} U_i\left(\theta_i + \frac{\pi}{4r}\right)\ket{\psi_{i-1}} \nonumber \\
    & \hspace{0.03cm} - r\bra{\psi_{i-1}}U_i\left(\theta_i - \frac{\pi}{4r}\right)^\dagger \cH_{i+1} U_i\left(\theta_i - \frac{\pi}{4r}\right) \ket{\psi_{i-1}} \nonumber \\
    &= r\left[ f\left(\bm{\theta} + \frac{\pi}{4r}e_i\right) - f\left(\bm{\theta} - \frac{\pi}{4r}e_i\right) \right].
\end{align}
That is, by shifting the parameter $\theta_i$ by $\pm \pi / 4r$ and measuring, one can exactly determine the gradient of $f$ with respect to $\theta_i$. While Eq.~\eqref{eq:param_shift} looks quite similar to the finite difference rule in Eq.~\eqref{eq:finite_difference}, it uses a finite, not infinitesimal, shift and is in fact exact.

A noteworthy gradient-descent method that uses a stochastic approach is \ac{SPSA}, which is used to obtain many of the results in Chapter~\secref{chap:5}. The basic premise of \ac{SPSA} is to perturb all directions of the gradient evaluation simultaneously in each iteration. Specifically, let $\bm{\Delta}^{(n)}$ be a random perturbation vector, typically following the Bernoulli distribution $\pm1$ with probability $1/2$. The stochastic perturbation gradient estimator is then given by
\begin{equation}
    \partial_i f(\bm{\theta}^{(n)}) \approx \frac{f(\bm{\theta}^{(n)} + \epsilon \bm{\Delta}^{(n)}) - f(\bm{\theta}^{(n)} - \epsilon \bm{\Delta}^{(n)})}{2\epsilon \Delta_i^{(n)}}.
\end{equation}
Compared to finite difference techniques, \ac{SPSA} requires only two evaluations to compute the update step at each iteration, making it much more suitable for \ac{NISQ} algorithms.

\begin{figure}[t]
    \centering
    \includegraphics[width=\textwidth]{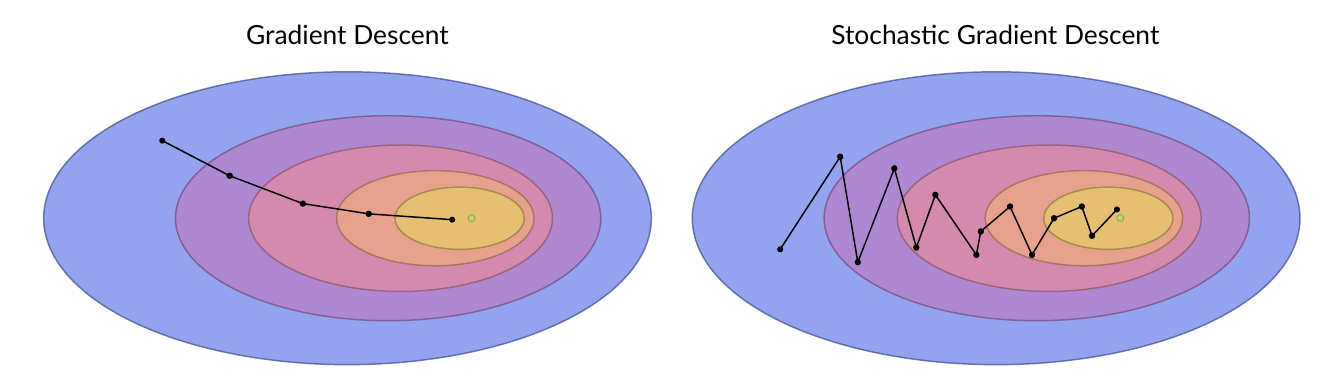}
    \caption{A schematic representation of gradient descent and stochastic gradient descent.}
    \label{fig:gradient_descent}
\end{figure}

A schematic representation of gradient descent and stochastic gradient descent can be seen in Fig.~\figref{fig:gradient_descent}.

There are many more techniques which can be used to calculate the gradient of an objective function on a quantum device; such as the \ac{BFGS} algorithm~\cite{Nocedal2006} (typically used to obtain statevector results in Chapters~\secref{chap:3},~\secref{chap:4} and~\secref{chap:5}), that utilises Hessian information for gradient descent~\cite{Huembeli2021}; \ac{SLSQP}~\cite{Kraft1988, Kraft1994} (used in Chapter~\secref{chap:4}), a quasi-Newton method that optimises successive second-order approximations of the objective function (via BFGS updates), with first-order approximations of the constraints; quantum natural gradient~\cite{Stokes2020}, that uses the Fisher information matrix; stochastic gradient descent~\cite{Sweke2020, Harrow2021}, that replaces the exact partial derivative with an estimator of the partial derivative, similar to \ac{SPSA}; quantum imaginary time evolution~\cite{McArdle2019}, invoking McLachlan’s variational principle~\cite{McLachlan1964}; and quantum analytic descent~\cite{Koczor2022}, which reconstructs local regions of the energy landscape and carries out classical optimisation in that region.

\subsection{Gradient-Free Approach} \label{sec:gradient_free}

One may prefer to employ an optimiser that does not rely on gradient information measured on a quantum computer, for a variety of technical reasons. There are methods based on evolutionary algorithms~\cite{Beyer2002}, which have been recently applied to \acp{VQA}~\cite{Zhao2020a,Anand2021}, that show comparable performance to gradient-based approaches. Reinforcement learning is another method, originally used to optimise the parameters of \ac{QAOA}~\cite{Garcia_saez2019,Khairy2019,Wauters2020,Yao2020a,Yao2020b}.

One such gradient-free approach that is employed in Chapter~\secref{chap:4} is \ac{GSA}~\cite{Tsallis1996, Xiang1997, Xiang2000}, a probabilistic algorithm that searches for a global minimum, rather than a local one, in contrast with optimisers in Sec.~\secref{sec:gradient_based}. \ac{GSA} simulates an annealing process, similar to annealing molten metal in metallurgy, to ﬁnd a global minimum of the objective function. A visiting distribution is employed to generate a trial jump distance $\Delta x^{(n)}$, which is proportional to a distorted Cauchy-Lorentz distribution
\begin{equation}
    P_{q_v}\left(\Delta x^{(n)}\right) \propto \frac{\left(T_{q_v}^{(n)}\right)^{-\frac{D}{3-q_v}}}{\left[ 1 + (q_v - 1)\frac{\left(\Delta x^{(n)}\right)^2}{\left(T_{q_v}^{(n)}\right)^\frac{2}{3-q_v}} \right]^{\frac{1}{q_v-1} + \frac{D-1}{2}}},
\end{equation}
where $n$ is the iteration, and $T_{q_v}^{(n)}$ is the artificial temperature, with both the temperature and visiting distribution controlled by the parameter $q_v$. The trial jump is automatically accepted if it is downhill, and possibly accepted according to an acceptance probability if it is uphill. A generalised Metropolis algorithm is used for the acceptance probability:
\begin{equation}
    p_{q_a} = \min\left\{1, \left(1 - (1 - q_a) \beta \Delta E\right)^\frac{1}{1-q_a}\right\},
\end{equation}
where $\beta \equiv 1 / T$, $\Delta E$ is the proposed energy difference, with the acceptance probability controlled by the parameter $q_a$. The artificial temperature is decreased according to
\begin{equation}
    T_{q_v}^{(n)} = T_{q_v}^{(0)}\frac{2^{q_v - 1} - 1}{(2 + n)^{q_v - 1} - 1}.
\end{equation}
The \ac{GSA} algorithm can be supplemented with a local optimiser, which activates either after each newly accepted position, or once the \ac{GSA} algorithm terminates upon meeting a criterion.

Lastly, sequential minimal optimisation is a technique used to obtain results as detailed in Chapters~\secref{chap:3} and~\secref{chap:4}, specifically by applying the method described by \citet{Nakanishi2020}. Consider a \ac{PQC} where; the parameters are independent of one another; is composed solely of fixed unitary gates and parametrised rotation gates; and the cost function is defined as a weighted sum of $K$ expectation values
\begin{equation}
    f(\bm{\theta}) = \sum_{i=0}^{K-1} w_k \bra{\psi_k} U^\dagger(\bm{\theta}) \cH_k U(\bm{\theta}) \ket{\psi_k},
\end{equation}
then, if $\bm{\theta}^{(n)}$ represents the parameters of the $n^\text{th}$ iteration, where $U_j^{(n)}(\theta_j)$ is the unitary $U(\bm{\theta})$ with all parameters fixed except $\theta_j$, and similarly for the cost function $f_j^{(n)}(\theta_j)$, then we have
\begin{align}
    f_j^{(n)}(\theta_j) &= \sum_{i=0}^{K-1} w_k \bra{\psi_k} U_j^{(n)\dagger}(\bm{\theta}) \cH_k U_j^{(n)}(\bm{\theta}) \ket{\psi_k} \nonumber \\
    &= \alpha_j^{(n)} \sin(\theta_j + \beta_j^{(n)}) + \gamma_j^{(n)},
    \label{eq:nft}
\end{align}
where $\alpha_j^{(n)}$, $\beta_j^{(n)}$ and $\gamma_j^{(n)}$ denote constants independent of $\theta_j$. Eq.~\eqref{eq:nft} signifies that $f_j^{(n)}(\theta_j)$ is simply a sine curve with period $2\pi$, where the constants can be determined by evaluating $f_j^{(n)}(\theta_j)$ at three different points $\theta_j$ (typically chosen to be $\theta_j^{(n)}$ and $\theta_j^{(n)} \pm \pi/2$). Then $\theta_j$ can be found by minimising $f_j^{(n)}(\theta_j)$, such that $\theta_j= \arg\min_{\theta_j} f_j^{(n)}(\theta_j)$. The optimisation is carried out sequentially, or randomly, over all indices $j$, until convergence. Note that $f(\bm{\theta})$ needs to be reevaluated every so often, since the minimum is calculated from the sine curve, and errors will accumulate.

\section{Theoretical Challenges}

One important definition in the analysis of \acp{PQC} is the concept of unitary $t$-designs. Following the formalism of \citet{Holmes2022a}, we define $\dU$ as the set of unitaries accessible by an ansatz $U$, and $\cU(n)$ the complete unitary group in which the ansatz is expressed, where $n$ is the number of qubits. Note that $\dU \subseteq \cU(n)$. A unitary $t$-design is a set of $K$ unitary operators $\{U_k\} \in \cU(N)$ that reproduce the first $t$ moments of the Haar measure~\cite{Dankert2009, McClean2018}, that is
\begin{equation}
    \frac{1}{K}\sum_{k=0}^{K-1} f(U_k) = \int_{\cU(N)} f(U) d_\mu U,
\end{equation}
where $f$ is any polynomial of degree at most $t$ in both the matrix elements of $U$ and $U^*$, and the integral is over the unitarily invariant Haar measure, with $d_\mu V$ being the volume metric. Thus, averaging over the $t$-design will yield the same result as averaging over the unitary group with respect to the Haar measure. 

$t$-designs are important to analyse, since \acp{PQC} that form an approximate 2-design will generally result in the occurrence of \acp{BP} in the optimisation landscape. Furthermore, \citet{Fontana2024} showed that for \acp{PQC} with observables that lie in their dynamical Lie algebra, the variance of the gradient of the cost function for 2-design scales inversely with the dimension of the dynamical Lie algebra, while \citet{Ragone2023} uncover the precise role that entanglement and locality play in the nature of \acp{BP}, and provide a unified theory for understanding \acp{BP} in deep \acp{PQC}.

\subsection{Expressibility}

The success of a \ac{VQA} heavily relies on selecting the appropriate ansatz tailored to the problem at hand. Beyond just its trainability --- how effectively the ansatz can be optimised --- another crucial aspect is expressibility. This pertains to the capability of a specific \ac{PQC} to produce a diverse range of quantum states~\cite{Sim2019, Holmes2022a}. Moreover, the expressibility of the circuit is intertwined with factors such as the number of \ac{PQC} layers, parameters, or entangling gates necessary to attain a desired level of accuracy.

A possible metric for the expressiblity of a \ac{PQC} (albeit state dependent) is defined as
\begin{equation}
    \cA_\dU(t) = \left\lVert \int_{\cU(n)} \left(V \rho V^{\dagger}\right)^{\otimes t} d_\mu V - \int_{\dU} \left(U \rho U^{\dagger}\right)^{\otimes t} dU \right\rVert_2,
\end{equation}
where $\rho$ is the initial state, $dU$ the uniform distribution over $\dU$, and $\lVert A \rVert_2 \equiv \Tr{A A^\dagger}$ denotes the Hilbert--Schmidt norm~\cite{Bharti2022}. A \ac{PQC} with small $\cA_\dU(t)$ is more expressible, such that $\cA_\dU(t) = 0$ denotes maximum expressibility, meaning it produces quantum states that align closely with the Haar distribution. The PQC uniformly samples from the entire Hilbert space, enabling it to approximate any possible state. This becomes particularly important when training the \ac{PQC} to model a specific quantum state without any prior knowledge. Thus, a \ac{PQC} with high expressiveness is more likely to accurately represent the target state.

\subsection{Entangling Capability}

Another quantifier for the expressiveness of a \ac{PQC} is entangling capability, which denotes the ability of a \ac{PQC} to create suitably entangled states. \citet{Sim2019} proposed the Meyer and Wallach measure~\cite{Meyer2002} to estimate a \ac{PQC}'s quality of prepared entangled states. This measure can be represented as
\begin{equation}
    Q(\ket{\psi}) = 2 \left( 1 - \frac{1}{n}\sum_{i=0}^{n-1} \Tr{\rho_i^2} \right),
\end{equation}
which condenses to the average purity of each qubit~\cite{Brennen2003}. $Q(\ket{\psi})$ was proven to be an entanglement monotone~\cite{Scott2004}, which is zero if the state $\ket{\psi}$ is a product state, and unity is reached for certain states, such as the $\GHZ$ state. The entangling capability of a \ac{PQC} is thus defined as the average $Q$ of states randomly sampled from the circuit:
\begin{equation}
    \text{Ent} = \frac{1}{|S|}\sum_{\bm{\theta}_i \in S} Q(\ket{\psi(\bm{\theta}_i)}),
\end{equation}
where $S = \{\bm{\theta}_i\}$ is the set of sampled circuit parameters~\cite{Bharti2022}.

\subsection{Reachability}

Another condition for a successful \ac{VQA} is quantifying the reachability of a \ac{PQC}, that is whether there exists a set of parameters $\bm{\theta}$ such that the state $\ket{\psi(\bm{\theta})}$ minimises the objective function. The reachability deficit is defined as
\begin{equation}
    f_R = \min_{\bm{\theta}} \bra{\psi(\bm{\theta})} f \ket{\psi(\bm{\theta})} - \min_{\ket{\psi} \in \cH} \bra{\psi} f \ket{\psi},
\end{equation}
where the first term is the minimum over all states that can be represented by the \ac{PQC}, whereas the second term on the right-hand side is the minimum over all states $\ket{\psi}$ of the Hilbert space $\cH$. The reachability deficit is non-negative, and $f_R = 0$ signifies that there is an optimal set of parameters $\bm{\theta^*}$ such that $\ket{\psi(\bm{\theta^*})}$ minimises the objective function $f$.

\subsection{Barren Plateaus} \label{sec:barren_plateaus}

\begin{figure}[t]
    \centering
    \includegraphics[width=\textwidth]{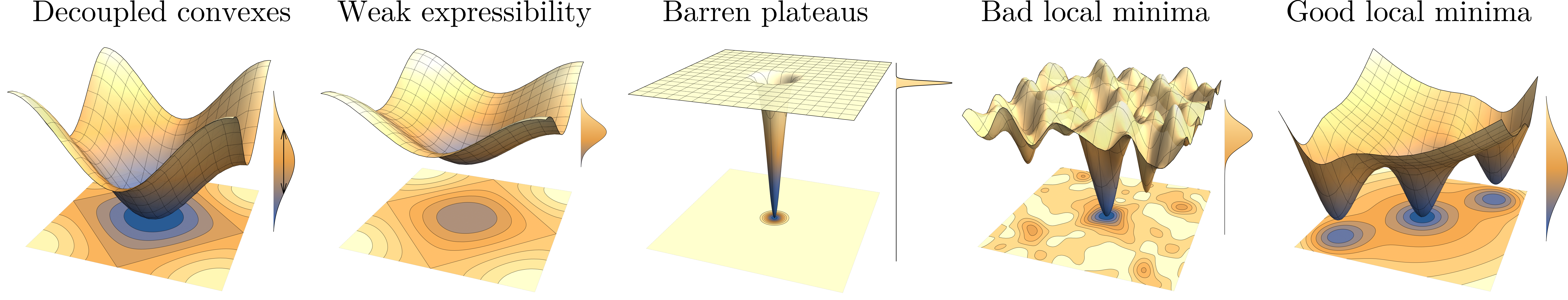}
    \caption[Plots of typical training landscapes.]{Plots of typical training landscapes, with the density of states cost value shown on the right side. The spread represents the landscape fluctuation $\sigma$. Although, the parameter space is inherently high-dimensional, it is presented in 2D here for ease of visualisation. Figure adapted from \citet{Zhang2024}.}
    \label{fig:bp}
\end{figure}

A \ac{PQC} of the form of Eq.~\eqref{eq:ala} becomes a unitary 2-design as the circuit depth increases polynomially with the number of qubits. \citet{McClean2018} proved that the expectation value of the gradient of the objective function, corresponding to randomly initialised \acp{PQC}, decays exponentially to zero as a function of the number of qubits; this is known as the \acf{BP} phenomenon. Certain assumptions or proofs can establish that specific ans\"{a}tze create approximate 2-designs, but demonstrating this in general remains difficult. Previous research has frequently relied on computing gradients and variances of a local observable with increasing system sizes to verify the existence of \acp{BP} when employing a particular ansatz~\cite{McClean2018, Skolik2021}. Furthermore, \citet{Cerezo2024} suggested that provable absence of \acp{BP} in a \ac{VQA} may imply classical simulability, however this analysis was only carried out for objective functions based on quantities such as that in Eq.~\eqref{eq:general_cost_function}.\footnote{Including the mixed state version.} On the other hand, it is expected to be hard to classically simulate different types of objective functions, such as those of thermal state preparation, as implemented in Chapter~\secref{chap:5}. Fig.~\figref{fig:bp} showcases different types of training landscapes.

In the context of the \ac{VQE}, the \ac{BP} problem can be formally characterised as follows. Consider a \ac{VQE} optimisation problem with cost function
\begin{equation}
    E(\bm{\theta}) = \ev{\cH}{\psi(\bm{\theta})},
\end{equation}
where $\cH$ is the Hamiltonian, and $\ket{\psi(\bm{\theta})}$ is a parametrised state given a set of parameters $\bm{\theta}$. The cost function exhibits a \ac{BP} if, $\forall~\theta_i \in \bm{\theta},~\exists~\epsilon > 0 \text{ and } b > 1$, such that
\begin{equation}
    P(|\partial_{\theta_i} E(\bm{\theta})| \geq \epsilon) \leq \cO\left(\frac{1}{\epsilon^2b^n}\right),
\end{equation}
which is a direct consequence of Chebyshev's inequality~\cite{Cerezo2021a}. This implies that the probability of the gradient being above a certain threshold, is upper-bounded by a number that decreases exponentially with the number of qubits $n$.

The emergence of \acp{BP} can stem from a number of sources, such as: circuits that are too expressive, irrespective of the operator or input state; global operators, even for shallow circuits; highly entangled initial states, even for shallow circuits and local operators; and hardware noise~\cite{Ragone2023}. \Acp{BP} pose a significant obstacle to the trainability of any \ac{PQC}, and as a result, those affected by this phenomenon are likely to struggle in properly training their parameters to find (near-)optimal solutions. Research by \citet{Arrasmith2021} indicates that even gradient-free approaches, which simulate gradient-based optimisation through local search methods, encounter similar difficulties. As a consequence, by following general guidelines that take heed of the above conditions, one can attempt to limit the hindrance caused by \acp{BP}.\footnote{Furthermore, one can look towards Section 6.1.2 of \citet{Tilly2022} and Section IV. A. of \citet{Bharti2022} for generic methods on tackling the \ac{BP} problem.} Appendix~\secref{app:bp} also features some qualitative analysis on the \ac{BP} problem for the Gibbs state preparation algorithm in Chapter~\secref{chap:5}.

\section{Tackling NISQ Devices}

As highlighted earlier in the introduction, the era of \ac{NISQ} devices consists of quantum computers that exhibit considerable noise, and coupled with the limited number of qubits, results in restricted use cases for such devices. This section will discuss potential methods and techniques of tackling the current restrictions within \ac{NISQ} devices.

\subsection{Shot Allocation and Pauli String Grouping} \label{sec:shots_and_grouping}

In any sampling experiment, the standard error is equal to $\epsilon = \sigma / \sqrt{N_s}$, where $\sigma$ is the population standard deviation, and $N_s$ is the experimental sample size, that is the number of shots. This means that the number of times an experiment needs to be repeated to achieve a given expected error $\epsilon$ goes as $\cO(1 / \epsilon^2)$. More specifically, when measurements are distributed optimally among the different Pauli strings, such that the variance is minimised with respect to a given precision $\epsilon$, the number of measurements required is upper-bounded by 
\begin{equation}
    N_s \leq \left( \frac{\sum_i w_i}{\epsilon} \right)^2,
\end{equation}
where $w_i$ are the weights of the Pauli strings of the Hamiltonian, with $\cP$ Pauli strings. For fermionic Hamiltonians, the number of Pauli strings is typically in the order of $\cP \sim \cO(n^4)$, where $n$ is the number of qubits, which would imply that the number of shots required for estimating the energy is
\begin{equation}
    N_s \sim \cO\left(\frac{n^4}{\epsilon^2}\right).
\end{equation}

Measurements of different Pauli operators (when not grouped) are independent and therefore uncorrelated, resulting in mean squared error for the energy estimate given a total of $\cP$ Pauli strings of
\begin{equation}
    \epsilon = \sqrt{\sum_{i=0}^{\cP-1} w_i^2\frac{\bV(P_i)}{N_i}} = \sqrt{\sum_{i=0}^{\cP-1} w_i^2\frac{1 - |\ev{P_i}{\psi}|^2}{N_i}},
\end{equation}
such that $N_i$ are the number of shots used to measure $P_i$ with weight $w_i$, $\sum_i N_i = N_s$, and we used
\begin{equation}
    \bV(P_i) = |\bra{\psi}P_i^2\ket{\psi}| - |\ev{P_i}{\psi}|^2 = 1 - |\bra{\psi}P_i\ket{\psi}|^2 \leq 1.
\end{equation}
\citet{Rubin2018} revealed that an optimal shot allocation is $N_i \propto |w_i|\sqrt{\bV(P_i)}$, while \citet{Arrasmith2020} demonstrated that given arbitrary states $\ket{\psi}$, one could then allocate according to $N_i \propto |w_i|$, which avoids accessing the variance, which might be hard to acquire.

Nevertheless, the naive approach of individually measuring each Pauli string will quickly approach a level of infeasibility. One technique of tackling this issue is partitioning the Hamiltonian based on its commutative sets. The idea of this method is that a group of Pauli strings are simultaneously diagonalisable through a unitary rotation if they form an Abelian group~\cite{Griffiths2005}. Consequently, the aim is to identify a set of generators for the Abelian group $\{\tau_i\}$, such that there exists a unitary $U$, where $U^\dagger \tau_i U$ is diagonal in the computational basis. Thus, finding $U$ allows the evaluation of the expectation value of Pauli strings which are generators within the group, by measuring in the computational basis. Pauli strings which are not generators of the group can be inferred from the same data.\footnote{For a review on grouping strategies, refer to Chapter 5 of \citet{Tilly2022}.}

\subsection{Error Mitigation} \label{sec:error_mitigation}

Presently, \ac{NISQ} devices are constrained by a finite quantity of qubits, around a hundred. Moreover, due to their inherent noise and short coherence time, the capacity for executing gate operations is restricted. A set of methodologies has been devised to alleviate noise impacts in quantum algorithms operating on \ac{NISQ} hardware. These methods have demonstrated the ability to lower noise levels in expectation value estimates, all without necessitating the extensive resources typically associated with error correction. While error mitigation was not the scope of the work included in this thesis, a brief overview of the most commonly applied techniques will be presented in this section.

The first commonly-employed methodology was proposed by \citet{Li2017} and \citet{Temme2017}, the zero-noise extrapolation technique. The idea of this technique is to let a quantum algorithm to operate at various noise levels, allowing the output to be extrapolated to acquire an expectation value with zero noise.

The second approach involves probabilistic error cancellation, which was introduced by \citet{Temme2017}. This method suggests estimating the expectation value of an operator by sampling from a collection of circuits that may contain errors. The idea behind this approach is that errors in individual circuits are random and may cancel each other out when averaged over a sufficiently large number of samples.

The third approach was proposed by \citet{Koczor2021} and \citet{Huggins2021}, and is known as exponential error suppression. Using different circuits, this method computes the value of a specific higher order moment of the outputs of a quantum state. Once the energy is minimised, the principal pure state of the noisy state generated by the \ac{PQC} exhibits energy that is exponentially close to the genuine ground state energy of the Hamiltonian.

\ac{SPAM} errors hold significance in \ac{NISQ} quantum computation as well. Mitigating their impact through straightforward measures is often crucial for attaining high accuracy in \acp{VQA}, especially for Pauli operators with significant weight. To address this form of noise, it is common practice to initially conduct calibration experiments on the quantum computer. These experiments yield a measurement calibration matrix, which correlates the likelihood of measuring each state with one another. This data is then utilised to mitigate \ac{SPAM} errors present in the outputs of the quantum computer. A type of \ac{SPAM} error mitigation algorithm is used in the \ac{VQA} of Chapter~\secref{chap:5} when running noisy simulations on IBM quantum computers.

The above are examples of commonly-employed error mitigation techniques, though many more exist.\footnote{For more detailed reviews on error mitigation, see Section V. A. of \citet{Bharti2022} and Section 8 of \citet{Tilly2022}.}

\subsection{Circuit Transpilation}

When translating a quantum circuit to a particular device layout, one must consider various factors such as the available quantum gates, the connectivity between qubits enabling two-qubit gate operations, and practical constraints like decoherence time, which limits the maximum circuit depth in terms of gate count. Consequently, the development of tools facilitating circuit simplification and efficient mapping of algorithms onto specific hardware configurations has become essential. These tools, often referred to as quantum compilers or transpilers~\cite{Chong2017}, serve to convert theoretical circuits into executable code for either simulators or actual quantum devices.

The available gates that can be implemented experimentally on a particular hardware platform are sometimes referred to as the native gate set. With a universal gate set $\cG \subseteq \text{SU}(d)$ (also called an instruction gate set), any unitary operation can be constructed efficiently. More formally, the Solovay--Kitaev theorem~\cite{Dawson2005} states that, given a universal set $\cG$, any unitary operation $U \in \cG$ can be approximated to within $\epsilon$-accuracy using a finite sequence of gates from $\cG$.\footnote{Although this is one of the most important theorems in quantum computation, it is an existence theorem; i.e., it does not provide the decomposition that it predicts. It also requires that the gate set contains the inverse of all gates.} This sequence scales logarithmically as $\cO(\log^c(1/\epsilon))$ where $c$ is a constant that depends on the theorem proof. For $d = 2^n$, this theorem guarantees that qubit quantum circuits can be decomposed using a finite gate sequence.

The generators of the Clifford group for qubits include the Clifford gates $H$, $S = \exp(-\imath \pi/4\sigma^z)$, and \textsc{CNOT}. According to the Gottesman--Knill theorem~\cite{Aaronson2004}, circuits composed solely of Clifford gates can be efficiently simulated using classical computers. These gates generate states known as stabiliser states, which can exhibit significant entanglement. However, not all unitary operations can be broken down into Clifford gates. To perform arbitrary quantum computations, a universal gate set is required. An example of such a set combines the Clifford gates with the $T = \exp(-\imath \pi/8\sigma^z)$ gate. The presence of $T$ gates is a necessary condition for achieving quantum advantage, as acknowledged by various studies~\cite{Amy2013, Gosset2013, Heyfron2018, Amy2018, Kissinger2020}. The computational challenge of classically simulating a quantum circuit intensifies with the inclusion of $T$ gates. Consequently, many algorithms aim to minimise the number of $T$ gates in quantum circuits to gauge their classical simulation complexity. 

Any individual quantum operation acting on a single qubit can be broken down into a series of rotational gates, represented by the gate sequence $U(\theta, \phi, \lambda) = R_z(\phi)R_y(\theta)R_z(\lambda)$, This suggests the use of rotational gates for single-qubit manipulation alongside at least one entangling gate, like a \textsc{CNOT} or \textsc{C}$Z$ gate, as the fundamental gate sets. Depending on the manufacturing technique of the quantum device, an inherent two-qubit gate implementation might be more appropriate. Instances include the utilisation of \textsc{C}$Z$ gates within tunable superconducting circuits~\cite{Krantz2019}, cross-resonance gates with fixed frequency superconducting qubits~\cite{Krantz2019, Kjaergaard2020}, and \textit{XX} gates in trapped ions~\cite{Haffner2008}.

Once the fundamental gate set is established, the subsequent stage involves breaking down the theoretical unitary circuit into this set. Directly translating all single- and two-qubit gates into the fundamental gate set could lead to a substantial circuit depth, thereby diminishing the efficiency of the decomposition process. Furthermore, determining the decomposition of gates that act on more than one qubit could generally present challenges. In addition to the conventional circuit decompositions mentioned previously, mathematical techniques may be necessary to comprehend and derive universal circuit reductions into specific, smaller components. One mathematical tool of significance is the ZX-calculus, which operates as a graphical language, associating quantum circuits with specific graph representations and establishing a set of rules for manipulating these graphs. Its utility spans from measurement-based quantum computation to quantum error correction.\footnote{For an extensive overview of the ZX-calculus and its diverse applications, refer to \citet{vandewetering2020}.} Other approaches use well-known artificial intelligence algorithms to find optimal circuit decompositions such as reinforcement learning~\cite{Zhang2020, Pirhooshyaran2021}, among other methods.

Once the quantum circuit has been decomposed and simplified into native gates, there remains a task specific to the hardware: mapping the resulting circuit to the particular qubit connectivity of the quantum device, commonly referred to as the qubit-mapping problem,\footnote{For a more in-depth review of the qubit-mapping problem, refer to Section V. B. 3. of \citet{Bharti2022}.} which is \ac{NP}-complete~\cite{Botea2018}. Typically, due to experimental constraints, not all qubits are interconnected in such a way that allows for the application of two-qubit gates between them. A straightforward method to address this challenge involves exchanging each qubit's state with its adjacent qubit --- utilising \textsc{SWAP} gates --- until a connected pair is found, executing the desired two-qubit operation, and then reverting the qubits' states back to their original configuration, effectively restoring the initial state with the intended two-qubit gate applied. However, this approach results in a notable increase in circuit depth, particularly for circuit topologies characterised by sparse qubit connectivity. Similarly to the previous section, circuit transpilation techniques present in \texttt{Qiskit} (IBM's quantum software development kit) are used when running the \ac{VQA} of Chapter~\secref{chap:5}.

%% file: chapter3/vqe.tex
\chapter{Variational Quantum Eigensolver for SU(\textit{N}) Fermions} \label{chap:3}

\epigraph{\textit{Particles not only collide, they compute.}}{--- \citet{LLoyd2006}}

\textit{Parts of this chapter are based on the published manuscript by \citet{Consiglio2022a}.}\\

The Fermi--Hubbard model, originally introduced to study the dynamics of electrons in solids~\cite{Hubbard1963}, is a paradigmatic example in addressing the physical properties of strongly interacting quantum many-body systems, ranging from superconductivity to quantum magnetism~\cite{Essler2005}. The Fermi--Hubbard model describes itinerant electrons sensing a local interaction. With cold atom quantum technology, the Fermi--Hubbard model can be studied with unprecedented control and flexibility of the system's physical conditions~\cite{Lewenstein2007,Mazurenko2017, Tarruell2018,Esslinger2010}. Despite the straightforward logic underlying the Fermi--Hubbard Hamiltonian dynamics, finding its ground-state remains a challenging problem in many-body physics, with numerous attempts made to solve it using the most advanced methods available~\cite{LeBlanc2015}. With no exception, different quantum algorithms have been devised to address the problem~\cite{Hensgens2017,Cade2020,Cai2020,Abrams1997, Dallaire_demers2016b,Bauer2016,Reiner2019,Uvarov2020}, mainly focusing on determining the ground-state via the application of the aforementioned \acp{VQA}.

In this chapter,  the \ac{VQE}~\cite{Peruzzo2014} is applied to SU($N$) Fermi--Hubbard models, describing strongly interacting fermions with $N$ spin-components. Interacting SU($N$) fermions play an important role in a variety of different contexts ranging from high-energy physics~\cite{Cherng2007,Rapp2007,chetcuti2023}, to specific situations in condensed matter physics~\cite{Keller2014,Kugel2015,Nomura2006,Arovas1999}. The higher symmetry accounts for a variety of novel phenomena, such as symmetry-protected topological phases, and Mott-insulator transitions at finite interaction values~\cite{Capponi2016}. Recently, the research scope on SU($N$) fermions has been substantially enlarged by cold atom quantum technology~\cite{Cazalilla2014,Sowinski2019}. Specifically, experiments with alkaline-earth and ytterbium atoms have simulated SU($N$) fermions~\cite{Pagano2014,Cappellini2014,Scazza2014} with $N$ as large as ten. SU($N$) cold atoms at the mesoscopic scale can provide a new platform for atomtronic circuits to widen the scope of current quantum simulation and access quantum devices with new specifications~\cite{Amico2021}. In particular, by using the logic of current-voltage characteristics of solid-state physics, an important goal of the atomtronics field is to eventually exploit matter-wave currents to probe quantum phases of matter. For SU($N$) fermions, such a programme has been initiated by studying the persistent current~\cite{Richaud2021, Chetcuti2022}.

\section{Simulating SU(\textit{N}) Fermions in a Fermi--Hubbard Model}

The prototypical Fermi--Hubbard Model, written in second quantisation for SU(2) fermions~\cite{Altland2010}, is
\begin{equation}
    \cH = -t \sum_{i}\sum_{s=\uparrow,\downarrow}\left( c_{i, s}^\dagger c_{i+1,s} + c_{i+1, s}^\dagger c_{i,s}  \right) + U \sum_{i} n_{i,\uparrow} n_{i,\downarrow},
    \label{eq:FH}
\end{equation}
where $c_{i, s}^\dagger$ ($c_{i, s}$) creates (annihilates) a fermion with spin-component $s$ at site $i$. The $s$-fermion number operator is $n_{i, s} = c_{i, s}^\dagger c_{i, s}$, and the total number operator on site $i$ is defined as $n_i = n_{i,\uparrow} + n_{i,\downarrow}$. Setting $U = 0$ reduces~\eqref{eq:FH} to the tight-binding model. Varying the ratio $U / t$ transitions the model from a conductor to an insulator, whereas in the case of $U / t \gg 1$ the model reduces to a set of isolated magnetic moments. The one-dimensional SU(2) Hubbard model is directly solvable using the Bethe ansatz~\cite{Lieb1968}. For $N > 2$, the Bethe ansatz integrability is preserved in the continuous limit of vanishing lattice spacing, with~\eqref{eq:FH} turning into the Gaudin--Yang--Sutherland model describing SU($N$) fermions with a delta interaction~\cite{Decamp2016, Chetcuti2022}; such a regime is achieved by~\eqref{eq:FH} in the dilute limit of small $N_p/L$, where $N_p$ denotes the total number of fermions in the model, i.e. $N_p = \sum_{i,s}n_{i,s}$. Another integrable regime of~\eqref{eq:FH} is obtained for $n_{i,s} = 1~\forall i, s$, and large repulsive values of $U \gg t$ for which the system is governed by the Lai--Sutherland model~\cite{Capponi2016, Chetcuti2022}.

\subsection{The SU(\textit{N}) Extended Fermi--Hubbard Model}

\begin{figure}[t]
    \centering
    \includegraphics[width=0.7\textwidth]{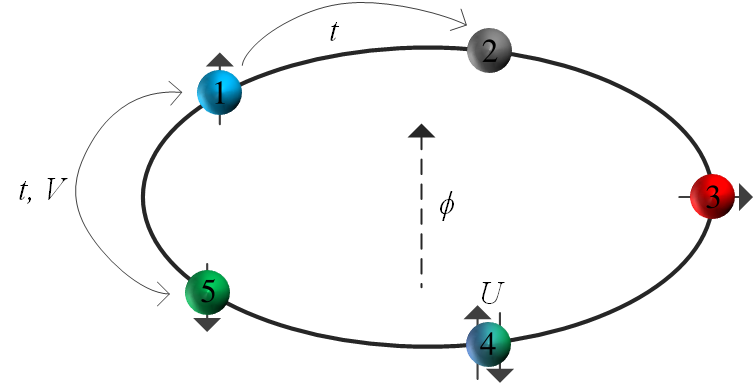}
    \caption[Example of an SU(3) 5-site Fermi--Hubbard model.]{Example of an SU(3) 5-site Fermi--Hubbard model. $t$ is the hopping term, $U$ is the on-site interaction, $V$ is the nearest-neighbour interaction, and $\phi$ is the magnetic flux. The colours (including up, right and down arrows) are meant to represent occupancy by spins of different components, with the arrows showing an example of how spins can hop between different sites. Site 1 consists of a spin up fermion, site 2 is empty, site 3 consists of a spin right fermion, site 4 consists of both a spin up and spin down fermion, and site 5 consists of a spin down fermion.}
    \label{fig:model}
\end{figure}

In this work, the physical system of interest is made of a set of SU($N$) fermions localised in a 1D ring lattice with $L$ sites,
\begin{align}
\label{eq:hubbard_model}
\cH = -t &\sum_{i=0}^{L-1}\sum_{s=0}^{N-1} \left(e^{\imath \frac{2 \pi \phi}{L}} c_{i, s}^\dagger c_{i+1,s} +\acs{hc} \right) + U\sum_{i=0}^{L-1} \sum_{s=0}^{N-1}\sum_{s'=s+1}^{N-1} n_{i, s}n_{i, s'} + V\sum_{i=0}^{L-1} n_{i} n_{i+1} ,
\end{align}
with the \ac{hc} denoting that of the preceding terms. The parameter $t$ is the hopping amplitude of a fermion between nearest-neighbour sites, the parameters $U$ and $V$ describe the on-site and density-density nearest-neighbour interaction, respectively, $n_{i, s} = c_{i, s}^\dagger c_{i, s}$ and $n_i = \sum_{s=0}^{N-1} n_{i,s}$. Furthermore, only non-negative values are considered for both $U$ and $V$.\footnote{Appendix~\secref{app:vqe} contains the generalisation of Eq.~\eqref{eq:hubbard_model} to a higher-dimensional long-range Fermi--Hubbard model.} A diagram for an SU(3) 5-site Fermi--Hubbard model is shown in Fig.~\figref{fig:model}.

For $V=0$, a superfluid to Mott insulator transition takes place for a finite value of the local repulsive interaction $U$ for $N>2$, whilst for $N = 2$ it was discovered that there is no finite $U$ Mott transition~\cite{Capponi2016,Xu2018}. Due to the interplay between $U$ and $V$, different quantum phases can be displayed. Specifically, the phase diagram involves a superfluid phase, a Mott phase, and a ``beat'' phase in which the particle occupation is modulated along the chain with a vanishing spin gap~\cite{Perez_romero2021}.

\subsection{Persistent Current}
 
The phase factor $e^{\imath \frac{2 \pi \phi}{L}}$ in Eq.~\eqref{eq:hubbard_model} takes into account the effective magnetic flux piercing the ring (see Fig.~\figref{fig:model}), which is able to impart a persistent current with the following form:
\begin{equation}
\label{eq:persistent_current}
I\left(\phi\right) = -\bra{\psi_0}\frac{\partial \cH}{\partial \phi}\ket{\psi_0} = \frac{2\pi\imath t}{L}\sum\limits_{i=0}^{L-1}\sum\limits_{s=0}^{N-1}\bra{\psi_{0}}e^{\imath\frac{2\pi\phi}{L}}c_{i,s}^{\dagger}c_{i+1,s} -\mathrm{h.c}.\ket{\psi_{0}},
\end{equation}
where $\ket{\psi_{0}}$ denotes the ground-state of Eq.~\eqref{eq:hubbard_model}. For the case of $V=0$, it has  been recently demonstrated that as a combination of the interaction, magnetic flux and spin correlations, the effective elementary flux quantum $\phi_{0}$ which fixes the persistent current periodicity, is observed to evolve from a single particle one to an extreme case of fractional flux quantum, in which one flux quantum is shared by all the fermions~\cite{Chetcuti2022}. Such a phenomenon reflects a type of attraction from repulsion: despite the repulsive interaction, spin correlations can lead to many-body states reacting as if they were bound states of fermions.

\subsection{Transforming the Hamiltonian}

To implement the \ac{VQE} on a \ac{NISQ} computer, it is crucial to represent the system described by Eq.~\eqref{eq:hubbard_model} in terms of qubits~\cite{McArdle2020,Cai2020,Dallaire_demers2016a,Cade2020}. To this end, the \ac{JW} transformation, originally devised for two-component fermions~\cite{Jordan1928, Shastry1986,Reiner2016}, is extended to the general SU($N$) case, as developed in \citet{Consiglio2022a}. $N$ sets of Pauli operators are introduced, one for every spin-component $s$, hereafter called colour, of the fermionic atom. The mapping assumes the following form:
\begin{equation}
\label{eq:map}
    c_{i,s}^\dagger=\left(\bigotimes_{j<n}\sigma_j^z\right) \otimes \sigma_n^+,~ c_{i,s}=\left(\bigotimes_{j<n} \sigma_j^z\right) \otimes \sigma_n^- ,~ (i, s) \xrightarrow{} n = i + sL,
\end{equation}
where $\sigma^\pm = (\sigma^x \pm \imath\sigma^y) / 2$ are the ladder operators, and $\ket{0}_n$ and $\ket{1}_n$ represent the absence and presence of a fermion of colour $s$ on site $i$, respectively. In this way, up to the Pauli string operator $\bigotimes_{j<n}\sigma_j^z$, a fermion of colour $s$ on site $i$ is mapped onto a spin-$1/2$ operator $\sigma_n^+$ acting on qubit $n=sL + i$. As a consequence, a system of $N_p$ SU($N$) fermions on $L$ sites is mapped onto $NL$ qubits.\footnote{It is straightforward to show that the mapping in Eq.~\eqref{eq:map} preserves fermionic commutation rules, independently of the chosen order of the colour $s$ appearing in the definition of $n$.} The Hamiltonian of Eq.~\eqref{eq:hubbard_model} is thus mapped to
\begin{align}
    \cH= &-t\sum\limits_{i, s} P_{i,s}\left(e^{\imath \frac{2 \pi \phi}{L}} \sigma_{sL + i}^+\sigma_{sL + i + 1}^- +\acs{hc}\right) \nonumber\\
    &+ \frac{U}{4}\sum\limits_{i, s < s'}\left( 1-\sigma_{sL + i}^z \right)\left( 1-\sigma_{s'L + i}^z \right)\nonumber \\
    &+ \frac{V}{4}\sum\limits_{i, s, s'} \left( 1-\sigma_{sL + i}^z \right)\left( 1-\sigma_{s'L + i + 1}^z \right),
\label{eq:hubbard_hamiltonian}
\end{align}
and
\begin{equation}
    I(\phi) = \frac{2\pi \imath t}{L}\sum\limits_{i, s} P_{i,s}\ev{e^{\imath \frac{2 \pi \phi}{L}} \sigma_{sL + i}^+\sigma_{sL + i + 1}^- -\acs{hc}}{\psi_0},
\end{equation}
where $P_{i,s}$ is the colour-dependent parity term given by
\begin{equation}
    P_{i,s} = \begin{cases}
        -1 ,~ \text{if } i = L - 1 \text{ and } N_s \text{ is odd}, \\
        +1 ,~ \text{otherwise},
        \end{cases}
\end{equation}
with $N_s$ being the number of fermions of colour $s$, such that $\sum_{i=0}^{L-1} n_{i,s} = N_s$.

\subsection{Introducing the VQE to the Model}

The aim of the \ac{VQE} is to find a set of parameters $\bm{\theta}$ able to obtain a quantum state $\ket{\psi(\bm{\theta})}$, that minimises the expectation value of the energy $E(\bm{\theta})$ of a Hamiltonian, through a shallow \ac{PQC}, $U(\bm{\theta})$, acting on an initial state $\ket{\psi}$. The \ac{PQC} is specifically constructed to be number-preserving \acf{HVA}, that is intended to be hardware-efficient on quantum devices that can natively employ parametrised \textsc{SWAP}-type gates~\cite{Ganzhorn2019,Sagastizabal2019}. Details of this construction can be found in Appendix~\secref{app:vqe}.

The minimisation is achieved through an adiabatically-assisted\footnote{"Adiabatic assistance is achieved via the magnetic flux $\phi$, by employing the previously optimized parameters from a neighbouring value of $\phi$ during the optimisation procedure.} quantum--classical loop of an optimiser using evaluations of the expectation value of the Hamiltonian~\cite{Garcia_saez2018}. The proposed \ac{PQC} utilises the extension of the \ac{JW} transformation to SU($N$) fermions, inspired by \citet{Cade2020} who adopted a number-preserving \ac{HVA} for SU(2) fermions. Parametrised $\imath$\textsc{SWAP} gates model the hopping terms between sites representing the same colour, while the interaction terms, as well as the on-site terms, between fermions of a different colour, are represented by \textsc{C}$R_z$ and $R_z$ gates, respectively. An example of such a \ac{PQC} is shown in Fig.~\figref{fig:circuit}, the details of which are discussed in Appendix~\secref{app:vqe}.

\begin{figure}[t]
	\centering
	\includegraphics[width=\textwidth]{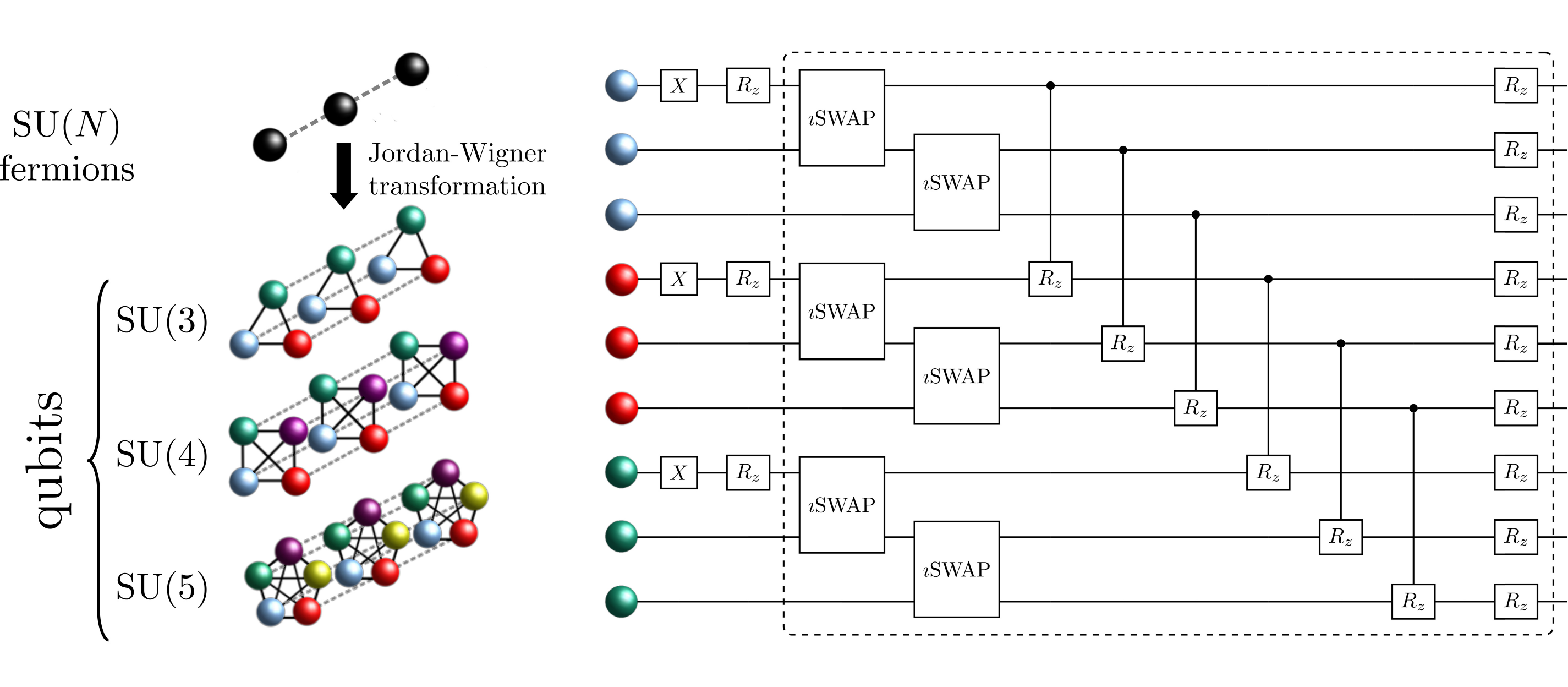}
	\caption[\acs{JW} mapping and \acs{PQC} used in a \acs{VQE} for a three-site SU(3) Fermi--Hubbard model.]{\ac{PQC} employed for the \ac{VQE}. (Left) Mapping a three-site linear SU($N$) Fermi--Hubbard Hamiltonian (black spheres) with only nearest-neighbour hopping and on-site interaction to an $N$-sided qubit prism of length $L$. The dashed grey lines represent the hopping term $t$ between same-colour qubits, and continuous black lines depict the on-site interaction $U$. (Right) \ac{PQC} for the three-site SU(3) Fermi--Hubbard Hamiltonian in the $N_p=3$ (number of fermions) sector starting from a state with all fermions on the same site (which are created in the circuit via initial $X$ gates). The variational parameters of the \ac{PQC} are contained within the $\imath$\textsc{SWAP}, C$R_z$ and $R_z$ gates. The dashed box corresponds to one \ac{HVA} layer. Note that the last sequence of entangling gates can be applied in two steps of parallel operations.}
	\label{fig:circuit}
\end{figure}

The simulation of the \ac{VQE} algorithm was carried out using the \texttt{Qibo} API~\cite{Efthymiou2022}, disregarding the noise of the quantum gates. The classical optimisation technique employed in the quantum--classical loop, in the case of statevector simulations (exact energy measurements), is the \ac{BFGS} method~\cite{Nocedal2006}, a gradient-based approach which involves the computation of the inverse Hessian matrix. On the other hand, for shot-based optimization, the \ac{NFT} method was used, which is a gradient-free sequential minimal optimiser~\cite{Nakanishi2020}. Finally, the results were benchmarked using the exact diagonalisation methods provided by the \texttt{QuSpin} software package~\cite{Weinberg2017, Weinberg2019}. The persistent current is also monitored, as given in Eq.~\eqref{eq:persistent_current}, that is accessed by means of a \ac{VQE} subroutine, after acquiring the ground-state.

\section{Results}

First, the results of the \ac{VQE} for commensurate ($N_p/L$ is an integer) and incommensurate ($N_p/L$ is not an integer) SU(3) models are discussed, these are found in Figs.~\figref{fig:3comm} and \figref{fig:incommensurate}, respectively, with different values of the parameters $U$ and $V$ corresponding to different physical regimes in the phase diagram of the system.

\subsection{SU(3)} \label{sec:SU3}

\begin{figure}[t]
	\centering
	\includegraphics[width=\textwidth]{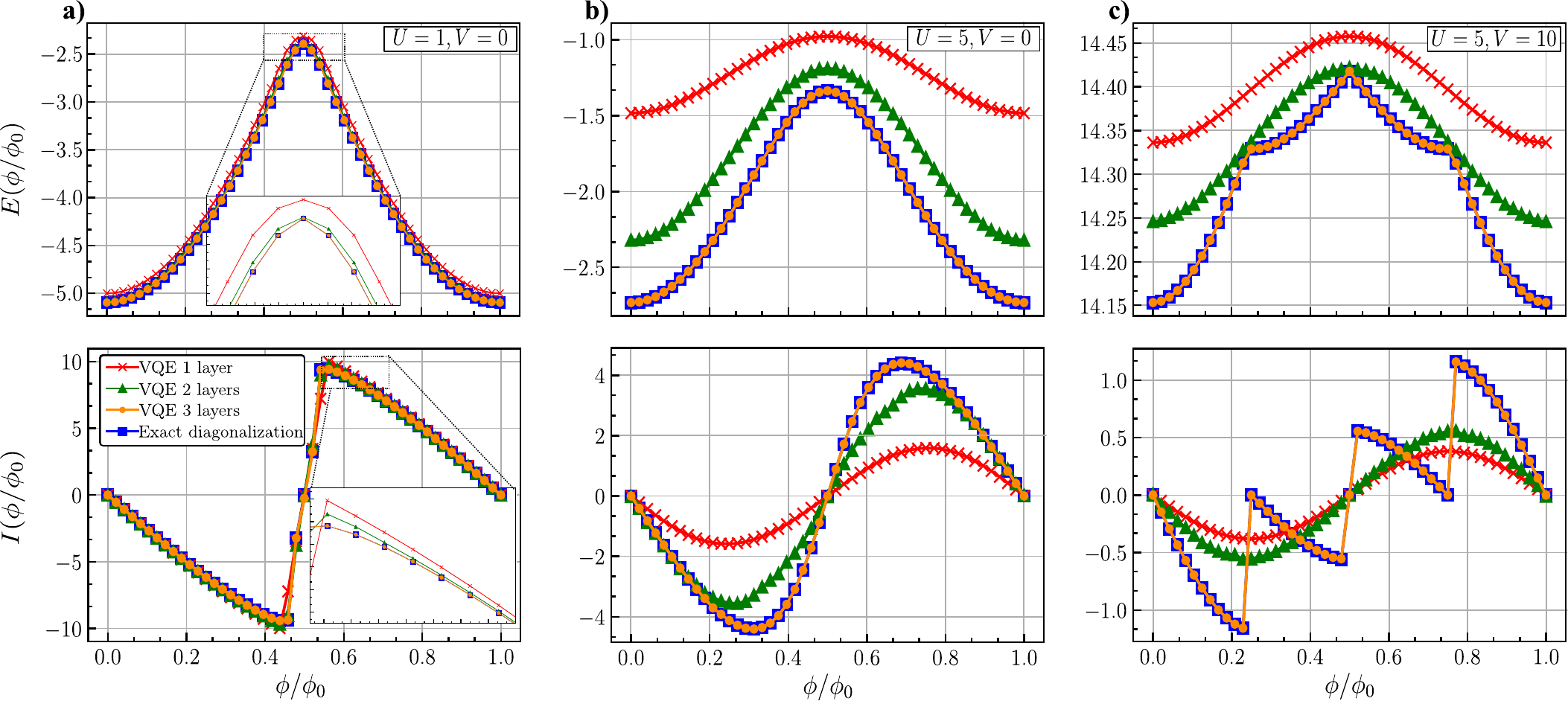}
	\caption[Ground-state energy and corresponding persistent current for a three-site SU(3) Fermi--Hubbard model.]{The ground-state energy $E_{0}(\phi)$ (top panel) and the corresponding persistent current $I(\phi)$ (bottom panel) for SU(3) fermions with different local $U$ and nearest-neighbour $V$ interactions, in the integer filling regime of the Fermi--Hubbard model, using exact energy measurements. The insets in the first column figures show zoomed in regions of their respective plots. The profile of the persistent current gives a clear indication between the \textbf{a)} superfluid, \textbf{b)} Mott and \textbf{c)} beat phases. Exact diagonalisation for $L=N_{p}=3$ is used to monitor the results obtained by the \ac{VQE}.}
	\label{fig:3comm}
\end{figure}

\begin{figure}[t]
    \centering
    \includegraphics[width=0.75\textwidth]{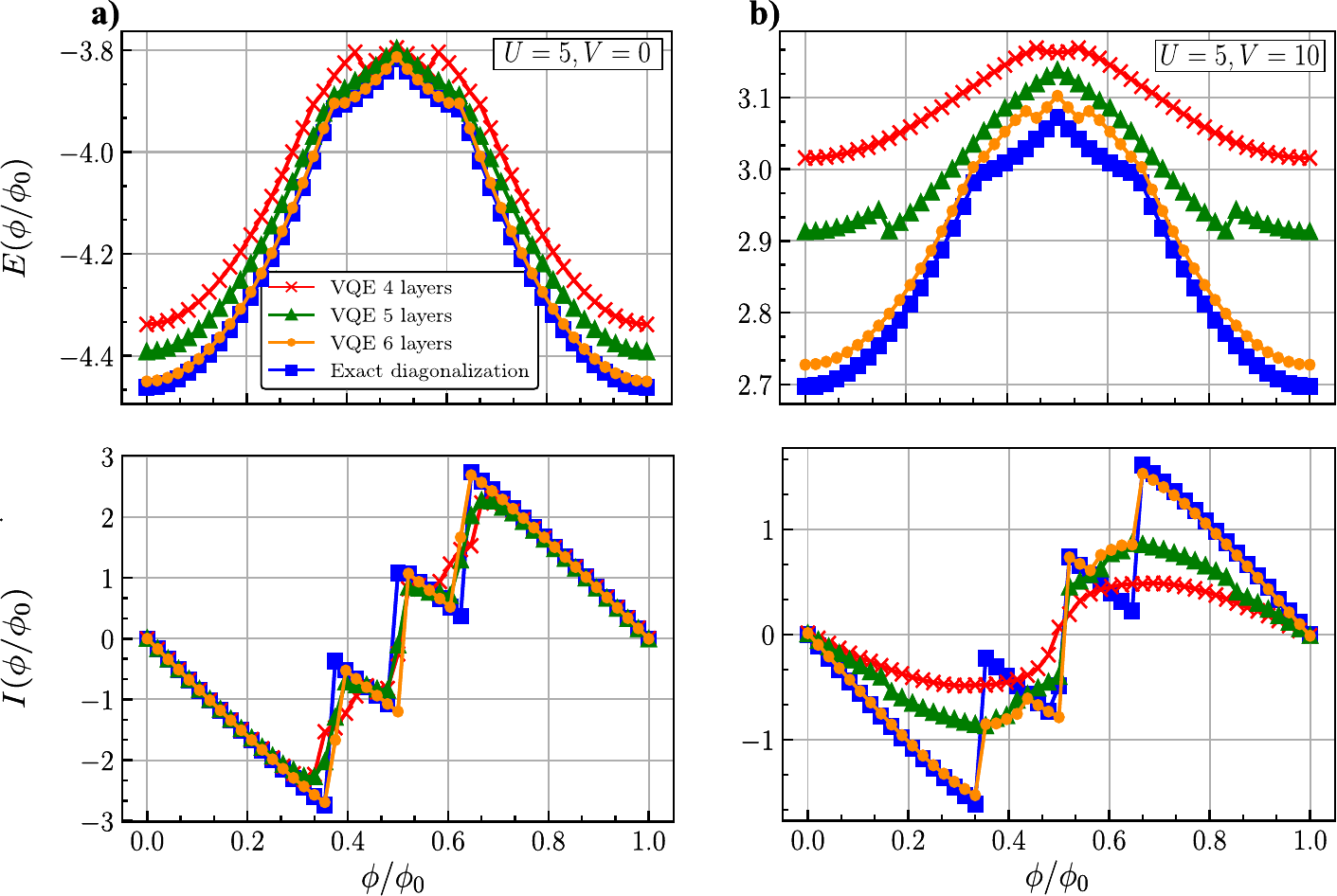}
    \caption[Ground-state energy and corresponding persistent current for a five-site SU(3) Fermi--Hubbard model.]{The ground-state energy $E_{0}(\phi)$ (top panel) and the corresponding persistent current $I(\phi)$ (bottom panel) for SU(3) fermions with different local $U$ and nearest-neighbour $V$ interactions, in the incommensurate regime of the Fermi--Hubbard model, using exact energy measurements. Exact diagonalisation for $L=5$ and $N_{p}=3$ is used to monitor the results obtained by \ac{VQE}.}
    \label{fig:incommensurate}
\end{figure}

In the lower panels of Figs.~\figref{fig:3comm} and \figref{fig:incommensurate} the persistent current given in Eq.~\eqref{eq:persistent_current} is monitored. In the commensurate case, the persistent current is observed to have a saw-tooth shape as in Fig.~\figref{fig:3comm} \textbf{a)}. Upon increasing the on-site interaction, the saw-tooth shape is smoothed out into a sinusoidal one due to the opening of the spectral gap, which indicates the onset of the Mott phase transition~\cite{Chetcuti2022} --- Fig.~\figref{fig:3comm} \textbf{b)}. The Panel \textbf{c)} of Fig.~\figref{fig:3comm} depicts a very interesting phenomenon. By increasing nearest-neighbour interactions through $V$, one goes from the Mott to the beat phase~\cite{Perez_romero2021}, where the latter is characterised by the modulation of the density arising out of effective attraction from repulsion. This in turn results in a hybrid form of the persistent current: a smoothed out yet fractionalised one. Figs.~\figref{fig:incommensurate} \textbf{a)} and \textbf{b)} display the persistent current against interaction for incommensurate systems. Upon increasing $U$, the persistent current fractionalises due to energy level crossings that occur to counteract the increase in $\phi$, resulting in the creation of spinon excitations in the ground-state~\cite{Chetcuti2022}. Upon increasing $V$, as in the lower panel of Fig.~\figref{fig:incommensurate} \textbf{b)}, the fractionalisation is enhanced due to the increased repulsion in the system.

At $\phi = 0$, the final optimal parameters of each case were determined by optimising the \ac{VQE} with the random initialisation of parameters. However, each subsequent instance of the \ac{VQE}, tasked with finding the ground-state energy of the model with the next iteration of $\phi$, was fed the optimal parameters from the previous iteration. This form of adiabatic assistance, along with the symmetry of the ground-state energy along the degeneracy point $\phi/\phi_{0} = 0.5$, offers a significant speed up in mapping out the ground-states of the extended Fermi--Hubbard Hamiltonian over the entire range of $\phi$, which is necessary for determining the persistent current.

Notice that the discontinuities of $I\left(\phi\right)$, due to spinon creation in the ground-state, are fully captured. Note that such a phenomenon is only captured by the \ac{VQE} if a sufficient number of layers are considered. This feature is consistent with the general theory signifying that the flux quantum fractionalisation is a genuine many-body correlation effect. Furthermore, for small $U$ and $V$, a shallow circuit estimates the ground-state with high accuracy. However, the correlations that are present for larger $U$ and $V$ necessitate the need for more \ac{PQC} layers to fully capture the ground-state of the model. The entanglement entropy of the final state is employed as an additional measure to gauge how closely the \ac{VQE} reproduces the desired ground-state. The entanglement entropy of a pure state of two systems $A$ and $B$ is defined as
\begin{equation}
    \cS(\rho_A) = -\Tr{\rho_A \log \rho_A} = -\Tr{\rho_B \log \rho_B} = \cS(\rho_B),
\end{equation}
where $\rho_A$ and $\rho_B$ are the reduced density matrices of the pure state $\ketbra{\psi_{AB}}$. In our case, $A$ is defined as the set of qubits $0$ to $\floor{n - 1}/2$, while $B$ is the set of qubits $\floor{n + 1}/2$ to $n - 1$, which equates to splitting the circuit in the middle.

\begin{figure}[t]
    \centering
    \includegraphics[width=0.75\textwidth]{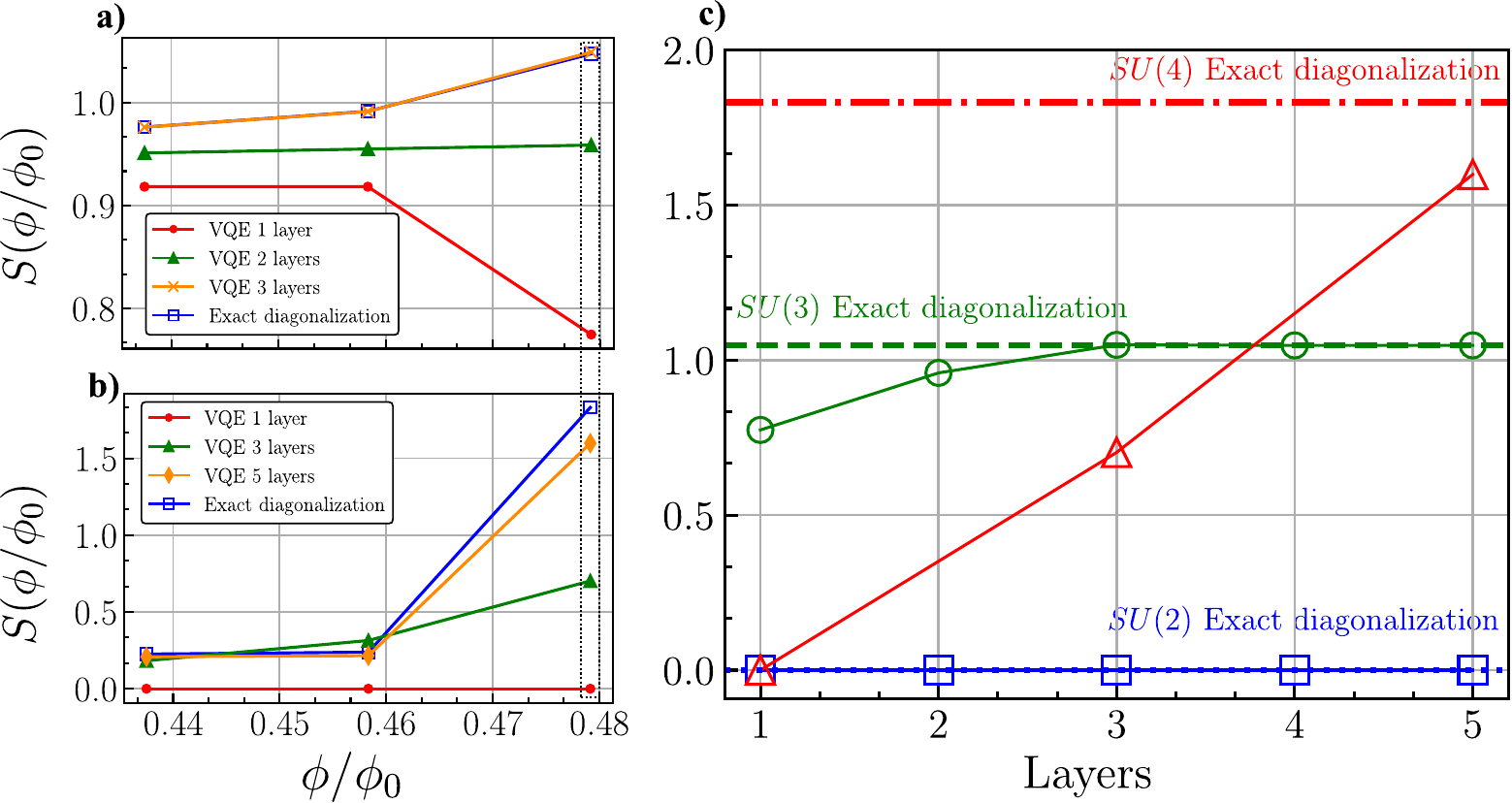}
    \caption[von Neumann Entropy of the Fermi--Hubbard model scaling with the number of layers in the \acs{VQE}.]{Values for the bipartite entropy $\cS(\phi)$ of half the qubit chain for \textbf{a)} SU(3) and \textbf{b)} SU(4) for an increasing number of \ac{PQC} layers for a Fermi--Hubbard model with $U=1$ and $V=0.5$ at commensurate fillings, namely, $N_p=N$, using exact energy measurements. \textbf{c)} By looking specifically at $\phi=0.5$, the SU(2) and SU(3) cases quickly reach the entropy of the exact diagonalisation state, while the SU(4) case still requires more correlations to be built up within the quantum circuit to fully approximate the ground-state (see Fig.~\figref{fig:su4}). Data was obtained using \texttt{quspin} via exact diagonalisation, after acquiring the ground-state through the \ac{VQE}.}
    \label{fig:entanglement}
\end{figure}

Note that, by increasing the number of layers of the \ac{PQC}, the entanglement in the ground-state increases, as shown in Fig.~\figref{fig:entanglement}. These results obtained using \texttt{quspin} via exact diagonalisation, after acquiring the ground-state through the \ac{VQE}. This layer-by-layer build-up of entanglement is crucial to capture the correlations of the target quantum state. It is only after the \ac{PQC} has enough depth to reach the target entanglement entropy, that the \ac{VQE} can start to approximate the target observable with high precision, as previously shown by \citet{Bravo_prieto2020}.

The number-preserving \ac{HVA} explores a subspace of the Hilbert space of size $\prod_{s=0}^{N-1}\binom{L}{N_s}$ for SU($N$) fermions in $L$ sites, where $N_s$ is the number of fermions of colour $s$, rather than the full Hilbert space of size $N^L$. Therefore, the number of layers needed to properly capture the ground-state energy will depend on the expressibility of the \ac{PQC} in covering the reduced subspace. In order to reduce the number of parameters needed for the \ac{VQE}, one could introduce a translationally-invariant \ac{PQC}, to account for the periodic nature of the fermionic chains studied. However, this \ac{PQC} would possibly require non-trivial and non-local quantum gates, which may not be suitable for implementation in \ac{NISQ} computers.

\subsection{SU(4)}

\begin{figure}[t]
    \centering
    \includegraphics[width=0.75\textwidth]{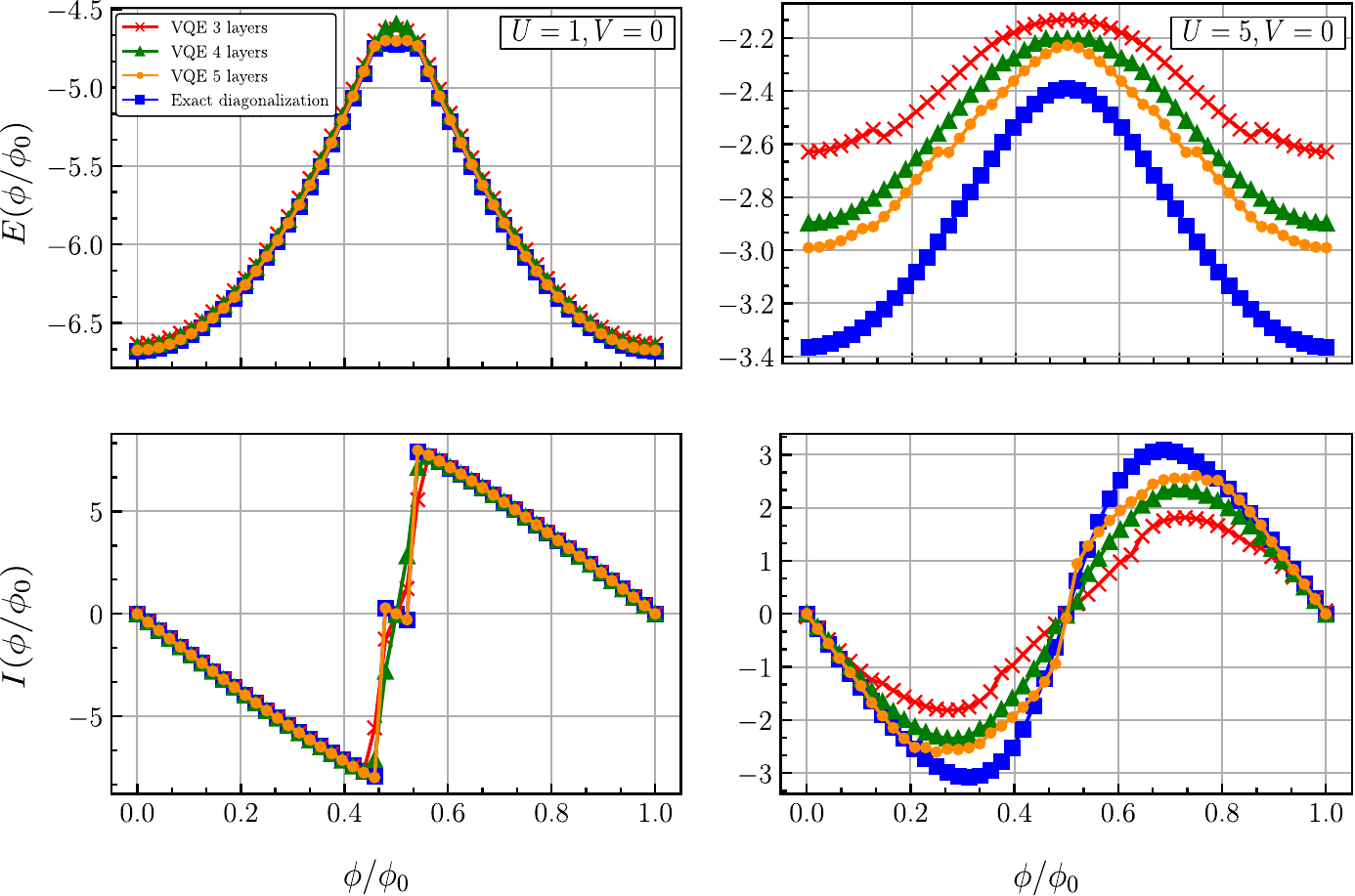}
    \caption[Ground-state energy and corresponding persistent current for a four-site SU(4) Fermi--Hubbard model.]{The ground-state energy $E_{0}(\phi)$ (top panel) and the corresponding persistent current $I(\phi)$ (bottom panel) for SU(4) fermions with different local $U$ and nearest-neighbour $V$ interactions, in the integer filling regime of the Fermi--Hubbard model, using exact energy measurements. Exact diagonalisation for $L=N_{p}=4$ is used to monitor the results obtained by \ac{VQE}.}
    \label{fig:su4}
\end{figure}

Now, the results of a commensurate SU(4) Fermi--Hubbard model, with matching parameters ($U$ and $V$) as in Sec.~\secref{sec:SU3}, are shown in Fig.~\figref{fig:su4}. For small values of $U$ and $V$ (left panels), five layers are enough to capture the ground-state energy, yet for higher interaction terms (right panels) the \ac{VQE} does not build up enough correlations to approximate the ground-state energy, whereas a higher number of layers does (see Fig.~\figref{fig:entanglement}).

\subsection{Shots}

Finally, the results of taking a finite number of shots during the optimisation procedure are presented. For the following simulations, after testing the performance of several optimisers available at~\cite{Qiskit_team2019b}, the \ac{NFT} optimiser~\cite{Nakanishi2020} was chosen. The reason being that this optimiser provided the best results, due to its consistency and analytical gradient-free optimisation technique, as recommended by \citet{Tilly2022}.

\begin{figure}[t]
    \centering
    \includegraphics[width=\textwidth]{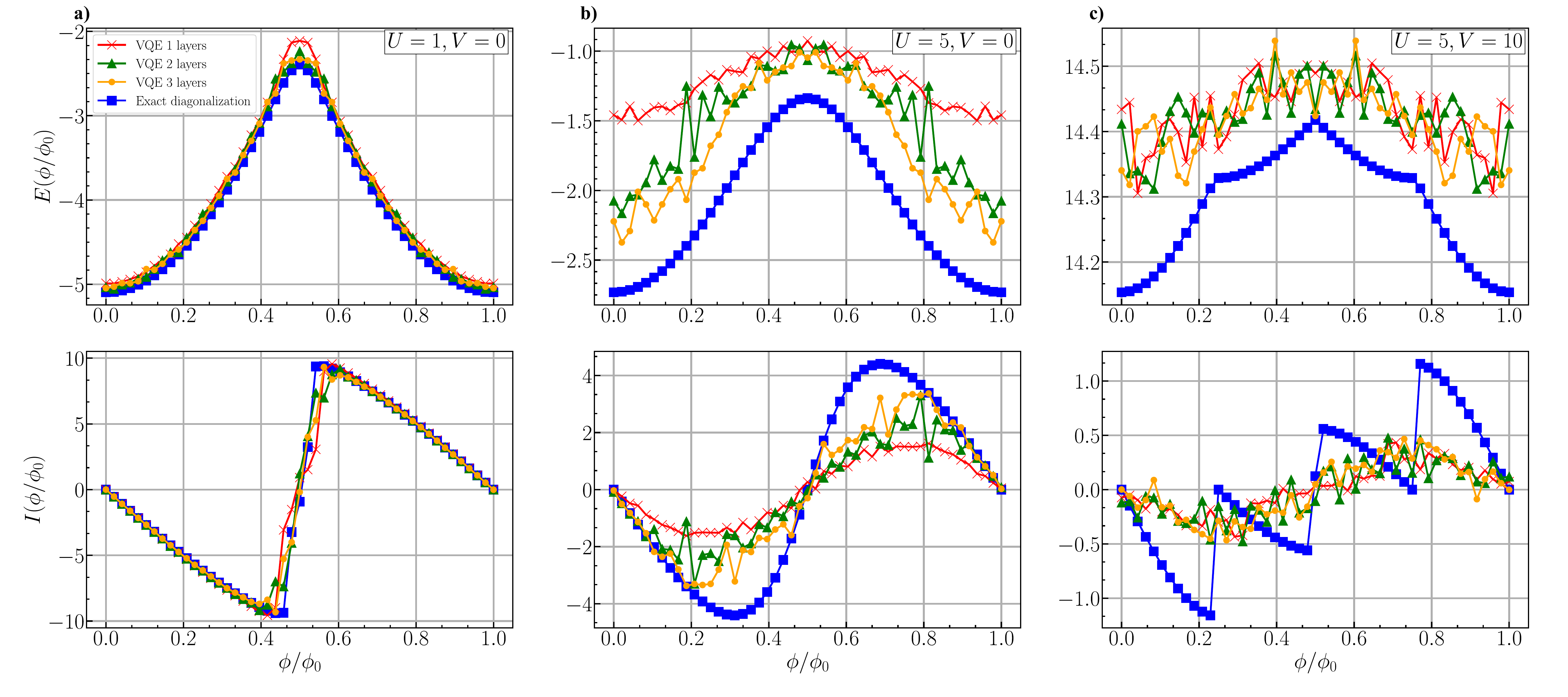}
    \caption[Ground-state energy and corresponding persistent current for a three-site SU(3) Fermi--Hubbard model (using shot-based optimization).]{The ground-state energy $E_{0}(\phi)$ (top panel) and the corresponding persistent current $I(\phi)$ (bottom panel) for SU(3) fermions with different local $U$ and nearest-neighbour $V$ interactions, in the integer filling regime of the Fermi--Hubbard model, using shot-based energy measurements. The profile of the persistent current gives a clear indication between the \textbf{a)} superfluid, \textbf{b)} Mott and \textbf{c)} beat phases. Exact diagonalisation for $L=N_{p}=3$ is used to monitor the results obtained by the \ac{VQE}.}
    \label{fig:shots_su3}
\end{figure}

In Fig.~\figref{fig:shots_su3}, a total of 65,536 function evaluations were carried out for each case. For small values of $U$ and $V$, similar to the results of exact energy measurements, even just one layer of the \ac{PQC} is sufficient to nearly capture the ground-state energies and corresponding persistent current, with three layers acquiring a close-to-perfect accuracy. However, in contrast with the statevector simulations, three layers are not enough to capture the ground-state energies for higher values of $U$ and $V$. Running a \ac{VQE} using shot-based energy measurements is undeniably more complex, with optimisers typically more prone to getting stuck in a \ac{BP} or local minima, as well as taking a significantly longer time to converge to the ground-state.

\begin{figure}[t]
    \centering
    \includegraphics[width=0.9\textwidth]{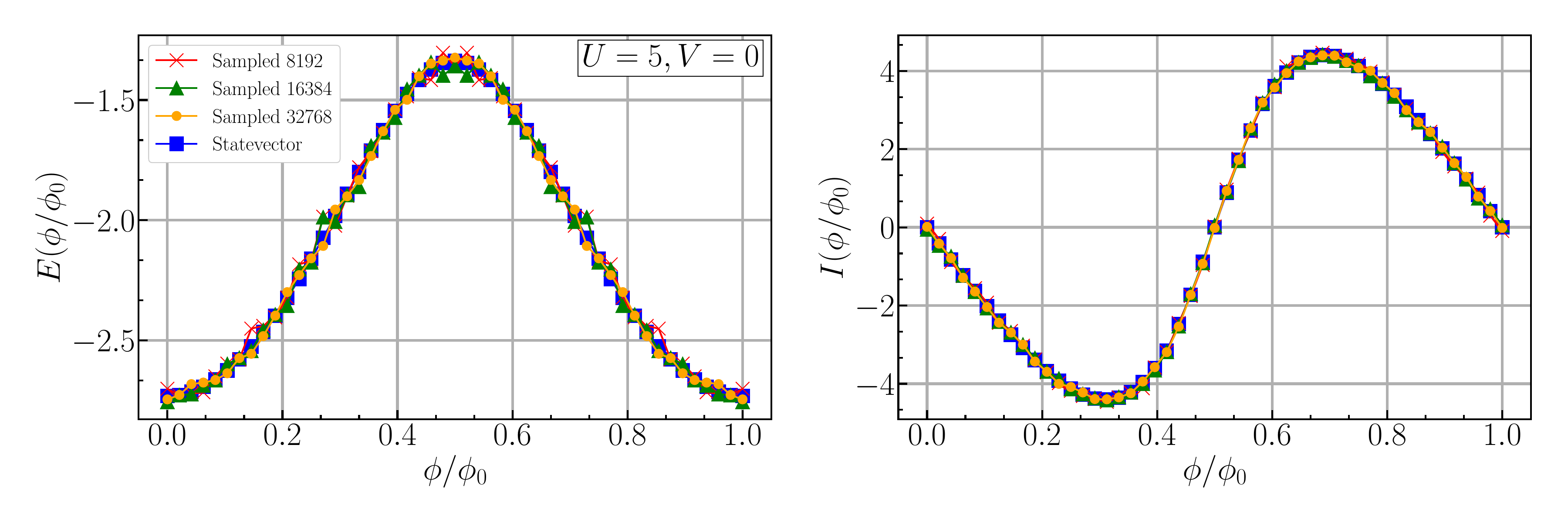}
    \caption[Ground-state energy and corresponding persistent current for a three-site SU(3) Fermi--Hubbard model (starting from statevector parameters).]{The ground-state energy $E_{0}(\phi)$ (left panel) and the corresponding persistent current $I(\phi)$ (right panel) for SU(3) fermions with $U=5$ and no nearest-neighbour interaction, in the integer filling regime of the Fermi--Hubbard model. The statevector results were obtained with exact energy measurements. On the other hand the shot-based profiles were obtained by reevaluating the ground-state energies using the parameters obtained via the statevector results.}
    \label{fig:shots_statevector}
\end{figure}

To ascertain that taking shots was not the problem, the statevector parameters of the ground-states of Fig.~\figref{fig:3comm} \textbf{b)} ($U = 5, V = 0$) are used to reevaluate the same ground-state energies, but instead using a finite number of shots. The results are shown in Fig.~\figref{fig:shots_statevector}. With just 8,192 shots, the ground-state energies and corresponding persistent current were obtained with sufficient accuracy. Using 16,384 and 32,768 shots resulted in even closer convergence to the statevector results.

\begin{figure}[t]
    \centering
    \includegraphics[width=0.9\textwidth]{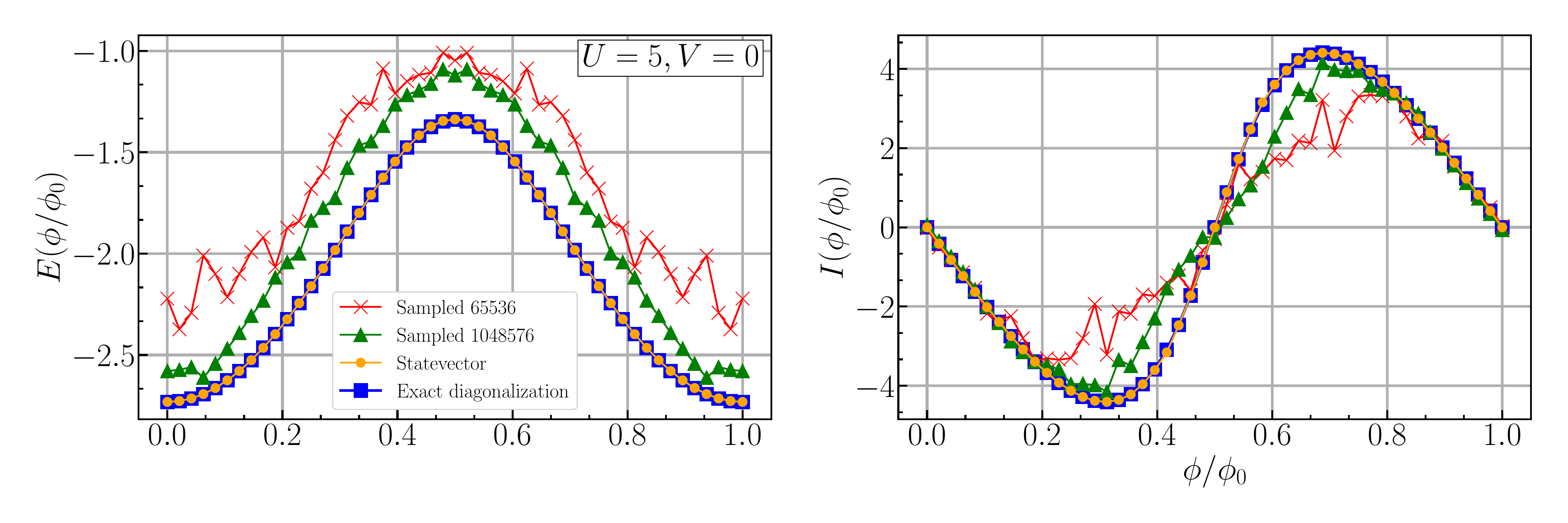}
    \caption[Ground-state energy and corresponding persistent current for a three-site SU(3) Fermi--Hubbard model (using a million function evaluations).]{The ground-state energy $E_{0}(\phi)$ (left panel) and the corresponding persistent current $I(\phi)$ (right panel) for SU(3) fermions with $U=5$ and no nearest-neighbour interaction, in the integer filling regime of the Fermi--Hubbard model. The statevector results were obtained with exact energy measurements. On the other hand the shot-based profiles (each 32,678 shots) were obtained by carrying out the optimisation procedure with a different number of function evaluations in the \ac{NFT} optimiser. Exact diagonalisation for $L=N_{p}=3$ is used to monitor the results obtained by the \ac{VQE}.}
    \label{fig:shots_fevs}
\end{figure}

To justify the argument that the optimisation procedure needs improvement, simulations of the case in Fig.~\figref{fig:3comm} \textbf{b)} ($U = 5, V = 0$) with three layers were conducted, using 1,048,576 function evaluations, and compared with the results from the one using 65,536 function evaluations, as shown in Fig.~\figref{fig:shots_fevs}. A better representation of the ground-state energies are obtained using a higher number of function evaluations, however, although close, the energies still do not match the statevector simulations, with deviations ranging between $3\%$ and $19\%$ in fidelity (left panel of Fig.~\figref{fig:shots_fevs}). Interestingly, the curvature of the energy values as a function of $\phi$, as represented by the persistent current, is well captured even with shot-based simulations over a wide interval of $\phi$ (right panel of Fig.~\figref{fig:shots_fevs}).

\section{Final Remarks}

In this work, multiple \ac{VQE} simulations were carried out on the SU($N$) Fermi--Hubbard model. As an important technical step, the \ac{JW} transformation was extended to encompass $N$-component fermions. This work provides a specific instance of a current-based quantum simulator: the physics of the system is probed by the current response to the gauge field. Specifically, a 1D ring lattice of the Fermi--Hubbard type was considered, while the persistent current was monitored.

The SU($N$) fermion-to-qubit mapping was then applied, using it to determine the ground-state energy of the Fermi--Hubbard Hamiltonian via the \ac{VQE}. The ground-state energy can be approached by utilising a number-preserving \ac{HVA}, similarly to what is done for the SU(2) case~\cite{Cade2020}. However, this \ac{HVA} is generalised to the SU($N$) fermion case, and optimised to minimise the depth and the number of gates and parameters of the \ac{PQC} needed to achieve the ground-state of the magnetic-flux-induced, extended, Fermi--Hubbard Hamiltonian.

In view of the recent interest for the SU($N$) Fermi--Hubbard model, motivated both by its experimental realisation with cold atoms, and its relevance in modelling mesoscopic quantum devices stemming from the rapidly developing atomtronics technology, this work opens this model for investigation via \acp{VQA}. 

The \texttt{Python} code used in obtaining the results is also publicly available at~\cite{Ramos_calderer2021}.

As a follow-up the work discussed in this Chapter, the \ac{VQE} was applied to topological quantum systems, namely the \ac{SSH} and Kitaev models, in \citet{Ciaramelletti2024}. Due to the quasi-degenerate ground-states occurring in the topological phase of the models, the standard \ac{VQE} procedure, while achieving the correct energy within machine-precision, is insufficient for obtaining the true ground-state. Consequently, we devised and implemented various add-on methods to the \ac{VQE} to attempt to address the degeneracy problem. The most optimal results were obtained by utilising a problem-inspired ansatz coupled with adiabatic assistance, starting from the non-topological phase of the model, and slowly moving towards the topological phase.

%% file: chapter4/vsv.tex
\chapter{Variational Separability Verifier} \label{chap:4}

\epigraph{\textit{Spooky action at a distance.}}{--- Einstein, when referring to entanglement}

\textit{Parts of this chapter are based on the published manuscript by \citet{Consiglio2022b}.}\\

\section{Entanglement} \label{sec:entanglement}

Entanglement is the principal defining feature of quantum mechanics~\cite{Einstein1935}, and a quantitative description of the phenomenon started with Bell's inequalities~\cite{Bell1964}. Unlike classical mechanics, entanglement is a type of non-local quantum correlations of a composite system, where each individual subsystem cannot be independently described. Consider two states, $\ket{\psi}_A$ and $\ket{\phi}_B$ lying in respective Hilbert spaces $\cH_A$ and $\cH_B$. If the state of the composite system is $\ket{\psi}_A \otimes \ket{\phi}_B$, then these states are said to be product states, or separable states. Mathematically, an entangled state is defined as a state which cannot be represented as a separable state, meaning that a general entangled state represented as
\begin{equation}
    \ket{\psi}_{AB} = \sum\limits_{i, j} c_{ij} \ket{i}_A \otimes \ket{j}_B,
\end{equation}
written in basis vectors $\{\ket{i}_A\}$ for $\cH_A$ and $\{\ket{j}_B\}$ for $\cH_B$, does not have vectors $\vec{c}_{i_{A}}$ and $\vec{c}_{j_{B}}$ such that $c_{ij} = c_{i_{A}} c_{j_{B}}$, which would give $\ket{\psi}_A = \sum_i c_{i_{A}}\ket{i}_A$ and $\ket{\phi}_B = \sum_j c_{j_{B}}\ket{j}_B$ \cite{Bengtsson2006}. A distinction must be made on what differentiates correlations of a quantum nature from those of a classical nature, and the \ac{LOCC}~\cite{Chitambar2014} paradigm enables us to distinguish between them. \Ac{LOCC} provides the necessary conditions to consider entanglement as a \textit{quantum resource}. It is fundamental to note that this is one of the main tools to show that quantum effects are indeed non-local.\footnote{Note that non-locality and entanglement are two separate concepts, as discussed by \citet{Bennett1999}.}

It is also important to identify states which exhibit no entanglement. Since using \ac{LOCC} with entangled states allows some tasks to be accomplished better than by \ac{LOCC} alone, then we can say that all non-separable states are entangled. It is fundamental to note that the entanglement between states does not increase under \ac{LOCC} operations. This is implied from the fact that a state $\rho$ can be created deterministically by \ac{LOCC} if, and only if, it is constructed from separable states. Moreover, given a state $\rho$, that can be transformed into the state $\sigma$ by \ac{LOCC}, then what can be achieved by $\sigma$ and \ac{LOCC} can just as well be accomplished by using $\rho$ and \ac{LOCC}.\footnote{A remarkable corollary to this statement is that $\rho$ has at least as much entanglement as $\sigma$~\cite{Plenio2007}.}

This begs the question of how one can go about measuring entanglement? Such a measure must satisfy a set of axioms, such as the nullification of the measure for separable states, invariance under local unitary operations, monotonicity under \ac{LOCC}, and non-increasing under \ac{CPTP} maps~\cite{Vedral1997}. To this end, many entanglement measures have been defined, such as distillable entanglement, entanglement cost, and the entanglement of formation~\cite{Plenio2007, Eltschka2014}. The difficulty in calculating these measures for generic states lies in their non-closed form, given that they involve computing of an extremum over a large Hilbert space, which becomes generally intractable as the system's dimension scales. On the other hand, the problem of measuring entanglement is more approachable for pure states, given that even for different entanglement classes, under \ac{SLOCC}~\cite{Eltschka2014}, different measures of entanglement can be computed, albeit with certain limitations. 

The simplest system that has been completely characterised is a two-qubit one. An interesting property for two qubits is that only two \ac{SLOCC} classes exist, one which contains the only type of entanglement possible, i.e. bipartite entanglement, and the other consists of fully separable states, i.e. non-entangled states. The relevant literature for three-, four-qubit pure states, and the multipartite case can be found in Refs.~\cite{Coffman2000, Cunha2019, Guhne2010, Dur2000},~\cite{Regula2014, Verstraete2002, Gour2010, Ghahi2016, Osterloh2016}, and~\cite{Guo2020, Eisert2001, Love2006, Guhne2010, Bengtsson2016} respectively. Only highly specific mixed states are known to have closed forms of multipartite entanglement measures, such as combinations of \ac{GHZ} and W states~\cite{Eltschka2008}, \ac{GHZ}-symmetric~\cite{Eltschka2012}, and \acf{X-MEMS}~\cite{Agarwal2013}, among others. \citet{Consiglio2021} also proved that any arbitrary teleportation protocol~\cite{Bennett1993}, using an arbitrary multipartite entanglement channel and Bell measurement scheme, requires an entanglement monotone termed localisable concurrence~\cite{Popp2005} as a quantum resource for achieving better-than-classical fidelity.

We can also relax the specificity of requiring an exact quantifier of entanglement, and focus on whether a particular state is certified to be entangled. This is known as the quantum separability problem, and is in fact deemed to be \ac{NP}-hard, even for the bipartite case~\cite{Gurvits2004, Ioannou2007, Gharibian2010}. Consequently, verifying that a state is \textit{fully} separable requires that the state is separable with respect to all bipartitions, which implies that the full separability problem is at least as hard as the bipartite separability problem. Thus, we look towards \acp{VQA} as a potential avenue for tackling the separability problem.

\begin{figure}[t]
    \centering
    \includegraphics[width=0.6\textwidth]{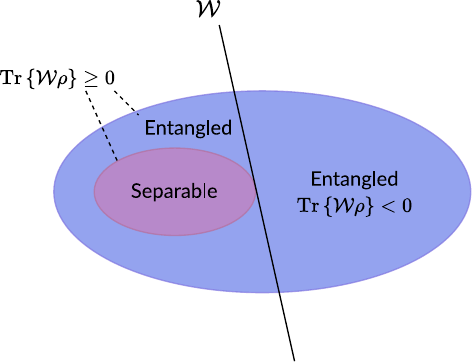}
    \caption[Schematic representation of an optimal entanglement witness.]{Schematic representation of an optimal entanglement witness, $\mathcal{W}$, where a state is guaranteed to be entangled if $\Tr{\cW\rho} < 0$, otherwise no conclusion can be drawn.}
    \label{fig:witness}
\end{figure}

To this end, a novel \ac{VQA} is introduced in this chapter: the \acf{VSV}, capable of determining the \acf{CSS} of an arbitrary quantum state (assuming it can be prepared on a quantum device), with respect to the \acf{HSD}~\cite{Bengtsson2006}. In addition, the \ac{HSD} induces an entanglement measure, denoted as the \ac{HSE}~\cite{Witte1999, Bengtsson2006}. If the \ac{VSV} finds a \ac{CSS} of a state $\rho$ with zero \ac{HSE}, then this signifies that the state $\rho$ is fully separable. Nevertheless, it is still up for debate whether the \ac{HSE} qualifies as an entanglement monotone,\footnote{~\citet{Ozawa2000} proved that the \ac{HSD} is non-contractive under \ac{CPTP} maps, however, it is not known whether \ac{HSE} also has this property.} however, it could in some cases: quantify the amount of entanglement present in specific states; behave as a separability witness; or else provide useful constructions of entanglement witnesses~\cite{Pandya2020}. A formal definition of a witness $\cW$ for a state $\rho$ is the following:
\begin{subequations}
\begin{align}
    &\Tr{\cW\rho} \geq 0~\text{for all separable}~\rho, \\
    &\Tr{\cW\rho} < 0~\text{for at least one entangled}~\rho.
\end{align}
\end{subequations}
This means that a state $\rho$ is entangled if $\Tr{\cW\rho} < 0$~\cite{Guhne2009}. A schematic representation of an entanglement witness is given in Fig.~\figref{fig:witness}.

The prospect of extending the \ac{VSV} to find the \ac{k-CSS} will be briefly mentioned, while for the time being, the current algorithm is designed to find the closest \textit{fully} separable state, i.e. the $n$-\ac{CSS}, where $n$ is the number of qubits in the system, which for the sake of brevity is referred to as the \ac{CSS}. Although only qubit systems are investigated, the \ac{VSV} can be adapted to find the \ac{CSS} for qudit systems, however, this requires an encoding of the $d$-level system onto a two-level system, as well as adjustments to the gates needed to generate ($k$-)separable qudit states.

\section{The Hilbert--Schmidt Distance Induces a Measure of Entanglement} \label{sec:HSD}

The \ac{HSD} between two quantum states $\DHS{\rho}{\sigma}$, is defined as
\begin{equation}
    \DHS{\rho}{\sigma} \equiv \Tr{\left( \rho - \sigma \right)^2},
    \label{eq:HSD}
\end{equation}
which is a non-monotonic Riemannian metric~\cite{Bengtsson2006}. Using this definition, an entanglement measure can be induced by the \ac{HSD}~\cite{Witte1999, Bengtsson2006}, denoted as the \ac{HSE},
\begin{equation}
    \EHS{\rho} \equiv \underset{\sigma \in \cC_k}{\min}~\DHS{\rho}{\sigma},
    \label{eq:EHS_min}
\end{equation}
with the \ac{CSS} defined as
\begin{equation}
    \rho_\text{CSS} \equiv \underset{\sigma \in \cC_k}{\arg\min}~\DHS{\rho}{\sigma}.
    \label{eq:CSS_min}
\end{equation}
where $\cC_k$ is the convex set of states for some degree $k$ of separability. To define it concretely, a rank $d$ state $\rho$ is said to be $k$-separable if it can be written as a convex sum of $k$-separable states, that is
\begin{equation}
    \rho = \sum_{i=0}^{d-1} p_i \ketbra{\Psi_i^k},
\end{equation}
where a $k$-separable pure state can be written as a tensor product of $k$-local states
\begin{equation}
    \ket{\Psi_i^k} = \bigotimes_{j=0}^{k-1} \ket{\psi_j},
\end{equation}
such that $\ket{\psi_j}$ are $k$-separable states on subsets of the $n$-qubit parties~\cite{Shang2018}. More specifically, states are called biseparable for $k = 2$, triseparable for $k = 3$, and up to fully separable for $k = n$. Although the \ac{k-CSS}~\eqref{eq:CSS_min} (and corresponding \ac{HSE}~\eqref{eq:EHS_min}) can be distinctly defined for all convex sets of $k$-separable states, as stated in Sec.~\secref{sec:entanglement}, the \ac{VSV} is designed to find the \ac{CSS} (and corresponding \ac{HSE}) of entangled states with respect to the set of \textit{fully} separable states.

While it is unclear whether the \ac{HSE} is a good entanglement measure, the \ac{HSD}, and the corresponding \ac{HSE}, are still useful quantities to investigate for quantum systems, since they are utilised in generalised Bell inequalities~\cite{Bertlmann2002, Silva2022}, while also providing insight into the geometry of entangled states~\cite{Bengtsson2006, Streltsov2010}. The \ac{HSD} is also relatively straightforward to evaluate on a quantum device, since
\begin{equation}
    \Tr{\left( \rho - \sigma \right)^2} = \Tr{\rho^2} + \Tr{\sigma^2} - 2\Tr{\rho\sigma},
    \label{eq:HSD_decomp}
\end{equation}
coupled with using quantum primitives such as the shallow destructive \textsc{SWAP} test to measure the overlap and purities of $\rho$ and $\sigma$.

On the other hand, it was shown by \citet{Streltsov2010} that distance measures, such as the Bures measure of entanglement~\cite{Bengtsson2006}, are directly related to the convex roof extension of the geometric measure of entanglement. The Bures measure therefore requires the calculation of the fidelity~\cite{Uhlmann1976} between a state and its \ac{CSS}. However, the fidelity is a harder quantity to evaluate on both a classical and quantum device~\cite{Witte1999, Brun2004, Garcia_escartin2013, Bartkiewicz2013, Cincio2018, Cerezo2021b} when compared with the simplicity of the \ac{HSD}. It is also interesting to mention that the \ac{HSD} is being used for tackling the quantum low-rank approximation problem~\cite{Ezzell2022}, with applications in principal component analysis and preparing arbitrary mixed states on quantum computers~\cite{Ezzell2023}.

\section{Framework of the Algorithm}  \label{sec:framework}

In this procedure, $\rho$ is designated as being the test state, and $\sigma$ as the trial state, which is the state that iteratively approaches the \ac{CSS}. As a result, one method is to prepare $s$ separable states on a quantum computer in the following way:
\begin{equation}
    \sigma(\bm{p}, \bm{\theta}, \bm{\phi}) = \sum_{i=0}^{s-1} p_i \ketbra{\psi (\bm{\theta}_i, \bm{\phi}_i)}{\psi(\bm{\theta}_i, \bm{\phi}_i)},
    \label{eq:CSS}
\end{equation}
where
\begin{subequations}
\begin{align}
    \bm{p} &= \left(
    \begin{array}{c}
        p_0 \\
        p_1 \\
        \vdots \\
        p_{s-1} \\
    \end{array}
    \right),~\sum\limits_{i=0}^{s-1} p_i = 1,~0 \leq p_i \leq 1, \\[1ex]
    \bm{\theta} &=
    \left( 
    \begin{array}{c}
        \bm{\theta}_0 \\
        \bm{\theta}_1 \\
        \vdots \\
        \bm{\theta}_{s-1} \\
    \end{array}
    \right) =
    \left( 
    \begin{array}{cccc}
        \theta_{0,0} & \theta_{0,1} & \cdots & \theta_{0,n-1} \\
        \theta_{1,0} & \theta_{1,1} & \cdots & \theta_{1,n-1} \\
        \vdots & \vdots & \ddots & \vdots \\
        \theta_{s-1,0} & \theta_{s-1,1} & \cdots & \theta_{s-1,n-1} \\
    \end{array} 
    \right),~0 \leq \theta_{ij} < \pi, \\[1ex]
    \bm{\phi} &= 
    \left(
    \begin{array}{c}
        \bm{\phi}_0 \\
        \bm{\phi}_1 \\
        \vdots \\
        \bm{\phi}_{s-1} \\
    \end{array}
    \right) =
    \left( 
    \begin{array}{cccc}
        \phi_{0,0} & \phi_{0,1} & \cdots & \phi_{0,n-1} \\
        \phi_{1,0} & \phi_{1,1} & \cdots & \phi_{1,n-1} \\
        \vdots & \vdots & \ddots & \vdots \\
        \phi_{s-1,0} & \phi_{s-1,1} & \cdots & \phi_{s-1,n-1} \\
    \end{array} 
    \right),~0 \leq \phi_{ij} < 2\pi,
\end{align}
\end{subequations}
totalling $s(2n + 1)$ parameters, where $n$ is the number of qubits and $s$ is the number of separable pure states chosen to prepare the trial state. Due to Carath\'eodory's theorem, $d \leq s \leq d^2$ separable pure states are required to describe a rank $d$ mixed state~\cite{Horodecki1997, Vedral1998, Streltsov2010}, where $d = 2^n$ is the dimension of the Hilbert space. However, in the simulations it was deduced that $s = d$ suffices to find the \ac{CSS} of all the test states. $\left\{ p_i,\ket{\psi(\bm{\theta}_i, \bm{\phi}_i)} \right\}$ represent the ensemble of separable pure states, which can be decomposed as
\begin{equation}
    \ket{\psi(\bm{\theta}_i, \bm{\phi}_i)} = \bigotimes_{j=0}^{n-1} \left( \cos\left(\theta_{ij}\right)\ket{0_j} + e^{\imath\phi_{ij}}\sin\left(\theta_{ij}\right)\ket{1_j} \right).
\end{equation}
A separable pure state can thus be generated on a quantum computer by applying a set of one-qubit gates to the all-zero state:
\begin{equation}
    \ket{\psi(\bm{\theta}_i, \bm{\phi}_i)} = \bigotimes_{j=0}^{n-1} U(\theta_{ij}, \phi_{ij}) \ket{0}_j,
    \label{eq:unitary}
\end{equation}
where
\begin{equation}
    U(\theta_{ij}, \phi_{ij}) = \left( \begin{array}{cc}
        \cos\left(\theta_{ij}\right) & -\sin\left(\theta_{ij}\right) \\[1ex]
        e^{\imath\phi_{ij}}\sin\left(\theta_{ij}\right) & e^{\imath\phi_{ij}}\cos\left(\theta_{ij}\right)
    \end{array} \right),
\end{equation}
which is equivalent to applying an $R_y(\theta_{ij})$ gate followed by an $R_z(\phi_{ij})$ gate.

\subsection{Measuring the Hilbert--Schmidt Distance} \label{sec:HSD_measure}

Preparing a mixed state on a quantum computer requires an additional $\ceil{\log_2 r}$ ancillary qubits, where $r$ is the rank of the mixed state~\cite{Benenti2009}. This, coupled with the issue of preparing an arbitrary separable state, leads us to find an alternative method for computing the overlap of the trial states with the test state. The solution is to individually supply separable (non-orthogonal) pure states of the separable mixed state to the quantum computer, evaluate the individual overlaps using the destructive \textsc{SWAP} test, and then classically mix them during the optimisation procedure. The benefit of this method is that the direct preparation of the \ac{CSS} on the quantum computer is not required, as only the computation of the \ac{HSD} is needed.

Suppose the trial state $\sigma$ is decomposed as in Eq.~\eqref{eq:CSS}, then the overlap between $\rho$ and $\sigma$ can be evaluated as
\begin{align}
    \Tr{\rho\sigma} &= \Tr{\rho \sum_{i=0}^{s-1} p_i \ketbra{\psi_i}{\psi_i}} = \sum_{i=0}^{s-1} p_i \Tr{\rho \ketbra{\psi_i}{\psi_i}} = \sum_{i=0}^{s-1} p_i \ev{\rho}{\psi_i}.
    \label{eq:overlap}
\end{align}
with the purity of $\sigma$ being computed as
\begin{align}
    \Tr{\sigma^2} &= \Tr{\sum_{i=0}^{s-1} p_i \ketbra{\psi_i}{\psi_i} \sum_{j=1}^{s-1} p_j
    \ketbra{\psi_j}{\psi_j}} \nonumber \\
    &= \sum_{i,j=0}^{s-1} p_i p_j \Tr{\ket{\psi_i}\braket{\psi_i}{\psi_j}\bra{\psi_j}} \nonumber \\
    &= \sum_{i,j=0}^{s-1} p_i p_j |\braket{\psi_i}{\psi_j}|^2 \nonumber \\
    &= \sum_{i=0}^{s-1} p_i^2 + \sum_{i \neq j}^{s-1} p_i p_j |\braket{\psi_i}{\psi_j}|^2 \nonumber \\
    &= \sum_{i=0}^{s-1} p_i^2 + 2\sum_{i<j}^{s-1} p_i p_j |\braket{\psi_i}{\psi_j}|^2.
    \label{eq:purity}
\end{align}
The purity of $\rho$ is trivially obtained assuming one can prepare it directly on a quantum computer; given that it is the test state.

\subsection{Measuring Overlaps on a Quantum Computer} \label{sec:SWAP}

Given Eqs.~\eqref{eq:overlap} and~\eqref{eq:purity}, one requires subroutines in the \ac{VSV} capable of measuring the purity and overlap of quantum states. Specifically, one can utilise the destructive \textsc{SWAP} test~\cite{Garcia_escartin2013, Cincio2018} to calculate these quantities. The concept of this procedure stems from the fact that measuring in the Bell basis determines the amount of correlations present between two systems, and it can be shown that it is equivalent to the non-destructive \textsc{SWAP} test~\cite{Buhrman2001}.\footnote{See Appendix~\secref{app:swap} for more information on \textsc{SWAP} tests.} The one-qubit gates necessary to generate the separable pure states of Eq.~\eqref{eq:unitary}, coupled with the two-depth circuit needed for the destructive \textsc{SWAP} test, results in a noticeably shallow circuit for the \ac{VSV}.

\subsection{Variational Optimisation}

Eqs.~\eqref{eq:overlap} and~\eqref{eq:purity} give us the means to compute the \ac{HSD} as in Eq.~\eqref{eq:HSD_decomp}, and by combining these with Eqs.~\eqref{eq:EHS_min} and~\eqref{eq:CSS_min}, one can devise a \ac{VQA} that is able to compute the \ac{CSS} and the corresponding \ac{HSE}.

The optimiser in this scenario is tasked with providing angles $\bm{\theta}$ and $\bm{\phi}$ to prepare the separable pure states, and probabilities $\bm{p}$ to classically mix them during post-processing, to generate a separable state $\sigma(\bm{p}, \bm{\theta}, \bm{\phi})$. One can notice that the quantum computer is only tasked with computing $\Tr{\rho^2}$ once at the beginning, and $\ev{\rho}{\psi(\bm{\theta}_i, \bm{\phi}_i)}$ and $|\braket{\psi(\bm{\theta}_i, \bm{\phi}_i)}{\psi(\bm{\theta}_j, \bm{\phi}_j)}|^2$ $\forall~i,j$, at each iteration. The probabilities $\bm{p}$ are only incorporated classically when computing the final results for calculating the purity $\Tr{\sigma(\bm{p}, \bm{\theta}, \bm{\phi})^2}$, and the overlap $\Tr{\rho \sigma(\bm{p}, \bm{\theta}, \bm{\phi})}$. As a result, the optimisation routine is split into a bilevel system, which is essentially a nested optimisation routine~\cite{Dempe2002}.

\begin{figure}[H]
    \centering
    \includegraphics[width=0.9\textwidth]{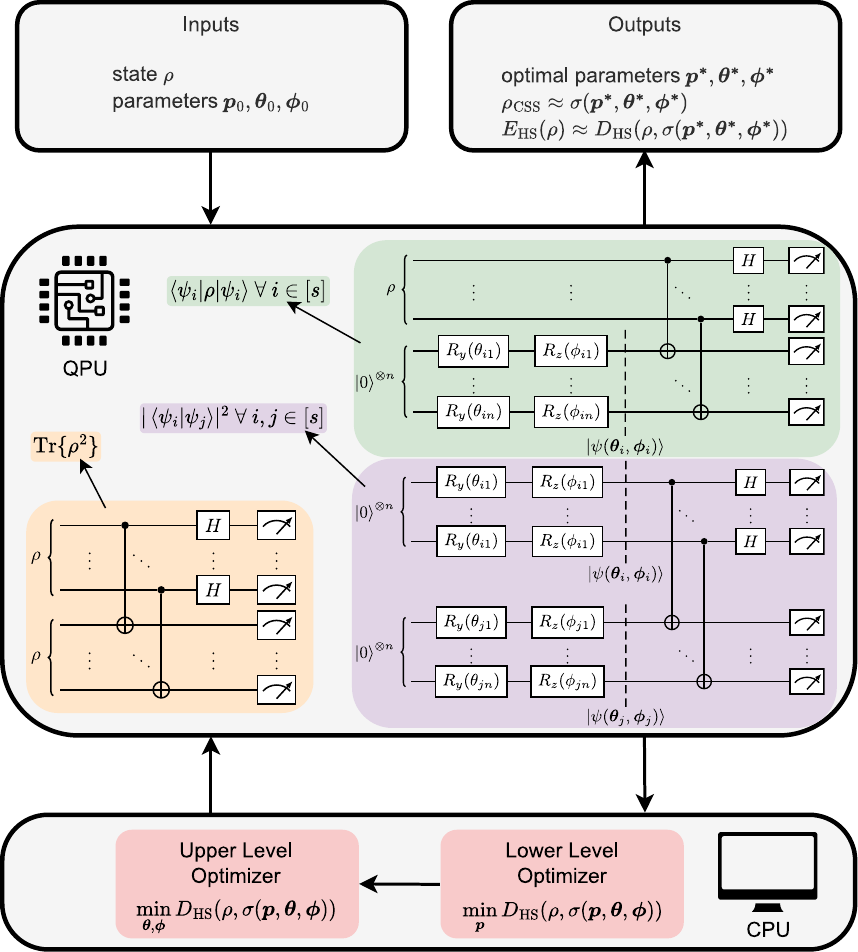}
    \caption[Diagrammatic representation of the \acs{VSV}.]{Diagrammatic representation of the \ac{VSV}. The inputs to the \ac{VSV} are the test state $\rho$, and initial parameters $\bm{\rho}_0, \bm{\theta}_0, \bm{\phi}_0$. The \ac{VSV} then computes the orange (bottom-left \ac{QPU}) circuit once, to evaluate $\Tr{\rho^2}$, with the green (top-right \ac{QPU}) and purple (bottom-right \ac{QPU}) circuits computed to evaluate the overlaps $\ev{\rho}{\psi_i}$ and $\braket{\psi_i}{\psi_j}~\forall~i,j \in [s] = \{0, 1, 2, \dots, s - 1\}$, which are used in Eqs.~\eqref{eq:overlap} and~\eqref{eq:purity}, respectively. The red (bottom \acs{CPU}) section is the classical feedback loop computed by the classical computer, consisting of first minimising the parameters $\bm{p}$, followed by proposing new parameters $\bm{\theta}$ and $\bm{\phi}$ to evaluate on the \ac{QPU}. After a sufficient number of iterations, the \ac{VSV} outputs optimal parameters $\bm{\rho^*}, \bm{\theta^*}, \bm{\phi^*}$, along with the computed \ac{CSS} and corresponding \ac{HSE}.}
    \label{fig:diagram}
\end{figure}

The structure of the bilevel system is as follows: at the beginning of every iteration, an upper-level optimiser selects the parameters $\bm{\theta}$ and $\bm{\phi}$, and proceeds to call the quantum computer to compute the overlaps in Eqs.~\eqref{eq:overlap} and \eqref{eq:purity}. The upper-level optimiser then launches a lower-level optimiser to obtain the parameters $\bm{p}$ to minimise Eq.~\eqref{eq:HSD_decomp}. The upper-level optimiser repeats this until a suitable convergence criterion is achieved. Ultimately, (near-)optimal parameters of the minimisation of the cost function~\eqref{eq:HSD} are obtained, such that
\begin{equation}
    \{\bm{p^*}, \bm{\theta^*}, \bm{\phi^*}\} = \underset{\bm{p}, \bm{\theta}, \bm{\phi}}{\arg\min}~\DHS{\rho}{\sigma(\bm{p}, \bm{\theta}, \bm{\phi})},
\end{equation}
such that the \ac{CSS} of $\rho$ would then be
\begin{equation}
    \rho_\text{CSS} = \sigma(\bm{p^*}, \bm{\theta^*}, \bm{\phi^*}),
\end{equation}
from which the \ac{HSE} of $\rho$ can be calculated as
\begin{equation}
    \EHS{\rho} = \DHS{\rho}{\sigma(\bm{p^*}, \bm{\theta^*}, \bm{\phi^*})}.
\end{equation}
A diagrammatic representation of the \ac{VSV} algorithm is shown in Fig.~\figref{fig:diagram}.

A limitation of the \ac{HSD}~\eqref{eq:HSD} is that it is in fact a global cost function, even when employing the destructive \textsc{SWAP} test for its computation, and consequently, will exhibit \acp{BP}. However, \citet{Ezzell2023} demonstrated that the \ac{HSD} can be faithfully reformulated to rely solely on local measurements, effectively mitigating the issue of \acp{BP}.

\subsection{Complexity Analysis} \label{sec:complexity}

The scalability of the \ac{VSV}, similar to other hybrid quantum--classical algorithms, is characterised from both the classical and quantum point of view. The algorithm prepares the test state $\rho$ (which may be unknown) and trial state $\sigma$ on a quantum computer. Additionally, the algorithm requires $n$ \textsc{CNOT} gates and $n$ Hadamard gates for carrying out the destructive \textsc{SWAP} test, which scale linearly with the number of qubits, and uses a constant circuit depth of two, assuming the quantum computer has a ladder-like connectivity.

In the current implementation of the \ac{VSV}, the number of parameters scale as $s(2n + 1)$ for determining the closest fully separable state of $\rho$, where $s$ is the number of (non-orthogonal) components representing the trial state $\sigma$. Unless $\rho$ is separable, the \ac{CSS} of $\rho$ was generally found to be of full rank. Thus, it was required that at least $s \geq d$, due to Carath\'eodory's theorem, although in all of the simulations $s = d$ sufficed to find the \ac{CSS} of an entangled state. This implied that an exponential scaling in the number of parameters was needed to determine the \ac{CSS} via this method. On the other hand, the number of parameters needed to define the \ac{CSS} may be reduced such that it scales sub-exponentially with the number of qubits, given a more elaborate scheme of supplying pure states. It should also be noted that finding the \ac{k-CSS} would require more complex ans\"atze, which are more likely to require a larger set of parameters to represent $k$-separable states.

\section{Results} \label{sec:VSV_results}

\begin{figure}[t]
    \centering
    \includegraphics[width=0.8\textwidth]{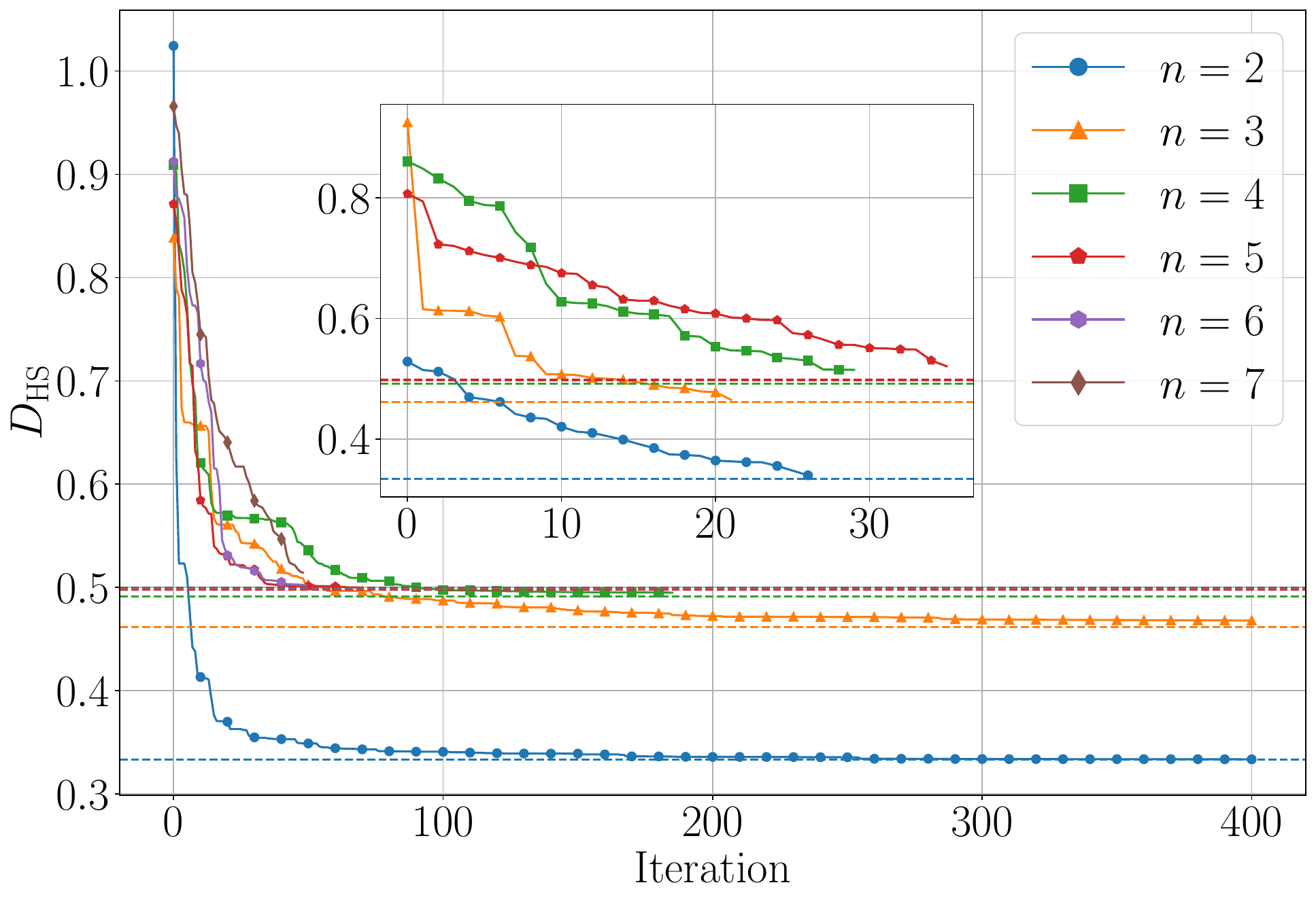}
    \caption[Plot of the convergence of \acs{HSD} with respect to the number of iterations for the statevector optimisation for $n$-qubit \acs{GHZ} states.]{Plot of the convergence of \ac{HSD} with respect to the number of iterations for the statevector optimisation for $n$-qubit \ac{GHZ} states. The solid lines represent the convergence for two to seven qubits (bottom-up). The dashed lines represent the analytical \ac{HSE} of the $n$-qubit \ac{GHZ} states (as given by Eq.~\eqref{eq:HSE_GHZ}). The inset figure represents the shot-based optimisation convergence with 8192 shots for two to five qubits (bottom-up). Note that there are less iterations for the shot-based cases (although there is the same number of function evaluations), since it is harder for the \ac{GSA} optimiser to find a strictly lower \ac{HSE} at each function evaluation.}
    \label{fig:GHZ_plot}
\end{figure}

The results involving the \ac{VSV} are presented in this section for both statevector and shot-based simulations, with all initial points being generated randomly. We show the results of the \ac{HSD} convergence using statevector optimisation for generalised \ac{GHZ} states, which are defined as
\begin{equation}
    \GHZ = \frac{1}{\sqrt{d}} \sum_{i = 0}^{d - 1} \ket{i}^{\otimes n},
\end{equation}
where $d$ is the dimension of each of the $n$ subsystems. The results are seen in Fig.~\figref{fig:GHZ_plot}, for \ac{GHZ} states ranging from two to seven qubits (that is $n = 2, \dots, 7$ and $d = 2$) --- with an inset similarly showing two to five qubits, but instead using shot-based optimisation with 8192 shots. The optimiser used in this instance is the \ac{GSA} algorithm~\cite{Tsallis1996, Xiang1997} from the \texttt{SciPy} library~\cite{dual_annealing}, for upper parameter ($\bm{\theta}$, $\bm{\phi}$) optimisation, with the \ac{SLSQP} algorithm~\cite{Kraft1988} for the lower parameter ($\bm{p}$) optimiser. The \ac{GHZ} states are specifically chosen so as to demonstrate the performance of the \ac{VSV}, and also since the \ac{CSS} is analytically known, as determined by \citet{Pandya2020}, with the analytical \ac{HSE} given by
\begin{equation}
    \label{eq:HSE_GHZ}
    \EHS{\GHZ_n} = \frac{2^n-2}{2^{n + 1} + 2^{3-n} - 4}.
\end{equation}

\ac{GHZ} states are the basis of numerous quantum communication protocols, such as quantum secret sharing~\cite{Hillery1999}, and quantum communication complexity reduction~\cite{Ho2022}, besides being the experimental workhorse for more refined tests of quantum non-locality with respect to two-qubit Bell states~\cite{Pan2000}. \ac{GHZ}-diagonal states may inherit their non-classical features from the former~\cite{Chen2012}, since they are typically a consequence of local damping channels acting on individual qubits. Thus, investigating the \ac{CSS} may provide us with a better understanding of the optimal strategies in the presence of noise, as well as insight into the amount of error that is admissible in these protocols, in order to retain a quantum advantage.

The following application of the \ac{VSV} is on two- and three-qubit \ac{X-MEMS}, which are $X$-states\footnote{$X$-states are density matrices where the only non-zero elements are the diagonal and the anti-diagonal elements when expressed in the computational basis.} that have the maximal amount of multipartite entanglement for a given linear entropy~\cite{Agarwal2013}. The general form of \ac{X-MEMS} is given by
\begin{equation}
    \tilde{X} = \left( 
    \begin{array}{cccccccc}
        f(\gamma) & & & & & & & \gamma \\
        & g(\gamma) & & & & & 0 & \\
        & & \ddots & & & \iddots & & \\
        & & & g(\gamma) & 0 & & & \\
        & & & 0 & 0 & & & \\
        & & \iddots & & & \ddots & & \\
        & 0 & & & & & 0 & \\
        \gamma^* & & & & & & & g(\gamma) \\
    \end{array} 
    \right),
\end{equation}
where
\begin{subequations}
\begin{align}
    f(\gamma) &= 
    \begin{cases}
        \frac{1}{N+1} & 0 \leq |\gamma| \leq \frac{1}{N+1}, \\
        |\gamma| & \frac{1}{N+1} \leq |\gamma| \leq \frac{1}{2},
    \end{cases} \\
    g(\gamma) &= 
    \begin{cases}
        \frac{1}{N+1} & 0 \leq |\gamma| \leq \frac{1}{N+1}, \\
        \frac{1 - 2|\gamma|}{N-1} & \frac{1}{N+1} \leq |\gamma| \leq \frac{1}{2},
    \end{cases}
\end{align}
\end{subequations}
and $N = 2^{n - 1}$. The \ac{GME} concurrence~\cite{Eltschka2014} was shown to be equal to $2|\gamma|$ for \ac{X-MEMS}~\cite{Agarwal2013}. The \ac{GME} concurrence is an extension of the bipartite concurrence, which is a measure related to the entanglement of formation~\cite{Hill1997, Wootters1998}, and in fact the \ac{GME} concurrence reduces to the bipartite concurrence for two-qubit states. 

\begin{figure}[t]
    \centering
    \includegraphics[width=0.8\textwidth]{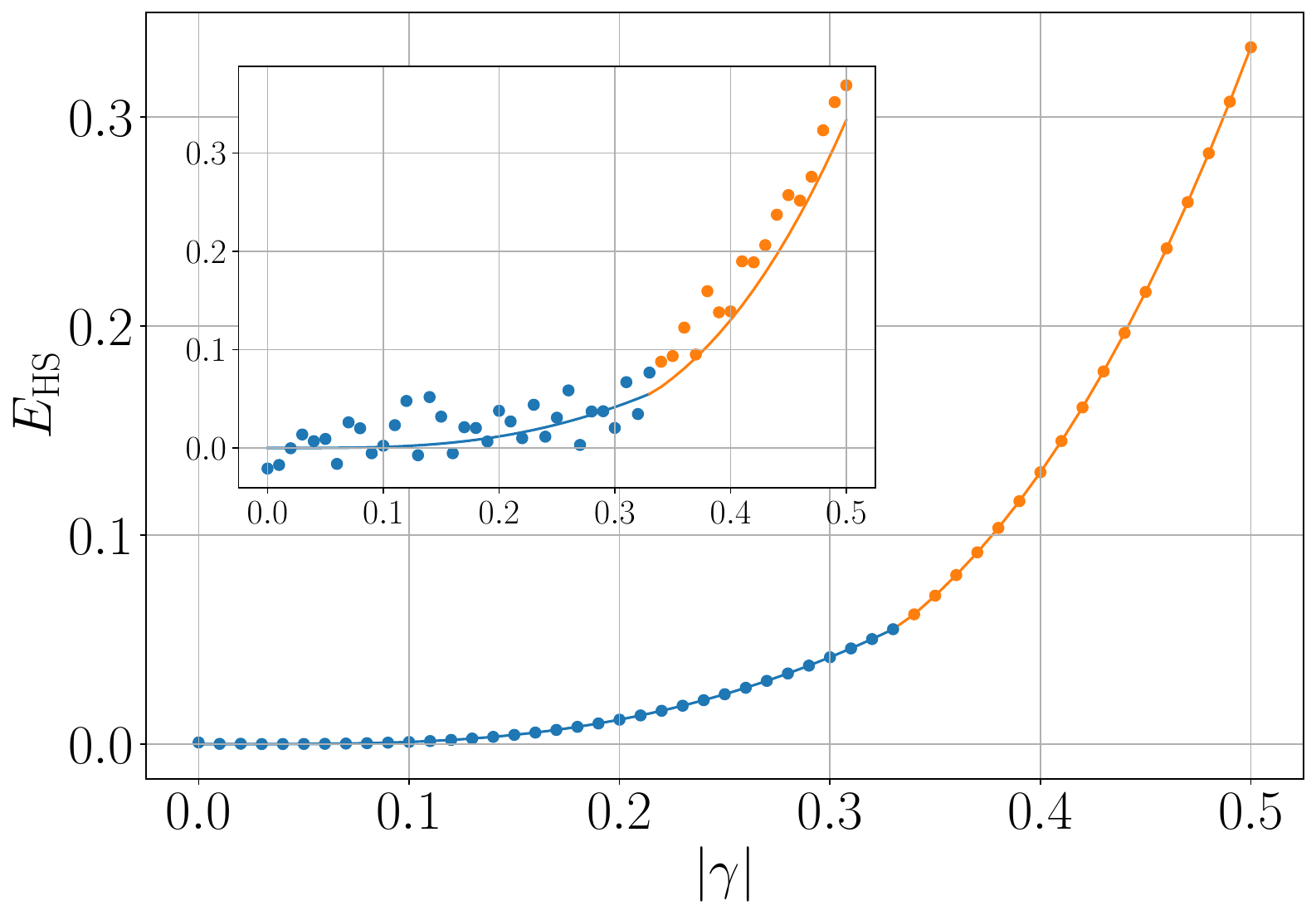}
    \caption[Plot of the results of the optimisation applied to two-qubit \acs{X-MEMS}.]{Plot of the results of the statevector optimisation applied to two-qubit \ac{X-MEMS}. The solid lines represent the analytical \ac{HSE} as a function of $|\gamma|$. The left blue segment represents the range between $0 \leq |\gamma| \leq 1/3$, while the right orange segment represents the range between $1/3 \leq |\gamma| \leq 1/2$. The inset figure represents the data acquired with shot-based optimisation using 8192 shots.}
    \label{fig:X-MEMS_2}
\end{figure}

\begin{figure}[t]
    \centering
    \includegraphics[width=0.8\textwidth]{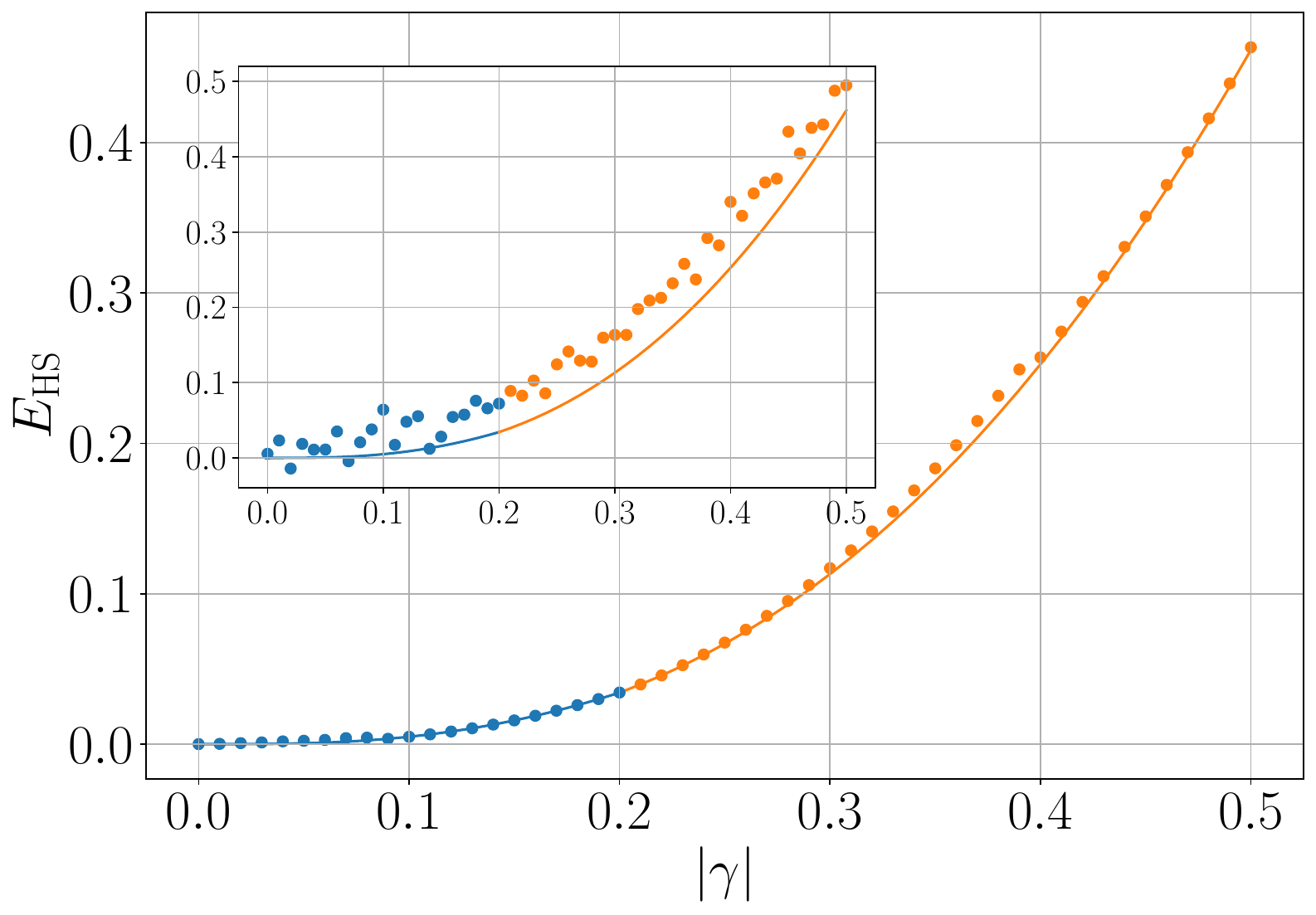}
    \caption[Plot of the results of the optimisation applied to three-qubit \acs{X-MEMS}.]{Plot of the results of the statevector optimisation applied to three-qubit \ac{X-MEMS}. The solid lines represent the analytical \ac{HSE} as a function of $|\gamma|$. The left blue segment represents the range between $0 \leq |\gamma| \leq 1/5$, while the right orange segment represents the range between $1/5 \leq |\gamma| \leq 1/2$. The inset figure represents the data acquired with shot-based optimisation using 8192 shots.}
    \label{fig:X-MEMS_3}
\end{figure}

Obtaining a relation between the \ac{HSE} and the \ac{GME} concurrence, may potential enable the \ac{HSE} to behave as an entanglement monotone for \ac{X-MEMS}, albeit it does not generally satisfy the conditions for being one~\cite{Bengtsson2006}. Figs.~\figref{fig:X-MEMS_2} and~\figref{fig:X-MEMS_3} show the analytical \ac{HSE} along with the values determined by the \ac{VSV}, for two- and three-qubit \ac{X-MEMS}, respectively. In the case of the inset figures, the \ac{NFT} optimiser~\cite{Nakanishi2020} was used as an upper-level optimiser, since it offered significant improvement over \ac{GSA} when evaluating the overlaps using shots rather than statevector calculations.

By inspection of the \ac{CSS} determined by the \ac{VSV}, we were able to surmise the form of the \ac{CSS} for $n$-qubit \ac{X-MEMS}, which is
\begin{equation}
    \left( 
    \begin{array}{cccccccc}
        \frac{a}{2} & & & & & & & d \\
        & \frac{b}{N - 1} & & & & & 0 & \\
        & & \ddots & & & \iddots & & \\
        & & & \frac{b}{N - 1} & 0 & & & \\
        & & & 0 & \frac{1 - a - b}{N - 1} & & & \\
        & & \iddots & & & \ddots & & \\
        & 0 & & & & & \frac{1 - a - b}{N - 1} & \\
        d^* & & & & & & & \frac{a}{2} \\
    \end{array} 
    \right),
\end{equation}
where $N = 2^{n-1}$, $0 \leq a, b \leq 1$, $a + b \leq 1$, and $0 \leq |d| \leq a/2$, to ensure a valid density matrix. The parameters $a$, $b$ and $d$ can be directly determined by using the \texttt{Mathematica} code developed by \citet{Consiglio2022c}. We determined that for $n \geq 3$, $a$ is a root of a quartic equation including $\gamma$, with $b$ and $d$ depending on both $a$ and $\gamma$.\footnote{More details on the \ac{CSS} of \ac{X-MEMS} can be found in Appendix~\secref{app:css}.}

In the case of two-qubit states, the \ac{HSE} and the concurrence are known to be equal for a restricted set of states~\cite{Mundarain2007, Ganardi2022}. With the \ac{VSV}, there is numerical evidence supporting the conjecture that the \ac{HSE} of two distinct, arbitrary, two-qubit states, $\rho$ and $\sigma$, are equal if, and only if, the concurrence of $\rho$ and $\sigma$ are also equal, that is
\begin{equation}
    C(\rho) = C(\sigma) \iff \EHS{\rho} = \EHS{\sigma},
\end{equation}
and similarly, that the \ac{HSE} of $\rho$ is greater than that of $\sigma$ if, and only if, the concurrence of $\rho$ is also greater than that of $\sigma$, that is
\begin{equation}
    C(\rho) > C(\sigma) \iff \EHS{\rho} > \EHS{\sigma}.
\end{equation}
Following this, we tested whether similar statements hold when comparing $\ac{GME}$ concurrence and \ac{HSE} for $(n \geq 3)$-qubit $X$-states. All of the simulations we carried out pointed towards the null hypothesis in both cases, meaning that equality does not hold:
\begin{equation}
    C_\text{GME}(\rho) = C_\text{GME}(\sigma) \centernot\iff \EHS{\rho} = \EHS{\sigma},
\end{equation}
nor does ordered inequality:
\begin{equation}
    C_\text{GME}(\rho) > C_\text{GME}(\sigma) \centernot\iff \EHS{\rho} > \EHS{\sigma}.
\end{equation}
This is not surprising, since the \ac{GME} concurrence characterises $n$-qubit entanglement. On the other hand, the \ac{HSE} seems to describe the departure of a state from a fully separable one, where in the case of three qubits, may also consist of biseparable states which are not detected by \ac{GME} concurrence.

\section{Final Remarks}

The \ac{VSV} is a novel \ac{VQA} that finds the \ac{CSS} of arbitrary quantum states, with respect to the \ac{HSD}, and obtains the \ac{HSE}. The \ac{VSV} has been applied to $n$-qubit \ac{GHZ} states, showing the convergence of the algorithm to the \ac{CSS}. It has also been applied to investigate \ac{X-MEMS}, producing a relation between the \ac{GME} concurrence and \ac{HSE}, as well as helping deduce the analytical form of the \ac{CSS} for $n$-qubit \ac{X-MEMS}. In this respect, the \ac{VSV} can be useful for shedding light on possible analytical forms of \acp{CSS} for arbitrary states. Further simulations on two-qubit states demonstrated that the equality (or ordering) of the concurrence between two states is present if, and only if, equality (or ordering) of the \ac{HSE} is also present. However, this relationship does not generally hold for the \ac{GME} concurrence and \ac{HSE} in the case of $X$-states with three or more qubits.

If the calculated \ac{HSE} of an entangled state is equal to zero, then we trivially acquire $\rho_\text{CSS} = \rho$, which immediately implies that the test state $\rho$ is fully separable, and hence, is not entangled. In either case, if one knows the \ac{CSS}, then an entanglement witness~\cite{Horodecki1996, Doherty2004, Guhne2009} can be defined as $\cW \equiv \rho - \rho_\text{CSS}$. For the most part however, one does not exactly acquire the \ac{CSS}, but rather a close approximation to it, say, $\sigma$. In this case, the optimal entanglement witness would be of the form $\cW \equiv \rho - \sigma - \max_{\ket{\psi} \in \cC_k} \ev{(\rho - \sigma)}{\psi}$~\cite{Pandya2020}.

The strength of the \ac{VSV} as a potential \ac{NISQ} algorithm stems from its simplicity. While the algorithm requires in general many calls to the quantum computer for evaluating overlaps, the destructive \textsc{SWAP} test, coupled with the ease of preparing separable states through one-qubit gates, results in a three-depth circuit which is remarkably tractable on a \ac{NISQ} device. It should be noted that the entire trial state(s) can be directly prepared on the quantum computer by utilising (up to $n$) ancillary qubits, which are then traced out after performing the relevant unitary gates~\cite{Benenti2009}, leading to less overlap computations at the cost of more qubits, as discussed in Sec.~\secref{sec:framework}. It is also noteworthy to mention that the evaluation of state overlaps could be improved by looking towards alternative methods, such as in Refs.~\cite{Flammia2011, Elben2019, Elben2020, Fanizza2020, Liu2022}.

Apart from the \ac{VSV}, there are alternative approaches to finding the \ac{CSS} of arbitrary states using other methods, such as neural networks~\cite{Girardin2022}, and an adaptive polytope approximation~\cite{Ohst2022}, among others. However, the work that inspired the variational approach to the problem at hand is the so-called \acf{QGA}~\cite{Brierley2016, Shang2018, Wiesniak2020, Pandya2020}. The implementation of the \ac{QGA} on a quantum device and related discussion can be found in Appendix~\secref{app:gilbert}.

The \texttt{Python} code for running the \ac{VSV} simulations, using the package \texttt{Qulacs}~\cite{Suzuki2021}, and the \texttt{Mathematica} code for determining the \ac{CSS} of \ac{X-MEMS}, can be found at~\cite{Consiglio2022b}.

%% file: chapter5/gibbs.tex
\chapter{Gibbs State Preparation} \label{chap:5}

\epigraph{\textit{If physical theories were people, thermodynamics would be the village witch.}}{--- \citet{Goold2016}}

\textit{Parts of this chapter are based on the published manuscripts by \citet{Consiglio2024a}, \citet{Consiglio2025} and \citet{Consiglio2024b}.}\\

When compared to the theory of quantum physics, thermodynamics is a much older science, dealing with the physics of heat, work and temperature. It is a phenomenological theory, invoked to understand and improve on the capacities of steam engines in the \nth{19} century~\cite{clausius1960}. Since then, thermodynamics has been linked with even more abstract physical notions, such as that of energy or entropy~\cite{Maxwell1872}. However, it was also proven that entanglement is (thermodynamically) costly for two-qubit pure states~\cite{Huber2015}, meaning that to generate entanglement, work must be carried out, ultimately associating the two formally related fields of entanglement and thermodynamics~\cite{Popescu1997, Brandao2008}. Since then, the study of the thermodynamics in quantum computation has been intensely investigated~\cite{Landauer1961, Bennett1982, Goold2016}, with the current objective of the field being the development of a framework governing the thermodynamics of quantum information.

An integral task in quantum state preparation is the generation of finite-temperature thermal states of a given Hamiltonian on a quantum computer. Indeed, Gibbs states\footnote{Also known as thermal states.} can be used for quantum simulation~\cite{Childs2018}, quantum machine learning~\cite{Kieferova2017, Biamonte2017}, quantum optimisation~\cite{Somma2008}, and the study of open quantum systems~\cite{Poulin2009}. In particular, combinatorial optimisation problems~\cite{Somma2008}, semi-definite programming~\cite{Brandao2016}, and training of quantum Boltzmann machines~\cite{Kieferova2017}, can be tackled by sampling from well-prepared Gibbs states.

The preparation of an arbitrary initial state is a challenging task in general, with finding the ground-state of a Hamiltonian being a \ac{QMA}-complete problem~\cite{Watrous2008}. Preparing Gibbs states, specifically at low temperatures, could be as hard as finding the ground-state of that Hamiltonian~\cite{Aharonov2013}. The first algorithms for preparing Gibbs states were based on the idea of coupling the system to a register of ancillary qubits, and letting the system and environment evolve under a joint Hamiltonian, simulating the physical process of thermalisation, such as in Refs.~\cite{Terhal2000, Poulin2009, Riera2012}, while others relied on dimension reduction~\cite{Bilgin2010}.

The algorithm for preparing Gibbs states proposed in this work can be placed in the category of \acp{VQA}, such as in Refs.~\cite{Wu2019, Chowdhury2020, Wang2021a, Warren2022, Foldager2022}, and similarly for preparing \ac{TFD} states~\cite{Zhu2020, Premaratne2020, Sagastizabal2021}. Variational ans\"atze based on multi-scale entanglement renormalisation~\cite{Sewell2022} and product spectrum ansatz~\cite{Martyn2019} have also been proposed in order to prepare Gibbs states.

Alternative algorithms prepare thermal states through quantum imaginary time evolution, such as in Refs.~\cite{Verstraete2004, Chowdhury2016, Zoufal2021, Gacon2021, Wang2023}, starting from a maximally mixed state, while others start from a maximally entangled state~\cite{Yuan2019}. \citet{Haug2022} proposed quantum-assisted simulation to prepare thermal states, which does not require a hybrid quantum--classical feedback loop. In addition, methods exist to sample Gibbs state expectation values, rather than prepare the Gibbs state directly, such as in quantum metropolis methods~\cite{Temme2011, Yung2012}, imaginary time evolution applied to pure states~\cite{Motta2020}, and random quantum circuits using intermediate measurements~\cite{Shtanko2023}.

Recent methods also propose using rounding promises~\cite{Rall2023}, fluctuation theorems~\cite{Holmes2022b}, pure thermal shadow tomography~\cite{Coopmans2023}, and minimally entangled typical thermal states for finite temperature simulations~\cite{Getelina2023}.

\begin{figure}[t]
    \centering
    \includegraphics[width=0.8\linewidth]{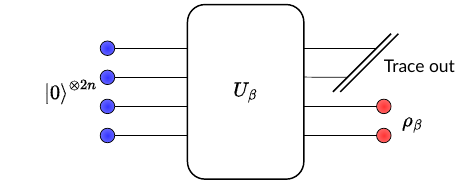}
    \caption[Diagrammatic representation of Gibbs state preparation.]{A diagram showcasing Gibbs state preparation on a quantum computer. Starting from the all-zero state in a $2n$-qubit register, and applying an (entangling) unitary $U_\beta$ that is dependent on the inverse temperature $\beta$, results in the mixed Gibbs state $\rho_\beta$ on the bottom $n$-qubit register when tracing out the top $n$ qubits.}
    \label{fig:gibbs_state_diagram}
\end{figure}

For a more detailed review on Variational Gibbs state preparation algorithms and their applications, such as in quantum Boltzmann machines, refer to \citet{Consiglio2025}. Fig.~\figref{fig:gibbs_state_diagram} shows the general constructive framework of preparing Gibbs states on a quantum computer, highlighting the need for ancillary qubits to prepare a mixed thermal state.

The goal of this work is to propose a \ac{VQA} that efficiently prepares Gibbs states on \ac{NISQ} computers, employing the generalised Helmholtz free energy as a (physically-motivated) objective function. This requires the evaluation of the von Neumann entropy~\cite{Bengtsson2006}, which is generally hard to obtain from a quantum register, as it is not an observable. In contrast to some of the currently employed \acp{VQA}~\cite{Riera2012, Wu2019, Chowdhury2020, Zhu2020, Wang2021a, Shtanko2023, Sagastizabal2021, Warren2022}, which employ truncated equations to approximate it, this work directly estimates the von Neumann entropy without any truncation, and with an error solely dependent on the number of shots, using sufficiently expressible ans\"atze capable of preparing the Boltzmann distribution. The \ac{VQA}, in fact, is composed of two ans\"atze: a heuristic, shallow one that prepares the Boltzmann distribution, for a given temperature; and another one, that may be designed with a problem-inspired approach which depends on the Hamiltonian, while being independent of the temperature.

\section{Variational Gibbs State Preparation} \label{sec:GSP}

Consider a Hamiltonian $\cH$, describing $n$ interacting qubits, the Gibbs state at inverse temperature $\beta \equiv 1 / k_\text{B} T$, where $k_\text{B}$ is the Boltzmann constant and $T$ is the temperature, is defined as
\begin{equation}
    \rho(\beta, \cH) = \frac{e^{-\beta \cH}}{\cZ(\beta, \cH)},
\end{equation}
where the partition function $\cZ(\beta, \cH)$ is
\begin{equation}
    \cZ(\beta, \cH) = \Tr{e^{-\beta \cH}} = \sum_{i=0}^{d - 1} e^{-\beta E_i}.
\end{equation}
Here the dimension $d = 2^n$, while $\{E_i\}$ are the eigenenergies of $\cH$, with $\{\ket{E_i}\}$ denoting the corresponding eigenstates, i.e. $\cH\ket{E_i} = E_i\ket{E_i}$. Physically speaking, the Gibbs state is the (thermal) equilibrium state at temperature $T$, of a system in the canonical ensemble. The canonical ensemble is a statistical framework that describes the possible states of a mechanical system in thermal equilibrium with a heat reservoir at a constant temperature. In the absence of external perturbations, a Gibbs state remains unchanged over time.

Fixing a Hamiltonian $\cH$ and inverse temperature $\beta$, for an arbitrary state $\rho$, one can define a generalised Helmholtz free energy as
\begin{equation}
    \cF(\rho) = \Tr{\cH \rho} - \beta^{-1}\cS(\rho),
    \label{eq:helmholtz_free_energy}
\end{equation}
where the von Neumann entropy $\cS(\rho)$ can be expressed in terms of the eigenvalues, $p_i$, of the state $\rho$,
\begin{equation}
    \cS(\rho) = -\sum_{i=0}^{d - 1} p_i \ln p_i.
    \label{eq:von_neumann}
\end{equation}
The Gibbs state is the unique state that minimises the free energy of $\cH$~\cite{Matsui1994}, and so a variational procedure can be put forward that takes Eq.~\eqref{eq:helmholtz_free_energy} as an objective function, such that the Gibbs state
\begin{equation}
    \label{eq:argmin_F}
    \rho(\beta, \cH) = \underset{\rho}{\arg\min}~\cF(\rho).
\end{equation}
Only when Eq.~\eqref{eq:argmin_F} is satisfied, $p_i = \exp\left(-\beta E_i\right)/\cZ(\beta, \cH)$ is the probability\footnote{Also known as the Boltzmann occupation probability.} of getting the eigenstate $\ket{E_i}$ from the ensemble $\rho(\beta, \cH)$.

\subsection{Framework of the Algorithm}

The difficulty in measuring the von Neumann entropy, defined by Eq.~\eqref{eq:von_neumann}, of a quantum state on a \ac{NISQ} device is typically the challenging part of variational Gibbs state preparation algorithms, as $\cS(\rho)$ is not an observable. With this in mind, the proposed \ac{VQA} is designed in a way such that it avoids the direct measurement of the von Neumann entropy on a quantum computer, by using a specifically constructed \ac{PQC}.

When preparing an $n$-qubit state starting from the input state $\ket{0}^{\otimes n}$, given that a quantum computer operates using only unitary gates, the final quantum state of the entire register will be pure (ignoring noise). As a result, in order to prepare an $n$-qubit Gibbs state on the system register, an $m \leq n$-qubit ancillary register is required. For example, in the case of the infinite-temperature Gibbs state, which is the maximally mixed state, $m = n$ qubits are required in the ancillary register to achieve maximal von Neumann entropy on the system register. In order to evaluate the von Neumann entropy, without any truncation, the entire Boltzmann distribution needs to be prepared on the ancillary register, hence, we set $m = n$, irrespective of the temperature.

We shall denote the ancillary register as $A$, while the preparation of the Gibbs state will be carried out on the system register $S$. The purpose of the \ac{VQA} is to effectively create the Boltzmann distribution on $A$, which is then imposed on $S$, via intermediary \textsc{CNOT} gates, to generate a diagonal mixed state in the computational basis. In the ancillary register, a \ac{PQC} capable of preparing such a probability distribution is chosen. Thus, the ancillary qubits are responsible for classically mixing in the probabilities of the thermal state, while also being able to access these probabilities via measurements in the computational basis. On the other hand, the system register will host the preparation of the Gibbs state, via a unitary rotation of computational basis states into the eigenenergy basis states, as well as the measurements of the expectation values of the desired Hamiltonian.

The specific design of the \ac{PQC} allows classical post-processing of simple measurement results, carried out on ancillary qubits in the computational basis, to determine the von Neumann entropy. A diagrammatic representation of the structure of the \ac{PQC} is shown in Fig.~\figref{fig:gibbs_circuit}. Note that while the \ac{PQC} of the algorithm has to have a particular structure --- a unitary acting on the ancillae and a unitary acting on the system, connected by intermediary \textsc{CNOT} gates --- it is not dependent on the choice of Hamiltonian $\cH$, inverse temperature $\beta$, or the other variational ans\"atze, employed within.

\subsection{Modular Structure of the PQC}

\begin{figure}[t]
    \centering
    \includegraphics[width=0.5\textwidth]{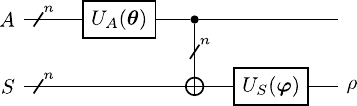}
    \caption[\acs{PQC} for Gibbs state preparation.]{\ac{PQC} for Gibbs state preparation, with systems $A$ and $S$ each carrying $n$ qubits. \textsc{CNOT} gates act between each qubit $A_i$ and corresponding $S_i$. $U_A(\bm{\theta})$ is responsible for preparing the Boltzmann Distribution, whilst $U_S(\bm{\varphi})$ is responsible for mapping computational basis states to the eigenstates of the Hamiltonian $\cH$.}
    \label{fig:gibbs_circuit}
\end{figure}

The \ac{PQC}, as shown in Fig.~\figref{fig:gibbs_circuit} for the \ac{VQA}, is composed of a unitary gate $U_A$ acting on the ancillary qubits, and a unitary gate $U_S$ acting on the system qubits, with \textsc{CNOT} gates in between. Note that the circuit notation we are using here means that there are $n$ qubits for both the system and the ancillae, as well as $n$ \textsc{CNOT} gates that act in parallel, and are denoted as
\begin{equation}
    \textsc{CNOT}_{AS} \equiv \bigotimes\limits_{i=0}^{n - 1}\textsc{CNOT}_{A_i S_i}.
\end{equation}

The parametrised unitary $U_A$ acting on the ancillae, followed by \textsc{CNOT} gates between the ancillary and system qubits, are responsible for preparing a probability distribution on the system. The parametrised unitary $U_S$ is then applied on the system qubits to transform the computational basis states into the eigenstates of the Hamiltonian.

A general unitary gate of dimension $d = 2^n$ is given by
\begin{equation}
    U_A = \left( \begin{array}{cccc}
        u_{0,0} & u_{0,1} & \cdots & u_{0,d-1} \\
        u_{1,0} & u_{1,1} & \cdots & u_{1,d-1} \\
        \vdots & \vdots & \ddots & \vdots \\
        u_{d-1,0} & u_{d-1,1} & \cdots & u_{d-1,d-1}
        \end{array} \right).
\end{equation}
Starting with the initial state of the $2n$-qubit register, $\ket{0}_{AS}^{\otimes 2n}$, the unitary gate $U_A$ is applied on the ancillae to obtain a quantum state $\ket{\psi}_A$, such that
\begin{equation}
    (U_A \otimes \dI_S) \ket{0}_{AS}^{\otimes 2n} = \ket{\psi}_{A} \otimes \ket{0}_{S}^{\otimes n},
\end{equation}
where
\begin{equation}
    \ket{\psi}_A = \sum_{i=0}^{d - 1} u_{i,0} \ket{i}_A,
\end{equation}
and $\dI_S$ is the identity operator acting on the system. The next step is to prepare a classical probability mixture on the system qubits, which can be done by applying \textsc{CNOT} gates between each ancilla and system qubit. This results in a state
\begin{align}
    \textsc{CNOT}_{AS} \left( \ket{\psi}_{A} \otimes \ket{0}_{S}^{\otimes n} \right) = \sum_{i=0}^{d - 1} u_{i,0} \ket{i}_A \otimes \ket{i}_S.
\end{align}
By then tracing out the ancillary qubits, we arrive at
\begin{align}
    \PTr{A}{\left( \sum_{i=0}^{d-1} u_{i,0}\ket{i}_A \otimes \ket{i}_S \right) \left( \sum_{j=0}^{d-1} u_{j,0}^* \bra{j}_A \otimes \bra{j}_S \right)} &= \sum_{i,j=0}^{d-1} u_{i,0}u_{j,0}^* \braket{i}{j} \ketbra{i}{j}_S \nonumber \\ 
    &= \sum_{i=0}^{d-1} |u_{i,0}|^2 \ketbra{i}{i}_S,
\end{align}
ending up with a diagonal mixed state on the system, with probabilities given directly by the absolute square of the entries of the first column of $U_A$, that is, $p_i = |u_{i,0}|^2$. If the system qubits were traced out instead, we would end up with the same diagonal mixed state, but on the ancillary qubit register:
\begin{align}
    \PTr{S}{\left( \sum_{i=0}^{d-1} u_{i,0}\ket{i}_A \otimes \ket{i}_S \right) \left( \sum_{j=0}^{d-1} u_{j,0}^* \bra{j}_A \otimes \bra{j}_S \right)} &= \sum_{i,j=0}^{d-1} u_{i,0}u_{j,0}^* \braket{i}{j} \ketbra{i}{j}_A \nonumber \\ 
    &= \sum_{i=0}^{d-1} |u_{i,0}|^2 \ketbra{i}{i}_A.
\end{align}
This implies that by measuring in the computational basis of the ancillary qubits, the probabilities $p_i = |u_{i,0}|^2$ can be determined, which can then be post-processed to determine the von Neumann entropy $\cS$ of the state $\rho$ via Eq.~\eqref{eq:von_neumann}.\footnote{Since the entropy of register $A$ is the same as that of register $S$.} As a result, since $U_A$ only serves to create a probability distribution from the entries of the first column, one can use a parametrised orthogonal (real and unitary) operator, thus requiring less gates and parameters for the ancillary ansatz.

The unitary gate $U_S$ then serves to transform the computational basis states of the system qubits to the eigenstates of the Gibbs state,\footnote{Since $U_S$ does not modify the occupation probabilities $p_i$, i.e., the spectrum of $\rho$.} such that
\begin{equation}
    \rho = U_S \left( \sum_{i=0}^{d-1} p_i \ketbra{i}{i}_S \right) U_S^\dagger = \sum_{i=0}^{d-1} p_i U_S \ketbra{i}{i}_S U_S^\dagger = \sum_{i=0}^{d - 1} p_i \ketbra{\psi_i}_S,
\end{equation}
where the expectation value $\Tr{\cH \rho}$ of the Hamiltonian can be measured. Ideally, at the end of the optimisation procedure, $p_i = \exp\left(-\beta E_i\right)/\cZ(\beta, \cH)$ and $\ket{\psi_i} = \ket{E_i}$, so that
\begin{equation}
    \rho(\beta, \cH) = \sum_{i=0}^{d - 1} \frac{e^{-\beta E_i}}{\cZ(\beta, \cH)} \ketbra{E_i}_S.
\end{equation}
The \ac{VQA} therefore avoids measuring the von Neumann entropy of a mixed state on a quantum computer, and instead transfers the task of post-processing computational basis measurement results to the classical computer.

\subsection{Objective Function}

Finally, the objective function of the \ac{VQA} can be defined, such that it minimises the generalised Helmholtz free energy~\eqref{eq:helmholtz_free_energy}, via the constructed \ac{PQC}, to obtain the Gibbs state
\begin{equation}
    \rho(\beta, \cH) = \underset{\bm{\theta}, \bm{\varphi}}{\arg\min} \ \cF\left(\rho\left(\bm{\theta}, \bm{\varphi}\right)\right) = \underset{\bm{\theta}, \bm{\varphi}}{\arg\min} \left( \Tr{\cH \rho_S(\bm{\theta}, \bm{\varphi})} - \beta^{-1}\cS\left(\rho_A(\bm{\theta})\right) \right).
    \label{eq:gibbs_state}
\end{equation}

It is noteworthy to mention that while the energy expectation depends on both sets of angles $\bm{\theta}$ (as $U_A$ is responsible for parametrising the Boltzmann distribution) and $\bm{\varphi}$ (as $U_S$ is responsible for parametrising the eigenstates of the Gibbs state), the calculation of the von Neumann entropy only depends on $\bm{\theta}$.

\begin{figure}[t]
    \centering
    \includegraphics[width=0.6\textwidth]{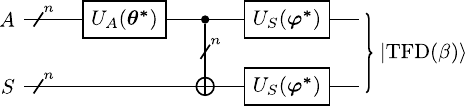}
    \caption[Optimised \acs{PQC} for \acs{TFD} state preparation.]{Optimised \ac{PQC} for \ac{TFD} state preparation, with systems $A$ and $S$ each carrying $n$ qubits. \textsc{CNOT} gates act between each qubit $A_i$ and corresponding $S_i$. $U_A(\bm{\theta^*})$ and $U_S(\bm{\theta^*})$ are similarly responsible as in Fig.~\figref{fig:gibbs_circuit}, with the addition of $U_S(\bm{\theta^*})$ in the ancillary register to prepare the \acs{TFD} state.}
    \label{fig:tfd_circuit}
\end{figure}

Furthermore, once the optimal parameters $\bm{\theta^*}$, $\bm{\varphi^*}$ are obtained to prepare the Gibbs state $\rho(\beta, \cH)$ on the system qubits $S$, one can place the same unitary $U_S$ with optimal parameters $\bm{\varphi^*}$ on the ancillary qubits to prepare the \ac{TFD} state on the entire qubit register, as shown in Fig.~\figref{fig:tfd_circuit}. A \ac{TFD} state~\cite{Wu2019, Zhu2020, Sagastizabal2021} is defined as
\begin{equation}
    \ket{\textrm{TFD}(\beta)} = \sum_{i=0}^{d - 1} \sqrt{\frac{e^{-\beta E_i}}{\cZ(\beta, \cH)}} \ket{i}_A \otimes \ket{i}_S~,
\end{equation}
and, tracing out either the ancilla or system register, yields the same Gibbs state on the other register.

\subsection{Alternative Implementations of the Algorithm}

\begin{figure}[t]
    \centering
    \includegraphics[width=0.5\textwidth]{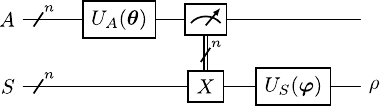}
    \caption[\acs{PQC} for Gibbs state preparation using mid-circuit measurements.]{\ac{PQC} for Gibbs state preparation using mid-circuit measurements, with systems $A$ and $S$ each carrying $n$ qubits. Classically-controlled \textsc{NOT} gates act between each qubit $A_i$ and corresponding $S_i$.}
    \label{fig:classical_gibbs_state}
\end{figure}

There are several adjustments that could be applied to the \ac{PQC} to modify the procedure. One specific example is replacing the intermediary \textsc{CNOT} gates with mid-circuit measurements and implementing classically controlled-\textsc{NOT} gates, since no subsequent unitary gates act on the control qubits, as shown in Fig.~\figref{fig:classical_gibbs_state}. This method admits a few benefits:
\begin{enumerate}
    \item Since the ancillary system, $A$, needs to be measured to compute the von Neumann entropy, utilising mid-circuit measurements followed by classically-controlling the system qubits, $S$, is a natural approach to the algorithm.
    \item The two registers $A$ and $S$ can be made fully distinct in terms of the device topology, as well as reducing the depth of the entire circuit, leading to less overall decoherence affecting the protocol.
    \item Once optimisation is carried out, the classically-controlled \textsc{NOT} gates can still be kept in the circuit, yet if the experimentalist ignores the measurement information (equivalent to tracing out), then there is no operational difference between preparing an ensemble of pure states and preparing a mixed state using quantum \textsc{CNOT} gates.
\end{enumerate}
The only downside is if the ancillary qubits are intended to be used again, such as when preparing the \ac{TFD} state. In this case, the optimisation for finding optimal parameters to prepare the Gibbs state can still be carried out using classically-controlled \textsc{NOT} gates. However, at the end of the optimisation procedure, the classically-controlled \textsc{NOT} gates can be replaced with \textsc{CNOT} gates followed by the optimised system unitary, with the same structure as in Fig.~\figref{fig:tfd_circuit}, to obtain the \ac{TFD} state.

The \ac{VQA} can be further be adapted so that an $U_A$ is replaced by a classical procedure that generates the probability distribution, that then prepares pure states of the Gibbs state ensemble $\{p_i, \ket{E_i}\}$, where $p_i = \exp(-\beta E_i) / \cZ(\beta, \cH)$ on the system qubits $S$. This procedure can be carried out by parametrising a classical probability distribution $p(\bm{\theta})$ by $\cO(\text{poly}(n))$ parameters $\bm{\theta}$, using methods such as Markov chains composed of a sequence of local stochastic matrices, among others. The probability distribution will output bit strings $\ket{i}$ that can be fed as a computational input state to the unitary $U_S$ that prepares the eigenstates $\ket{E_i}$ of the Hamiltonian. By reducing the number of qubits and eliminating the requirement for intermediary \textsc{CNOT} gates, this process may result in a less expressible probability distribution function because entanglement is not used as a resource.

Furthermore, if the parametrisation of the probability distribution corresponds with the output distribution of a known unitary circuit, of a sufficiently shallow depth and expressibility, then the optimisation can be carried using only the classical subroutine of sampling bit strings from the probability distribution $p(\bm{\theta})$, and feeding them to $U_S$ to compute the free energy. Once the \ac{VQA} is trained, $U_A$ can be introduced with the optimised parameters $\bm{\theta^*}$, to prepare the mixed Gibbs on the quantum computer. Nevertheless, finding such a parametrisation, that corresponds to a shallow, yet expressible enough unitary, is a non-trivial task.

\section{Performance of the VQA on the Ising Model} \label{sec:results_ising}

In this Section, the performance of the \ac{VQA} is assessed for preparing the Gibbs state of a transverse-field Ising model, defined as
\begin{equation}
    \cH = -\sum_{i=0}^{n-1} \sigma_i^x \sigma_{i+1}^x - h \sum_{i=0}^{n-1} \sigma_i^z.
    \label{eq:Ising}
\end{equation}
The Ising Hamiltonian is a widely-investigated model~\cite{Franchini2017}, and here only one relevant property for implementing a problem-inspired ansatz for $U_S$ is reported. The Hamiltonian in Eq.~\eqref{eq:Ising} commutes with the parity operator $\cP=\bigotimes_{i=0}^{n - 1} \sigma^z_i$. As a consequence, the eigenstates of $\cH$ have definite parity, and so will the eigenstates of $\rho(\beta, \cH)$.

The Uhlmann--Josza fidelity~\cite{Uhlmann2011}, defined as $F\left(\rho, \sigma\right)=\left(\Tr{\sqrt{\sqrt{\rho}\sigma\sqrt{\rho}}}\right)^2$, is used as a figure of merit for assessing the performance of the \ac{VQA}, since it describes how `close' the prepared state is to the Gibbs state, and it is also the most commonly-employed measure of distinguishability. However, other measures having different interpretations can be used. One such measure is the trace distance~\cite{Nielsen2010}, which has the property that if its value between the two states is bounded by $\epsilon$, expectation values computed on the effectively prepared state, differ from those taken on the Gibbs state by an amount that is, at most, proportional to $\epsilon$~\cite{Holmes2022b}. Another choice is the relative entropy~\cite{Nielsen2010}, which characterises the distinguishability between the two states as the unexpected result that arises when an event, that is not possible with the true Gibbs state, occurs~\cite{Vedral2002}.

A simple, linearly entangled \ac{PQC} for the unitary $U_A$ is chosen, with parametrised $R_y(\theta_i)$ gates, and \textsc{CNOT}s as the entangling gates. This ansatz is hardware efficient and is sufficient to produce real amplitudes for preparing the probability distribution.\footnote{Note that the use of entangling gates is required, as otherwise the \ac{PQC} will not be able to prepare arbitrary probability distributions, including the Boltzmann distribution of the Ising model. A proof by contradiction of this is given in Appendix~\secref{app:entangling_gates}.}

\begin{figure}[t]
    \centering
    \includegraphics[width=\textwidth]{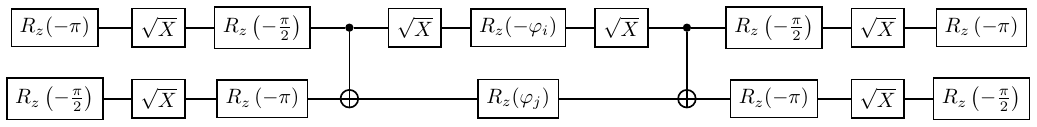}
    \caption[Decomposed parity-preserving gate.]{Decomposed parity-preserving gate, $R_p$, defined in Eq.~\eqref{eq:R_p}, written in basis gates of IBM quantum computers.}
    \label{fig:R_p}
\end{figure}

\begin{figure}[t]
    \centering
    \includegraphics[width=\textwidth]{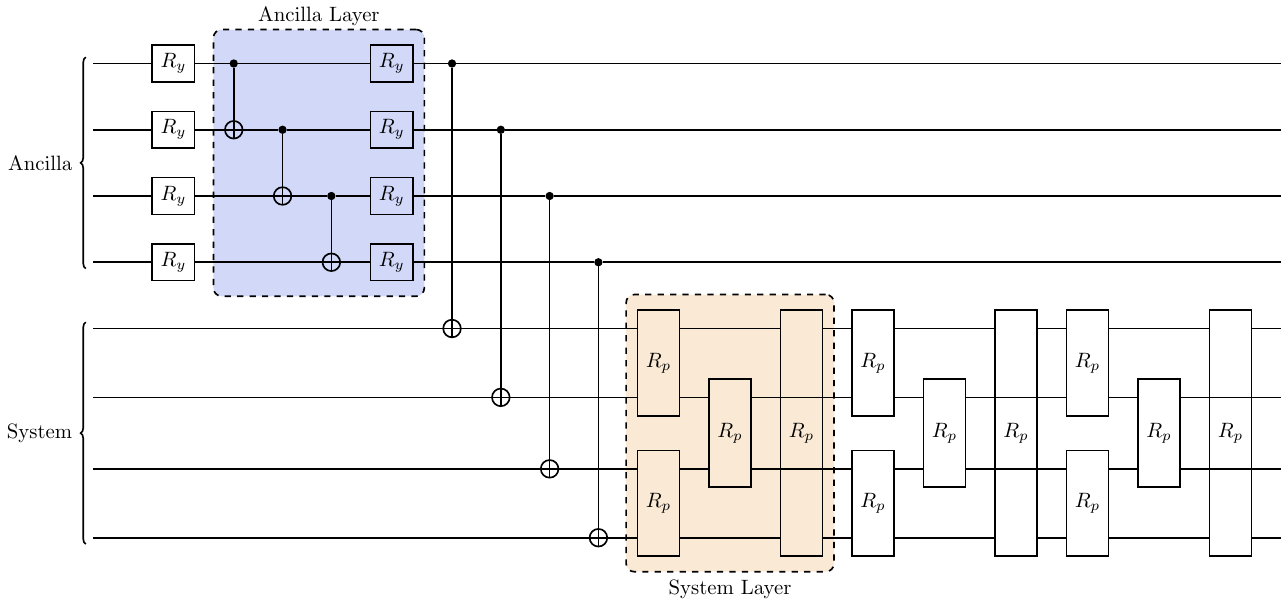}
    \caption[Example of an eight-qubit Gibbs state preparation \acs{PQC} for the Ising model.]{Example of an eight-qubit \ac{PQC}, consisting of one ancilla layer acting on a four-qubit register, and three ($n - 1$) system layers acting on another four-qubit register. Each $R_y$ gate is parametrised with one parameter $\theta_i$, while each $R_p$ gate has two parameters $\varphi_i$ and $\varphi_j$. The $R_p$ gate is defined in Eq.~\eqref{eq:R_p}. Note that the intermediary \textsc{CNOT} gates, as well as the $R_p$ gates acting on qubits two and three, and on qubits one and four of the system, can be carried out in parallel, respectively.}
    \label{fig:pqc_4qubits}
\end{figure}

For the unitary $U_S$, a parity-preserving \ac{PQC} is chosen, with the gates being $R_{xy}(\varphi_i) \equiv \exp(-\imath \varphi_i (\sigma^x \otimes \sigma^y)/2)$ followed by $R_{yx}(\varphi_j) \equiv \exp(-\imath \varphi_j (\sigma^y \otimes \sigma^x)/2)$ gates, employed in a brick-wall structure. If the two gates are combined, which we denote as $R_p(\varphi_i, \varphi_j)$, we get
\begin{align}
    R_p(\varphi_i, \varphi_j) &= R_{yx}(\varphi_j) R_{xy}(\varphi_i) \nonumber \\
    &=
    \left(
    \begin{array}{cccc}
     \cos \left(\frac{\varphi_i +\varphi_j }{2}\right) & 0 & 0 & \sin \left(\frac{\varphi_i +\varphi_j }{2}\right) \\[1ex]
     0 & \cos \left(\frac{\varphi_i -\varphi_j }{2}\right) & -\sin \left(\frac{\varphi_i -\varphi_j }{2}\right) & 0 \\[1ex]
     0 & \sin \left(\frac{\varphi_i -\varphi_j }{2}\right) & \cos \left(\frac{\varphi_i -\varphi_j }{2}\right) & 0 \\[1ex]
     -\sin \left(\frac{\varphi_i +\varphi_j }{2}\right) & 0 & 0 & \cos \left(\frac{\varphi_i +\varphi_j }{2}\right) \\
    \end{array}
    \right),
    \label{eq:R_p}
\end{align}
which can be decomposed into two \textsc{CNOT} gates, six $\sqrt{X}$ gates and ten $R_z$ gates. One layer of the unitary acting on the system qubits consists of a brick-wall structure, composed of an even-odd sublayer of $R_p$ gates, followed by an odd-even sublayer of $R_p$ gates. The decomposed unitary is shown in Fig.~\figref{fig:R_p}. An example of a \ac{PQC}, split into a four-qubit ancillary register, and a four-qubit system register, is shown in Fig.~\figref{fig:pqc_4qubits}. Table~\tabref{tab:scaling} shows the scaling of the \ac{VQA} assuming a closed ladder connectivity, for $n > 2$.

\begin{table}[t]
\centering
\begin{tabular}{|l|l|l|} 
    \hline
    \# of parameters & $n(l_A + 1) + 2nl_S$ & $\cO(n(l_A + l_S))$ \\ 
    \hline
    \# of \textsc{CNOT} gates & $(n - 1)l_A + 2nl_S + n$ & $\cO(n(l_A + l_S))$ \\
    \hline
    \# of $\sqrt{X}$ gates & $2n(l_A + 1) + 6nl_S$ & $\cO(n(l_A + l_S))$ \\ 
    \hline
    Circuit depth & $(n - 1)l_A + 2Pl_S + 1$ & $\cO(nl_A + l_S)$ \\ 
    \hline
\end{tabular}
\caption[\acs{PQC} scaling of the Gibbs state preparation algorithm for the Ising model.]{Scaling of the \ac{VQA} assuming a closed ladder connectivity for the Ising model, with $n > 2$, and $l_A$ and $l_S$ are the number of ancilla ansatz and system ansatz layers, respectively, and $P$ is 2 when $n$ is even and 3 when $n$ is odd. The depth only counts \textsc{CNOT} gates.}
\label{tab:scaling}
\end{table}

\subsection{Statevector Results} \label{sec:statevector_results}

\begin{figure}[t]
    \centering
    \includegraphics[width=\textwidth]{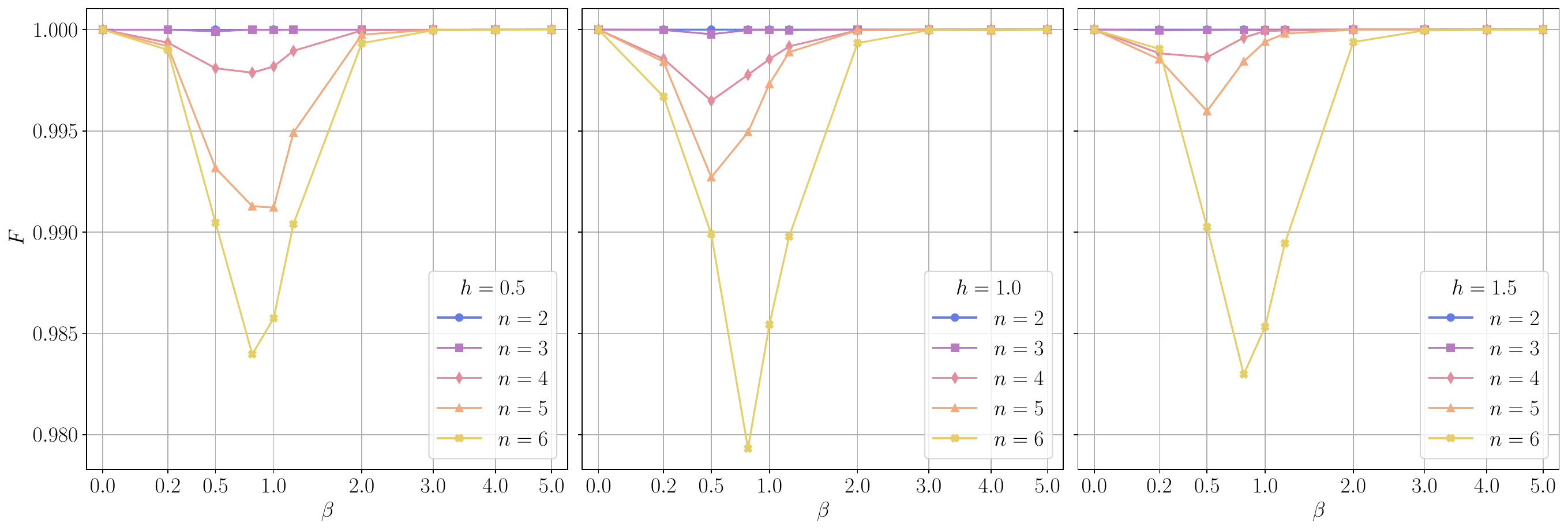}
    \caption[Fidelity of the Gibbs state preparation algorithm via statevector simulations for the Ising model.]{Fidelity $F$ of the obtained state via statevector simulations (using \ac{BFGS}) with the exact Gibbs state, vs inverse temperature $\beta$, for two to six qubits of the Ising model with $h = 0.5, 1.0, 1.5$. A total of 100 runs are made for each point, with the optimal state taken to be the one that maximises the fidelity. Lines are for guiding the eyes.}
    \label{fig:statevector_fidelity}
\end{figure}

Fig.~\figref{fig:statevector_fidelity} shows the fidelity of the prepared mixed state when compared with the exact Gibbs state of the Ising model with $h = 0.5, 1.0, 1.5$, respectively, across a range of temperatures for system size between two to six qubits. The \ac{VQA} was carried out using statevector simulations with the \ac{BFGS} optimiser~\cite{Nocedal2006}. One layer for the ancilla ansatz was used, and $n - 1$ layers for the system ansatz, with the scaling highlighted in Table~\tabref{tab:scaling_2}. The number of layers was heuristically chosen to satisfy, at most, a polynomial scaling in quantum resources, while achieving a fidelity higher than 0.98 in statevector simulations. Furthermore, in order to alleviate the issue of getting stuck in local minima, the optimiser was embedded in a Monte Carlo framework, that is, taking multiple random initial positions and carrying out a local optimisation from each position --- which is called a `run' --- and finally taking the global minimum to be the minimum over all runs. 

\begin{table}[t]
\centering
\begin{tabular}{|l|l|l|} 
    \hline
    \# of parameters & $2n^2$ & $\cO(n^2)$ \\ 
    \hline
    \# of \textsc{CNOT} gates & $2n^2 - 1$ & $\cO(n^2)$ \\
    \hline
    \# of $\sqrt{X}$ gates & $2n(3n - 1)$ & $\cO(n^2)$ \\ 
    \hline
    Circuit depth & $(2P + 1)n - 2P$ & $\cO(n)$ \\
    \hline
\end{tabular}
\caption[Specific \acs{PQC} scaling of the Gibbs state preparation algorithm for the Ising model.]{Scaling of the \ac{VQA} assuming a closed ladder connectivity, for the Ising model, with $n > 2$, $l_A = 1$, $l_S = n - 1$, and $P$ is 2 when $n$ is even and 3 when $n$ is odd, obtained from Tab.~\tabref{tab:scaling}. The depth only counts \textsc{CNOT} gates.}
\label{tab:scaling_2}
\end{table}

A total of 100 runs of \ac{BFGS} per $\beta$ were carried out to verify the reachability of the \ac{PQC}, with Fig.~\figref{fig:statevector_fidelity} showcasing the maximal fidelity achieved for each $\beta$ out of all runs. The results show that, indeed, the \ac{VQA} is able to reach a very high fidelity $F \gtrsim 0.98$ for up to six-qubit Gibbs states of the Ising model. In the case of the extremal points, that is $\beta \rightarrow 0$ and $\beta \rightarrow \infty$, the fidelity reaches one, for all investigated system sizes. 

For $\beta \rightarrow 0$, the problem reduces to maximising the von Neumann entropy, i.e., achieving the maximally mixed state, where the problem becomes independent of the energy, and therefore, independent of the prepared eigenstates by $U_S$ (since the von Neumann entropy depends solely on $U_A(\bm{\theta})$ as in Eq.~\eqref{eq:gibbs_state}). For $\beta \rightarrow \infty$, the problem reduces to minimising the energy, which is equivalent to finding the ground-state of the Hamiltonian, and so, independent of $U_A$ (or rather $U_A$ is tasked with preparing the all-zero state, i.e. $U_A$ becomes the identity gate). This is in contrast with intermediary temperatures $\beta \sim 1$, where the fidelity decreases with the number of qubits. This could be attributed to the fact that at intermediate temperatures, a substantial amount of Boltzmann probabilities are necessary for evaluating the von Neumann entropy with a high precision, and similarly, a large number of eigenstates contribute to preparing the Gibbs state, resulting in a dip in the fidelity at those temperatures.

\subsection{Noisy Simulation Results} \label{sec:noisy_results}

The next step was to carry out noisy simulations of the \ac{VQA}, including both shot and hardware noise simulations. The noise model of \texttt{ibmq\_guadalupe}~\cite{IBM_compute_resources} was used with the Ising model with $h = 0.5$, similarly with one layer for the ancilla ansatz and $n - 1$ layers for the system ansatz. However, it must be noted that in this case, the scaling of the algorithm does not follow Table~\tabref{tab:scaling_2}, due to the fact that \texttt{ibmq\_guadalupe} does not have a closed ladder connectivity. As a result, transpilation was carried out by the \texttt{Qiskit} transpiler~\cite{Li2018}. Apart from this, since it is difficult to optimise a noisy objective function via \ac{BFGS}, an optimiser that accommodates noisy measurements was chosen: \ac{SPSA}~\cite{Spall1992}.

\begin{figure}[t]
    \centering
    \includegraphics[width=0.8\textwidth]{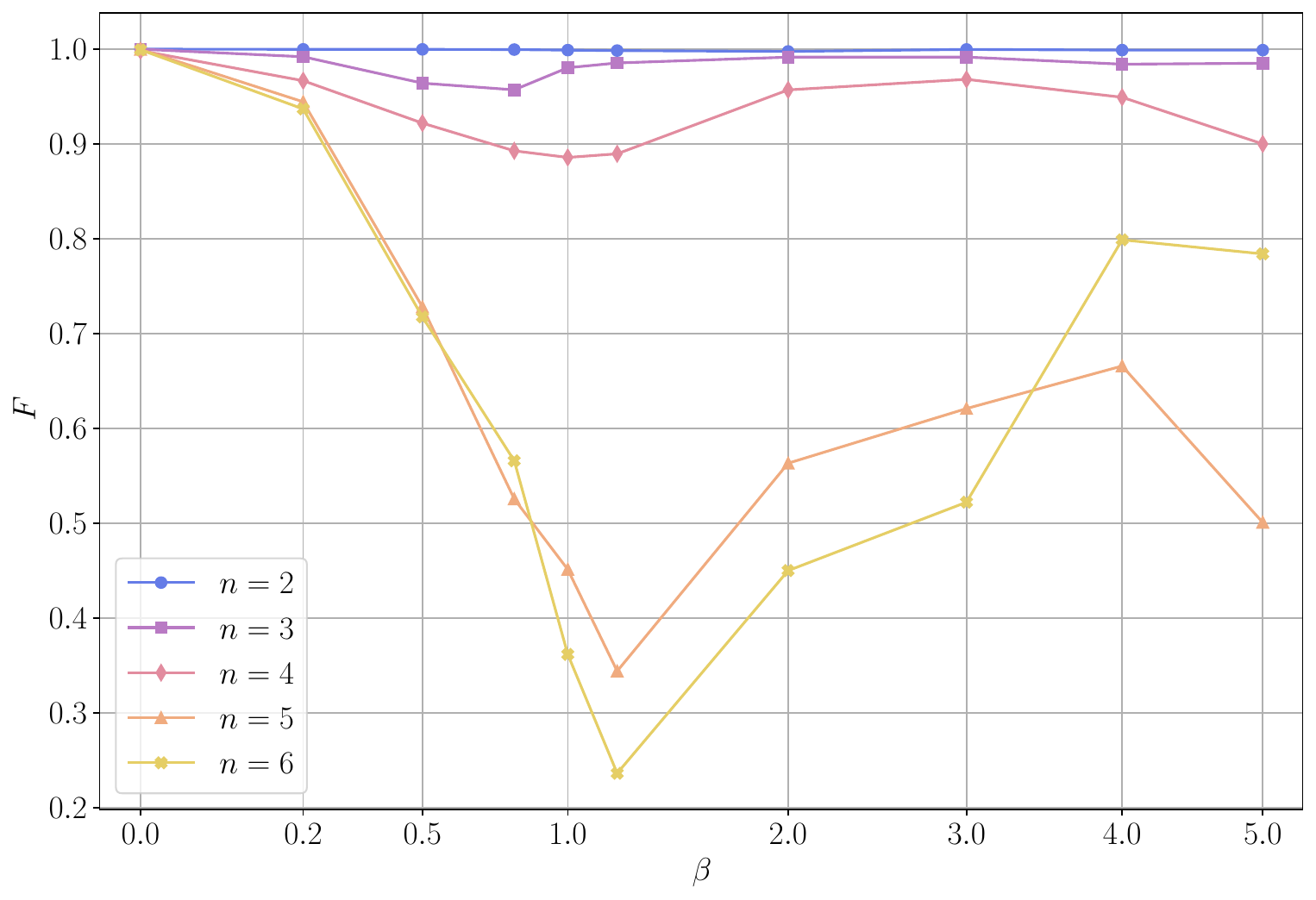}
    \caption[Fidelity of the Gibbs state preparation algorithm via noisy simulations for the Ising model.]{Fidelity $F$ of the obtained state via noisy simulations (using \ac{SPSA}) of \texttt{ibmq\_guadalupe} with the exact Gibbs state, vs inverse temperature $\beta$, for two to six qubits of the Ising model with $h = 0.5$. A total of ten runs are made for each point, with the optimal state taken to be the one that maximises the fidelity. Lines are for guiding the eyes.}
    \label{fig:noisy_fidelity}
\end{figure}

Using \ac{SPSA}, ten runs were carried out for each $\beta$, while the number of iterations was taken to be $100n$ for each run, with only two measurements at each iteration to estimate the gradient in a random direction, i.e. $200n$. As a consequence, a total of $2000n$ function evaluations were used to obtain the fidelity for each $\beta$ shown in Fig.~\figref{fig:noisy_fidelity} (with an extra 50 function evaluations at each run to calibrate the hyperparameters of \ac{SPSA}). Similar to the number of layers, the choice of the number of iterations was heuristically chosen so that, at most, the scaling is linear, while still retaining a fidelity greater than 0.95 for the two- and three-qubit noisy simulation cases. On the other hand, the fidelity for the five- and six-qubit cases drops dramatically in the mid-to-high $\beta$ range, and the Gibbs state is only faithfully prepared for low $\beta$. This can be attributed to the level of noise present in the optimisation procedure, with the transpiled circuits going well beyond the quantum volume of \texttt{ibm\_guadalupe}. There are also some points which obtain a fidelity worse for five qubits than six qubits, which could be due to the larger depth acquired by an odd number of qubits in the system ansatz, as highlighted in Table~\tabref{tab:scaling_2}.

To measure the energy expectation value of the Ising model $\Tr{\cH\rho}$, the Ising Hamiltonian needs to be split into its constituent Pauli strings, whose number scales linearly with the number of qubits as $2n$. However, the $\sigma^x\sigma^x$ terms can be grouped together, as well as the $\sigma^z$ terms, and measured simultaneously, reducing the number of measurement circuits to two. Each circuit was also measured with 1024 shots. The \texttt{M3} package~\cite{Nation2021} was also utilised to perform basic error mitigation. It operates by using a matrix-free preconditioned iterative-solution method to mitigate measurement error that does not form the full assignment matrix, or its inverse. A summary of the optimisation scaling is shown in Table~\tabref{tab:scaling_3}.

\begin{table}[t]
\centering
\begin{tabular}{|l|l|l|}
    \hline
    \# of iterations for each run & $100n$ & $\cO(n)$ \\
    \hline
    \# of function evaluations for each run & $200n$ & $\cO(n)$ \\
    \hline
    \# of circuits per function evaluation & $2$ & $\cO(1)$ \\
    \hline
    \# of circuit evaluations for each run & $400n$ & $\cO(n)$ \\
    \hline
    \# of shots for each circuit evaluation & $1024$ & $\cO(1)$ \\
    \hline
\end{tabular}
\caption{Scaling of \acs{SPSA} for noisy simulations and on quantum hardware for the Ising model.}
\label{tab:scaling_3}
\end{table}

The performance of \acp{VQA} is heavily impacted by the presence of noise-induced \acp{BP}~\cite{Wang2021b}. While a full analysis of \acp{BP} for Gibbs state preparation is beyond the scope of this thesis, which aims to provide an alternative approach to preparing Gibbs states variationally, and avoid any estimations of the entropy using Taylor expansions or other truncations, brief analyses as starting points are still discussed, for future works. In particular, analysis on the error of estimating the entropy is carried out in Section~\secref{sec:entropy_estimation}, while the implications of \acp{BP} on the \ac{VQA} are presented in Appendix~\secref{app:bp}. 

\subsection{IBM Quantum Device Results} \label{sec:device_results}

\begin{figure}[t]
    \centering
    \includegraphics[width=0.8\textwidth]{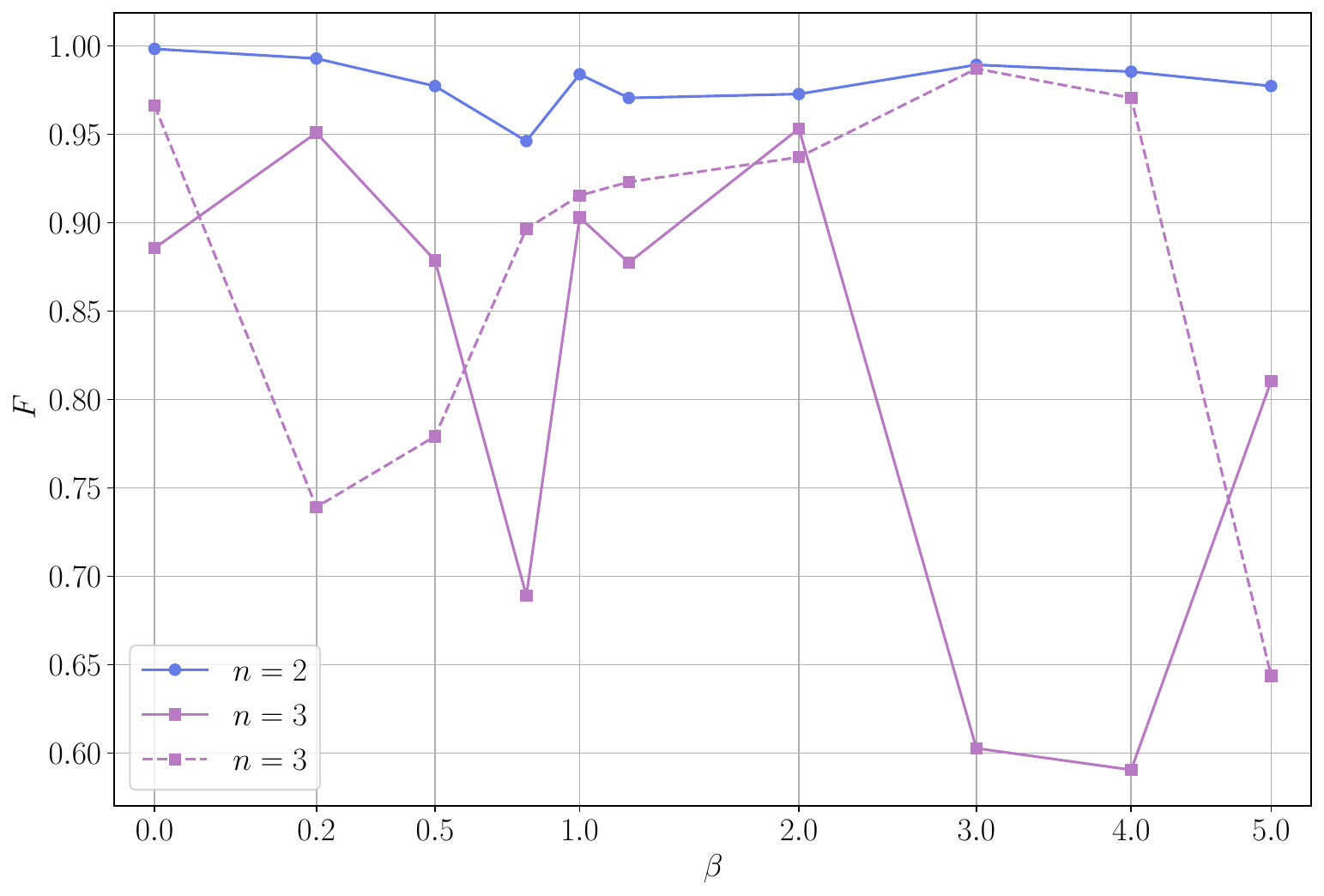}
    \caption[Fidelity of the Gibbs state preparation algorithm on \texttt{IBM\_nairobi} for the Ising model.]{Fidelity $F$ of the obtained state (using \ac{SPSA}) running directly on \texttt{ibm\_nairobi} with the exact Gibbs state, vs inverse temperature $\beta$, for two and three qubits of the Ising model with $h = 0.5$. The dashed line represents the run with no $R_p$ gate between non-adjacent qubits in the system layers. One run is carried out for $n = 2$ and $n = 3$ for the dashed line, and two runs for $n = 3$ for the solid line. Lines are for guiding the eyes.}
    \label{fig:ibm_nairobi}
\end{figure}

Finally, the \ac{VQA} was carried out on an actual quantum device. While noisy simulations were ran using the calibration data of \texttt{ibmq\_guadalupe} --- since it has access of up to 16 qubits --- the actual hardware used for the two- and three-qubit Gibbs state preparation was \texttt{ibm\_nairobi}, due to its accessibility. Fig.~\figref{fig:ibm_nairobi} displays the fidelity of Gibbs states obtained by running on IBM quantum hardware~\cite{IBM_compute_resources}, specifically \texttt{ibm\_nairobi}. Similarly to the noisy simulations, \ac{SPSA} was used; however, this time, with only one run for each $\beta$ in the case $n = 2$, and two runs in the case $n = 3$, with $100n$ iterations and 1024 shots. The Gibbs states were obtained by taking the optimal parameters from the optimisation carried out on \texttt{ibm\_nairobi}, and reconstructing the obtained mixed state on a classical computer.

The solid lines in Fig.~\figref{fig:ibm_nairobi} represent the two- and three-qubit results. At all points, the two-qubit Gibbs state shows excellent fidelity. On the other hand, the three-qubit Gibbs state is remarkably reproduced at certain temperatures, while it is lacking at others. Since \texttt{ibm\_nairobi} does not have a closed ladder connectivity, several \textsc{SWAP} gates are necessary for carrying out transpilation. In an attempt to reduce the number of \textsc{SWAP} gates, another run was carried out at each $\beta$, where the $R_p$ gate acting on non-adjacent qubits in the system layers was removed, with the result shown in the dashed line of Fig.~\figref{fig:ibm_nairobi} (note that this also resulted in less parameters and depth of the \ac{PQC}). A considerable improvement in fidelity is achieved at the points where fidelity was lacking in the previous case. Since the available running time on the quantum device was limited, the number of runs was too low to determine the reason as to why the Gibbs state was not achieved with a higher fidelity. Nevertheless, comparing the results of Fig.~\figref{fig:ibm_nairobi}, with the statevector results in Fig.~\figref{fig:statevector_fidelity} and with the noise-simulated results in Fig.~\figref{fig:noisy_fidelity}, it can be concluded that limited connectivity, combined with device noise, severely hampers the effectiveness of the \ac{VQA}.

\begin{figure}[t]
    \centering
    \includegraphics[width=0.7\textwidth]{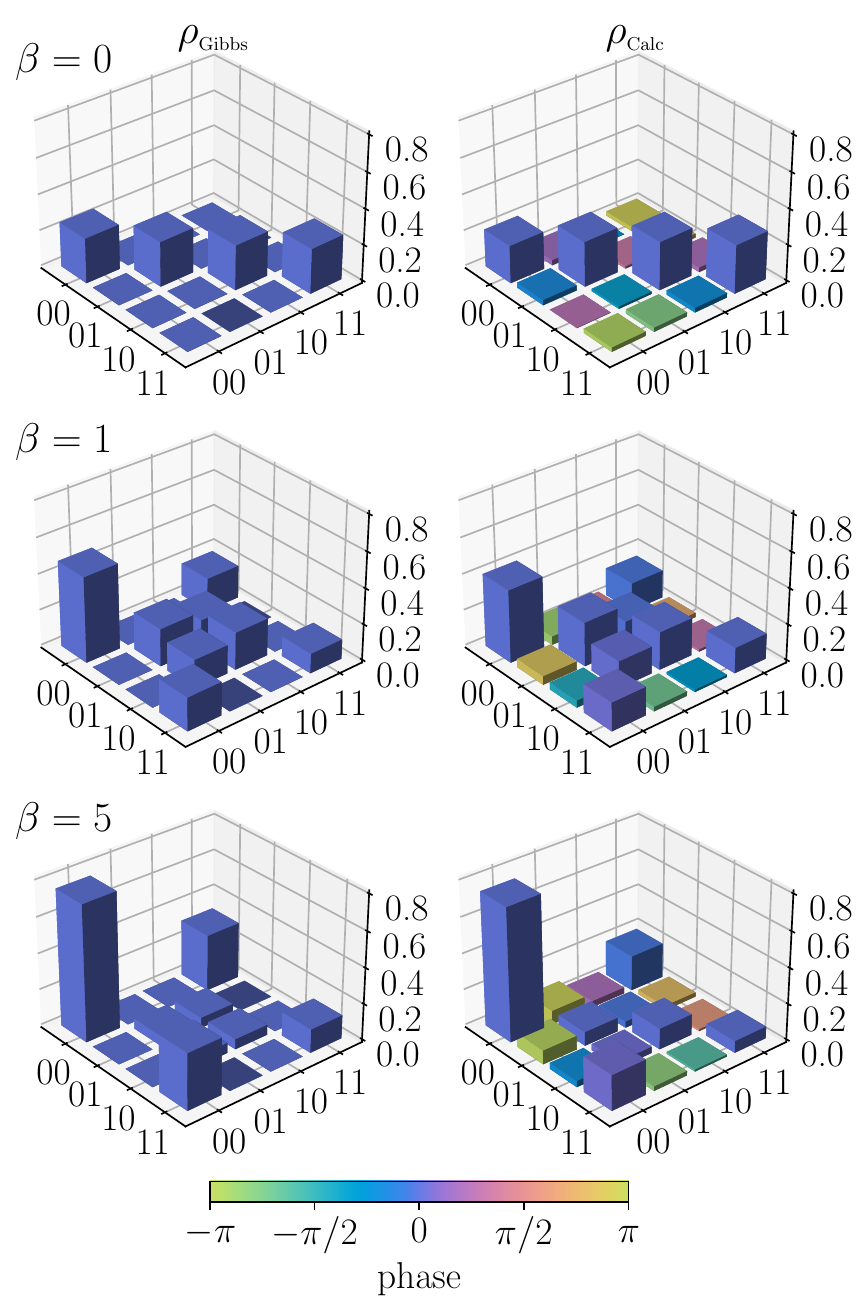}
    \caption[3D bar plot of the two qubit results from \texttt{ibm\_nairobi}.]{3D bar plot of the two qubit results from \texttt{ibm\_nairobi} for $\beta = 0, 1, 5$, of the Ising model with $h = 0.5$. The analytical Gibbs states are shown in the left column, while the tomographically obtained Gibbs States are shown in the right column.}
    \label{fig:bar3d}
\end{figure}

In addition, quantum state tomography for the two-qubit case was carried out on \texttt{ibm\_nairobi}, with 1024 shots, for the cases of $\beta = 0, 1, 5$, where the fidelities obtained were 0.992, 0.979, and 0.907, respectively. A 3D bar plot of the tomographic results can be seen in the right column Fig.~\figref{fig:bar3d}, which can be compared with the analytical form of the Gibbs state on the left column. The largest discrepancies can be witnessed in the off-diagonal terms, which increase as $\beta$ increases, as well as showcasing symptoms of amplitude damping. This could be attributed to the thermal relaxation and dephasing noise present in the quantum devices, leading to an overall decoherence in the Gibbs state.

\section{Performance of the VQA on the XY Model} \label{sec:XY}

In this Section, the performance of the \ac{VQA} on the XY model~\cite{Lieb1961} is assessed. The XY model is defined as
\begin{equation}
    \cH = -\sum_{i=0}^{n-1} \left( \frac{1 + \gamma}{2} \sigma_i^x \sigma_{i+1}^x + \frac{1 - \gamma}{2} \sigma_i^y \sigma_{i+1}^y \right) - h \sum_{i=0}^{n-1} \sigma_i^z,
    \label{eq:XY}
\end{equation}
where the anisotropy parameter $0 \leq \gamma \leq 1$ defines the \textit{XX} model ($\gamma=0$), the Ising  ($\gamma=1$) and the XY model ($0<\gamma<1$), with the last two belonging to the same universality class. Similar to the Ising model~\eqref{eq:Ising}, the XY model is parity-preserving, and so the $R_p$ gate~\eqref{eq:R_p} is used in the system unitary $U_S$ to determine its eigenstates.

\begin{figure}[t]
    \centering
    \includegraphics[width=\textwidth]{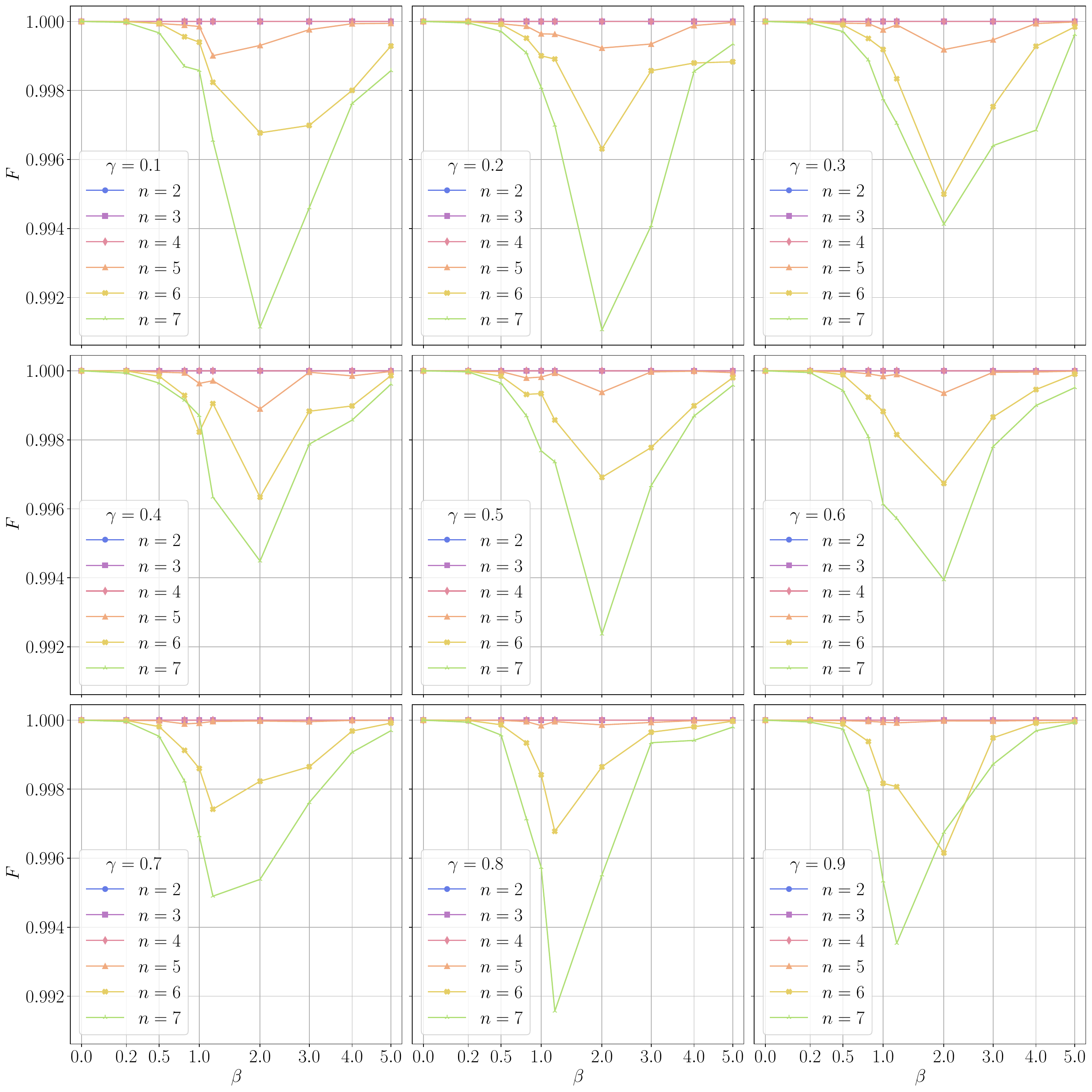}
    \caption[Fidelity of the Gibbs state preparation algorithm via statevector simulations for the XY model.]{Fidelity $F$, of the obtained state via statevector simulations (using \ac{BFGS}) with the exact Gibbs state, vs inverse temperature $\beta$, for two to seven qubits of the XY model with $\gamma$ between 0.1 and 0.9, and $h = 0.5$. A total of ten runs are made for each point, with the optimal state taken to be the one that maximises the fidelity. Lines are for guiding the eyes.}
    \label{fig:statevector_fidelity_XY}
\end{figure}

Fig.~\figref{fig:statevector_fidelity_XY} shows the fidelity of the generated mixed state when compared with the exact Gibbs state of the XY model with $h = 0.5$, and $\gamma$ ranging from 0.1 to 0.9 in steps of 0.1, across a range of temperatures for system sizes between two to seven qubits. The \ac{VQA} was carried out using statevector simulations with the \ac{BFGS} optimiser~\cite{Nocedal2006}. We used $n - 1$ layers for both the ancilla ansatz and for the system ansatz, with the scaling highlighted in Table~\tabref{tab:scaling_XY}. Similar to the Ising model, the number of layers was chosen to satisfy, at most, a polynomial scaling in quantum resources (but linear in depth), while achieving a fidelity higher than 0.99. Furthermore, in order to alleviate the issue of getting stuck in local minima, the number of runs was chosen to be ten, with the global minimum taken to be the minimum over all runs.

\begin{table}[t]
\centering
\begin{tabular}{|l|l|l|} 
    \hline
    \# of parameters & $n(3n - 2)$ & $\cO(n^2)$ \\ 
    \hline
    \# of \textsc{CNOT} gates & $3n^2 - 2n$ & $\cO(n^2)$ \\
    \hline
    \# of $\sqrt{X}$ gates & $2n(4n - 3)$ & $\cO(n^2)$ \\ 
    \hline
    Circuit depth & $3P(n - 1) + 1$ & $\cO(n)$ \\
    \hline
\end{tabular}
\caption[Specific \acs{PQC} scaling of the Gibbs state preparation algorithm for the XY model.]{Scaling of the \ac{VQA} assuming a closed ladder connectivity for the XY model, for $n > 2$, with $l_A = n - 1$, and $l_S = n - 1$, and $P$ is 2 when $n$ is even and 3 when $n$ is odd, obtained from~\tabref{tab:scaling}. Circuit depth is calculated on the depth of \textsc{CNOT} gates.}
\label{tab:scaling_XY}
\end{table}

A total of ten runs of \ac{BFGS} per $\beta$ were carried out to verify the reachability of the \ac{PQC}, with Fig.~\figref{fig:statevector_fidelity} showcasing the maximal fidelity achieved for each $\beta$ out of all runs. The results show that, indeed, our \ac{VQA}, is able to reach a very high fidelity $F > 0.99$ for up to seven-qubit Gibbs states of the XY model. In the case of the extremal points, that is $\beta \rightarrow 0$ and $\beta \rightarrow \infty$, the fidelity reaches unity, for all investigated system sizes. The same dip in fidelity around intermediary temperatures can be seen in the results of the XY model.

\section{Performance of the VQA on the XXZ Model} \label{sec:results_heisenberg}

In this Section, to further explore the feasibility of the \ac{VQA} for few-body thermal state preparation on NISQ devices, its performance is assessed on a more complex, interacting system: the Heisenberg model. The Heisenberg XXZ model is defined as
\begin{equation}
    \cH = -\frac{1}{4}\sum_{i=0}^{n-1} \left( \sigma_i^x \sigma_{i+1}^x + \sigma_i^y \sigma_{i+1}^y + \Delta \sigma_i^z \sigma_{i+1}^z \right) - h\sum_{i=0}^{n-1} \sigma_i^z.
\end{equation}
At variance with the Ising model investigated in Section~\secref{sec:results_ising}, the so-called XXZ model in a transverse magnetic field is an interacting Hamiltonian once mapped into spinless fermions. The phase diagram is much more complex and exhibits a paramagnetic-to-ferromagnetic phase transition at $h=\left(1-\Delta\right)/2$~\cite{Franchini2017}.

\begin{figure}[t]
    \centering
    \includegraphics[width=\textwidth]{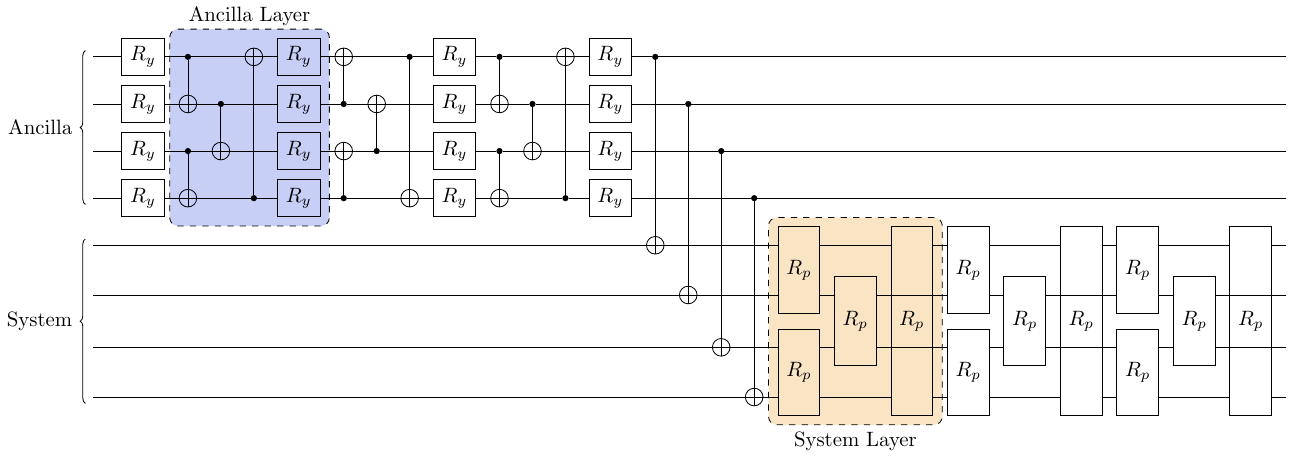}
    \caption[Example of an eight-qubit Gibbs state preparation \ac{PQC} for the XXZ model.]{Example of an eight-qubit \ac{PQC}, consisting of three ($n - 1$) ancilla layers acting on a four-qubit register, and three ($n - 1$) system layers acting on another four-qubit register. Each $R_y$ gate is parametrised with one parameter $\theta_i$, while each $R_p$ gate has two parameters $\varphi_i$ and $\varphi_j$. The $R_p$ gate is defined in Eq.~\eqref{eq:R_p}. Note that the intermediary \textsc{CNOT} gates, as well as the $R_p$ gates acting on qubits two and three, and on qubits one and four of the system, can be carried out in parallel, respectively.}
    \label{fig:pqc_4qubits_heisenberg}
\end{figure}

The Heisenberg model also commutes with the parity operator $\cP=\bigotimes_{i=0}^{n - 1} \sigma^z_i$. As such, $U_S$ is the same as in Section~\secref{sec:results_ising}. On the other hand, an alternating-layered ansatz for $U_A$ is used, with parametrised $R_y(\theta_i)$ gates, and \textsc{CNOT} gates as the entangling gates. $U_A$ was changed to attempt to obtain high fidelity results, while attaining a circuit depth that is independent of $n$. Once again, this ansatz is hardware efficient and is sufficient to produce real amplitudes for preparing the Boltzmann distribution. The fidelity is employed as a figure of merit for quantifying the performance of the algorithm. In this case, the scaling of $U_A$ and $U_S$ are given in Table~\tabref{tab:scaling_4}. Moreover, an example of a \ac{PQC}, split into a four-qubit ancilla register, and a four-qubit system register, is shown in Fig.~\figref{fig:pqc_4qubits_heisenberg}.

\begin{table}[t]
\centering
\begin{tabular}{|l|l|l|} 
    \hline
    \# of parameters & $n(l_A + 1) + 2nl_S$ & $\cO(n(l_A + l_S))$ \\ 
    \hline
    \# of \textsc{CNOT} gates & $nl_A + 2nl_S + n$ & $\cO(n(l_A + l_S))$ \\
    \hline
    \# of $\sqrt{X}$ gates & $2n(l_A + 1) + 6nl_S$ & $\cO(n(l_A + l_S))$ \\ 
    \hline
    Circuit depth & $Pl_A + 2Pl_S + 1$ & $\cO(l_A + l_S)$ \\ 
    \hline
\end{tabular}
\caption[\acs{PQC} scaling of the Gibbs state preparation algorithm for the XXZ model.]{Scaling of the \ac{VQA} assuming a closed ladder connectivity for the XXZ model, with $n > 2$, and $l_A$ and $l_S$ are the number of ancilla ansatz and system ansatz layers, respectively, and $P$ is 2 when $n$ is even and 3 when $n$ is odd. The depth only counts \textsc{CNOT} gates.}
\label{tab:scaling_4}
\end{table}

\subsection{Statevector Results}

\begin{figure}[t]
    \centering
    \includegraphics[width=\textwidth]{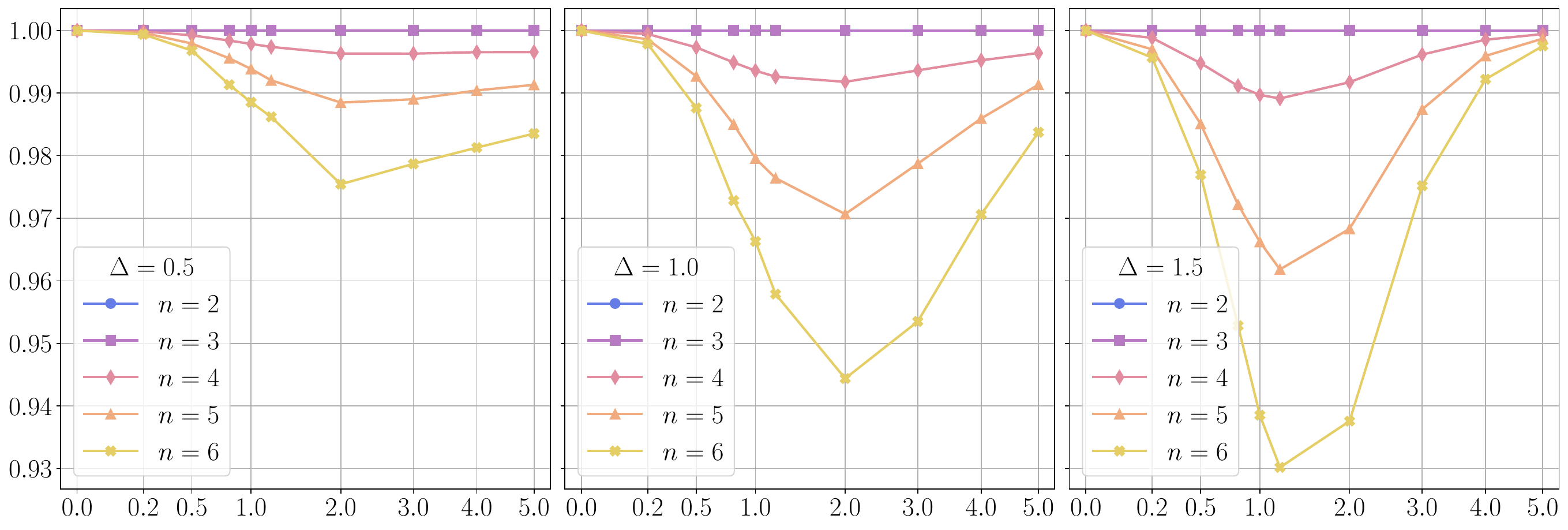}
    \caption[Fidelity of the Gibbs state preparation algorithm via statevector simulations for the XXZ model.]{Fidelity $F$ of the obtained state via statevector simulations (using \ac{BFGS}) with the exact Gibbs state, vs inverse temperature $\beta$, for two to six qubits of the XXZ model with $h = 0.5$ and $\Delta = -0.5, 0.0, 0.5$. A total of 100 runs are made for each point, with the optimal state taken to be the one that maximises the fidelity. Lines are for guiding the eyes.}
    \label{fig:heisenberg_statevector}
\end{figure}

Fig.~\figref{fig:heisenberg_statevector} shows the fidelity of the prepared mixed state when compared with the exact Gibbs state of the XXZ model with $h = 0.5$ and $\Delta = -0.5, 0.0, 0.5$, respectively, across a range of temperatures for system sizes between two to six qubits. The specific values of $\Delta$ and $h$ were chosen so as to represent different quantum phases in the zero-temperature regime~\cite{Franchini2017}. The \ac{VQA} was carried out using statevector simulations with the \ac{BFGS} optimiser~\cite{Nocedal2006}. $n - 1$ layers are used for both the ancilla ansatz and for the system ansatz, with the scaling highlighted in Table~\tabref{tab:scaling_5}. The number of layers was heuristically chosen to satisfy, at most, a polynomial scaling in quantum resources, while achieving a fidelity higher than 0.98 in statevector simulations. Furthermore, in order to alleviate the issue of getting stuck in local minima, the optimiser is embedded in a Monte Carlo framework.

\begin{table}[t]
\centering
\begin{tabular}{|l|l|l|} 
    \hline
    \# of parameters & $3n^2 - 2n$ & $\cO(n^2)$ \\ 
    \hline
    \# of \textsc{CNOT} gates & $3n^2 - 2n$ & $\cO(n^2)$ \\
    \hline
    \# of $\sqrt{X}$ gates & $8n^2 - 6n$ & $\cO(n^2)$ \\ 
    \hline
    Circuit depth & $3Pn - 3P + 1$ & $\cO(n)$ \\
    \hline
\end{tabular}
\caption[Specific \acs{PQC} scaling of the Gibbs state preparation algorithm for the XXZ model.]{Scaling of the \ac{VQA} assuming a closed ladder connectivity for the XXZ, with $n > 2$, $l_A = n - 1$, $l_S = n - 1$, and $P$ is 2 when $n$ is even and 3 when $n$ is odd, obtained from Tab.~\tabref{tab:scaling_4}. The depth only counts \textsc{CNOT} gates.}
\label{tab:scaling_5}
\end{table}

Similar to the statevector results of the Ising model in Section~\secref{sec:statevector_results}, very high fidelities $F > 0.98$ are obtained for a number of qubits ranging from two to six, across a broad range of temperatures of the Heisenberg XXZ model. It must be noted that the paramagnetic-to-ferromagnetic transition point lies at $\Delta = 0$ for $h = 0.5$, resulting in the non-interacting \textit{XX} model, achieving a much better performance. The same dip in fidelity at intermediary temperatures reappears at around $\beta \sim 1$ for all the plots in Fig.~\figref{fig:heisenberg_statevector}.

\subsection{Shot-Based Results}

\begin{figure}[t]
    \centering
    \includegraphics[width=\textwidth]{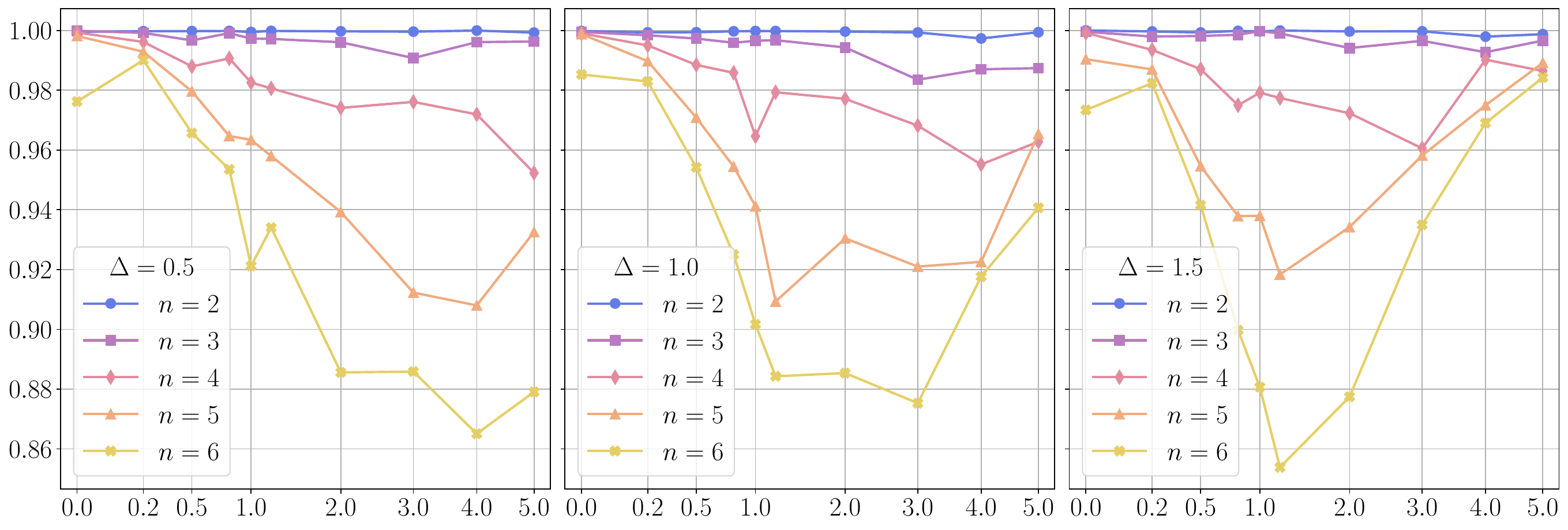}
    \caption[Fidelity of the Gibbs state preparation algorithm via shot-based simulations for the XXZ model.]{Fidelity $F$ of the obtained state via shot-based simulations (using \ac{SPSA}) with the exact Gibbs state, vs inverse temperature $\beta$, for two to six qubits of the XXZ model with $h = 0.5$ and $\Delta = -0.5, 0.0, 0.5$. A total of 100 runs are made for each point, with the optimal state taken to be the one that maximises the fidelity. Lines are for guiding the eyes.}
    \label{fig:heisenberg_shots}
\end{figure}

The next step was to carry out shot-based simulations for the XXZ model with $h = 0.5$ and $\Delta = -0.5, 0.0, 0.5$, respectively, as shown in Fig.~\figref{fig:heisenberg_shots}. Using \ac{SPSA}, ten runs were carried out for each $\beta$, while the number of iterations was taken to be $100n$ for each run, with $2n$ function evaluations at each iteration to estimate the gradient in $n$ random directions, i.e. $200n^2$. Similar to the number of layers, the choice of the number of function evaluations was heuristically chosen so that the scaling is polynomial. The number of commuting sets of Pauli strings of the XXZ model is three. Furthermore, each circuit was also measured with 1024 shots. and the \texttt{M3} package~\cite{Nation2021} was also utilised to perform error mitigation. Table~\tabref{tab:scaling_6} shows a summary of the optimisation scaling.

\begin{table}[t]
\centering
\begin{tabular}{|l|l|l|}
    \hline
    \# of iterations for each run & $100n$ & $\cO(n)$ \\
    \hline
    \# of function evaluations for each run & $200n^2$ & $\cO(n^2)$ \\
    \hline
    \# of circuits per function evaluation & $3$ & $\cO(1)$ \\
    \hline
    \# of circuit evaluations for each run & $600n^2$ & $\cO(n^2)$ \\
    \hline
    \# of shots for each circuit evaluation & $1024$ & $\cO(1)$ \\
    \hline
\end{tabular}
\caption{Scaling of \acs{SPSA} for shot-based simulations for the XXZ model.}
\label{tab:scaling_6}
\end{table}

Naturally, the choice of optimiser, along with the finite number of measurements used to reconstruct both the von Neumann entropy and the expectation value, are shown to affect the performance of the \ac{VQA}. Nevertheless, while the results of Fig.~\figref{fig:heisenberg_shots} exhibit a lower, albeit relatively high fidelity $F \gtrsim 0.93$, the \ac{VQA} shows notable promise in being able to produce Gibbs states of complex interacting Hamiltonians, such as the Heisenberg XXZ model.

\section{Preparing the Boltzmann Distribution using the Grover--Rudolph Ansatz}
\label{sec:boltzmann_distribution}

This Section explores the method for preparing the Boltzmann distribution of the XY model in greater detail. We apply the \ac{VQA} to the XY Hamiltonian utilising the \ac{GR} ansatz~\cite{Grover2002} to prepare the Boltzmann distribution, highlighting how the symmetries of the model can be utilised to reduce the number of variational parameters in the circuit. We then use a parity-preserving ansatz to diagonalise the Hamiltonian. For the sake of completeness, a thorough derivation of the exact solution of the XY model is considered in Appendix~\secref{app:xy}, which, due to the quadratic nature of the Hamiltonian in the fermionic representation, can be achieved both for finite and infinite system sizes.

\subsection{The Grover--Rudolph Ansatz} \label{sec:GR_ansatz}

Grover and Rudolph~\cite{Grover2002} initially described a procedure for preparing log-concave probability density functions on a quantum computer. However, it was later proven that this approach does not achieve a quantum speed-up in general~\cite{Herbert2021}. Nevertheless, it is still interesting to utilise the \ac{GR} ansatz as a proof-of-principle application of exploiting symmetries of a model to reduce the number of parameters needed for preparing its Boltzmann distribution.

To prepare the Boltzmann distribution of the XY model, we take the approach of utilising a variational approach for the \ac{GR} ansatz. In this case, an $n$-qubit ansatz requires $2^n - 1$ parameters, and $2^n - 1$ parametrised $R_y$ gates, most of which are multi-controlled. As a result, since the \ac{GR} ansatz serves to create a probability distribution from the entries of the first column, it utilises only parametrised orthogonal (real unitary) gates. Note that an extended algorithm was proposed by \citet{Benenti2009} to obtain an optimal purification of an $n$-qudit state.

\begin{figure}[t]
    \centering
    \includegraphics[width=\textwidth]{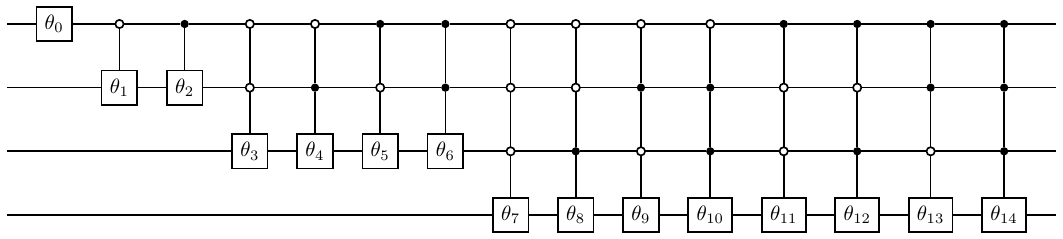}
    \caption[\acs{GR} ansatz for preparing an arbitrary probability distribution embedded in a real quantum state.]{\ac{GR} ansatz for preparing an arbitrary probability distribution embedded in a real quantum state. Each gate is represented an $R_y$ gate with its respective parameter $\theta_i$. Black dots represent the control qubit activating on $\ket{1}$, while the white dots represent the control qubit activating on $\ket{0}$.}
    \label{fig:GR_circuit}
\end{figure}

Let us consider the case of $n = 4$, with Fig.~\figref{fig:GR_circuit} showing the circuit. Each box represents the $R_y$ gate with the relevant parameter, and the black and white dots represent the control qubits (acting on 1 and 0, respectively). For example, the gate with $R_y(\theta_9)$ acts only on qubit 3 if qubit 0 is in the state $\ket{0}$, qubit 1 is in the state $\ket{1}$, and qubit 2 is in the state $\ket{0}$. Since there are $2^k$ possible binary controls for a $k$-qubit controlled gate, one can see how the circuit scales exponentially, both in number of gates and number of parameters.

The action of the circuit $U$ in Fig.~\figref{fig:GR_circuit} acting on the all-zero four-qubit state results in
\begin{equation}
    \label{eq:U}
    U\ket{0000} = \sum_{i=0000}^{1111} \sqrt{p_i} \ket{i}.
\end{equation}
As a result, when measuring the quantum state in Eq.~\eqref{eq:U} in the computational basis, we end up with the following probabilities for each configuration:
{\allowdisplaybreaks
\begin{subequations}\label{eq:ration}
\begin{align}
    \cos^2\left(\frac{\theta_0}{2}\right) \cos^2\left(\frac{\theta_1}{2}\right) \cos^2\left(\frac{\theta_3}{2}\right) \cos^2\left(\frac{\theta_7}{2}\right) &= p_{0000}, \\
    \cos^2\left(\frac{\theta_0}{2}\right) \cos^2\left(\frac{\theta_1}{2}\right) \cos^2\left(\frac{\theta_3}{2}\right) \sin^2\left(\frac{\theta_7}{2}\right) &= p_{0001}, \\
    \cos^2\left(\frac{\theta_0}{2}\right) \cos^2\left(\frac{\theta_1}{2}\right) \sin^2\left(\frac{\theta_3}{2}\right) \cos^2\left(\frac{\theta_8}{2}\right) &= p_{0010}, \\
    \cos^2\left(\frac{\theta_0}{2}\right) \cos^2\left(\frac{\theta_1}{2}\right) \sin^2\left(\frac{\theta_3}{2}\right) \sin^2\left(\frac{\theta_8}{2}\right) &= p_{0011}, \\
    \cos^2\left(\frac{\theta_0}{2}\right) \sin^2\left(\frac{\theta_1}{2}\right) \cos^2\left(\frac{\theta_4}{2}\right) \cos^2\left(\frac{\theta_9}{2}\right) &= p_{0100}, \\
    \cos^2\left(\frac{\theta_0}{2}\right) \sin^2\left(\frac{\theta_1}{2}\right) \cos^2\left(\frac{\theta_4}{2}\right) \sin^2\left(\frac{\theta_9}{2}\right) &= p_{0101}, \\
    \cos^2\left(\frac{\theta_0}{2}\right) \sin^2\left(\frac{\theta_1}{2}\right) \sin^2\left(\frac{\theta_4}{2}\right) \cos^2\left(\frac{\theta_{10}}{2}\right) &= p_{0110}, \\
    \cos^2\left(\frac{\theta_0}{2}\right) \sin^2\left(\frac{\theta_1}{2}\right) \sin^2\left(\frac{\theta_4}{2}\right) \sin^2\left(\frac{\theta_{10}}{2}\right) &= p_{0111}, \\
    \sin^2\left(\frac{\theta_0}{2}\right) \cos^2\left(\frac{\theta_2}{2}\right) \cos^2\left(\frac{\theta_5}{2}\right) \cos^2\left(\frac{\theta_{11}}{2}\right) &= p_{1000}, \\
    \sin^2\left(\frac{\theta_0}{2}\right) \cos^2\left(\frac{\theta_2}{2}\right) \cos^2\left(\frac{\theta_5}{2}\right) \sin^2\left(\frac{\theta_{11}}{2}\right) &= p_{1001}, \\
    \sin^2\left(\frac{\theta_0}{2}\right) \cos^2\left(\frac{\theta_2}{2}\right) \sin^2\left(\frac{\theta_5}{2}\right) \cos^2\left(\frac{\theta_{12}}{2}\right) &= p_{1010}, \\
    \sin^2\left(\frac{\theta_0}{2}\right) \cos^2\left(\frac{\theta_2}{2}\right) \sin^2\left(\frac{\theta_5}{2}\right) \sin^2\left(\frac{\theta_{12}}{2}\right) &= p_{1011}, \\
    \sin^2\left(\frac{\theta_0}{2}\right) \sin^2\left(\frac{\theta_2}{2}\right) \cos^2\left(\frac{\theta_6}{2}\right) \cos^2\left(\frac{\theta_{13}}{2}\right) &= p_{1100}, \\
    \sin^2\left(\frac{\theta_0}{2}\right) \sin^2\left(\frac{\theta_2}{2}\right) \cos^2\left(\frac{\theta_6}{2}\right) \sin^2\left(\frac{\theta_{13}}{2}\right) &= p_{1101},\\
    \sin^2\left(\frac{\theta_0}{2}\right) \sin^2\left(\frac{\theta_2}{2}\right) \sin^2\left(\frac{\theta_6}{2}\right) \cos^2\left(\frac{\theta_{14}}{2}\right) &= p_{1110}, \\
    \sin^2\left(\frac{\theta_0}{2}\right) \sin^2\left(\frac{\theta_2}{2}\right) \sin^2\left(\frac{\theta_6}{2}\right) \sin^2\left(\frac{\theta_{14}}{2}\right) &= p_{1111}.
\end{align}
\end{subequations}
}
If one knows beforehand the discrete probability distribution, then, by taking specific ratios of expressions in Eqs.~\eqref{eq:ration}, one can relate the parameters $\theta_i$ to ratios of probabilities, e.g., 
\begin{align}
    \frac{p_{1111}}{p_{1110}}=\tan^2\left(\frac{\theta_{14}}{2}\right)\implies \theta_{14}=2 \arctan\left( \sqrt{\frac{p_{1111}}{p_{1110}}}\right).
\end{align}
By iterating this procedure for all $\theta$, one can readily obtain
\begin{equation}
    \theta_m = 2\arctan\left(\sqrt{\frac{\sum_{i=0}^{k-1} p_{2^k l + i + k}}{\sum_{j=0}^{k-1} p_{2^k l + j}}}\right),
\end{equation}
where $k = n - \floor{\log_2 m}$ and $l = m - 2^{\floor{\log_2 m}} + 1$, with $m$ ranging from $0, \dots, 2^{n - 1} - 1$. Therefore, a variational procedure is not necessary to load a known discrete probability distribution onto a quantum state.

\subsection{Applying the Grover--Rudolph Ansatz to the XY Model}

To illustrate an instance of the Boltzmann distribution preparation algorithm, via the \ac{GR} ansatz, where the degeneracies of the energy levels allow for a reduction of the number of variational parameters needed in the \ac{PQC}, let us restrict to a few-body scenario, specifically the $n = 4$ XY chain. This allows us to better keep track of the protocol relying on the analytic expressions of the eigenenergies.

\begin{figure}[t]
    \centering
    \includegraphics[width=\textwidth]{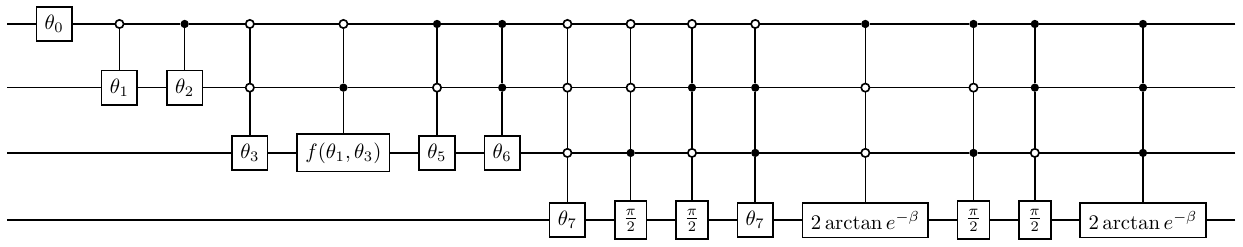}
    \caption[Four-qubit \acs{GR} ansatz with substituted parameters.]{Four-qubit \ac{GR} ansatz with substituted parameters (refer to Fig.~\figref{fig:GR_circuit}), specifically using Eqs.~\eqref{eq:first}~-~\eqref{eq:last}, with $f(\theta_1, \theta_3)$ defined in Eq.~\eqref{eq:f}. The ansatz results in 8 less parameters from an initial total of 15. Furthermore, combinations of multi-controlled gates with equal parameters, such as the $\pi/2$ gates, may possibly simplify the circuit even further.}
    \label{fig:GR_circuit_substituted}
\end{figure}

Since the XY model is non-interacting, the spectrum is given by the single-particle energies. The model is also symmetric: when the eigenenergies are calculated with the modes ordered lexicographically, we can assign the first half of the computational basis states to cover $\left\{p_n^+\right\}$, with the latter half to cover $\left\{p_n^-\right\}$. The eigenenergies are:
\begin{subequations}
\begin{align}
      E_0^+ &= \frac{-\sqrt{\gamma ^2+2 h \left(h-\sqrt{2}\right)+1}-\sqrt{\gamma ^2+2 h
   \left(h+\sqrt{2}\right)+1}}{\sqrt{2}}, \\
    E_1^+ &= \frac{\sqrt{\gamma ^2+2 h \left(h-\sqrt{2}\right)+1}-\sqrt{\gamma ^2+2 h
	   \left(h+\sqrt{2}\right)+1}}{\sqrt{2}}, \\
    E_2^+ &= E_3^+ = E_4^+ = E_5^+ = 0, \\
    E_6^+ &= \frac{-\sqrt{\gamma ^2+2 h
   \left(h-\sqrt{2}\right)+1}+\sqrt{\gamma ^2+2 h \left(h+\sqrt{2}\right)+1}}{\sqrt{2}}, \\
    E_7^+ &= \frac{\sqrt{\gamma ^2+2 h \left(h-\sqrt{2}\right)+1}+\sqrt{\gamma ^2+2 h
   \left(h+\sqrt{2}\right)+1}}{\sqrt{2}}, \\
    E_0^- &= -1-\sqrt{\gamma ^2+h^2},~
    E_1^- = 1-\sqrt{\gamma ^2+h^2}, \\
    E_2^- &= E_3^- = -E_4^- = -E_5^- = -h, \\
    E_6^- &= -1+\sqrt{\gamma ^2+h^2},~
    E_7^- = 1+\sqrt{\gamma ^2+h^2}.
\end{align}
\end{subequations}
We start to look at the Boltzmann coefficients generated by the eigenenergies in the negative parity sector. In the case of 
\begin{align}
    \frac{p_{1111}}{p_{1110}} &= \tan^2\left(\frac{\theta_{14}}{2}\right) = e^{-\beta\left(E_{7}^- - E_{6}^-\right)} = e^{-2\beta} \implies \theta_{14} = 2\arctan\left(e^{-\beta}\right).
    \label{eq:first}
\end{align}
Next we find that
\begin{align}
    \frac{p_{1101}}{p_{1100}} &= \tan^2\left(\frac{\theta_{13}}{2}\right) = e^{-\beta\left(E_{5}^- - E_{4}^-\right)} = 1 \implies \theta_{13} = \frac{\pi}{2},
\end{align}
and
\begin{align}
    \frac{p_{1011}}{p_{1010}} &= \tan^2\left(\frac{\theta_{12}}{2}\right) = e^{-\beta\left(E_{3}^- - E_{2}^-\right)} = 1 \implies \theta_{12} = \frac{\pi}{2},
\end{align}
due to the pairs of degeneracies. Furthermore,
\begin{align}
    \frac{p_{1001}}{p_{1000}} = \tan^2\left(\frac{\theta_{11}}{2}\right) = e^{-\beta\left(E_{1}^- - E_{0}^-\right)} = e^{-2\beta} \implies \theta_{11} = 2\arctan\left(e^{-\beta}\right).
\end{align}
Now looking at the Boltzmann coefficients given by the eigenenergies in the positive parity sector:
\begin{align}
    \frac{p_{0101}}{p_{0100}} = \tan^2\left(\frac{\theta_{9}}{2}\right) = e^{-\beta\left(E_{5}^+ - E_{4}^+\right)} = 1 \implies \theta_{9} = \frac{\pi}{2},
\end{align}
and
\begin{align}
    \frac{p_{0011}}{p_{0010}} = \tan^2\left(\frac{\theta_{8}}{2}\right) = e^{-\beta\left(E_{3}^+ - E_{2}^+\right)} = 1 \implies \theta_{8} = \frac{\pi}{2}.
\end{align}
Moreover, while we cannot determine the exact value of $\theta_{10}$ and $\theta_7$, due to their dependency on $\gamma$ and $h$, we can still relate them through
\begin{align}
    \frac{p_{0111}}{p_{0110}}= \frac{p_{0001}}{p_{0000}} 
    \implies \tan^2\left(\frac{\theta_{10}}{2}\right)= \tan^2\left(\frac{\theta_7}{2}\right) 
    \implies \theta_{10}= \theta_7.
\end{align}
Finally, we also have that
\begin{align}
    \frac{p_{0101}}{p_{0011}} &= \frac{\cos^2\left(\frac{\theta_{4}}{2}\right)}{\sin^2\left(\frac{\theta_{3}}{2}\right)} \tan^2\left(\frac{\theta_{1}}{2}\right) = e^{-\beta\left(E_{5}^+ - E_{3}^+\right)} = 1 \nonumber \\
    &\implies \theta_{4} = 2\arccos\left(\frac{\sin\left(\frac{\theta_{3}}{2}\right)}{\tan\left(\frac{\theta_{1}}{2}\right)}\right).
    \label{eq:last}
\end{align}
Thus, we have managed to eliminate 8 parameters from a total of 15. This may be due to the fact that we have $n$ single-particle energy levels in both the positive and the negative parity subspace. However, in the negative parity subspace, the sum of the energies of the $\pi$- and 0-modes always adds up to 2, which imposes an additional constraint. This result may hint towards the fact that for the XY model composed of $n$ particles, the number of parameters needed to characterise its Boltzmann distribution is $2n - 1$. Fig.~\figref{fig:GR_circuit_substituted}, represents the \ac{GR} ansatz in Fig.~\figref{fig:GR_circuit}, with the reduced number of parameters obtained from Eqs.~\eqref{eq:first}~-~\eqref{eq:last}, with
\begin{equation}
    \label{eq:f}
    f(\theta_1, \theta_3) = 2\arccos\left(\frac{\sin\left(\frac{\theta_{3}}{2}\right)}{\tan\left(\frac{\theta_{1}}{2}\right)}\right).
\end{equation}

\subsection{Results}

We apply the Gibbs state preparation algorithm to the XY model once again, this time using the \ac{GR} ansatz in Fig.~\figref{fig:GR_circuit_substituted} as $U_A$, with $U_S$ once again the brick-wall structure of the parity-preserving gates $R_p$~\eqref{eq:R_p}.

\begin{figure}[t]
    \centering
    \includegraphics[width=\textwidth]{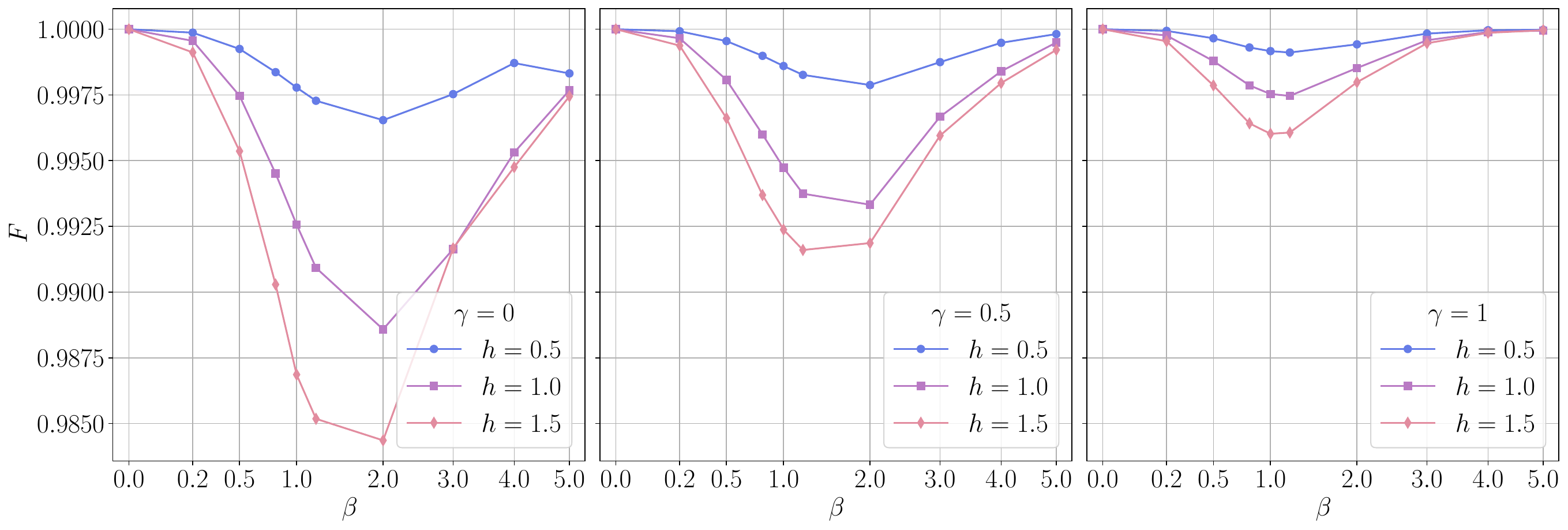}
    \caption[Results of applying the \acs{GR} ansatz to the XY model.]{Fidelity $F$ between the obtained statevector simulations (using \ac{BFGS}) and the exact Gibbs state, vs inverse temperature $\beta$, for the XY model with $h = 0.5, 1, 1.5$ and $\gamma=0, 0.5, 1$ for four qubits. A total of 100 runs are made for each point, with the optimal state taken to be the one that maximises the fidelity. Lines are for guiding the eyes.}
    \label{fig:fidelity_plot}
\end{figure}

Fig.~\figref{fig:fidelity_plot} shows the fidelity of the prepared Gibbs state with the theoretical Gibbs state of the four-qubit XY model at $h = 0.5, 1, 1.5$ and $\gamma=0, 0.5, 1$, across a broad range of temperatures. Three layers of both $U_A$ and $U_S$ were used for all simulations. The results show that the fidelity, while dipping at intermediary temperatures, still achieves values greater than 0.98 for all models and temperatures. Interestingly, the results for $\gamma=0$ appear to be slightly less optimal. This may be due to the fact that the \textit{XX} model belongs to a different universality class than the XY and Ising model, exhibiting an additional symmetry: the conservation of the total magnetisation.

The final fidelities obtained for the Ising model ($\gamma = 1$) are also similar to the ones obtained in Sec.~\secref{sec:statevector_results}, when using a hardware-efficient \ac{PQC} as $U_A$. However, the convergence of the optimiser towards the Boltzmann distribution was typically faster for the reduced \ac{GR} ansatz when compared to the hardware-efficient one, due to the smaller probed Hilbert space.

\section{Error Analysis of Entropy Estimation} \label{sec:entropy_estimation}

The \ac{VQA} for preparing Gibbs states requires generating a probability distribution on a quantum computer. In general, preparing a probability distribution faithfully requires an exponential number of shots, and particularly free fermion distributions can be hard to learn~\cite{Nietner2023}. However, let us look at estimating the entropy using the \ac{ML} estimator, rather than focusing on the generation of the probability distribution. The \ac{ML} estimator was shown to have a bias and variance that in general decreases as $\cO(M^{-1})$ for $M \gg B$~\cite{Paninski2003}, where $M$ is the number of shots and $B$ the number of bins. The outcome of one shot of a quantum circuit can be described by a multinomial distribution $\vec{p}$, where $p_i$ is the probability of observing a bit string $i$. Given $M$ shots, filling $B$ bins, the \ac{ML} estimator~\cite{Roulston1999, Paninski2003} of the von Neumann entropy is given by
\begin{equation}
    \cS_\text{ML} = -\sum_{i=0}^{B-1} q_i \log q_i,
\end{equation}
were $q_i = n_i / M$, such that $n_i$ represents the number of times the bit string $i$ appears given $M$ shots. The variance of the entropy can be easily computed as
\begin{equation}
    \bV(\cS_\text{ML}) = \sum_{i=0}^{B-1} (1 + \log q_i)^2 \bV(q_i),
\end{equation}
where $\bV(q_i) \approx q_i (1 - q_i) / M$. In fact, it can be shown that for all $M$, and all possible distributions, the variance of the \ac{ML} estimator for entropy is bounded above as
\begin{equation}
    \bV(\cS_\text{ML}) \leq \frac{(\log M)^2}{M},
\end{equation}
as proven by \citet{Antos2001}, and that the probability that the estimated entropy is more than $\epsilon$ away from the true entropy, decreases exponentially in the order of $\cO(M\epsilon^2/(\log M)^2)$, or written as a Chebyshev inequality:
\begin{equation}
    P(|\cS_\text{ML} - \bE(\cS_\text{ML})| > \epsilon) \leq 2e^{-\frac{M \epsilon^2}{2(\log(M))^2}}.
\end{equation}
It is important to note that this bound is not particularly tight, and it is independent of $B$ and the probability distribution. Moreover, the \ac{ML} estimator was proven to be negatively biased everywhere, such that
\begin{equation}
    \bE_{\vec{p}}(\cS_\text{ML}) \leq \cS(\vec{p}),
\end{equation}
where $\bE_{\vec{p}}$ denotes the conditional expectation given $\vec{p}$, and that equality is only achieved when $\cS(\vec{p}) = 0$, meaning the distribution is supported on a single point.\footnote{Or in the case of Boltzmann distributions for $\beta \rightarrow \infty$.} In the case of $M \gg B$, the Miller-Madow bias correction~\cite{Roulston1999, Paninski2003} gives that
\begin{equation}
    \cS(\vec{p}) = \cS_\text{ML}(\vec{p}) + \frac{B - 1}{2M} + \cO(M^{-1}).
\end{equation}
This bias correction is particularly relevant for simulations involving a small number of qubits.

Looking at \ac{NISQ} and near-term quantum algorithms, given that the Hilbert space of qubits grows as $B = 2^n$, we can reasonably assume that $M \ll B$ as soon as $n > 20$, corresponding to more than a million possible output bit-strings when measuring qubits. As a result, we need to look towards entropy estimation techniques when we are in a heavily undersampled regime. Bayesian inference is a typically employed method in these situations. While learning a probability distribution might generally require an exponential number of samples~\cite{Nietner2023}, computing functionals of such distributions might not. As a result, the \ac{NSB} estimator employs Bayesian inference to obtain both the entropy and its a posteriori standard deviation.  We utilise the Python package \texttt{ndd}~\cite{Marsili2023} to compute the $\cS_\text{NSB}$, while referring interested readers to \citet{Nemenman2002} and \citet{Nemenman2011} for the details.

\begin{figure}[!ht]
    \centering
    \includegraphics[width=\textwidth]{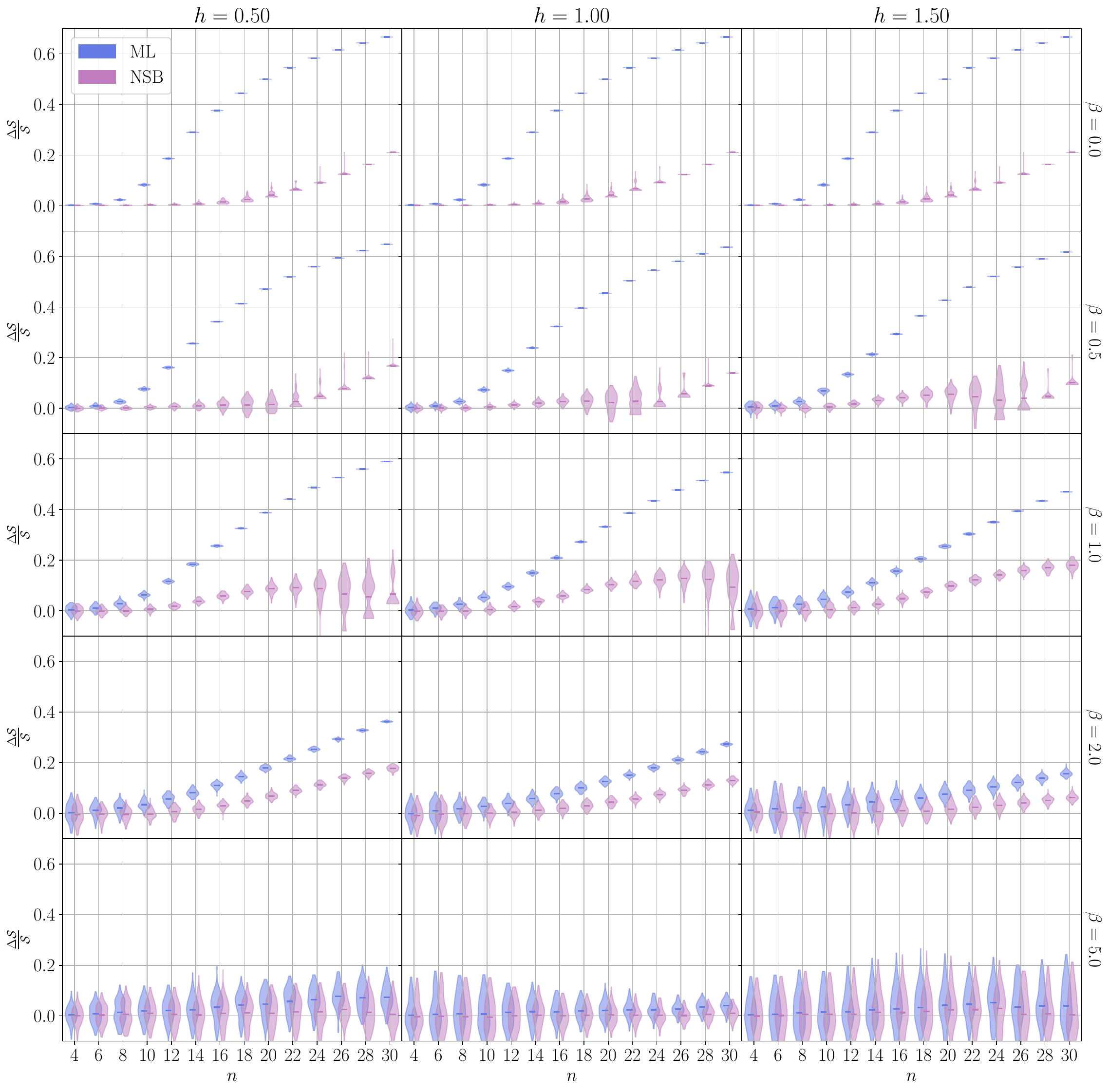}
    \caption[Violin plots of the relative error (bias) in entropy estimation as a function of the number of qubits, using the \acs{ML} and \acs{NSB} estimators.]{Violin plots of the relative error $\Delta\cS/\cS$ (bias) in entropy estimation as a function of the number of qubits $n$, using the \ac{ML} (blue) and \ac{NSB} (purple) estimators. The columns represent the Ising model with $h = 0.5, 1.0, 1.5$, and the rows represent $\beta = 0.0, 0.5, 1.0, 2.0, 5.0$, respectively. Each violin plot is obtained by calculating the entropy of $100$ samples taking $M = 1024$ shots. The mean is shown as the small flat horizontal lines, with the area around each mean representing the distribution of bias. The violin plots at each $n$ are shifted slightly from each tick to allow for better visualisation. Notice how, at each instance the \ac{NSB} estimator achieves a much lower bias than the \ac{ML} estimator, especially in the low $\beta$ regime.}
    \label{fig:entropy}
\end{figure}

Fig.~\figref{fig:entropy} shows the results of using the \ac{ML} and \ac{NSB} estimators for the Ising model. In particular we show how the relative error (bias) of the entropy scales as the number of qubits increases. Each violin plot in Fig.~\figref{fig:entropy} is obtained by calculating the entropy of $100$ samples taking $M = 1024$ shots.

As one can expect, for a number of qubits $n$ such that $2^n \ll M$, both \ac{ML} and \ac{NSB} estimators obtain an average bias close to zero, for all values of $\beta$. There is a transition region where $2^n \sim M$, where the bias, particularly for low values of $\beta$, starts to increase. In the region where $2^n \gg M$, the \ac{ML} estimator is only valid in the large $\beta$ regime, since there is usually a finite number of non-zero probabilities which is much smaller than $M$. It is important to note that although the error is increasing linearly with the number of qubits, the number of shots $M$ needed to reduce the bias to a constant error increases exponentially as a function of the number of qubits $n$. Specifically, in the region $2^n \gg M$, the \ac{ML} estimator reaches the upper bound of $\log(M)$, and so for $\beta = 0$, $\Delta\cS_\text{ML} = \log(2^n) - \log(M) = \log(2^n/M)$.

On the other hand, the \ac{NSB} estimator obtains a much lower bias, at the cost of a slightly higher standard deviation. While the behaviour of the \ac{NSB} estimator is hard to surmise given that exact diagonalisation results allowed us to analyse until $n = 30$, there are a few instances where for intermediate and high $\beta$ the estimator seems to flatten, or even decrease as $n$ increases. Proving that for a particular $\beta$, the \ac{NSB} estimator\footnote{Or any other entropy estimator for the matter.} reliably acquires a bias that scales as $\cO(\text{poly}(\log(n)))$, would mean that using a number of shots $M$ that scales polynomially would achieve a constant additive error, implying scalability in entropy estimation.

\section{Final Remarks}

The preparation of a thermal equilibrium state of a quantum many-body system on a \ac{NISQ} device was addressed, with the uniqueness of the Gibbs state, being the state that minimises the Helmholtz free energy, exploited to provide a faithful objective function for minimisation. The proposed \ac{VQA} consists of splitting the \ac{PQC} into two parametrised unitaries, one acting on an ancillary register, and one on a system register. The former is tasked with determining the weights of the Boltzmann distribution, corresponding to a given temperature, while the latter performs the rotation from the computational basis to the energy basis of a given Hamiltonian.

The \ac{VQA} was benchmarked by preparing the Gibbs state of the transverse field Ising model, obtaining fidelities $F \simeq 1$ for system sizes up to six qubits in statevector simulations, across a broad range of temperatures, with a slight dip at intermediate ones. Moreover, the \ac{VQA} was tested on the XY and Heisenberg XXZ models with a transverse field, similarly obtaining fidelities $F \simeq 1$ in statevector simulations. However, performance on current \ac{NISQ} devices, investigated both by noisy simulations and real-hardware execution on IBM devices, showed a degradation in the results of the \ac{VQA} with increasing system size. This may have been caused by the limited connectivity and the noise present in the device. Nevertheless, executing the \ac{VQA} on \ac{NISQ} devices still provides an improvement upon the recent developments in variational Gibbs state preparation.\footnote{See, e.g., \citet{Sagastizabal2021}.}

It is important to observe that the structure of the \ac{VQA} does not depend on the specific Hamiltonian to be tackled, nor on any prior knowledge of its spectrum.\footnote{Although this could possibly help improve the algorithm as seen in Sec.~\secref{sec:boltzmann_distribution}.} For example, the structure of the unitary $U_S(\bm{\varphi})$ could be adjusted in order to match some specific features of the eigenstates (if these are known), or the parametrised unitary $U_A(\bm{\theta})$ could be replaced by a deterministic procedure, such as the one reported by \citet{Sannia2023}, if the probabilities of the Boltzmann distribution are known. 

However, even without requiring any such knowledge, the `Hamiltonian-agnostic' variational approach gives an effective way to prepare Gibbs states of arbitrary quantum many-body systems on a quantum computer, providing an advancement over previous methods, especially thanks to the modular structure of the \ac{PQC}. This could significantly contribute to both performing quantum thermodynamical experiments on quantum computers, as well as faithfully preparing Gibbs states to be used in a great variety of computational tasks.

In fact, many modular elements of the \ac{VQA} have the capacity to be significantly improved. While the scope of this work was to provide a proof-of-concept \ac{VQA} for preparing Gibbs states by directly estimating the entropy without any truncation, potential avenues for future research are mentioned. In particular, more robust error mitigation techniques, such as those present in the \texttt{mitiq} library~\cite{Larose2022}, could be implemented. 

The \ac{GR} ansatz approach for the preparation of an arbitrary probability distribution function has also been investigated. By exploiting the symmetries of the XY model, a reduction of the number of parameters in the \ac{GR} ansatz was achieved. Although this approach was only applied to a small system, the results hint towards the fact that a sub-exponential number of variational parameters --- and possibly even gates, by combining similar multi-controlled gates~\cite{Mottonen2006} --- may be necessary for the preparation of a many-body Boltzmann distribution, by exploiting the degeneracies in the energy spectrum of Hamiltonians.

Choosing and/or developing the correct entropy estimation technique would also go a long way in minimising the error present in determining the von Neumann entropy. Furthermore, \acp{BP} in deep \acp{PQC} and noisy devices are also a considerable challenge to address. In Appendix~\secref{app:bp}, the requirements needed to investigate \acp{BP} for Gibbs state preparation are qualitatively discussed, as well as the \acp{PQC} that stem from the particular structure of the procedure. Lastly, it is worth noting that the choice of optimiser plays a pivotal role in the performance of the \ac{VQA}, particularly in the presence of noise. An in-depth analysis of various optimisers applied to the \ac{VQA} could significantly enhance its reliability.

The Python code for running the statevector simulations, using \texttt{Qulacs}~\cite{Suzuki2021}, and the noisy simulations, as well as the Runtime Program, using \texttt{Qiskit}~\cite{Qiskit_team2019a}, can be found at~\cite{Consiglio2023}.

%% file: chapter6/conclusion.tex
\chapter{Conclusion} \label{chap:6}

\acresetall

\epigraph{\textit{All quantumists should appreciate that our field can fulfil its potential only through sustained, inspired effort over decades. If we pay that price, the ultimate rewards will more than vindicate our efforts.}}{---~\citet{Preskill2018}}

\section{Outcomes of the Thesis}

In this thesis, several \acp{VQA} were investigated and applied to relevant problems, while showcasing results and discussing future work and improvements. 

The first of which was the \ac{VQE} applied to the SU($N$) Hubbard model in Chapter~\secref{chap:3}, where an extension of \ac{JW} encoding was applied, while the persistent current, a mesoscopic quantity, was used to probe quantum phase transitions. The \ac{PQC} utilised in the \ac{VQE} was a number-preserving \ac{HVA}, which ensured that spin configurations were preserved to find specific ground-states of the Hubbard model. Statevector results showed that the \ac{VQE} performed significantly well in the low-interaction regime (small $U$ and $V$), while for high-interaction (large $U$ and $V$), more layers were required to fully resolve the shape of the energy curve with respect to the varying magnetic flux. Optimisation with sampling was also a highly challenging task, and the results revealed that more research needs to be undertaken to optimise the procedure of finding the ground-state of the SU($N$) Hubbard models on \ac{NISQ} devices. A peer-reviewed publication of this work can be found in \citet{Consiglio2022a}.

The second study discussed in Chapter~\secref{chap:4} was the \ac{VSV}, a \ac{VQA} that finds the \ac{CSS} of arbitrary quantum states with respect to the \ac{HSD}. This enabled the characterisation of entanglement, as well as provided the possibility of constructing entanglement witnesses. The \ac{VSV} has been applied to \ac{GHZ} states, and compared with the performance of the \ac{QGA}, showing a notable improvement. It has also been applied to \ac{X-MEMS}, producing a relation between the \ac{GME} concurrence and \ac{HSE} measure induced by the \ac{HSD}. Further simulations carried out on two-qubit $X$-states demonstrated that while the equality (or majorisation) of the concurrence of two $X$-states immediately implies a similar equality (or majorisation) of the \ac{HSE}, this does not hold in general for the \ac{GME} concurrence and \ac{HSE} in the case of $X$-states with three or more qubits. A peer-reviewed publication of this work can be found in \citet{Consiglio2022b}.

Chapter~\secref{chap:5} investigates the third \ac{VQA}, which involved the preparation of Gibbs states on quantum devices. A novel design of a \ac{PQC} enables the von Neumann entropy to be easily computed after post-processing measurements carried out on ancillary qubits, while the Gibbs state is simultaneously prepared on system qubits by minimising the free energy. Statevector results showed that the algorithm correctly found the Gibbs states of the Ising model across a broad range of temperatures. Peer-reviewed publications of this work can be found in \citet{Consiglio2024a}, \citet{Consiglio2025}, and \citet{Consiglio2024b}.

\section{Future Prospects}

All the works discussed in this thesis have the potential for enhancement and expansion. This section will outline and discuss possible improvements.

In principle, most \acp{VQA} can be ``easily'' adapted by incorporating general enhancements to the structure of the underlying algorithms, particularly regarding noise mitigation. Several techniques were covered in Sec.~\secref{sec:error_mitigation}, and will not be reiterated here. Instead, we will explore potential avenues for further research and applications of the \acp{VQA} discussed in this thesis.

Regarding the work presented in Chapter~\secref{chap:3}, it would be interesting to consider SU($N$) generalisation of other two-component fermion-to-qubit mappings, such as the Bravyi--Kitaev~\cite{Bravyi2002}, Ball--Verstraete--Cirac~\cite{Ball2005, Verstraete2005}, among others, as well as various custom encodings.\footnote{It is important to note that alternative encodings may require non-trivial gates in the \ac{PQC} to maintain symmetry-preserving operations, unless such constraints are moved to the cost function, which can in turn make the optimisation much more difficult.}

With regards to the work in Chapter~\secref{chap:4}, future research could focus on enhancing the \ac{VSV} to determine the \ac{k-CSS} of an arbitrary quantum state, which is not straightforward, since even the closest biseparable state for three-qubit states is significantly challenging to implement. The issue lies in properly expressing a biseparable state in a variational form.\footnote{Since the number of pure state bipartitions scales as $2^{n-1} - 1$ for an $n$-qubit state.} The set of all biseparable states is generated by the convex hull of all pure state bipartitions. This, coupled with the fact that an $n$-qubit state can have rank up to $2^n$ makes the extension of the \ac{VSV} to determine the \ac{k-CSS} extremely non-trivial. The question of determining the \ac{k-CSS}, along with improvements to the performance of the \ac{VSV}, remains an open problem.

Concerning the work in Chapter~\secref{chap:5}, the algorithm can be adapted to operate outside the scope of just Gibbs state preparation, in particular for applications such as partition function evaluation and Hermitian matrix diagonalisation. 

The free energy of a Gibbs state is related to its partition function by
\begin{equation}
    \cF_\beta = -\frac{1}{\beta}\ln \cZ_\beta.
\end{equation}
Rearranging, we get
\begin{equation}
    \cZ_\beta = e^{-\beta \cF_\beta}.
\end{equation}
Thus, if we have computed the free energy, then we can also compute the partition function. Given an error in the free energy of $\Delta \cF_\beta = \sqrt{\ev*{\cF_\beta^2} - \ev{\cF_\beta}^2}$, then the resulting error in the partition function is 
\begin{equation}
    \Delta \cZ_\beta \approx \beta \cZ_\beta \Delta \cF_\beta.
\end{equation}
Since the Helmholtz free energy typically scales inversely as a function of $\beta$, then the error of the partition function should decrease as a function of $\beta$, implying that the evaluation of the partition function is more accurate at lower temperatures.

The system unitary is responsible for transforming the computational basis states --- input in the system register --- into the eigenstates of the thermal state, which coincide with the eigenstates of the Hamiltonian. As a result, the system unitary is an approximation to the diagonalising unitary of the Hermitian operator $\cH$, that is
\begin{equation}
    \cH \approx U_S D U_S^\dagger,
\end{equation}
where $D = \text{diag}(p_1, p_2, \dots, p_d)$. Furthermore, since
\begin{equation}
    p_i = \frac{e^{-\beta E_i}}{\cZ_\beta},
\end{equation}
then equivalently,
\begin{equation}
    E_i = \cF_\beta - \beta^{-1}\ln p_i.
\end{equation}
Assuming an error of $\Delta p_i$ for each $p_i$, and $\Delta \cF_\beta$ for $\cF_\beta$, we get
\begin{equation}
    \Delta E_i \approx \sqrt{\left(\Delta \cF_\beta \right)^2 + \left(\frac{\Delta p_i}{\beta p_i}\right)^2},
\end{equation}
and so we can compute all the eigenvalues (and their errors) of a Hamiltonian.

Due to the insufficient capabilities of current quantum technologies in carrying out prototypical quantum algorithms, some of which were developed in the 1990s and mathematically proven to achieve a form of quantum advantage, brought about the invocation of \acp{VQA}. Many scientists thought (and still do) that a hybrid approach may lead to quantum advantage, possibly by addressing simpler problems than initially envisioned. Although a simple Google search might suggest that quantum advantage has been in principle achieved, I am of the belief that we have not yet truly reached this milestone, since it has only been demonstrated in limited and specifically-constructed cases, particularly through the use of quantum computers developed by the likes of Google, IBM, and other corporations. Thus, the question remains, can \acp{VQA} perhaps offer a (sufficient and feasible) solution to a hard classical problem? 

Much research has been poured into the understanding of the main villain of \acp{VQA}, the \ac{BP}, and recently, \citet{Cerezo2024} have put forward a thorough analysis and lengthy discussion on this very problem. They present many important points, but to discuss and analyse all of them would result in another thesis, therefore I will focus on just one particularly relevant point. Trying to purposefully avoid \acp{BP} typically results in an algorithm that is classically simulable, rendering the entire point moot. However, there are a few caveats to this, namely that the work focused on objective functions which are simple expectation values of quantum operators, and may in fact not hold for \acp{VQA} built upon a quantum primitive, such as Gibbs state preparation, which one expects is hard to simulate classically. Furthermore, while provable absence of \acp{BP} does not equate to quantum advantage, it can still be the case that a classically equivalent algorithm, while in theory scalable, still operates on a timeline that is realistically intractable. Thus, discovering a quantum algorithm that, despite scaling exponentially, results in a realistic and feasible timeline whilst still being faster than its classical counterpart, will still be useful to investigate.

This indicates to me, in line with \citet{Cerezo2024}, that the \ac{VQA} research community needs to approach the framework in innovative ways, whilst avoiding the temptation to simply plug in a \ac{VQA} wherever a new objective function is concerned. While \acp{VQA} may not herald a new era of attaining exponential quantum advantage, it may still pave the way towards novel quantum technologies that have beneficial and valuable applications.

%% file: appendixA/vqe_appendix.tex
\chapter{Derivation and Implementation of the SU(\textit{N}) Fermi--Hubbard Model} \label{app:vqe}

\textit{Parts of this appendix are based on the published manuscript by \citet{Consiglio2022a}.}\\

The details of the SU($N$) magnetic-flux-induced Hubbard Hamiltonian shown in Eq.~\eqref{eq:hubbard_model} are defined and derived explicitly here. First we show how we obtained the qubit Hamiltonian from the SU($N$) \ac{JW} transformation, followed by extending the Hamiltonian to cater for both long-range hopping and interaction terms.

\section{Obtaining the Qubit Hamiltonian via the SU(\textit{N}) Mapping}

Starting with some preliminary calculations, notice that $\sigma_n^z \sigma_n^\pm = \mp \sigma_n^\pm$ and $\sigma_n^\pm \sigma_n^z = \pm \sigma_n^\pm$. Looking at only the nearest-neighbour hopping terms, and using the \ac{JW} transformation given in Eq.~\eqref{eq:map}, we find that
\begin{align}
    c^\dagger_{i, s}c_{i+1, s} = \left[\left(\bigotimes_{j<n}\sigma_j^z\right) \otimes \sigma_n^+\right] \left[\left(\bigotimes_{k<n+1}\sigma_k^z\right) \otimes \sigma_{n+1}^-\right] = \sigma_n^+\sigma_n^z\sigma_{n+1}^- = \sigma_n^+\sigma_{n+1}^-,
\end{align}
and trivially, $c^\dagger_{i+1, s}c_{i, s} = \sigma_n^-\sigma_{n+1}^+$. Note that if $i = L - 1$, then $i + 1 = 0\mod L$, meaning that
\begin{align}
    c^\dagger_{L-1, s}c_{0, s} 
    &= \left[\left(\bigotimes_{j<sL + L - 1}\sigma_j^z\right) \otimes \sigma_{sL + L - 1}^+\right] \left[\left(\bigotimes_{k<sL}\sigma_k^z\right) \otimes \sigma_{sL}^-\right] \nonumber \\
    &= \left(\bigotimes_{j=sL}^{sL + L - 2}\sigma_j^z\right)\sigma_{sL + L - 1}^+\sigma_{sL}^- \nonumber \\
    &= \left(\bigotimes_{j=sL + 1}^{sL + L - 2}\sigma_j^z\right)\sigma_{sL + L - 1}^+\sigma_{sL}^- \nonumber \\
    &= \sigma_{sL}^-\left(\bigotimes_{j=sL + 1}^{sL + L - 2}\sigma_j^z\right)\sigma_{sL + L - 1}^+,
\end{align}
and similarly for the \acl{hc} terms. This results in a parity term, made up of $(L-2)$ $\sigma^z$ terms, which is $-1$ at the ``looping'' index $i=L-1$ only when the number of spins of colour $s$, $N_s$, is odd, otherwise it is $+1$. Thus, in the final Hamiltonian we can replace the $\sigma^z$ terms with
\begin{equation}
    P_{i,s} = \begin{cases}
        -1,~\text{if } i = L - 1 \text{ and } N_s \text{ is odd,} \\
        +1,~\text{otherwise.}
        \end{cases}
\end{equation}
Now, since $n_{i, s} = c^\dagger_{i, s}c_{i, s}$, then
\begin{equation}
    n_{i, s} = \left[\left(\bigotimes_{j<n}\sigma_j^z\right) \otimes \sigma_n^+\right] \left[\left(\bigotimes_{k<n}\sigma_k^z\right) \otimes \sigma_n^-\right] = \sigma_n^+ \sigma_n^-,
\end{equation}
which can be rewritten as
\begin{equation}
    n_{i, s} = \dfrac{1 - \sigma_n^z}{2}.
\end{equation}
Therefore, looking at the interaction terms of $U$, we have
\begin{equation}
    n_{i, s}n_{i, s'} = \left(\dfrac{1 - \sigma_{sL + i}^z}{2}\right)\left(\dfrac{1 - \sigma_{s'L + i}^z}{2}\right),
\end{equation}
while for the interaction terms of $V$,
\begin{align}
    n_i n_{i + 1} = \left(\sum_s n_{i, s}\right) \left(\sum_{s'} n_{i+1, s'}\right) = \sum_{s, s'} \left(\dfrac{1 - \sigma_{sL + i}^z}{2}\right)\left(\dfrac{1 - \sigma_{s'L + i + 1}^z}{2}\right).
\end{align}
Thus, starting from our original fermionic Hamiltonian~\eqref{eq:hubbard_model}, and applying the SU($N$) \ac{JW} transformation~\eqref{eq:map}, we end up with the following representation:
\begin{align}
    \cH= &-t\sum\limits_{i, s} P_{i,s}\left(e^{\imath \frac{2 \pi \phi}{L}} \sigma_{sL + i}^+\sigma_{sL + i + 1}^- + \acs{hc}\right) \nonumber\\
    &+ \frac{U}{4}\sum\limits_{i, s < s'}\left( 1-\sigma_{sL + i}^z \right)\left( 1-\sigma_{s'L + i}^z \right) \nonumber \\
    &+ \frac{V}{4}\sum\limits_{i, s, s'} \left( 1-\sigma_{sL + i}^z \right)\left( 1-\sigma_{s'L + i +1}^z \right),
\end{align}
which is exactly the derived qubit Hamiltonian~\eqref{eq:hubbard_hamiltonian}.

\section{Extending the Hamiltonian}

The Hubbard Hamiltonian in Eq.~\eqref{eq:hubbard_model} can be generalised to incorporate symmetric long-range hopping and interaction terms, transforming into 
\begin{align}
\label{eq:hubbard_model1}
 \cH = -&\sum_{i=0}^{L-1} \sum_{s=0}^{N-1} \sum_{r=1}^{R_t} t_{r} \left(e^{\imath \frac{2 \pi \phi}{L}}  c_{i, s}^\dagger c_{i+r,s} +\acs{hc} \right) \nonumber \\ 
 +&\ U\sum_{i=0}^{L-1}\sum_{s=0}^{N-1}\sum_{s'=s+1}^{N-1} n_{i, s}n_{i, s'} \nonumber \\ 
 +&\ \sum_{i=0}^{L-1}\sum_{r=1}^{R_V} V_{r} n_{i} n_{i+r},
\end{align}
where $t_{r}$ is the hopping amplitude of a fermion between sites at a distance $r$, $U$ is the on-site interaction, and $V_{r}$ is the interaction between fermions at a distance $r$. $R_V$ and $R_t$ represent the range of the interacting $V$ and hopping $t$ terms, respectively, which can be at most $\floor{L/2}$, due to periodic boundary conditions.

Given that the SU($N$) fermion-to-qubit mapping is independent of the Hamiltonian parameters, Eq.~\eqref{eq:hubbard_model1} (which includes both site-dependent long-range hopping and interaction terms) can also be mapped to a qubit Hamiltonian, although with different geometry and interaction patterns:
\begin{align}
    \cH = &-\sum\limits_{i, s, r} t_r \left(e^{\imath \frac{2 \pi \phi}{L}} \sigma_{sL + i}^+\sigma_{sL + i + r}^- + \acs{hc} \right) \bigotimes_{j=sL + i + 1}^{sL + i + r - 1} \sigma_j^z \nonumber \\
    &+ \frac{U}{4}\sum\limits_{i, s < s'}\left( 1-\sigma_{sL + i}^z \right)\left( 1-\sigma_{s'L + i}^z \right)\nonumber\\
    &+ \sum\limits_{i,r, s, s'} \frac{V_{r}}{4} \left( 1-\sigma_{sL + i}^z \right)\left( 1-\sigma_{s'L + i + r}^z \right).
\label{eq:mapr}
\end{align}

\section{Implementation of the VQE}

Determining the ground-state energy of the Hubbard model requires the preparation of a suitable initial state, and a \ac{PQC} able to modify the quantum state, with observable measurements efficiently calculating the expectation value on the ground-state. As the number of fermions for each colour is conserved, the spin configuration of the initial state, i.e. the number of qubits in the $\ket{1}$ state, must be preserved with regards to the mapping. As a consequence, the gates in the \ac{PQC} must be in the form of Pauli terms present in the Hubbard Hamiltonian. This results in a form of \ac{HVA}, which can also be hardware-efficient if the quantum architecture supports the native implementation of \textsc{SWAP}-type gates, significantly reducing the depth of the \ac{PQC}~\cite{Ganzhorn2019,Sagastizabal2019}.

Note that the ladder operators $\sigma^+$, $\sigma^-$, of the transformed hopping terms can be rewritten in terms of $\sigma^x$ and $\sigma^y$ Pauli operators (ignoring qubit indices and $\sigma^z$ operators without loss of generality) as follows:
\begin{align}
\label{eq:hopping}
	&\left(e^{\imath\frac{2\pi\phi}{L}}\sigma^+\sigma^- + e^{-\imath \frac{2 \pi \phi}{L}}\sigma^-\sigma^+\right) \nonumber \\ 
    &= e^{\imath \frac{2 \pi \phi}{L}}\frac{(\sigma^x+\imath \sigma^y)}{2}\frac{(\sigma^x-\imath \sigma^y)}{2} + e^{-\imath \frac{2 \pi \phi}{L}}\frac{(\sigma^x-\imath \sigma^y)}{2}\frac{(\sigma^x+\imath \sigma^y)}{2} \nonumber \\ 
    &= \frac{1}{4}\left( e^{\imath\frac{2\pi\phi}{L}} \left(\sigma^x\sigma^x - \imath \sigma^x\sigma^y + \imath \sigma^y\sigma^x + \sigma^y\sigma^y\right) + e^{-\imath\frac{2\pi\phi}{L}}\left(\sigma^x\sigma^x + \imath \sigma^x\sigma^y - \imath \sigma^y\sigma^x + \sigma^y\sigma^y\right) \right) \nonumber \\ 
    &= \frac{1}{4} \left[ \left(e^{\imath\frac{2\pi\phi}{L}} + e^{-\imath\frac{2\pi\phi}{L}}\right)\left(\sigma^x\sigma^x + \sigma^y\sigma^y\right) - \imath\left(e^{\imath\frac{2\pi\phi}{L}} - e^{-\imath\frac{2\pi\phi}{L}}\right)\left(\sigma^x\sigma^y - \sigma^y\sigma^x\right) \right] \nonumber \\ 
    &= \frac{1}{2}\left[\cos\left(\frac{2\pi\phi}{L}\right)\left(\sigma^x\sigma^x + \sigma^y\sigma^y\right) + \sin\left(\frac{2\pi\phi}{L}\right)\left(\sigma^x\sigma^y - \sigma^y\sigma^x\right) \right].
\end{align}
We take the $(\sigma^x\sigma^x + \sigma^y\sigma^y)/2$ exponential operators to represent the sub-layer consisting of hopping terms, such that
\begin{equation}
        e^{-\imath\frac{\theta}{2}\left(\sigma^x\sigma^x + \sigma^y\sigma^y\right)} =
        \left( \begin{array}{cccc}
        1 & 0 & 0 & 0 \\
        0 & \cos(\theta) & -\imath\sin(\theta) & 0 \\
        0 & -\imath\sin(\theta) & \cos(\theta) & 0 \\
        0 & 0 & 0 & 1
        \end{array} \right) =
    \vcenter{\hbox{\includegraphics[width=0.17\textwidth]{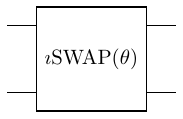}}}
\end{equation}
Similarly, for the interaction terms of $V$, we directly take the $(1-\sigma^z)(1-\sigma^z)/4$ terms in exponential form,
\begin{equation}
    e^{\imath\frac{\theta}{4}\left(1-\sigma^z\right)\left(1-\sigma^z\right)} =
    \left( \begin{array}{cccc}
    1 & 0 & 0 & 0 \\
    0 & 1 & 0 & 0 \\
    0 & 0 & e^{-\imath\theta} & 0 \\
    0 & 0 & 0 & e^{\imath\theta}
    \end{array} \right) =
    \vcenter{\hbox{\includegraphics[width=0.13\textwidth]{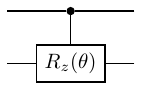}}}
\end{equation}
Finally, for the on-site interaction terms representing $U$, we simply take the individual exponential terms of $(1-\sigma^z)/2$ to represent the sub-layer, so that
\begin{equation}
    e^{\imath\frac{\theta}{2}\left(1-\sigma^z\right)} =
    \left( \begin{array}{cc}
		e^{-\imath\theta} & 0 \\
		0 & e^{\imath\theta}
	\end{array} \right) = \vcenter{\hbox{\includegraphics[width=0.13\textwidth]{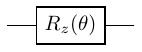}}}
\end{equation}
The individual terms were taken, rather than the combined $(1-\sigma^z)(1-\sigma^z)/4$ acting on the same site, to obtain a larger circuit expressibility. Now, to generate a specific spin configuration, $X$ gates can be used as an initial state, to create spins at particular sites with specific colours. In the simulations, created spins were followed by an initial sub-layer of parametrised $R_z$ gates (which were omitted where no spin was created since they would not affect the overall quantum state).

All of the proposed quantum gates are number-preserving, which is crucial for ascertaining that the \ac{VQE} searches for the ground-state of the SU($N$) Hubbard model in a specific spin configuration. An assembled \ac{HVA} for a 3-site SU(3) nearest-neighbour Hubbard model is presented in Fig.~\figref{fig:circuit}. This makes the \ac{PQC} more complex when compared with hardware-efficient ones. However, a symmetry-preserving ansatz provides certain advantages, such as potentially speeding up optimisation and reducing the risk of \acp{BP}, due to the restriction in the size of the Hilbert space~\cite{Tilly2022}.

\subsection{Circuit Complexity}

Here we give a brief analysis of basis gate decompositions of the \ac{HVA} of Fig.~\figref{fig:circuit}, where we assume that the native gate set is composed of any one-qubit gate and \textsc{CNOT} gates, with full qubit connectivity. For the sake of simplicity, we determine the effective circuit depth of a single ansatz layer by counting \textsc{CNOT} gates, assuming that the gate times and errors of one-qubit gates are significantly less than those of a typical \textsc{CNOT} gate~\cite{Willsch2017}. A single ansatz layer consists of an entangling sub-layer incorporating the hopping terms ($\imath$\textsc{SWAP} gates), followed by the other entangling sub-layer containing the interaction terms (C$R_z$ gates), and then a one-qubit sub-layer composed of $R_z$ gates.

The hopping sub-layer consists of $N(L - 1)$ $\imath$\textsc{SWAP} gates and the interaction sub-layer is composed of $L(N - 1)$ C$R_z$ gates. Each $\imath$\textsc{SWAP} gate introduces three \textsc{CNOT} gates~\cite{Vatan2004}, while each interaction term consisting of a C$R_z$ gate decomposes into two \textsc{CNOT} gates~\cite{Mottonen2006}. Therefore, the total number of decomposed \textsc{CNOT} gates is equal to $5NL - 3N - 2L$. Looking at the circuit depth, the hopping sub-layer garners a depth of $3(L - 1)$, with the interaction sub-layer achieving a depth of $2(N - 1)$. Hence, the effective depth of the \ac{PQC} is equal to $2N +3L - 5$. Now counting the number of parameters, the hopping sub-layer consists of $N(L - 1)$ parameters, while the interaction sub-layer contains $L(N - 1)$ parameters --- since the C$R_z$ gates acting from each colour to the next can be done in parallel --- with each one-qubit sub-layer having $NL$ parameters. Thus, each ansatz layer incorporates $3NL - N - L$ parameters, along with the initial layer of one-qubit $R_z$ gates acting after the $X$ gates (used to create a spin), introducing an extra $N_p$ parameters at the start of the circuit. As a specific example, referring to Fig.~\figref{fig:circuit} using one layer, we obtain a total of 24 parameters.

In each instance of the \ac{VQE}, we considered an initial state starting in the computational basis equating to the spin configuration of the intended model. In fact, this approach allows for the simple notion of error detection to be implemented: by checking the Hamming weight\footnote{The number of 1's in the bit-string output.} of the output and confirming that it is conserved with respect to each individual spin colour. If a non-conserved binary string is outputted, then one can be certain that an error occurred during computation and the result of the \ac{VQE} can be disregarded.

\subsection{Measurement}

By decomposing the Hubbard Hamiltonian given in Eq.~\eqref{eq:mapr} into Pauli strings, we end up with the following:
\begin{align}
\cH = -\ &\sum\limits_{i, s, r}t_r\left(e^{\imath\frac{2\pi\phi}{L}} \sigma_{sL + i}^+\sigma_{sL + i + r}^- + \acs{hc} \right)\bigotimes_{j=sL + i + 1}^{sL + i + r - 1}\sigma_j^z \nonumber\\
+\ &\dfrac{U}{4}\sum\limits_{i, s<s'}\sigma_{sL + i}^z\sigma_{s'L + i}^z +\frac{1}{4}\sum\limits_{i, s, s', r} V_r\sigma_{sL + i}^z\sigma_{s'L + i + r}^z\nonumber\\
-\ &\frac{1}{4}\left[ N\lambda_{R_V}(\vec{V}) + (N - 1)U \right] \sum\limits_{i, s} \sigma_{sL + i}^z\nonumber\\
+\ &\dfrac{NL}{8}\left[ N\lambda_{R_V}(\vec{V})+ (N - 1)U \right],
\end{align}
with
\begin{align}
    \lambda_{R_V}(\vec{V}) = \sum_{r = 1}^{R_V} g_L(r) V_r, \\
    \vec{V} = \left\{V_1, V_2, \dots, V_{R_V}\right\},
\end{align}
and
\begin{equation}
    g_L(r) = 
    \begin{cases}
        2 \text{ if } r < \frac{L}{2}, \\
        1 \text{ if } r = \frac{L}{2}.
    \end{cases}
\end{equation}
Note that $\lambda_R = \sum_{r = 1}^{R} g_L(r)$ counts the number of edges in a circulant graph~\cite{Weisstein2021} having edges of up to distance $R$, and at most $\floor{L/2}$, representative of the layout of a 1D ring lattice.

Thus, the Hubbard Hamiltonian consists of $NL\lambda_{R_t}/2$ $\sigma^x\sigma^x$, $\sigma^y\sigma^y$, $\sigma^x\sigma^y$ and $\sigma^y\sigma^x$ terms each, $NL[N(1 + \lambda_{R_V}) - 1]/2$ $\sigma^z\sigma^z$ terms, $NL$ $\sigma^z$ terms, and one constant term, adding to a total of $NL[4\lambda_{R_t} + N(1 + \lambda_{R_V}) + 1]/2$ terms, which equates to $3NL(N + 3)/2$ for nearest-neighbour hopping and interaction models as in Eq.~\eqref{eq:hubbard_hamiltonian}. This means that naively taking separate energy measurements for each term in the Hamiltonian would prove to be severely infeasible. One solution to this problem is incorporating commuting sets of observables and measuring them in parallel~\cite{Jena2019, Zhao2020b}. Immediately, it can be observed that the $\sigma^z\sigma^z$ and $\sigma^z$ terms can be measured in parallel, since they are already diagonal operators (and thus commuting) in the computational basis. The simplest approach is to take the original form of $(1 - \sigma_i^z)(1 - \sigma_j^z)/4 = \ket{11}\bra{11}_{ij}$ for the $\sigma^z\sigma^z$ terms and $(1 - \sigma_i^z)/2 = \ket{1}\bra{1}_{i}$ for the $\sigma^z$ terms, as given directly in Eq.~\eqref{eq:hubbard_hamiltonian}. 

For nearest-neighbour models, hopping terms can be divided into three sets: \textbf{a)} the even-odd hopping terms, i.e. the terms in Eq.~\eqref{eq:hopping} acting on qubits 0-1, 2-3, 4-5, \dots \textbf{b)} the odd-even hopping terms, i.e. qubits 1-2, 3-4, 5-6, \dots, and lastly \textbf{c)} the closed hopping terms acting on qubits 0-$(L-1)$, along with the $\sigma^z$ terms in between. To measure the hopping terms given in Eq.~\eqref{eq:hopping}, a unitary operator that diagonalises pairs of qubits in the hopping basis is given by
\begin{equation}
	\left( \begin{array}{cccc}
	    1 & 0 & 0 & 0 \\
		0 & \frac{e^{-\imath\frac{\pi\phi}{L}}}{\sqrt{2}} & \frac{e^{\imath\frac{\pi\phi}{L}}}{\sqrt{2}} & 0 \\[1.8ex]
		0 & -\frac{e^{-\imath\frac{\pi\phi}{L}}}{\sqrt{2}} & \frac{e^{\imath\frac{\pi\phi}{L}}}{\sqrt{2}} & 0 \\[1.5ex]
		0 & 0 & 0 & 1
	\end{array} \right),
\end{equation}
where $\phi$ is the magnetic flux. This unitary operator is responsible for transforming the hopping terms in Eq.~\eqref{eq:hopping} to the diagonal basis $\ket{01}\bra{01} - \ket{10}\bra{10}$. It is significant to note that applying sets of this transformation before measuring effectively increases the circuit depth by a further three \textsc{CNOT} gates~\cite{Vatan2004}.

This implies that only four sets of measurements are needed to calculate the expectation value of the Hamiltonian containing only nearest-neighbour hopping. It is also significant to note that if the number of spins for each colour is odd, and the number of sites is even, then one can ignore the $\sigma^z$ terms in between the closed hopping terms, due to parity symmetry. This will enable the closed hopping terms to fit in with the set of even-odd hopping terms, further reducing the total size of the set of measurements to three.

\subsection{Classical Optimisation}

The \ac{BFGS} method~\cite{Nocedal2006} is the classical optimisation technique used in all of the statevector simulations. It was specifically chosen due to its relatively quick and accurate convergence, requiring a moderate number of iterations to arrive at the ground-state energy, when compared with other optimisation algorithms. For all the simulations, the threshold for convergence was defined to be a tolerance value of $10^{-5}$ for the gradient norm between one iteration and the next. However, it is important to note that the \ac{BFGS} algorithm would not be ideal if shot-based simulations were performed instead of taking exact energy measurements. For this reason, carrying out a simulation either on a classical computer or on a quantum computer, and sampling measurement outcomes, would require the use of a more sophisticated optimisation technique, such as \ac{NFT}~\cite{Nakanishi2020}, to take into consideration the effect of the inherent statistical noise in the measurement results.

In the case of shot-based simulations, the \ac{BFGS} algorithm requires sampling of both the gradient and the curvature of the parameter space, meaning that both the Jacobian and the Hessian need to be computed for each shot. The classical complexity also scales with the cube of the number of parameters of the variational ansatz. On the other hand, the \ac{NFT} algorithm is a gradient-free optimiser, which only optimises one parameter at a time. Due to the analytical landscape of each parameter being a sinusoidal function, \ac{NFT} uses this fact to perform sinusoidal fitting, which has constant time complexity. Consequently, at each iteration only three measurement are required for each parameter, achieving a linear scaling on the number of measurements.

\subsection{Computational Complexity Analysis}

In this section we compile the computational complexity scaling of both the \ac{PQC}, as well as of the optimisers used in the simulations. Table~\tabref{tab:cs} highlights the computational complexity scaling of the \ac{PQC}, while Table~\tabref{tab:ca} outlines the computational complexity analysis of \ac{VQE} procedure using the different optimisers mentioned in this work.

\begin{table}[t]
\centering
\begin{tabular}{||c c c||} 
    \hline
    \textsc{CNOT} Gates & Circuit Depth & Parameters \\
    \hline
    \hline
    $5NL - 3N - 2L$ & $2N + 3L - 5$ & $3NL - N - L$ \\ 
    \hline
    $\cO(NL)$ & $\cO(N + L)$ & $\cO(NL)$ \\ 
    \hline
\end{tabular}
\caption[Computational complexity scaling of the \acs{PQC}.]{Computational complexity scaling of the \ac{HVA}, where we highlight the number of decomposed \textsc{CNOT} gates, associated circuit depth, and the number of parameters in the ansatz.}
\label{tab:cs}
\end{table}

\begin{table}[t]
\centering
\begin{tabular}{||c c c c||}
    \hline
    optimiser & Type & Measurement Complexity & Classical Complexity \\
    \hline
    \hline
    \acs{BFGS} & Second-order & $M(g_1 + g_2) $ & $\cO(p^3) = \cO(N^3L^3)$ \\ 
    \hline
    \acs{NFT} & Gradient-free & $\cO(p) = \cO(NL)$ & $\cO(1)$ \\ 
    \hline
\end{tabular}
\caption[Computational complexity analysis of classical optimisers.]{Computational complexity of the different optimisers~\cite{Tilly2022}, used in this work, where we highlight the type of optimiser, and its corresponding measurement complexity, denoting the number of measurements needed per iteration, as well as its classical complexity, representing the complexity of the optimiser for each iteration. $M$ denotes the number of shots taken per iteration, $g_1$ and $g_2$ represent the cost of evaluating the first-order and second-order gradients, respectively, and $p$ denotes the number of parameters in the circuit.}
\label{tab:ca}
\end{table}

%% file: appendixB/swap_test_appendix.tex
\chapter{Measuring Overlaps on a Quantum Computer} \label{app:swap}

\textit{Parts of this appendix are based on the published manuscript by \citet{Consiglio2022b}.}\\

In Section~\secref{sec:HSD_measure}, we decomposed the \ac{HSD} between two states into a sum of their purities and overlap. To measure these quantities, we require the use of either the \textsc{SWAP} test, or the destructive \textsc{SWAP} test.

\section{\textsc{SWAP} Test}

\begin{figure}[t]
    \centering
    \includegraphics[width=0.5\textwidth]{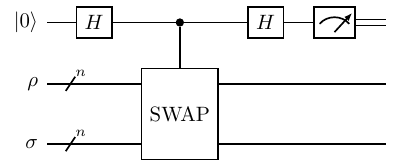}
    \caption[\textsc{SWAP} test for determining the overlap between two quantum states.]{\textsc{SWAP} test for determining the overlap between two quantum states. The multi-qubit controlled-\textsc{SWAP} gate in this case swaps the two $n$-qubit states $\rho$ and $\sigma$.}
    \label{fig:swap_test}
\end{figure}

Consider two $n$-qubit states, $\rho$ and $\sigma$, to measure the overlap between them one can employ the \textsc{SWAP} test~\cite{Buhrman2001}, which is capable of determining by how much two quantum states differ. Fig.~\figref{fig:swap_test} gives the quantum circuit needed for the \textsc{SWAP} test. The total number of qubits required in the circuit is $2n + 1$, where the extra qubit is utilised by an ancilla to perform the indirect measurement.

To show that this procedure estimates the overlap between two states, let us use two pure states $\ket{\psi}$ and $\ket{\varphi}$ for simplicity. The input for the circuit is then given by the state $\ket{0}\ket{\psi}\ket{\varphi}$. After the first Hadamard gate, the state transforms into
\begin{equation}
    \frac{1}{\sqrt{2}}\left(\ket{0}\ket{\psi}\ket{\varphi} + \ket{1}\ket{\psi}\ket{\varphi} \right).
\end{equation}
The controlled-\textsc{SWAP} gate then changes the state to
\begin{equation}
    \frac{1}{\sqrt{2}}\left(\ket{0}\ket{\psi}\ket{\varphi} + \ket{1}\ket{\varphi}\ket{\psi} \right).
\end{equation}
Finally the second Hadamard gate results in the state
\begin{equation}
    \frac{1}{2}\left(\ket{0}\ket{\psi}\ket{\varphi} + \ket{1}\ket{\psi}\ket{\varphi} + \ket{0}\ket{\varphi}\ket{\psi} - \ket{1}\ket{\varphi}\ket{\psi} \right),
\end{equation}
which can be rewritten as
\begin{equation}
    \frac{1}{2}\ket{0}\left(\ket{\psi}\ket{\varphi} + \ket{\varphi}\ket{\psi}\right) + \frac{1}{2}\ket{1}\left(\ket{\psi}\ket{\varphi} - \ket{\varphi}\ket{\psi} \right).
\end{equation}
Measuring the first qubit gives a result of 0 with probability
\begin{equation}
    p_0 = \frac{1}{4}\left(\bra{\psi}\bra{\varphi} + \bra{\varphi}\bra{\psi}\right)\left(\ket{\psi}\ket{\varphi} + \ket{\varphi}\ket{\psi}\right) = \frac{1}{2} + \frac{1}{2}|\braket{\psi}{\varphi}|^2.
\end{equation}
Therefore, the overlap between $\ket{\psi}$ and $\ket{\varphi}$ can be calculated as
\begin{equation}
    |\braket{\psi}{\varphi}|^2 = 2p_0 - 1,
\end{equation}
implying that if $M$ measurements are taken with each outcome yielding either 0 or 1, with the measurement result denoted by $M_i$, then
\begin{equation}
    p_0 \approx  1 - \frac{1}{M}\sum_{i=0}^{M-1} M_i,
\end{equation}
such that
\begin{equation}
    |\braket{\psi}{\varphi}|^2 \approx 1 - \frac{2}{M}\sum_{i=0}^{M-1} M_i,
\end{equation}
with equality occurring as $M \xrightarrow{} \infty$. Note that this result can be easily extended to mixed states $\rho$ and $\sigma$~\cite{Kobayashi2003}, where instead we would have
\begin{equation}
    \Tr{\rho\sigma} \approx 1 - \frac{2}{M}\sum_{i=0}^{M-1} M_i.
\end{equation}
The purity can be simply evaluated by instead supplying two copies of either $\rho$ or $\sigma$ in the procedure.

\begin{figure}[t]
    \centering
    \includegraphics[width=0.5\textwidth]{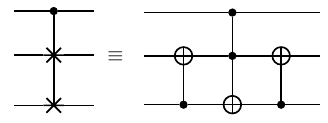}
    \caption{Controlled SWAP gate decomposition.}
    \label{fig:cswap}
\end{figure}

Nonetheless, the drawback of this procedure is twofold: it requires the addition of an ancillary qubit; and it also requires considerable circuit depth, scaling linearly with the number of qubits. This is due to the fact that a multi-qubit controlled-\textsc{SWAP} gate is composed of individual controlled-\textsc{SWAP} gates between every qubit pair of $\rho$ and $\sigma$. A controlled-\textsc{SWAP} gate in turn can be decomposed into a Toffoli gate interposed between two \textsc{CNOT} gates, as shown in Fig.~\figref{fig:cswap}. The Toffoli gate can then be decomposed into eight one-qubit gates and six \textsc{CNOT} gates \cite{Shende2008}.

As a consequence, we look towards alternative methods of calculating the overlap between two quantum states, which ideally avoid the issues present in the \textsc{SWAP} test. The destructive \textsc{SWAP} test offers solutions to both of these shortcomings, by requiring no ancillary qubit, and also having a constant depth of just two gates.

\section{Destructive \textsc{SWAP} Test}

\begin{figure}[t]
    \centering
    \includegraphics[width=0.4\textwidth]{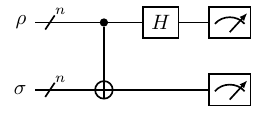}
    \caption[Destructive \textsc{SWAP} test for determining the overlap between two quantum states.]{Destructive \textsc{SWAP} test for determining the overlap between two quantum states. The bundled \textsc{CNOT} gate here actually represents \textsc{CNOT} gates between each pair of qubits of $\rho$ and $\sigma$, with the bundled Hadamard gate equating to performing a Hadamard gate on each qubit of $\rho$. This effectively reduces to a Bell basis transformation for each pair of qubits of $\rho$ and $\sigma$.}
    \label{fig:destructive_swap_test}
\end{figure}

The concept of this procedure stems from the fact that measuring in the Bell basis determines the amount of correlation present between two systems. Once again, given two $n$-qubit states, $\rho$ and $\sigma$, we now employ the destructive \textsc{SWAP} test~\cite{Garcia_escartin2013, Cincio2018}, which is shown in Fig.~\figref{fig:destructive_swap_test}. 

Each pair of qubits of $\rho$ and $\sigma$ are measured in the Bell basis, by putting a \textsc{CNOT} gate followed by a Hadamard on the control qubit, and then measured. Each pair of measurements are post-processed using the vector $\vec{c} = (1, 1, 1, -1)$, which corresponds to summing the probabilities of getting 00, 01 and 10, and subtracting the probability of getting 11. This is equivalent to measuring the expectation value of a C$Z$ operator. This results in obtaining the overlap $\Tr{\rho\sigma}$, as long as the qubits are rearranged as $R_1S_1R_2S_2\dots R_nS_n$, where $R_i$ and $S_i$ denote the subsystems of $\rho$ and $\sigma$, respectively. This results in a linear scaling in post-processing, given that we do not directly compute $\vec{c}\cdot\vec{p}$, where $\vec{p}$ is the probability vector, but rather binning the paired measurement outcomes into a $+1$ and $-1$ bin, and then averaging the results~\cite{Cincio2018}.

%% file: appendixC/css_appendix.tex
\chapter{Deriving the Closest Separable State for \textit{X}-MEMS} \label{app:css}

\textit{Parts of this appendix are based on the published manuscript by \citet{Consiglio2022b}.}\\

For the sake of conciseness, the general form of \ac{X-MEMS} is restated here:
\begin{equation}
    \tilde{X} = \left( 
    \begin{array}{cccccccc}
        f(\gamma) & & & & & & & \gamma \\
        & g(\gamma) & & & & & 0 & \\
        & & \ddots & & & \iddots & & \\
        & & & g(\gamma) & 0 & & & \\
        & & & 0 & 0 & & & \\
        & & \iddots & & & \ddots & & \\
        & 0 & & & & & 0 & \\
        \gamma^* & & & & & & & g(\gamma) \\
    \end{array} 
    \right),
\end{equation}
where
\begin{subequations}
\label{eq:fg}
\begin{align}
    f(\gamma) &= 
    \begin{cases}
        \frac{1}{N+1} & 0 \leq |\gamma| \leq \frac{1}{N+1}, \\
        |\gamma| & \frac{1}{N+1} \leq |\gamma| \leq \frac{1}{2},
    \end{cases} \\
    g(\gamma) &= 
    \begin{cases}
        \frac{1}{N+1} & 0 \leq |\gamma| \leq \frac{1}{N+1}, \\
        \frac{1 - 2|\gamma|}{N-1} & \frac{1}{N+1} \leq |\gamma| \leq \frac{1}{2},
    \end{cases}
\end{align}
\end{subequations}
and $N = 2^{n - 1}$.

\section{Two-qubit \textit{X}-MEMS}

The analytical form of the \ac{CSS} for two-qubit \ac{X-MEMS} was derived using analytical optimisation with the \ac{KKT} conditions~\cite{Tabak1971}, since the numerical results by the \ac{VSV} hinted towards a \ac{CSS} of the form
\begin{equation}
    \left(
    \begin{array}{cccc}
    \frac{a}{2} & 0 & 0 & d \\
    0 & b & 0 & 0 \\
    0 & 0 & 1 - a - b & 0 \\
    d^* & 0 & 0 & \frac{a}{2}
    \end{array}
    \right),
\end{equation}
where $0 \leq a, b \leq 1$, $a + b \leq 1$, and $0 \leq |d| \leq a/2$, to ensure a valid density matrix. For this state to be separable, then the concurrence must be equal to zero, meaning that
\begin{equation}
    |d|^2 \leq b(1 - a - b).
    \label{eq:no_conc}
\end{equation}
Optimising this problem results in parameters 
\begin{subequations}
\begin{align}
    a &= \frac{1}{9}\left(
    7 - \sqrt{1 + 36 |\gamma|^2}
    \right), \\
    b &= \frac{1 + 12 |\gamma|^2+\sqrt{1 + 36 |\gamma|^2}}{6 \sqrt{1 + 36|\gamma|^2}}, \\
    d &= \frac{\gamma}{3}  \left(1 + \frac{2}{\sqrt{1 + 36 |\gamma|^2}}\right),
\end{align}
\end{subequations}
with an \ac{HSE} of
\begin{equation}
    \EHS{\tilde{X}} = \frac{2}{27} \left(1 + 18 |\gamma|^2-\sqrt{1 + 36 |\gamma|^2}\right),
\end{equation}
for $0 \leq |\gamma| \leq \frac{1}{3}$, and
\begin{subequations}
\begin{align}
    a &= \frac{1}{3} \left(1 + 4|\gamma| - \sqrt{1 - 4|\gamma| + 8|\gamma|^2}\right), \\
    b &= \frac{1}{6} \left(3 - 6|\gamma| + \frac{3 - 12|\gamma| + 16 |\gamma|^2}{\sqrt{1 - 4|\gamma| + 8|\gamma|^2}}\right), \\
    d &= \frac{\gamma  \left(2 - 4|\gamma| + \sqrt{1 - 4|\gamma| + 8|\gamma|^2}\right)}{3 \sqrt{1 - 4|\gamma| + 8|\gamma|^2}},
\end{align}
\end{subequations}
with an \ac{HSE} of
\begin{equation}
    \EHS{\tilde{X}} = \frac{2}{3} \left(1 - 4|\gamma| + 6|\gamma|^2 + (2|\gamma| - 1)\sqrt{1 - 4|\gamma| + 8|\gamma|^2} \right),
\end{equation}
for $\frac{1}{3} \leq |\gamma| \leq \frac{1}{2}$. 

\section{Three-qubit \textit{X}-MEMS}

The analytical form of the \ac{CSS} for three-qubit \ac{X-MEMS} was similarly derived using analytical optimisation with the \ac{KKT} conditions~\cite{Tabak1971}, since the numerical results by the \ac{VSV} hinted towards a \ac{CSS} of the form
\begin{equation}
    \left(
    \begin{array}{cccccccc}
    \frac{a}{2} & 0 & 0 & 0 & 0 & 0 & 0 & d \\
    0 & \frac{b}{3} & 0 & 0 & 0 & 0 & 0 & 0 \\
    0 & 0 & \frac{b}{3} & 0 & 0 & 0 & 0 & 0 \\
    0 & 0 & 0 & \frac{b}{3} & 0 & 0 & 0 & 0 \\
    0 & 0 & 0 & 0 & \frac{1 - a - b}{3} & 0 & 0 & 0 \\
    0 & 0 & 0 & 0 & 0 & \frac{1 - a - b}{3} & 0 & 0 \\
    0 & 0 & 0 & 0 & 0 & 0 & \frac{1 - a - b}{3} & 0 \\
    d^* & 0 & 0 & 0 & 0 & 0 & 0 & \frac{a}{2}
    \end{array}
    \right),
    \label{eq:three_qubit_CSS}
\end{equation}
where similarly, $0 \leq a, b \leq 1$, $a + b \leq 1$, and $0 \leq |d| \leq a/2$, to ensure a valid density matrix. Since we require the state to be separable, we first need to certify that the \ac{GME} concurrence is zero, which results in the same condition as in~\eqref{eq:no_conc}. However, the \ac{GME} concurrence only assures us that the state does not have tripartite entanglement in this scenario, yet we still need to ensure that the state has no bipartite entanglement. It can be immediately seen from the density matrix~\eqref{eq:three_qubit_CSS} that the concurrence between pairs of individual qubits is zero --- since the partial trace with respect to each qubit results in a two-qubit diagonal matrix. 

What is left to check is the entanglement between each qubit and the rest, and so, we look to the negativity~\cite{Zyczkowski1998}, although the negativity does not capture bound entangled states~\cite{Horodecki1996, Peres1996, Horodecki1998}. However, we conjecture that the state in~\eqref{eq:three_qubit_CSS} is not bound entangled, since the numerical \ac{HSE}, determined by the \ac{VSV}, coincides with the \ac{HSE} of~\eqref{eq:three_qubit_CSS}, as shown in Fig.~\figref{fig:X-MEMS_3}. Due to the uniqueness of the \ac{CSS}~\cite{Pandya2020}, this implies that the fully separable state determined by the \ac{VSV}, is exactly the same state as~\eqref{eq:three_qubit_CSS}, meaning that it is fully separable.

The partial transpose of~\eqref{eq:three_qubit_CSS} with respect to each qubit is equivalent, and the negativity is non-zero if the partial transpose has negative eigenvalues. Now, the partial transpose of~\eqref{eq:three_qubit_CSS} has only one eigenvalue which can be negative, for some parameters $a$, $b$ and $d$, which is
\begin{equation}
    \frac{1}{6}\left( 1 - a - \sqrt{(1 - a - 2b)^2 + 36|d|^2} \right),
    \label{eq:condition}
\end{equation}
meaning we need to ascertain that condition~\eqref{eq:condition} is non-negative, for~\eqref{eq:three_qubit_CSS} to be separable. This condition can instead be rewritten and shown that it is equivalent to
\begin{equation}
    |d|^2 \leq \frac{b(1 - a - b)}{9},
    \label{eq:no_neg}
\end{equation}
which interestingly encompasses condition~\eqref{eq:no_conc} as well, resulting in an analytical optimisation problem using the \ac{KKT} conditions with only constraint~\eqref{eq:no_neg} needed to insure separability. For the sake of brevity, the parameters $a$, $b$ and $d$, as well as the \ac{HSE} are not provided here, since they are in terms of a root of a quartic equation and so the explicit equations would be too cumbersome to present. Nevertheless, they can be reproduced using the \texttt{Mathematica} code available at:~\cite{Consiglio2022c}.

\section{\textit{n}-qubit \textit{X}-MEMS}

While the \ac{CSS} of $n$-qubit \ac{X-MEMS} is not explicitly derived in general, we conjecture that it is of the form
\begin{equation}
    \left( 
    \begin{array}{cccccccc}
        \frac{a}{2} & & & & & & & d \\
        & \frac{b}{N - 1} & & & & & 0 & \\
        & & \ddots & & & \iddots & & \\
        & & & \frac{b}{N - 1} & 0 & & & \\
        & & & 0 & \frac{1 - a - b}{N - 1} & & & \\
        & & \iddots & & & \ddots & & \\
        & 0 & & & & & \frac{1 - a - b}{N - 1} & \\
        d^* & & & & & & & \frac{a}{2} \\
    \end{array} 
    \right),
\end{equation}
where $N = 2^{n-1}$, $0 \leq a, b \leq 1$, $a + b \leq 1$, and $0 \leq |d| \leq a/2$, to ensure a valid density matrix. This form is also invariant under the symmetries imposed by \ac{X-MEMS}, which is a necessary condition for the \ac{CSS} as shown in \citet{Pandya2020}. The symmetries of the \ac{CSS} consist of any qubit permutations from the set $\{2, \dots, n\}$, and the condition that the first element and the last element of the main diagonal must be equal. The parameter $a$ in the \ac{CSS} seems to be a root of a quartic equation in general, with parameters $b$ and $d$ depending on both $a$ and $\gamma$. The condition for zero negativity also encompasses the cases for $(k \geq 3)$-separability, which is equivalent to
\begin{equation}
    |d|^2 \leq \frac{b(1 - a - b)}{(N - 1)^2},
\end{equation}
presumably related to the hierarchy of mixed state entanglement~\cite{Eltschka2014}.

\begin{figure}[t]
    \centering
    \includegraphics[width=0.7\textwidth]{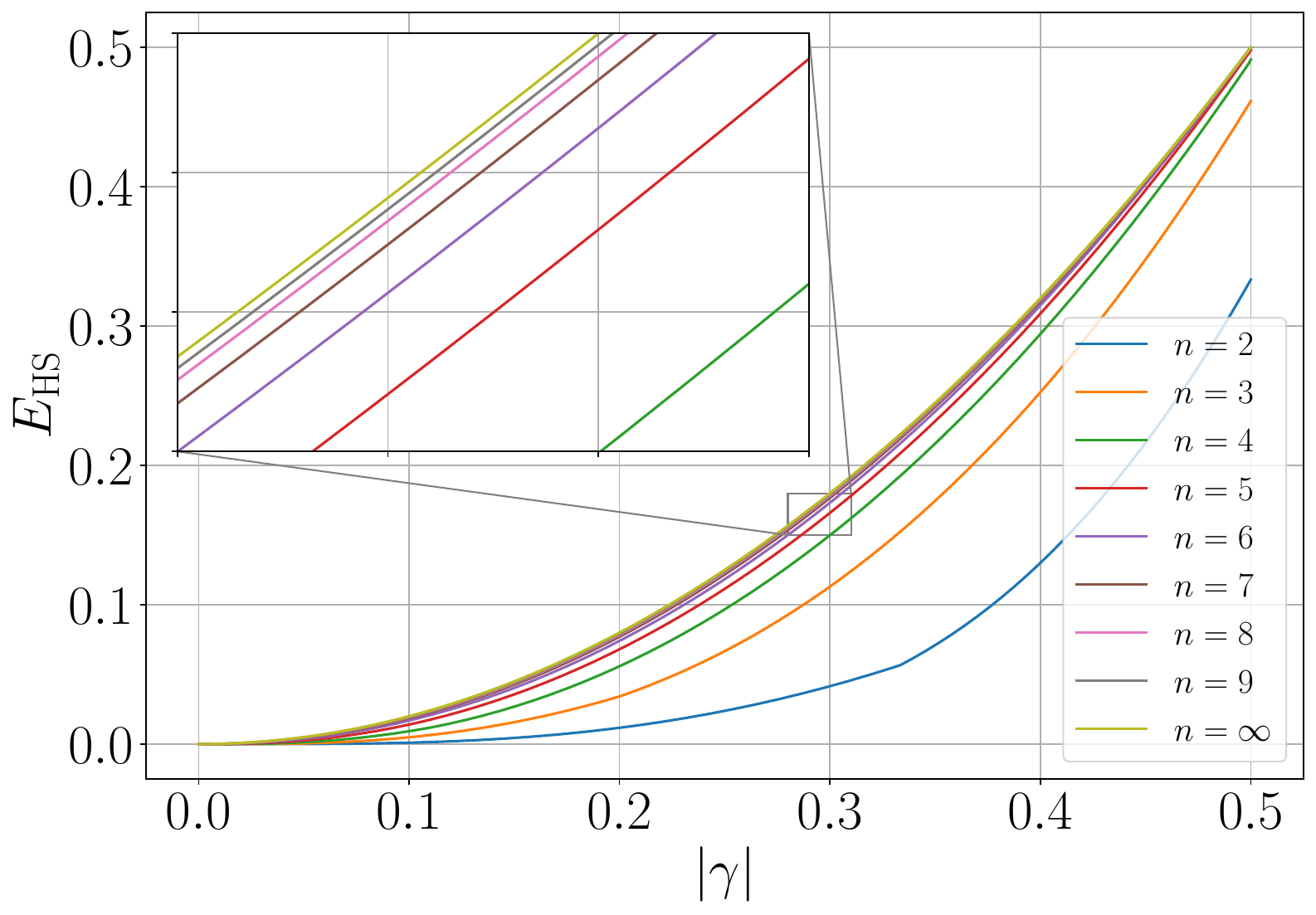}
    \caption[Analytical \acs{HSE} of \acs{X-MEMS}.]{Analytical \ac{HSE} of two- to nine-qubit \ac{X-MEMS} as a function of $|\gamma|$ (bottom-up). The infinite-qubit case (top-most line) represents the analytical upper bound of $2|\gamma|^2$ for the \ac{HSE} of \ac{X-MEMS}.}
    \label{fig:analytical_X-MEMS}
\end{figure}

Fig.~\figref{fig:analytical_X-MEMS} shows the analytical \ac{HSE} of two- to nine-qubit \ac{X-MEMS} as a function of $|\gamma|$. Each function for the \ac{HSE} is piece-wise continuous, with the domains divided at the point $|\gamma| = 1/(N + 1)$, as is similarly given by the definition of the \ac{X-MEMS} in Eqs.~\eqref{eq:fg}. \Ac{X-MEMS} possess an \ac{HSE} of zero at $|\gamma| = 0$ and an \ac{HSE} of
\begin{equation}
    \EHS{\GHZ_n} = \frac{2^n-2}{2^{n + 1} + 2^{3-n} - 4},
\end{equation}
at $|\gamma| = 1/2$, since the $n$-qubit \ac{X-MEMS} correspond to the $n$-qubit \ac{GHZ} states at those points~\cite{Pandya2020}. It is also interesting to note the limit of the functions in Fig.~\figref{fig:analytical_X-MEMS}, corresponding to the infinite-qubit $\ac{X-MEMS}$, which is equal to
\begin{equation}
    \lim\limits_{n \rightarrow \infty} \EHS{\tilde{X}_n} = 2|\gamma|^2,
\end{equation}
meaning that the \ac{HSE} for \ac{X-MEMS} is bounded above.

%% file: appendixD/gilbert_appendix.tex
\chapter{Direct Implementation of the Quantum Gilbert Algorithm on a Quantum Computer} \label{app:gilbert}

\textit{Parts of this appendix are based on the published manuscript by \citet{Consiglio2022b}.}\\

In this appendix, we describe the direct implementation of the \acf{QGA}~\cite{Gilbert1966} on a quantum device. Given an initial guess in the form of a pure product state $\sigma_0 = \op{\psi_0}$, after the $n^\text{th}$ success, the new \ac{CSS} $\rho_n$ to the test state $\rho$ is given iteratively as
\begin{equation}
    \rho_{n} = p_n \rho_{n-1} + (1 - p_n)\sigma_n,
\end{equation}
which equates to
\begin{equation}
    \rho_{n} = \sum_{i=0}^{n} \left(\prod_{j=i+1}^{n} p_j\right) \left(1-p_i\right) \sigma_i,
\end{equation}
where $\sigma_i = \op{\psi_i}$ is a pure product state and $0 \leq p_i \leq 1$ such that
\begin{equation}
    \sum_{i=0}^{n} \left(\prod_{j=i+1}^{n} p_j\right) \left(1-p_i\right) = 1.
\end{equation}
Note that this definition entails that $\rho_0 \equiv \sigma_0$. The implementation of a \ac{QGA} can thus be described as follows~\cite{Pandya2020}:
\begin{enumerate}[a)]
    \item Input data: the state to be tested $\rho$ and any pure product state $\sigma_0$.
    \item Output data: the closest state found after $n$ successes $\rho_n$, and lists of values of $\DHS{\rho}{\rho_i} = \Tr{(\rho - \rho_i)^2}$, $p_i$ and $\sigma_i$.
\end{enumerate}
\begin{enumerate}[1)]
    \setcounter{enumi}{-1}
    \item Calculate the value of $\DHS{\rho}{\sigma_0} \equiv \DHS{\rho}{\rho_0}$ and add it to the list.
    \item Increase the counter of trials $c_t$ by 1. Generate a random pure product state $\sigma_1$, hereafter called the first trial state.
    \item Run a preselection for the trial state by checking a value of a linear functional. If it fails, go back to point 1.
    \item In the case of successful preselection, find the minimum of $\Tr{(\rho - \rho_n)^2}$  with respect to $p_n$, which in the case of the first trial state is $\Tr{[\rho - p_1\rho_0 - (1 - p_1)\sigma_1]^2}$ with respect to $p_1$.
    \item If the minimum occurs for $p_1  \in [0, 1]$, then $\rho_1 = p_1\rho_0 + (1 - p_1)\sigma_1$ is saved to the list as the new \ac{CSS}, along with the value of $\DHS{\rho}{\rho_1}$, $p_1$ and $\sigma_1$, increase the success $c_s$ counter value by 1.
    \item Go to step 1), now generating a new random pure product state $\sigma_2$, such that after a successful preselection, one determines a new $p_2$, and consequently $\rho_2$, and then saves $\DHS{\rho}{\rho_2}$, $p_2$ and $\sigma_2$. Continue in this fashion until a specified \texttt{HALT} criterion is met.
\end{enumerate}

\begin{figure}[t]
    \centering
    \includegraphics[width=0.7\textwidth]{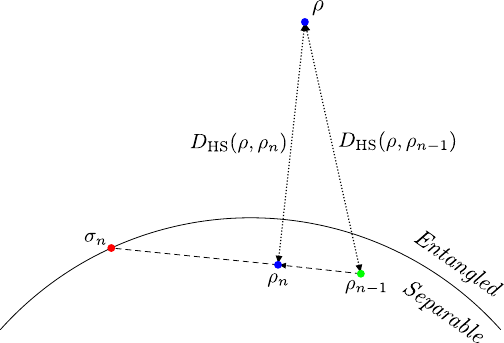}
    \caption[Visualisation of the \acs{QGA}.]{Visualisation of the $n^\text{th}$ iteration of the \ac{QGA}. A random pure product state $\sigma_n$ is generated that satisfies the preselection criterion. The next step is to find $\rho_n = p_n\rho_{n-1} + (1 - p_n)\sigma_n$, $p_n \in [0, 1]$, such that $\DHS{\rho}{\rho_n} < \DHS{\rho}{\rho_{n-1}}$. Figure adapted from \citet{Pandya2020}.}
    \label{fig:iteration}
\end{figure}

Fig.~\figref{fig:iteration} provides a visual representation of the $n^\text{th}$ iteration of the algorithm presented above, while the preselection criterion for the $n^\text{th}$ trial state is given by
\begin{equation}
    \Tr{(\sigma_n - \rho_{n-1})(\rho - \rho_{n-1})} > 0,
    \label{eq:preselection}
\end{equation}
which can be interpreted as checking whether the geometric angle between the new trial state and the test state, from the current \ac{CSS} is less than $\pi/2$, which ensures that the \ac{HSD} is monotonically decreasing as a function of the number of iterations.

The \ac{HSD} between the test state $\rho$ and the fully separable mixed state $\rho_n$ reduces to
\begin{align}
    \DHS{\rho}{\rho_n} &= \DHS{\rho}{\sum_{i=0}^{n} \left(\prod_{j=i+1}^{n} p_j\right) \left(1-p_i\right)\sigma_i} \nonumber \\
    &= \Tr{\left[\rho - \sum_{i=0}^{n} \left(\prod_{j=i+1}^{n} p_j\right) \left(1-p_i\right)\sigma_i \right]^2} \nonumber \\
    &= \Tr{\rho^2} + \Tr{\left[\sum_{i=0}^{n} \left(\prod_{j=i+1}^{n} p_j\right) \left(1-p_i\right)\sigma_i \right]^2} \nonumber \\
    & \hspace{0.5cm} - 2\Tr{\rho\sum_{i=0}^{n} \left(\prod_{j=i+1}^{n} p_j\right) \left(1-p_i\right)\sigma_i} \nonumber \\
    &= \Tr{\rho^2} + 2\sum_{i < k}^{n}\left(\prod_{j=i+1}^{n} p_j\right)\left(\prod_{l=k+1}^{n} p_l\right) \left(1-p_i\right) \left(1-p_k\right) \Tr{\sigma_i\sigma_k} \nonumber \\ & \hspace{0.5cm} + \sum_{i=0}^{n}\left(\prod_{j=i+1}^{n} p_j^2\right) \left(1-p_i\right)^2 - 2\sum_{i=0}^{n} \left(\prod_{j=i+1}^{n} p_j\right) \left(1-p_i\right)\Tr{\rho\sigma_i}.
\end{align}
This entails that the overlap between every new trial state $\sigma_n$ and every previous trial state $\sigma_0, \dots, \sigma_{n-1}$, as well as with the test state $\rho$, must be calculated at every iteration.

Initially, the overlap between the test state $\rho$ and the initial guess $\sigma_0$ is calculated to obtain $\Tr{\rho\sigma_0}$. Following this, the first random trial state $\sigma_1$ is drawn, and the overlaps with $\rho$ and $\sigma_0$ are calculated so that
the preselection criterion \eqref{eq:preselection} can be checked. This is carried out at the $n^\text{th}$ iteration by checking that
\begin{equation}
    \Tr{\rho_{n-1}^2} + \Tr{\rho\sigma_n} > \Tr{\rho\rho_{n-1}} + \Tr{\rho_{n-1}\sigma_n},
\end{equation}
where
\begin{align}
    \Tr{\rho_n^2} &= \Tr{(p_n\rho_{n-1} + (1 - p_n)\sigma_n)^2} \nonumber \\ 
    &= p_n^2\Tr{\rho_{n-1}^2} + (1 - p_n)^2 + 2p(1 - p_n)\Tr{\rho_{n-1}\sigma_n},
\end{align}
which denotes the purity of each new \ac{CSS} when calculated iteratively (which starts at 1 due to the initial trial state $\sigma_0$ being a pure state), and
\begin{equation}
    \Tr{\rho\rho_n} = p_n\Tr{\rho\rho_{n-1}} + (1 - p_n)\Tr{\rho\sigma_n},
\end{equation}
which denotes the overlap of each new \ac{CSS} with the test state when calculated iteratively. Thus, the cost function for the $n^\text{th}$ iteration is defined as follows:
\begin{align}
    \DHS{\rho}{p_n\rho_{n-1} + (1 - p_n)\sigma_n} &= \Tr{[\rho - p_n\rho_{n-1} - (1 - p_n)\sigma_n]^2} \nonumber \\
    &= \Tr{\rho^2} + p_n^2\Tr{\rho_{n-1}^2} + (1 - p_n)^2 \nonumber \\ & \hspace{0.5cm} + 2p_n(1 - p_n)\Tr{\rho_{n-1}\sigma_n} \nonumber \\ & \hspace{0.5cm} - 2p_n\Tr{\rho\rho_{n-1}} - 2(1 - p_n)\Tr{\rho\sigma_n}.
\end{align}

%% file: appendixE/entangling_gates.tex
\chapter{Necessity of Entangling Gates in Preparing the Boltzmann Distribution} \label{app:entangling_gates}

\textit{Parts of this appendix are based on the published manuscript by \citet{Consiglio2024a}.}\\

In Section~\secref{sec:results_ising}, we specified that we required entanglement in the ancillary register to be able to prepare the Boltzmann distribution of the Ising model. While we only used one layer of a hardware-efficient ansatz, we concluded that at least one entangling layer is necessary for preparing the Boltzmann distribution of the Ising model, and we will show this by considering the converse. Suppose the ancilla ansatz is only composed of local $R_y$ gates, then we get
\begin{align}
    \bigotimes_{i=0}^{n-1}R_y(\theta_i)\ket{0}_i &= \bigotimes_{i=0}^{n-1} \left[ \cos\left(\frac{\theta_i}{2}\right)\ket{0}_i + \sin\left(\frac{\theta_i}{2}\right)\ket{1}_i\right] \nonumber \\
    &= \sum_{i=0}^{d-1} \prod_{j \in S_{i=0}} \cos\left(\frac{\theta_j}{2}\right) \prod_{k \in S_{i=1}} \sin\left(\frac{\theta_k}{2}\right) \ket{i} \nonumber \\
    &= \sum_{i=0}^{d-1} \lambda_i \ket{i},
\end{align}
where $S_{i=0} \equiv \left\{j~|~j = 0~\forall~\textrm{bits}~j \in i \right\}$, that is, the set of bits in $i$ which are equal to $0$, and similarly defined for $S_{i=1}$, and $\ket{i}$ is the computational basis state. This implies that
\begin{equation}
    \prod_{j \in S_{i=0}} \cos\left(\frac{\theta_j}{2}\right) \prod_{k \in S_{i=1}} \sin\left(\frac{\theta_k}{2}\right) = \lambda_i.
\end{equation}
Now, without loss of generality, consider these specific cases, where we take the first four amplitudes:
\begin{subequations}
\begin{align}
    &\prod_{j=0}^{n-1} \cos\left(\frac{\theta_j}{2}\right) = \lambda_0, \\
    &\prod_{j=0}^{n-2} \cos\left(\frac{\theta_j}{2}\right) \sin\left(\frac{\theta_{n-1}}{2}\right) = \lambda_1, \\
    &\prod_{j=0}^{n-3} \cos\left(\frac{\theta_j}{2}\right) \cos\left(\frac{\theta_{n-1}}{2}\right) \sin\left(\frac{\theta_{n-2}}{2}\right)  = \lambda_2, \\
    &\prod_{j=0}^{n-3} \cos\left(\frac{\theta_j}{2}\right) \sin\left(\frac{\theta_{n-2}}{2}\right) \sin\left(\frac{\theta_{n-1}}{2}\right) = \lambda_3.
\end{align}
    \label{eq:lambdas}
\end{subequations}
Combining Equations~\eqref{eq:lambdas} results in
\begin{subequations}
\begin{align}
     \frac{\lambda_0}{\lambda_1} &= \frac{\cos\left(\frac{\theta_{n-1}}{2}\right)}{\sin\left(\frac{\theta_{n-1}}{2}\right)}, \\
    \frac{\lambda_0}{\lambda_2} &= \frac{\cos\left(\frac{\theta_{n-2}}{2}\right)}{\sin\left(\frac{\theta_{n-2}}{2}\right)}, \\
    \frac{\lambda_0}{\lambda_3} &= \frac{\cos\left(\frac{\theta_{n-2}}{2}\right) \cos\left(\frac{\theta_{n-1}}{2}\right)}{\sin\left(\frac{\theta_{n-2}}{2}\right) \sin\left(\frac{\theta_{n-1}}{2}\right)},
\end{align}
\label{eq:thetas}
\end{subequations}
and finally, combining the Equations~\eqref{eq:thetas} implies that
\begin{equation}
    \frac{\lambda_0}{\lambda_1}\frac{\lambda_0}{\lambda_2} = \frac{\lambda_0}{\lambda_3} \implies \frac{\lambda_0}{\lambda_1 \lambda_2} = \frac{1}{\lambda_3} \implies \lambda_0 \lambda_3 = \lambda_1 \lambda_2.
\end{equation}
Since after optimisation we want that $\lambda_i = \sqrt{p_i} = \exp(-\beta E_i / 2) / \sqrt{\cZ}$, then
\begin{equation}
    \lambda_0 \lambda_3 = \lambda_1 \lambda_2 \implies \frac{e^\frac{-\beta E_0}{2} e^\frac{-\beta E_3}{2}}{\cZ} = \frac{e^\frac{-\beta E_1}{2} e^\frac{-\beta E_2}{2}}{\cZ}.
\end{equation}
Applying logs to both sides and simplifying, we get
\begin{equation}
    E_0 + E_3 = E_1 + E_2,
    \label{eq:constraint}
\end{equation}
which is not in general true for the Ising model. The above reasoning can be adjusted to obtain further constraints in the manner of Eq.~\eqref{eq:constraint} on the spectrum of $\cH$. If entangling gates are not utilised in the preparing a probability distribution, then only probability distributions that satisfy Eq.~\eqref{eq:constraint} --- and other constraints obtained similarly --- can be prepared.

%% file: appendixF/barren_plateau.tex
\chapter{Barren Plateau Analysis of Gibbs State Preparation} \label{app:bp}

\textit{Parts of this appendix are based on the published manuscript by \citet{Consiglio2024a}.}\\

By following the analysis carried out in \citet{Cerezo2021a}, we qualitatively discuss the trainability of our \ac{VQA} for preparing Gibbs States (see Chapter~\secref{chap:5}). We can decompose the generalised Helmholtz free energy cost function as
\begin{align}
    \cF(\rho_S) &= \Tr{\cH \rho_S} - \beta^{-1} \cS(\rho_A) \nonumber \\
    &= \Tr{\cH U_S \rho_S' U_S^\dagger} + \beta^{-1} \sum_{i=0}^{d-1} \Tr{O_i U_A \ketbra{0}_{A}^{\otimes n} U_A^\dagger} \ln \Tr{O_i U_A \ketbra{0}_{A}^{\otimes n} U_A^\dagger},
    \label{eq:cost_function}
\end{align}
where $\rho_S' = \PTr{A}{V \ketbra{0}_{AS}^{\otimes 2n} V^\dagger}$, $V = \textsc{CNOT}_{AS}(U_A \otimes \dI_S)$ and $O_i = \ketbra{i}$. Now, \citet{Cerezo2021a} only considers cost functions of the form
\begin{equation}
    C = \Tr{O U \rho U^\dagger},
    \label{eq:cerezo_cost_function}
\end{equation}
where $\rho$ is an arbitrary quantum state of $n$ qubits, $O$ is any operator, and $U$ is an alternating layered ansatz. Given that our cost function in Eq.~\eqref{eq:cost_function} is not in the form of Eq.~\eqref{eq:cerezo_cost_function}, because of the logarithm in the von Neumann entropy, our comparison should be taken solely as a qualitative discussion on the possibility of \acp{BP}.

Now with reference to Eq.~\eqref{eq:cost_function}, we have that $\cH$ is 2-local, while $O_i$ is 1-local, and both $U_A$ and $U_S$ are alternating layered ans\"atze. Theorem 2 of \citet{Cerezo2021a} gives a lower bound on the variance of the gradient of the cost function as a function of the number of layers, and as such, the trainability of the \ac{PQC}. If the number of layers $l = \cO(\log(n))$, then the variance vanishes no faster than polynomially, hence making the \ac{PQC} trainable. If the number of layers $l = \cO(\text{poly}(\log(n)))$, then the variance vanishes faster than polynomially, but no faster than exponentially, settling in a transition region in between trainable and untrainable. Finally, if the number of layers $l = \cO(\text{poly}(n))$, then the variance vanishes exponentially, resulting in an untrainable circuit.

In the case of $\beta \rightarrow \infty$, our cost function equates directly to Eq.~\eqref{eq:cerezo_cost_function} (since the \ac{VQA} effectively reduces to a \ac{VQE}, and thus we require $l = \cO(\log(n))$ for our circuit to be trainable. On the other hand, in the case of $\beta \rightarrow 0$, the cost function simplifies to maximising the von Neumann entropy, which is acquiring the maximally mixed state. While we cannot directly relate the von Neumann entropy as a cost function with Eq.~\eqref{eq:cerezo_cost_function}, we have numerically seen that preparing the maximally mixed state is a relatively straightforward task. Nevertheless, analysis on the trainability of utilising the von Neumann entropy as the (or part of the) cost function should be sought to be able to detect the presence of \acp{BP}.

Now for any finite $\beta > 0$, the problem of determining whether a \ac{BP} is possible for the generalised free energy is out of the scope of this thesis. Nevertheless, we can possibly surmise that, given $U_A$ and $U_S$ being alternating layered ans\"atze, with $\cH$ being a 2-local Hamiltonian consisting of traceless operators, then using a number of layers for both $U_A$ and $U_S$ that scales at most as $l = \cO(\log(n))$, might result in a \ac{PQC} that is trainable. This would hold if Theorem 2 of \citet{Cerezo2021a} also holds for cost functions in the form of Eq.~\eqref{eq:cost_function}.

%% file: appendixG/xy.tex
\chapter{Solving the XY Model} \label{app:xy}

Let us consider a 1D spin-$1/2$ system with periodic boundary conditions exhibiting nearest-neighbour interactions of Heisenberg type in the XY plane, subject to a uniform magnetic field in the transverse $z$ direction. The XY spin-$1/2$ Hamiltonian~\cite{Lieb1961} is given by
\begin{equation}
    \cH = -\sum_{i=0}^{n-1} \left( \frac{1 + \gamma}{2} \sigma_i^x \sigma_{i+1}^x + \frac{1 - \gamma}{2} \sigma_i^y \sigma_{i+1}^y \right) - h \sum_{i=0}^{n-1} \sigma_i^z,
\end{equation}
which is exactly Eq.~\eqref{eq:XY} rewritten here for convenience. Turning to the spin ladder operators $\sigma^{\pm}=(\sigma^x\pm i \sigma^y)/2$, results in
\begin{align}
    \cH&=-\sum_{i=0}^{n-1}\left[\sigma^+_i \sigma^-_{i+1}+\sigma^-_i \sigma^+_{i+1}+\gamma\left(\sigma^+_i \sigma^+_{i+1}+\sigma^-_i \sigma^-_{i+1}\right)\right]-h\sum_{i=0}^{n-1}\left( \sigma_i^+ \sigma_i^- - \sigma_i^- \sigma_i^+ \right)\nonumber\\
    &=-\sum_{i=0}^{n-1}\left[\sigma^+_i \sigma^-_{i+1}+\sigma^-_i \sigma^+_{i+1}+\gamma\left(\sigma^+_i \sigma^+_{i+1}+\sigma^-_i \sigma^-_{i+1}\right)\right]-2h\sum_{i=0}^{n-1}\sigma^+_i\sigma^-_i + nh\dI,
\end{align}
where the commutator $\comm{\sigma_i^+}{\sigma_i^-}=\sigma_i^z$ and anticommutator $\acomm{\sigma_i^+}{\sigma_i^-}=\dI$ have been used, and we can rescale the Hamiltonian by the constant term $nh$ to remove it. Applying the \ac{JW} transformation:
\begin{equation}
	c_i^\dagger = \left(\bigotimes_{j=0}^{i-1} \sigma_j^z\right) \otimes \sigma_i^+,~c_i = \left(\bigotimes_{j=0}^{i-1} \sigma_j^z\right) \otimes \sigma_i^-,
\end{equation}
the Hamiltonian in the fermionic representation reads
\begin{align}
	\label{eq:XY_fermionic}
	\cH&=-\sum_{i=0}^{n-2}\left[\left(c_i^{\dagger}c_{i+1}+\gamma c_i^{\dagger}c_{i+1}^{\dagger}\right)+ \text{h.c.}\right]-\cP\left[\left(c_{n-1}^{\dagger}c_0+c_{n-1}^{\dagger}c_0^{\dagger}\right)+\text{h.c}\right]-2h\sum_{i=0}^{n-1}c^{\dagger}_ic_i,
\end{align}
where h.c. denotes the hermitian conjugate of the preceding terms, and the middle term on the right-hand side accounts for the periodic boundary conditions, by defining the parity operator $\cP=\bigotimes_{i=0}^{n-1}\sigma_i^z$. The parity operator $\cP$ has eigenvalues $\pm 1$, denoting, respectively, the positive and negative parity subspace. Hence, introducing the projectors on the positive and negative parity subspaces $\cP^{\pm}=\left(1\pm \cP\right)/2$, Eq.~\eqref{eq:XY_fermionic} can be cast in a direct sum of disjoint subspaces
\begin{align}
    \label{eq:XY_direct_sum}
    \cH=\cH^+ + \cH^-,
\end{align}
where $\cH^{\pm}=\cH\cP^{\pm}$, and
\begin{align}
    \cH^{\pm}=-\sum_{i=0}^{n-1}\left[\left(c_i^{\dagger}c_{i+1}+\gamma c_i^{\dagger}c_{i+1}^{\dagger}\right)+ \text{h.c.}\right]-2h\sum_{i=0}^{n-1}c^{\dagger}_ic_i,
\end{align}
where antiperiodic boundary conditions, $c_{n}=-c_0$, hold for $\cH^+$, and periodic boundary conditions, $c_{n}=c_0$, hold for $\cH^-$.\footnote{While parity effects may become relevant for finite-size systems~\cite{Damski2014, Franchini2017}, in the thermodynamic limit $n \rightarrow \infty$, they can be neglected being of order $\cO(1/n)$.} In the following Sections, Eq.~\eqref{eq:XY_direct_sum} will be diagonalised in the respective parity subspaces for finite size systems.

\section{Positive Parity Subspace Diagonalisation}

Due to the translational invariance of the system, we can perform a Fourier transformation of the fermionic operators,
\begin{equation}
    c_i^{\dagger}=\frac{e^{\frac{\imath \pi}{4}}}{\sqrt{n}}\sum_{k\in\cK^+}e^{\imath kn}c_k^{\dagger},~c_i=\frac{e^{\frac{-\imath \pi}{4}}}{\sqrt{n}}\sum_{k\in\cK^+}e^{-\imath kn}c_k,
\end{equation}
where the momenta $k\in\cK^+$ are chosen in order to satisfy the antiperiodic boundary conditions for $\cH^+$, and $\cK^+=\left\{-\frac{n-1}{n}\pi+j\frac{2\pi}{n}~|~j=0, 1, \dots, n-1\right\}$. The prefactor $e^{\frac{\imath \pi}{4}}$ has been introduced so as to end up with a real Hamiltonian, reading
\begin{equation}
\label{eq:XY_bogo}
\cH^+=\sum_{k\in\cK^+}\left[\left(h-\cos k\right)~ \left(c_k^{\dagger}c_k-c_{-k}c_{-k}^{\dagger}\right)-\gamma \sin k\left(c_k^{\dagger}c_{-k}^{\dagger}+c_{-k}c_{k}\right)\right].
\end{equation}
The spinless, quadratic Hamiltonian in Eq.~\eqref{eq:XY_bogo} is finally diagonalised via a Bogoliubov transformation:
\begin{equation}
   c_k=\cos\left(\frac{\theta_k}{2}\right)\gamma_k-\sin\left(\frac{\theta_k}{2}\right)\gamma^{\dagger}_{-k},~\tan \theta_k=\frac{\gamma\sin k}{h-\cos k},
\end{equation}
where $\gamma_k$ is another quasi-particle operator\footnote{Not to be confused with the anisotropy parameter $\gamma$.} yielding
\begin{equation}
    \cH^+=\sum_{k\in\cK^+}\epsilon_k\left(2\gamma_k^{\dagger}\gamma_k-1\right),
\end{equation}
where the single-particle energy spectrum is given by
\begin{align}
    \label{eq:single_particle_energy}
    \epsilon_k=\sqrt{(h-\cos k)^2 + \gamma^2\sin^2 k}.
\end{align}
The ground-state of $\cH^+$ is given by the vacuum of the $\gamma_k$ fermions, with energy
\begin{align}
    E_0^+=-\sum_{k\in\cK^+}\epsilon_k.
\end{align}
The full spectrum of $\cH^+$ is derived by adding an even number of fermions to the ground-state:
\begin{equation}
    E^+_k=E_0^+ + \sum_{k \in P_\text{even}(\cK^+)} \epsilon_k,
\end{equation}
where $P_\text{even}(\cK^+)$ is the subset of the power set of $\cK^+$ with an even number of terms.

\section{Negative Parity Subspace Diagonalisation}

The diagonalisation of the model in the negative parity subspace follows closely that in the positive one. In order to fulfil the periodic boundary condition for the negative parity subspace Hamiltonian $\cH^-$, the Fourier transformation is carried over $k\in \cK^-=\left\{-\pi + (j + 1) \frac{2\pi}{n}~|~j= 0, 1, \dots, n-1\right\}$. The Fourier transformation of $\cH^-$ reads
\begin{align}
	\cH^-&=\sum_{k\neq 0,\pi \in\cK^-}\left[\left(h-\cos k\right)~ \left(c_k^{\dagger}c_k+c_{-k}^{\dagger}c_{-k}\right)-\gamma \sin k\left(c_k^{\dagger}c_{-k}^{\dagger}+c_{-k}c_{k}\right)\right]\nonumber\\
    & \hspace{1.1em} +\left(h-1\right)\left(c_0^{\dagger}c_0-c_{0}c_{0}^{\dagger}\right)+\left(h+1\right)\left(c_{\pi}^{\dagger}c_{\pi}-c_{\pi}c_{\pi}^{\dagger}\right),
\end{align}
where the 0-mode is occupied, whereas the $\pi$-mode is empty in the ground-state of $\cH^-$. By performing a Bogoliubov rotation similar to the one used for $\cH^+$, we find that the ground-state energy of $\cH^-$ is given by
\begin{equation}
     E_0^-=-\sum_{k\in\cK^-}s(k) \epsilon_k,
\end{equation}
where
\begin{equation}
    s(k) = \begin{cases} 
      -1, & \text{if } k = 0, \\
      1, & \text{otherwise}.
   \end{cases}
\end{equation}
Similarly to $\cH^+$, the full spectrum of $\cH^-$ is derived by adding an even number of fermions to the ground-state
\begin{equation}
    E^-_k=E_0^- + \sum_{k \in P_\text{even}(\cK^-)} s(k) \epsilon_k,
\end{equation}
where $P_\text{even}(\cK^-)$ is the subset of the power set of $\cK^-$ with an even number of terms.

\section{Gibbs Density Matrix}

Due to the direct sum structure of the Hamiltonian in Eq.~\eqref{eq:XY_direct_sum}, the thermal state of the spin system can be cast in a sum form:
\begin{align}
    \rho(\beta, \cH)=\sum_{i=0}^{\frac{d}{2} - 1} \left(\frac{e^{-\beta E^+_i}}{\mathcal{Z}(\beta, \cH)}\ketbra{E_i^+}+\frac{e^{-\beta E^-_i}}{\mathcal{Z}(\beta, \cH)}\ketbra{E_i^-}\right),
\end{align}
where $\ket{E_i^{\pm}}$ are the eigenstates of $\cH^{\pm}$ and $p_i=e^{-\beta E_i^{\pm}}/\mathcal{Z}(\beta, \cH)$ are the Boltzmann weights of the Gibbs distribution. The latter is composed by
\begin{align}
    \label{eq:pdfG}
    \left\{p_i\right\}=\left\{p_i^+\right\}\cup\left\{p_i^-\right\}.
\end{align}
Given the non-interacting nature of the Hamiltonian and its set of symmetries, it is evident that there is a high degree of degeneracy in the Hamiltonian spectrum. In the following Sections we will analytically determine how many distinct $p_i$'s appear in Eq.~\eqref{eq:pdfG}, in order to minimise the number of necessary variational parameters in the \ac{GR} ansatz (which is examined in Sec.~\secref{sec:GR_ansatz}), capable of reproducing the Boltzmann distribution.

\section{Degeneracies in the Energy Spectrum}

Let us start by identifying the number of distinct energy levels in the positive parity subspace. Due to the translational symmetry, the single-particle energy spectrum is invariant under momentum inversion $k \rightarrow -k$, as evident from Eq.~\eqref{eq:single_particle_energy} being an even function of $k$: $\epsilon_k=\epsilon_{-k}$.

Clearly, the state with zero fermions: the ground-state, is non-degenerate. The next set of energies we analyse is that with two fermions added to the ground-state, that is $E_i^{\left(2\right)}=E_0^+ + \epsilon_k + \epsilon_q$, where $k \neq q \in \mathcal{K}^+$.
The number of 2-fermion energy levels is given by the binomial $\binom{n}{2}$ of which $n/2$
energy levels are non-degenerate. This can be readily seen, as degenerate levels occur in the form 
\begin{align}
    \label{eq:2deg}
    \epsilon_{-k}+\epsilon_{-q}=\epsilon_{k}+\epsilon_{q}=\epsilon_{k}+\epsilon_{-q}=\epsilon_{-k}+\epsilon_{q},
\end{align}
with $k \neq q$. On the other hand, non-degenerate levels occur in the form
\begin{align}
    \label{eq:2nondeg}
    \epsilon_{k}+\epsilon_{-k}.
\end{align}
Hence, non-degenerate levels in the two-fermion excitation spectrum are given by how many distinct $k$'s can be selected in Eq.~\eqref{eq:2nondeg}. This number is obtained by observing that the momenta $k\in\mathcal{K^+}$ are symmetric around 0, and hence the number of ways to choose one momentum $k$ out of $n/2$ momenta is trivially given by $n/2$. Moreover, as Eqs.~\eqref{eq:2deg} give rise to the same energy, the degeneracy is always four-fold in this particle sector. To summarise, out of the $\binom{n}{2}$ energy levels, there are $\binom{n/2}{2}$ four-fold degenerate energy levels, with the remaining $n/2$ being non-degenerate.

Moving to the four-fermion excited states, one can proceed in a similar way by identifying, out of the $\binom{n}{4}$ four-fermion excitation energy levels, as non-degenerate levels being those with energy
\begin{equation}
    \label{eq:4nondeg}
    \epsilon_{k}+\epsilon_{-k}+\epsilon_{q}+\epsilon_{-q},
\end{equation}
where $k\neq q$. Hence, the non-degenerate levels of the four-fermion excited states are $\binom{n/2}{2}$, corresponding to the number of distinct sets $\left(k,q\right)$ that can be selected from the available momenta in the positive subset of $\mathcal{K^+}$. The degrees of degeneracy in the four-particle sector are $1, 4, 16$ occurring, respectively, $\binom{n/2}{2}, \binom{n/2}{1} \binom{n/2-1}{2},\binom{n/2}{4}$ times.

Generalising the above procedure to an arbitrary $p$-fermion energy subspace, for a chain of length $n$, one obtains that out of the $\binom{n}{p}$ $p$-fermion excitation energy levels
\begin{itemize}
    \item the number of 1-fold degenerate levels is $\dbinom{\frac{n}{2}}{\frac{p}{2}}$
    \item the number of 4-fold degenerate levels is $\dbinom{\frac{n}{2}}{\frac{p}{2}-1}\dbinom{\frac{n}{2}-\left(\frac{p}{2}-1\right)}{2}$
    \item the number of 16-fold degenerate levels is $\dbinom{\frac{n}{2}}{\frac{p}{2}-2}\dbinom{\frac{n}{2}-\left(\frac{p}{2}-2\right)}{4}$
    \item the number of 64-fold degenerate levels is $\dbinom{\frac{n}{2}}{\frac{p}{2}-3}\dbinom{\frac{n}{2}-\left(\frac{p}{2}-3\right)}{6}$
    \item the number of $4^q$-fold degenerate levels is $\dbinom{\frac{n}{2}}{\frac{p}{2}-q}\dbinom{\frac{n}{2}-\left(\frac{p}{2}-q\right)}{2q}$
\end{itemize}
Clearly, the maximum degree of degeneracy in the $p$-fermion energy subspace is $4^q \leq \binom{n}{p}$, with $q$ being a non-negative integer representing the order of four-fold degeneracies. The procedure just described determines the degree of degeneracy, and the number of energy levels in each degenerate subspace of $\cH^+$ can be straightforwardly applied also to the negative parity subspace, which will not be reported here to avoid redundancy.

%% file: addendum/addendum.tex
\chapter*{Related Research} \label{add:related_research}
\addcontentsline{toc}{chapter}{Related Research}

The research carried out in this thesis is considerably related to the field of quantum communications and \ac{QIP}, specifically, that of \ac{QST}, entanglement generation and the theory of entanglement resources. In fact, preceding this work, the study of the role of localisable entanglement~\cite{Verstraete2004, Popp2005} in quantum teleportation protocols was undertaken by~\citet{Consiglio2021}. In this manuscript, I served as the primary and corresponding author, responsible for devising the plan of action, obtaining results and writing the manuscript.

Concurrently with this thesis, multiple studies were undertaken to investigate \ac{QST}: the process by which a quantum state is transferred from one location to another without physically moving the quantum system itself~\cite{Cirac1997}. This is an essential operation in the field of \ac{QIP} and quantum computing~\cite{Nielsen2010}, enabling the communication of quantum information over distances.

The first of these was ``Quantum map approach to entanglement transfer and generation in spin chains''~\cite{Lorenzo2022}, a chapter contributed to the Springer volume ``Entanglement in Spin Chains: Theory and Quantum Technology Applications''. This study focused on entanglement generation and transfer. My role in this work primarily involved deriving the results presented in Section 7, which concentrated on four-qubit entanglement generation in the spin chain. Specifically, I derived values for the concurrence, the geometric mean of the \ac{GME} concurrence, as well as the three-tangle and four-tangle, across all partitions.

The second study was ``Quantum transfer of interacting qubits''~\cite{Apollaro2022}, which explored the challenging task of $n$-qubit interacting \ac{QST} in \ac{QIP}. My role in this work involved analysing the three- and four-qubit interacting \ac{QST}, as detailed in in Section 5, including the standard derivation and the envelope of the average fidelity.

The third study, ``Entangled States are Harder to Transfer than Product States''~\cite{Apollaro2023}, demonstrated that the \ac{QST} of entangled states yields a lower average fidelity compared to the \ac{QST} of product states at fixed transition amplitudes. My role in this work was to investigate \ac{QST} fidelity as a function of entanglement, for systems ranging from two to four qubits, as detailed in Section 3. In particular, for three qubits, we looked towards quantities labelled as invariant polynomials, which are capable of identifying the class of entanglement. The case for four qubits was more challenging, due to the existence of an infinite number of entanglement classes under \ac{SLOCC}.

Finally, ``Distribution of Fidelity in Quantum State Transfer Protocols''~\cite{Lorenzo2024}, extends on the work from the previous manuscript, by investigating the probability distribution functions of the transfer fidelity of one- and two-qubit states. My role in this work involved analysing the results for two-qubit \ac{QST} as detailed in Section IV B.